\documentclass[12pt]{elsarticle}

\usepackage{graphicx, import}
\usepackage{amsfonts,textgreek}
\usepackage{mathtools}
\usepackage{dsfont}
\usepackage{multicol}
\usepackage{bbm}
\usepackage{url}
\usepackage[dvipsnames]{xcolor}
\usepackage[colorlinks=true,allcolors=blue]{hyperref}
\usepackage[nameinlink,noabbrev,capitalize]{cleveref}
\usepackage{xfrac}
\usepackage{caption}
\usepackage{subcaption}
\usepackage{xspace}
\usepackage{stmaryrd}
\usepackage{enumerate}
\usepackage{soul}

\usepackage[utf8]{inputenc}
\usepackage{wrapfig,bm}
\usepackage{lineno}
\usepackage{siunitx} 
\usepackage{caption}
\usepackage{psfrag}
\usepackage{upgreek}
\usepackage{isomath}
\usepackage{latexsym}
\usepackage{array}
\usepackage{multirow}
\usepackage{booktabs}
\usepackage{colortbl}
\usepackage{verbatim}
\usepackage{graphicx}
\usepackage{epstopdf,overpic}
\usepackage{float}
\usepackage{soul}
\usepackage{color}
\usepackage{dirtytalk}
\usepackage{todonotes}
\usepackage{textcomp}
\usepackage{algorithm, algpseudocode, lmodern, mathtools}
\usepackage{tikz}
\usetikzlibrary{shapes.geometric, arrows}

\usepackage{amssymb}
\usepackage{geometry}
\usepackage{amsmath}
\usepackage{mathrsfs}  
\usepackage{graphicx}
\usepackage{titlesec}
\usepackage{indentfirst}
\usepackage{bigints}
\usepackage{breqn}
\usepackage[export]{adjustbox}
\usepackage{pgfplots}
\usepackage{enumitem}
\usetikzlibrary{plotmarks,spy,calc}
\newlength{\figureheight}
\newlength{\figurewidth}
\usepackage{epstopdf,overpic}
\usepackage{multirow}
\usepackage[normalem]{ulem}
\usepackage{tabularx}

\newcommand{\eref}[1]{Eq.~(\ref{#1})}
\newcommand{\erefs}[1]{Eqs.~(\ref{#1})}
\newcommand{\fref}[1]{Figure~\ref{#1}}
\newcommand{\frefs}[1]{Figures~\ref{#1}}
\newcommand{\tref}[1]{Table~\ref{#1}}
\newcommand{\sref}[1]{Section~\ref{#1}}

\usetikzlibrary{calc,fit,intersections,shapes}
\usetikzlibrary{plotmarks,spy,calc}

\pgfdeclareradialshading{ballshading}{\pgfpoint{-10bp}{10bp}}
 {color(0bp)=(gray!40!white); 
 color(9bp)=(gray!75!white);
 color(18bp)=(gray!70!black); 
 color(25bp)=(gray!50!black); 
 color(50bp)=(black)}

\pgfplotsset{compat=1.18} 
\begin{document}
\begin{frontmatter}

\title{Design of thermal meta-structures made of functionally graded materials using isogeometric density-based topology optimization}

\author[label0]{Chintan Jansari}
\author[label1,label3]{St\'ephane P.A. Bordas}
\author[label4]{Marco Montemurro}
\author[label2]{Elena Atroshchenko\corref{cor1}}
\ead{e.atroshchenko@unsw.edu.au}

\cortext[cor1]{Corresponding author}
\address[label0]{Shaping Matter Lab, Faculty of Aerospace Engineering, Delft University of Technology, Delft, Netherlands.}
\address[label1]{Institute of Computational Engineering, Faculty of Sciences, Technology and Medicine, University of Luxembourg, Luxembourg City, Luxembourg.}
\address[label3]{Clyde Visiting Fellow, Department of Mechanical Engineering, The University of Utah, Salt Lake City, Utah, United States.}
\address[label4]{Universit\'{e} de Bordeaux, Arts et M\'{e}tiers Institute of Technology, CNRS, INRA, Bordeaux INP, HESAM Universit\'{e}, I2M UMR 5295, F-33405 Talence, France}
\address[label2]{School of Civil and Environmental Engineering, University of New South Wales, Sydney, Australia.}


\begin{abstract}
The thermal conductivity of Functionally Graded Materials (FGMs) can be efficiently designed through topology optimization to obtain thermal meta-structures that actively steer the heat flow. Compared to conventional analytical design methods, topology optimization allows handling arbitrary geometries, boundary conditions and design requirements; and producing alternate designs for non-unique problems. Additionally, as far as the design of meta-structures is concerned, topology optimization does not need intuition-based coordinate transformation or the form invariance of governing equations, as in the case of transformation thermotics. We explore isogeometric density-based topology optimization in the continuous setting, which perfectly aligns with FGMs. In this formulation, the density field, geometry and solution of the governing equations are parameterized using non-uniform rational basis spline entities. Accordingly, the heat conduction problem is solved using Isogeometric Analysis. We design various 2D \& 3D thermal meta-structures under different design scenarios to showcase the effectiveness and versatility of our approach. We also design thermal meta-structures based on architected cellular materials, a special class of FGMs, using their empirical material laws calculated via numerical homogenization. 
\end{abstract}

\end{frontmatter}
%
%
%
%
\section{Introduction}
\label{sec:Introduction}
\subsection{Thermal metamaterials and meta-structures}
\label{Sec:Intro thermal metamaterials}
\par By controlling heat flow, the thermal analogues of electrical devices can be created, such as resistors, capacitors, inductors, diodes and transistors. In addition, new thermal detection-anti-detection, computing and communication devices can be developed. Despite its significance, storing and steering heat is not an easy task as there are multiple modes of heat transfer and the heat transfer processes are intrinsically less ordered than ballistic/wave transport. In recent years, thermal metamaterials have emerged as a tool to manipulate and control heat transfer~\cite{yang_controlling_2021,dai_designing_2021,kadic_metamaterials_2013,peralta_brief_2020}. Thanks to their architected structures, thermal metamaterials can achieve thermal properties, which are difficult to find in natural materials.  
\par Traditionally, these metamaterials are mainly designed by the analytical methods, such as transformation thermotics~\cite{Fan2008,chenCloakCurvilinearlyAnisotropic2008,Guenneau2012} and the scattering cancellation method~\cite{Han2014,xu_ultrathin_2014}. Using these methods, several thermal metamaterials have already been proposed and experimentally demonstrated, such as thermal cloak, thermal concentrator, thermal rotator and thermal expander~\cite{Guenneau2012,narayana2012heat,schittnyExperimentsTransformationThermodynamics2013b,yang_controlling_2021}. However, these design methods are only efficient with limited regular geometries under specific design conditions. They face difficulties handling any arbitrary design scenario mainly due to their analytical nature. Transformation thermotics also requires an intuition-based coordinate transformation, which is not easy to devise. Moreover, often deduced thermal property distributions are (extremely) anisotropic and heterogeneous; and therefore difficult to manufacture. 
\par To overcome the above-mentioned limitations, a numerical method could be an effective alternative. As designing metamaterials/meta-structures is essentially an inverse problem, one can exploit structural optimization methods as a design tool. At first, Dede~\textit{et~al.}~\cite{dede_simulation_2010,dede_thermal-composite_2014} designed thermal composites for heat flux shielding, focusing and reversal using topology optimization. Following it, several other articles focusing on the usage of  optimization to design thermal meta-structures~\cite{peraltaOptimizationbasedDesignHeat2017,fachinottiOptimizationbasedDesignEasytomake2018, ALEKSEEV2019,fujiiOptimizingStructuralTopology2019,sha2020,sha_topology-optimized_2022,XU2023,NAKAGAWA2023123964}. Hirasawa~\textit{et~al.}~\cite{HIRASAWA2022123093} experimentally demonstrated optimization-based thermal cloaking meta-structures. Similarly, we also aspire to utilize the structural optimization for designing thermal meta-structures to achieve versatility and flexibility in handling arbitrary geometries and design conditions.
\par In our earlier works~\cite{jansari2022design,jansari2024design}, we used shape optimization and (a more flexible) topology optimization to design thermal meta-structures. Both methods focus on distributing natural constituent materials at the macroscale to produce an apparent effect of required anisotropy and heterogeneity for thermal metamaterials/meta-structures. Though these methods are effective in their respective design spaces, their design spaces are limited by a few discrete conductivities. At times, to achieve an apparent anisotropy and heterogeneity, the methods can produce designs with intricate material distributions that are challenging to manufacture as observed for thermal camouflages (without regularization), as discussed in~\cite{jansari2024design}. In this paper, we contemplate enlarging the design spaces by including functionally graded materials in the design. We could think of this as distributing the constituents at a smaller scale rather than the macroscale. Correspondingly, we propose thermal meta-structures made of functionally graded materials. 

\subsection{Functionally Graded Materials}
\label{sec:Intro Functionally Graded Materials (FGMs)}
\par Functionally Graded Materials (FGMs) is a new class of materials characterized by the gradual variation of properties across the volume through variation in structure, microstructure or composition~\cite{niino_functionally_1987,KOIZUMI1997}. With this definition, the functionally graded porous/lattice structures or Architected Cellular Materials (ACMs), with structural variations occurring at a smaller scale than the scale of observation, can also be included in the family of FGMs. In FGMs, smooth gradations in properties can be beneficial for reducing residual stresses or stress concentration and therefore, interfacial separation and cracking. Recent advances in additive manufacturing made the manufacturing of FGMs easier compared to expensive and time-consuming conventional manufacturing processes~\cite{kumar_recent_2022,li_review_2020,SUAREZAFANADOR2022}. From a constructive perspective, the material gradation can be accurately tailored to design desired heterogeneity, which is often a primary requirement of a thermal meta-structure. Therefore, utilizing an FGM in thermal meta-structure design is a straightforward logical approach. 
\subsection{Topology optimization}
\label{sec:Intro Topology optimization}
\par Topology optimization is a structural optimization method that focuses on optimizing the connectivity, shape, and placement of voids within a given design domain. Topology optimization has several variants such as (1) density-based approach~\cite{bendsoe_optimal_1989,zhou_coc_1991,mlejnek_aspects_1992}, (2) level-set method~\cite{allaire_level-set_2002,allaire_structural_2004,wang2003level} (3) phase field approach~\cite{bourdin_design-dependent_2003} and (4) evolutionary algorithm approach~\cite{xie_simple_1993} (5) Moving Morphable Component (MMC)/Moving Morphable Void (MMV)/Geometric Projection (GP) approach~\cite{Guo2014Doing,zhang2016new,ZHANG2017Explicit,NORATO2015geometry}. For more detailed information about all approaches, interested readers can refer to the following articles~\cite{rozvanyCriticalReviewEstablished2009,sigmundTopologyOptimizationApproaches2013,vandijkLevelsetMethodsStructural2013,munkTopologyShapeOptimization2015,wein2020review,li2024comprehensive}. Among different methods, density-based and level-set methods have remained the most popular ones. The density-based approaches consider the continuous material formulation and behave as a sizing problem in terms of density~\cite{bendsoe_optimal_1989,zhou_coc_1991,mlejnek_aspects_1992,bendsoe_material_1999}. Commonly, they employ a penalization scheme, such as the Solid Isotropic Material with Penalization (SIMP) and the Rational Approximation for Material Properties (RAMP), that penalizes the intermediate densities to achieve the discrete designs. However, the continuous framework of the density-based approach without penalization perfectly aligns with FGMs whose peculiarity is a continuous gradation in composition/structure.
 
\par As an early work on topology optimization of FGMs, Paulino~\textit{et~al.}~\cite{paulino_design_2005} proposed so-called FGM-SIMP formulation to design FGM structures. They employed the continuous approximation of material distribution (CAMD) for density field parameterization and corresponding nodal densities as the design variables. Later, Almeida \textit{et~al.}~\cite{almeida_layout_2010} studied the effect of global and local level gradation in the topology optimization of FGMs. The level-set method is also explored in the context of FGMs by Xia \textit{et~al.}~\cite{xia_simultaneous_2008}. Even functionally graded cellular/porous structures are designed using topology optimization in the following articles~\cite{liu_functionally_2018,montemurro_thermal_2022,li_optimal_2018,li_topology_2018}. Taheri and his collaborators published several articles~\cite{taheri_simultaneous_2014,taheri_thermo-elastic_2014,taheri_isogeometric_2017} on a fully isogeometric structural optimization for optimizing eigen-frequency, thermoelastic stress and compliance, respectively. In their fully isogeometric structural optimization approaches, density, geometry and solution field were parameterized using the same non-uniform rational b-splines (NURBS) basis functions and accordingly, Isogeometric Analysis (IGA)~\cite{hughes2005isogeometric,cottrell2009isogeometric} was employed to solve the boundary value problems. 
\par In this article, we too exploit fully NURBS-based density topology optimization with the NURBS-parameterized density, geometry and solution fields as  in~\cite{taheri_simultaneous_2014}. The NURBS-parameterised density provides quite a few advantages over element-wise densities or nodal densities as utilized in~\cite{qian_topology_2013,wang_efficient_2015,taheri_simultaneous_2014}; getting inherent filtering effect against the checker-boarding issue, obtaining smoother material distributions, straight forward calculation of derivatives of density, providing complete cost-effective analytical sensitivities. The fully NURBS-based topology optimization formulation holds the potential to build an integrated design-analysis-optimization model with its tightly integrated design field-to-geometry and geometry-to-solution-field mappings. Moreover, IGA offers advantages over the classical finite element method (FEM) by enabling the exact representation of conic geometries, handling  higher inter-element continuity and providing higher efficiency for higher-order elements~\cite{hughes2005isogeometric,cottrell2009isogeometric}. The detailed literature related to isogeometric optimization approaches can be found in the review articles~\cite{wangStructuralDesignOptimization2018,gaoComprehensiveReviewIsogeometric2020}. 

\subsection{Contribution of the present study}
\label{sec:Intro Contribution of the present study}
In the present study, we design thermal meta-structures made of FGMs using isogeometric density topology optimization. In summary, the key features of this paper are:
\begin{itemize}
    \item \textbf{Design of thermal meta-structures made of FGMs and ACMs}: In this article, we designed thermal meta-structures made of FGMs. We explored both analytical and experimental homogenization models for thermal conductivity. We also showcased thermal meta-structures made of ACMs using an empirical material law established on numerical homogenization results. Following it, the reconstruction of an entire cellular structure based on the optimization results is presented too.
    \item \textbf{Versatile and more flexible design tool than the conventional analytical methods}: In our article, we showcased that the proposed approach can effectively  design thermal meta-structures with arbitrary geometries, boundary conditions, design constraints/regularizations and objective functions. It can also generate alternative designs  for the  design problems which lacks uniqueness such as thermal cloak probelm~\cite{calderon_inverse_1980,uhlmann_electrical_2009,Greenleaf2003nonuniqueness} with simple modifications.
    \item \textbf{Fully NURBS-based formulation}: By using a fully NURBS-based topology optimization, the proposed method 
    provides inbuilt filtering, smoother material distributions, straightforward calculation of density gradient and closed-form sensitivities, higher inter-element continuity, exact geometric representation (for conic geometries) and higher numerical efficiency for the higher-order elements.
    It also has the potential for building an integrated design-analysis-optimization model.
    \item \textbf{Verification and comparison}: We verified our topology optimization-based tool by designing various 2D \& 3D thermal meta-structures found in literature such as thermal cloaks, thermal concentrators, thermal rotators, thermal cloaked sensors, multi-functional thermal meta-structures (thermal cloak concentrator), and multi-directional thermal meta-structure (thermal horizontal concentrator- vertical cloak). As the most of literature results are achieved using a different methodology (and does not involve FGMs), it is difficult to provide a quantitative comparison with our results. Yet, we tries to offer a qualitative comparison for some special cases of thermal cloaks and thermal concentrators. 
 
\end{itemize}

\par The remainder of the paper is organized as follows: \sref{sec:Boundary value problem} describes the boundary value problem and numerical formulation to solve it. \sref{sec:Material models} describes various models to calculate the effective thermal conductivity of FGMs that are utilized in our work. The optimization problem and sensitivity analysis are explained in \sref{sec:Optimization problem}. In~\sref{sec:Thermal cloak}, thermal cloaks are designed in various design scenarios to showcase the effectiveness of the proposed method. \sref{sec:Other thermal metamaterials} explores other thermal meta-structures and corresponding different objective functions. \sref{sec:Reconstruction of ACMs} briefly covers the reconstruction process of Architected Cellular Materials (ACMs) designed using topology optimization. Lastly, \sref{sec:Conclusions} presents the conclusions and some prospects of the current work.

%
%
%
%
\section{Boundary value problem}
\label{sec:Boundary value problem}
\subsection{Problem description}
\label{sec:Problem description}
\begin{figure}[!htbp]
\centering
\includegraphics[width=4.6 in]{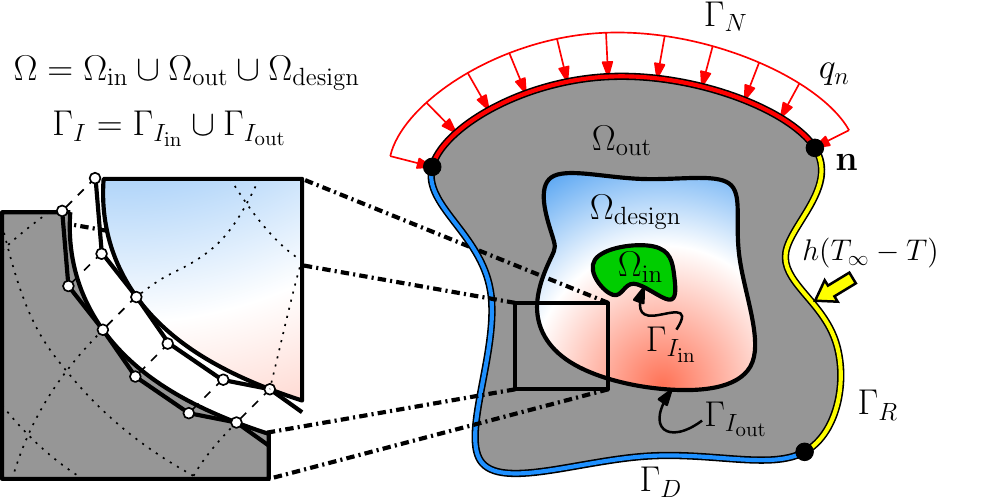}
\caption {Domain description of the boundary value problem. $\Omega_{\rm design}$ represents a design region where thermal meta-structure is optimized, $\Omega_{\rm in}~\&~\Omega_{\rm out}$ are the inside and outside regions with respect to $\Omega_{\rm design}$, respectively. $\Omega$=$\Omega_{\rm in}\cup\Omega_{\rm out}\cup\Omega_{\rm design}$. $\Gamma=\partial \Omega=\Gamma_D\cup\Gamma_N \cup\Gamma_R$. The solid black line shows an explicitly defined interfaces $\Gamma_{I_{\rm in}}$ and $\Gamma_{I_{\rm out}}$, $\Gamma_{I}=\Gamma_{I_{\rm in}}\cup\Gamma_{I_{\rm out}}$. The square is shown in detail highlighting the matching of control points of connecting patches at the interface $\Gamma_I$.} 
\label{fig:BVP domain}
\end{figure}
Let us consider an FGM-based thermal meta-structure distributed across $\mathrm{\Omega}_{\rm design}$, which is embedded in the homogeneous isotropic domain $\mathrm{\Omega} \in\mathbb{R}^{d}, d \in \{{2,3}\}$. The inside and outside regions are denoted as $\mathrm{\Omega}_{\rm in}$ and $\mathrm{\Omega}_{\rm out}$, respectively. $\mathrm{\Omega}=\mathrm{\Omega}_{\rm in} \cup \mathrm{\Omega}_{\rm design} \cup \mathrm{\Omega}_{\rm out}$. The external boundary $\mathrm{\Gamma}=\partial\mathrm{\Omega}$ is decomposed into three parts $\mathrm{\Gamma}_D $,  $\mathrm{\Gamma}_N$ and $\mathrm{\Gamma}_R$, $\mathrm{\Gamma}=\mathrm{\Gamma}_D \cup \mathrm{\Gamma}_N \cup \mathrm{\Gamma}_R$. On $\mathrm{\Gamma}_D$, $\mathrm{\Gamma}_N$ and $\mathrm{\Gamma}_R$, the Dirichlet, Neumann and Robin boundary conditions are applied, respectively. Moreover, internal boundaries $\mathrm{\Gamma}_{I_{\rm in}}$ \& $\mathrm{\Gamma}_{I_{\rm out}}$ separate $\mathrm{\Omega}_{\rm in}$ and $\mathrm{\Omega}_{\rm out}$ from
$\mathrm{\Omega}_{\rm design}$, respectively. Internal boundaries are denoted as $\mathrm{\Gamma}_I$, $\mathrm{\Gamma}_I=\mathrm{\Gamma}_{I_{\rm in}} \cup \mathrm{\Gamma}_{I_{\rm out}}$. The 2D domain description is shown in \fref{fig:BVP domain}. For the given arrangement, the steady-state heat conduction boundary value problem with an internal heat generation $q_b$ in the temperature field $T$ is given as: 
\begin{subequations}
\begin{align} 
\nabla \cdot \left( \boldsymbol{\kappa}(v)\nabla T\right) + q_b &= 0, \quad &&\textrm{in}\quad \mathrm{\Omega}, \label{eq:Laplace equation}\\
 T &= T_D, \quad &&\textrm{on}\quad \mathrm{\Gamma}_D, \label{eq:Boundary conditions a}\\
  (\boldsymbol{\kappa}(v)\nabla T)\cdot \mathbf{n} &= q_n, \quad &&\textrm{on} \quad \mathrm{\Gamma}_N, \label{eq:Boundary conditions b}\\
  (\boldsymbol{\kappa}(v)\nabla T)\cdot \mathbf{n} &= h(T_{\infty}-T), \quad &&\textrm{on} \quad \mathrm{\Gamma}_R, \label{eq:Boundary conditions c}
\end{align}
\label{eq:Heat conduction BVP}%
\end{subequations}
where $\nabla$ is the gradient operator, $q_n$ is the flux applied on $\mathrm{\Gamma}_N$, $T_D$ is the prescribed temperature on $\mathrm{\Gamma}_D$, $\mathbf{n}$ is the unit normal on the boundary, $\boldsymbol{\kappa}$ is the thermal conductivity matrix, $h$ is the heat transfer coefficient, $T_{\infty}$ is the bulk temperature. 
\par As our thermal meta-structure is made of FGM, the (macroscopic) thermal conductivity matrix varies point-wise. The thermal conductivity variation is governed by the variation of compositions/structures at a smaller scale. These microstructure variations are homogenized using an appropriate homogenization law to calculate an effective macroscopic property. Possibly, the effective macroscopic property can be defined as a function of several microstructure parameters. For a fixed type of unit cell, the microstructure parameters could be reduced to the volume fractions/relative densities of the constituents. In our case, we only consider FGMs made of two constituents (or one constituent in the case of an FGM characterized by structural variation with porosity being the other constituent) with a fixed-type unit cell. Therefore, the effective thermal conductivity is written as a function of the relative density $v$ of filler material/porosity, \textit{i.e.}, $v=V_{\rm}/V_{0}$, $V$ being the volume of filler material/porosity in the unit cell and $V_0$ being the total volume of the unit cell. The relative density of the other constituent will be $1-v$. Consequently, the current work focuses on optimizing this relative density distribution. Various models to calculate the effective thermal conductivity are presented in \sref{sec:Material models}. Other regions $\Omega_{\rm in}$ and $\Omega_{\rm out}$ are also assumed to be filled with homogeneous isotropic materials. Note that the materials constituting the unit cell of the FGM are isotropic. Nevertheless, the equivalent homogeneous material replacing the unit cell at the macroscopic scale exhibits a cubic syngony behaviour. That being the case, the conductivity matrix can be easily defined as $\boldsymbol{\kappa}(v)=\kappa(v) \mathbf{I}_d$ with $\mathbf{I}_d$ being an identity matrix in $\mathbb{R}^{d}$. The conductivities of homogeneous materials are included in the functional form via constant functions.
\par Across the internal interfaces $\mathrm{\Gamma}_{I_{\rm in}}$ and $\mathrm{\Gamma}_{I_{\rm out}}$, the temperature and normal flux are assumed to be continuous. If the connected regions at $\mathrm{\Gamma}_{I}$ are denoted by indices 1 and 2 locally, the interface boundary conditions are written as: 
\begin{subequations}\label{eq:Interface conditions}%
\begin{align} 
\left\llbracket T\right\rrbracket & \coloneqq T^1-T^2 = 0, \quad &\textrm{on} \quad \mathrm{\Gamma}_I, \label{eq:Interface conditions a}\\ \mathbf{n}\cdot \left\llbracket \boldsymbol{\kappa}(v)\nabla T\right\rrbracket &\coloneqq \mathbf{n}^1\cdot \boldsymbol{\kappa}^1(v^1)\nabla T^1 + \mathbf{n}^2\cdot \boldsymbol{\kappa}^2(v^2)\nabla T^2 =0,  \quad &\textrm{on} \quad \mathrm{\Gamma}_I, \label{eq:Interface conditions b} 
\end{align}
\end{subequations}
where 
$\llbracket\cdot\rrbracket$ is the jump operator and  $\mathbf{n}=\mathbf{n}^1=-\mathbf{n}^2$. 
\subsection{Solution of the boundary value problem using IGA}
\label{sec:Solution of the boundary value problem using IGA}
\par To solve the boundary value problem, the strong form described in \eref{eq:Heat conduction BVP} is transformed into the weak form using the standard Bubnov-Galerkin formulation. The interface conditions mentioned in \eref{eq:Interface conditions} are also incorporated in the weak formulation using Nitsche's method~\cite{Nguyen2014,HU2018}. At last, the modified Bubnov-Galerkin weak formulation is given as follows: Find $
T^h \in \mathscr{T}^h \subseteq \mathscr{T} = \big\lbrace T \in \mathbb{H}^1 ({\mathrm{\Omega}}),   T= T_D \hspace{0.15cm} \textrm{on} \ {\mathrm{\Gamma}}_D \big\rbrace
$ such that $\forall S^h \in \mathscr{S}^h_0 \subseteq \mathscr{S}_0 = \left\lbrace S \in \mathbb{H}^1 ({\mathrm{\Omega}}), S=0 \hspace{0.15cm} \textrm{on} \ {\mathrm{\Gamma}}_D \right\rbrace$,   
\begin{equation}
a(T^h,S^h,v) = \ell(S^h),
\label{eq:weak_form}
\end{equation}
with
\begin{subequations}
\begin{multline}
a(T^h,S^h,v) = \int _{\mathrm{\Omega}} (\nabla  S^h)^{\rm T} \boldsymbol{\kappa}(v)\nabla T^h d\mathrm{\Omega} - \int _{\mathrm{\Gamma}_I} \left(\mathbf{n}\cdot \{\boldsymbol{\kappa}(v)\nabla S^h\}\right)^{\rm T}\llbracket T^h \rrbracket~d\mathrm{\Gamma}  \\- \int _{\mathrm{\Gamma}_I} \llbracket S^h  \rrbracket^{\rm T} \left(\mathbf{n}\cdot\{\boldsymbol{\kappa}(v)\nabla T^h\}\right)  d\mathrm{\Gamma} + \int _{\mathrm{\Gamma}_I} \beta~\llbracket S^h  \rrbracket^{\rm T} \llbracket T^h  \rrbracket~d\mathrm{\Gamma} + \int _{\mathrm{\Gamma}_{R}} h(S^h)^{\rm T}~T^h~d\mathrm{\Gamma}, 
\label{eq:weak_form_a}%
\end{multline}
\begin{equation}
 \ell{(S^h)} =  \int _{\mathrm{\Omega}} (S^h)^{\rm T} q_b~  d\mathrm{\Omega} +\int _{\mathrm{\Gamma}_{N}} (S^h)^{\rm T} q_n~d\mathrm{\Gamma} + \int _{\mathrm{\Gamma}_{R}} (S^h)^{\rm T}hT_{\infty}~d\mathrm{\Gamma},
 \label{eq:weak_form_b}%
\end{equation}
\end{subequations}
where $\beta$ is the stabilization parameter and $\{\cdot\}$ is the averaging operator defined as $\{\theta\}=\gamma\theta^1 + (1-\gamma)\theta^2$ with $\gamma$ being the averaging parameter ($0<\gamma<1$). for the current work, $\beta=1\times 10^{12}$ and $\gamma=0.5$. In the literature~\cite{Nguyen2014,HU2018}, it is reported that the large stabilization parameter might cause ill-conditioning of the system, but we did not face any conditioning issues for our boundary value problem.  The verification of the accuracy of Nitsche’s method is performed in our earlier work on
shape optimization~\cite{jansari2022design}. 
\begin{figure}[!htbp]
\centering
\includegraphics[width=3.8in]{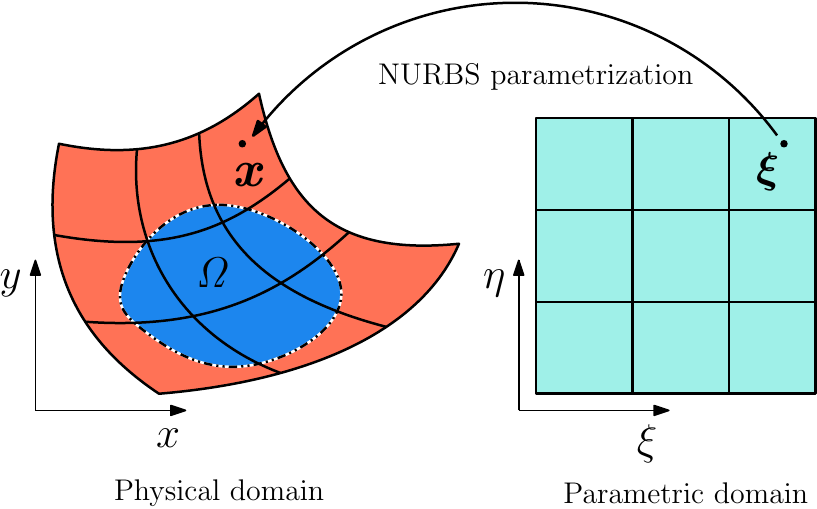}
\caption {Parametrization of a point from the parametric domain to a point in the physical domain using NURBS basis functions.} 
\label{fig:NURBS parameterization}
\end{figure} 
\par As we are exploiting IGA for approximation, the geometry, test function and trial function are parameterized using NURBS basis functions. If $n$ NURBS $N_{i}$,~$i=1,2,...,n$ are employed to discretize the weak form, these approximations are written as:
\begin{equation} \label{eq:NURBs_approx}
    \boldsymbol{x}(\boldsymbol{\xi}) = \sum _{i=1}^{n}\mathbf{X}_{i}N_{i}(\boldsymbol{\xi}), \quad 
    T^h(\boldsymbol{\xi}) = \sum_{i=1}^{n}T_{i}N_{i}(\boldsymbol{\xi}), \quad \textrm{and} \quad 
     S^h(\boldsymbol{\xi}) = \sum_{i=1}^{n}S_{i}N_{i}(\boldsymbol{\xi}),
\end{equation}
where $\boldsymbol{x}$ is a physical point in $ \Omega$, and $\boldsymbol{\xi}$ is the corresponding parametric point as shown in \fref{fig:NURBS parameterization}, $\mathbf{X}_i$ is the $i^{th}$ control point and $T_i$ \& $S_{i}$ are corresponding temperature \& arbitrary temperature. By substituting these approximations in \eref{eq:weak_form}, we can obtain a linear matrix system as follows:
\begin{equation}
\label{eq:Linear matrix system}
\mathbf{K} \mathbf{T} = \mathbf{F},
\end{equation}
where $\mathbf{K}$ is the global stiffness matrix, $\mathbf{F}$ is the global force vector and $\mathbf{T}$ is the vector of temperature at control points. A detailed derivation of matrix formulation is given in \ref{sec:Appendix A}.

\section{Models for effective thermal conductivity of functionally graded materials}
\label{sec:Material models}
\par For analysis of FGMs using numerical methods such as FEM, an accurate model of the material property gradation is crucial to ensure numerical accuracy. In earlier works of FEM, material properties were assumed to remain constant across each element. With element-wise constant properties, FEM requires a fine mesh and thus higher computational effort to represent a smooth gradation. To overcome this limitation, Santare \textit{et~al.}~\cite{Santare2000} proposed a graded finite element approach to analyze heterogeneous materials like FGMs. In this approach, the material properties are sampled directly at the integration points using explicit functions. Later, Kim \textit{et~al.} gave a generalized isoparametric FEM formulation~\cite{kimIsoparametricGradedFinite2002}, where material properties are sampled at nodes of finite elements. The properties inside the elements were interpolated using the same shape functions as geometry and solution field. A generalized IGA formulation was also proposed by Tahiri \textit{et~al.}~\cite{taheri_isogeometric_2017}. Other works on numerical analysis of FGMs can be found in~\cite{Minutolo2009,VALIZADEH2013,KOU2007,Reddy2000,chinosi2007approximation}. As we are working on optimizing the relative density, the thermal conductivity is directly sampled at integration points using the relative density values similar to~\cite{Santare2000}. 
\par The effective thermal conductivity tensor as a function of the relative density of the FGMs at the macroscopic scale is assessed through a dedicated homogenization method. Several homogenization models have been proposed to estimate the effective thermal conductivity~\cite{pietrak_review_2015} for heterogeneous materials. Generally, these models stem from different analytical and numerical homogenization techniques. Analytical homogenization schemes are mostly based on micro-mechanics models and may be valid for specific composite materials~\cite{pietrak_review_2015,nguyen_effective_2016,ngo_thermal_2016}. Their accuracy critically depends on parameters such as matrix-filler conductivities; dimensions, shape, orientation and dispersion pattern of filler materials. On the other hand, numerical homogenization schemes are more general and based on finite element simulations of Representative Volume Element (RVE)/unit-cell~\cite{matt2007effective,yue2010modeling,yvonnet2008numerical}. Numerical schemes face difficulties in the realistic modelling of material behaviour, interface-boundary conditions and fillers. Often, numerical schemes are computationally expensive due to the need for a large number of simulations across various configurations of RVEs and a fine mesh in each simulation to capture the features of fillers. 
\par
\begin{figure}[!htbp]
    \centering
    \setlength\figureheight{1\textwidth}
    \setlength\figurewidth{1\textwidth}
    \includegraphics[width=0.47\textwidth]{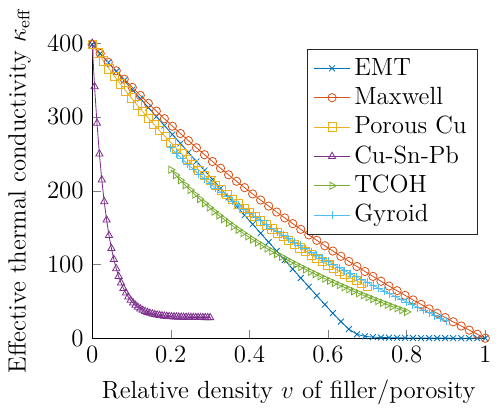}
    \includegraphics[width=0.52\textwidth]{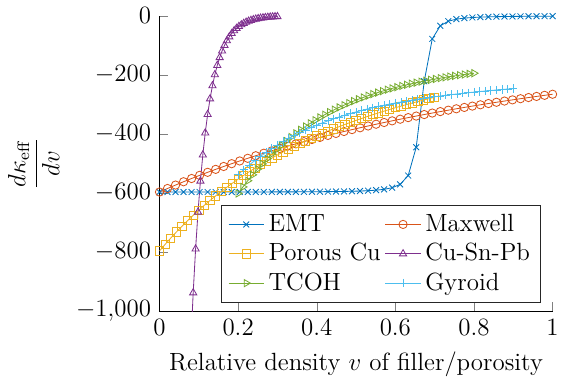}
 \caption{Effective thermal conductivity $\kappa_{\rm eff}$ and its derivative with respect to the relative density $v$ for all models shown in \tref{table: effective thermal conductivity models}.}
 \label{fig:effective thermal conductivity}
\end{figure}

 In this article, to demonstrate the universality across several types of material laws, we exploit a total of six material models, covering three distinct types~: micro-mechanics models, empirical models based on experimental data, and empirical models based on numerical homogenization of graded lattice structures. The discussed models and the chosen constituent materials are detailed in the next few paragraphs. Formul\ae{} to find the effective thermal conductivity $\kappa_{\rm eff}$, limits on the relative density values, \textit{i.e.}, $v$-range ([$v_{\rm min},v_{\rm max}$]) and corresponding limits on the effective thermal conductivity, \textit{i.e.}, $\kappa_{\rm eff}$-range ([$\kappa_{\rm min},\kappa_{\rm max}$]) of the considered models are reported in \tref{table: effective thermal conductivity models}. We also plot the effective thermal conductivity $\kappa_{\rm eff}$ and its derivative with respect to the relative density $v$ in \fref{fig:effective thermal conductivity}. 
\par Regarding analytical models, we explored two well-known models named Effective Medium Theory (EMT) and Maxwell model~\cite{ngo_thermal_2016,pietrak_review_2015}. We consider copper ($\kappa_{\rm copper}=398$~W/mK) and  Polydimethylsiloxane (PDMS) ($\kappa_{\rm PDMS}=0.27$~W/mK) as the constituents. By considering one higher conductivity material and one lower conductivity material, we aim to exploit a wider range of thermal conductivity for design and hence a larger design space. From the manufacturing perspective, very limited research can be found on FGMs made of highly contrasting thermal conductivities. Therefore, it is difficult to comment on the manufacturability of these FGMs. We also apply purely theoretical limits on relative density $v$, $v_{\rm min}=0$ \& $v_{\rm max}=1$.
\newcolumntype{L}[1]{>{\raggedright\let\newline\\\arraybackslash\hspace{0pt}}m{#1}}
\newcolumntype{C}[1]{>{\centering\let\newline\\\arraybackslash\hspace{0pt}}m{#1}}
\newcolumntype{R}[1]{>{\raggedleft\let\newline\\\arraybackslash\hspace{0pt}}m{#1}}
\renewcommand{\arraystretch}{1.5}
\begin{table} [!htbp]  
\begin{center}
\scalebox{0.97}{
\begin{tabular}{ | C{6.5em} | C{17.8em} | C{3.5em} | C{4.5em} |}
\hline

Model name
 & Effective thermal conductivity ($\kappa_{\rm eff}$) & $v$ - range & $\kappa_{\rm eff}$ - range (in W/mK)\\
 \hline
 \hline
 \multicolumn{4}{|C{34em}|}{Micromechanics models \linebreak($\kappa_m$ is thermal conductivity of a matrix material, $\kappa_i$ is thermal conductivity of a filler; here, $\kappa_m=\kappa_{\rm copper}=398$~W/mK and $\kappa_i=\kappa_{\rm PDMS}=0.27$~W/mK)} \\
 \hline
 \hline
Effective Medium Theory~(EMT)~\cite{ngo_thermal_2016}
 & $\kappa_{\rm eff} = \dfrac{1}{4} \left(\tau+\sqrt{\tau^2+8\kappa_i \kappa_m}\right)$ \quad with $\tau =(3v-1)\kappa_i+\left(3(1-v)-1\right)\kappa_m$ & [0, 1] & [0.27, 398]\\
\hline
Maxwell~\cite{ngo_thermal_2016} & $\kappa_{\rm eff}=\kappa_m\dfrac{2\kappa_m+\kappa_p-2v(\kappa_m-\kappa_p)}{2\kappa_m+\kappa_p+v(\kappa_m-\kappa_p)}$& [0, 1] & [0.27, 398]\\
  \hline
   \hline
 \multicolumn{4}{|C{34em}|}{Empirical models based on experimental data ($\kappa_{m}=\kappa_{\rm copper}=398$~W/mK)} \\
 \hline
 \hline
  Lotus type porous copper~\cite{ogushi_measurement_2004} & $\kappa_{\rm eff}=\kappa_{m}\dfrac{1-v}{1+v}$  & [0, 0.7] & [70.24, 398]\\
  \hline
  Cu-Sn-Pb composite~\cite{mercuri_infrared_2022}  & $\kappa_{\rm eff}=\kappa_{m}\left(a e^{bv}+c e^{dv}\right))$ \linebreak $a=9.34008\times10^{-1}$, $b=-2.81400\times10^{1}$, $c=7.08923 \times10^{-2}$, $d=1.14783\times10^{-3}$ & [0, 0.3]& [28.31, 399.95]\\
  \hline
   \hline
 \multicolumn{4}{|C{34
 em}|}{Empirical models based on numerical homogenization of graded lattice structure (of copper), ($\kappa_{m}=\kappa_{\rm copper}=398$~W/mK)} \\
 \hline
 \hline
  Truncated CubOctaHedron (TCOH)~\cite{montemurro_thermal_2022} & $\kappa_{\rm eff}=\kappa_{m}\sum\limits_{i=1}^{7} C_i v^i$, \linebreak $C_1=0.4231$, $C_2=0.1236$, $C_3=0.0933$, $C_4=0.0902$, $C_5=0.0899$, $C_6=0.0899$, $C_7=0.0899$ & [0.2,~0.8]& [36.01, 228.52]\\
  \hline
  Gyroid~\cite{montemurro_thermal_2022} & $\kappa_{\rm eff}=\kappa_{m}\sum\limits_{i=1}^{7} C_i v^i$, \linebreak $C_1=0.5934$, $C_2=0.1119$, $C_3=0.0631$, $C_4=0.0583$, $C_5=0.0578$, $C_6=0.0577$, $C_7=0.0577$ & [0.2,~0.9]& [7.13, 397.96]\\
  \hline
\end{tabular}}
\end{center}
\caption{Various effective thermal conductivity models considered for the FGMs.} 
\label{table: effective thermal conductivity models}
\end{table}
\par Next, we explore the empirical models of thermal conductivity based on experimental data. We refer to the effective thermal conductivity data of lotus-type porous copper and composite of Cu-Sn-Pb taken from Ogushi \textit{et~al.}~\cite{ogushi_measurement_2004} and Mercuri \textit{et~al.}~\cite{mercuri_infrared_2022}, respectively. For the lotus-type porous copper, the empirical model is already established in~\cite{ogushi_measurement_2004}, and we are utilizing the same model. As for Cu-Sn-Pb, we extracted the data from~\cite{mercuri_infrared_2022} and fitted a curve using MATLAB curve fitting toolbox~\cite{MATLAB:curvefittingtoolbox}. The limits on relative density $v$ are also imported from the reference articles.
\par At last, we explore the numerical homogenization results of graded ACMs~\cite{pan_design_2020,montemurro_thermal_2022,li_optimal_2018,li_topology_2018}. ACMs are often used in thermal applications due to their improved heat transfer due to their high surface-to-mass ratio. ACMs have a predefined unit cell, whose geometric features are directly linked to the relative density. Therefore, their numerical homogenization can be performed offline prior to the optimization, and its results can be used as the empirical model for effective thermal conductivity. Nonetheless, to ensure the accuracy of the numerical homogenization during optimization, the scale separation hypothesis must be satisfied. The scale separation hypothesis~\cite{BERTOLINO2022twoscale} assumes that the characteristic dimensions of the periodic
unit cells in the lattice are much smaller than the characteristic dimensions of the entire
structure \& phenomena. At last, the multiscale structure can be reconstructed based on optimized density distribution through the reconstruction process~\cite{montemurro_thermal_2022,wu_topology_2021}. More details about ACMs and their design using optimization can be found in~\cite{pan_design_2020,wu_topology_2021}.
\par Some of the commonly used unit cells for ACMs are Strut-like lattices and Triply Periodic Minimal Surface (TPMS) lattices. We chose two unit-cells called Truncated CubOctaHedron (TCOH) \& Thin Walled Gyroid. Their numerical homogenization data and their empirical models published in Montemurro \textit{et~al.}~\cite{montemurro_thermal_2022} are utilized. Both models are presented in terms of the relative density of porosity. For TCOH, $v_{\rm min}=0.2$ \& $v_{\rm max}=0.8$ are taken considering the manufacturing constraints. As for Gyroid, $v_{\rm min}=0$ \& $v_{\rm max}=0.97$ are possible as the feasible limits mentioned in Li~\textit{et~al.}~\cite{li_optimal_2018}. Nevertheless, we take slightly conservative limits of $v_{\rm min}=0.2$ and $v_{\rm max}=0.9$ into account.

%
%
%
%
\section{Optimization problem}
\label{sec:Optimization problem}
\subsection{Density field parameterization}
\label{sec:Optimization problem description}
\par As pointed out in the introduction section, we use the NURBS basis functions to parameterize the density field. The parameterized density field using $m$ NURBS basis functions $R_i$ can be given as:
\begin{equation}\label{eq:Density parameterization}
    v(\boldsymbol{\xi})=\sum_{i=1}^m R_i(\boldsymbol{\xi})v_i,
\end{equation}
with
\begin{equation}\label{eq:Density gradient}
    \nabla v(\boldsymbol{\xi})=\sum_{i=1}^m \nabla R_i(\boldsymbol{\xi})v_i,
\end{equation}
where $v_i$ is the relative density corresponding to the $i^{th}$ control point. The density values at control points are utilized as the design variables. Note that the weights associated with the control points are predefined (based on the geometry and coordinates of control points). These values do change during optimization, as the weights are not considered as the design variables. 
\par Often, due to the NURBS having support spanning multiple knot spans, the NURBS-based parameterization provides an inherent filtering effect that can prevent checker-boarding issues. In addition, the gradients of the density field would be given by a straightforward formula, as shown in \eref{eq:Density gradient}, which could simplify the employment of density-based restrictions in the optimization problem.
In the case of the element-wise density field, one of the reasons for checker-boarding is the inter-element discontinuity. This discontinuity also makes the calculation of density field gradient difficult. Alternatively, the nodal density field (with the so-called continuous approximation of material distribution using Lagrange basis functions) provides limited $C^0$ inter-element continuity. It is well-known that the higher-order elements are quite effective against checker-boarding. However, the higher-order Lagrange elements cannot be straightforwardly implemented due to their non-negativity property. Hence, the NURBS-based density parameterization is quite beneficial compared to other common alternatives.
\par As evident from \eref{eq:Density parameterization}, the choice of density parameterization defines the design freedom. A finer NURBS mesh for density means more design variables, and thereby a larger optimization problem and higher associated computational cost. For this reason, it is beneficial to decouple both design and solution parameterizations. By decoupling, we ensure satisfactory solution accuracy by refining solution parameterization without increasing the size of the optimization problem. For future, it would also be interesting to decouple geometry and solution field parameterizations in addition to design and solution field parameterizations~\cite{LIAN2017}. Geometric Independent Field approximaTion (GIFT), proposed by Atroshchenko \textit{et~al.}\cite{atroshchenko_weakening_2018}, offers such independence between geometry and solution field parameterizations in IGA framework. By decoupling the geometry and solution fields, we can exploit locally refined splines such as T-splines and PHT-splines for solution field parameterization~\cite{jansari2022adaptive}, to take advantage of local refinement, without disturbing the geometry parameterization.
\subsection{Optimization problem}
\par Next, we formulate the optimization problem based on the given design parameterization. Suppose $\mathbf{V}=[v_1\hspace{0.7em}v_2\hspace{0.7em}...\hspace{0.7em}v_{N_{\rm var}}]^{\rm T}$ is the vector of the $N_{\mathrm{var}}$ design variables, and $J$ is the objective function. In the most general case, $N_{\text{var}}$ can be different from the number of control points of the design mesh $m$ in \eref{eq:Density parameterization} depending on applied symmetry. The isogeometric density topology optimization problem for a thermal meta-structure can be defined as:
\begin{equation}
\label{eq:optimization problem}
\min_{\mathbf{V} \in  \mathbb{R}^{N_{\mathrm{var}}} }~J(T^h,v),
\end{equation}
with
\begin{subequations}
\begin{alignat}{1}
&J : \mathbb{R}^{N_{\mathrm{var}}}  \rightarrow \mathbb{R},\\
&J : \mathbf{V} \rightarrow J(T^h(\mathbf{V}),v(\mathbf{V})),
\end{alignat}
\label{eq:obj fun}
\end{subequations}
\noindent\textrm{such that the following constraints are satisfied,}
\begin{align}
&\text{Equality constraints:}\quad &h_i(T^h(\mathbf{V}),v(\mathbf{V}))= 0, \quad &i=1,2,...,N_h, \label{eq:equality constraint}\\
&\text{Inequality constraints:}\quad &g_j(T^h(\mathbf{V}),v(\mathbf{V}))\leq 0, \quad &i=1,2,...,N_g, \label{eq:Inequality constraint}\\
&\text{Box constraints:}\quad &v_{i,\rm min}\leq~v_{i}\leq~v_{i,\rm max} \quad &i=1,2,...,N_{\mathrm{var}} ,  \label{eq:Box constraints}
\end{align} 
where $N_h$ and $N_g$ are the number of equality constraints and inequality constraints, respectively. $v_{i,\rm min}$ and $v_{i,\rm max}$ are lower and
upper bounds of the design variable $v_{i}$. For our numerical examples, we will have at least two equality constraints related to the boundary value problem as given below,
\begin{align}
    &\text{Equality constraint:}\quad &a(T^h,S^h,v)= \ell(S^h), \forall S^h \in \mathscr{S}^h_0 \quad &\textrm{in} \quad \Omega, \label{eq:equality constraint_1}\\
&\text{Equality constraint:}\quad &T= T_D \quad &\textrm{on} \quad \Gamma_D,   \label{eq:equality constraint_2}
\end{align}
However, these equality constraints are satisfied explicitly by solving the linear system for evaluating temperature $T$ at each optimization iteration. 
\par The given optimization problem is then solved using Sequential Quadratic programming (SQP) algorithm. A nonlinear mathematical programming technique like SQP has sophisticated step selection and constraint handling strategies, in addition to optimized speed and efficiency. As we implement the methodology in MATLAB, its ‘fmincon’ optimization toolbox with the ‘sqp’ subroutine is directly used for straightforwardness~\cite{MATLAB:OptimizationToolbox}. Also, since SQP algorithm is a gradient-based algorithm, both objective function and constraint sensitivities with respect to design variables are required to update the values of the design variables at each iteration. The required sensitivities are calculated using the adjoint method~\cite{chenSensitivityAnalysisHeat2004} and fed to the algorithm. The next subsection will outline the procedure to evaluate the objective function and constraint sensitivities.  

%
%
%
\subsection{Sensitivity analysis}
\label{sec:Sensitivity analysis}
The adjoint method~\cite{chenSensitivityAnalysisHeat2004} to calculate sensitivity is described in this subsection. In the most general case, we define the objective function $J(T,v)$ of a thermal meta-structure as a sum of a domain integral and a surface integral as follows:
\begin{equation}
J(T,v)= \int _{\Omega_{J}} J_b(T,v)~d\Omega + \int _{\Gamma_{J}} J_s(T,v)~d\Gamma,
\label{eq:Objective fun definition}
\end{equation}
where ${\Omega_{J}}$ is the domain where the domain integral is calculated, and ${\Gamma_{J}}$ is the boundary where the surface integral is calculated. 
\par In order to find the objective function sensitivities, the Lagrangian $\mathcal{L}$ is defined as:
\begin{equation}\label{eq:Langrangian 1}
    \mathcal{L}:\underbrace{\mathbb{H}^1(\mathbb{R}^2)}_T \times \underbrace{\mathbb{H}^1(\mathbb{R}^2)}_{P_J} \times \underbrace{\mathbb{H}^1(\mathbb{R}^2)}_v \times \underbrace{\mathbb{H}^1(\mathbb{R}^2)}_{\lambda} \rightarrow \mathbb{R},
\end{equation} with
\begin{equation}\label{eq:Langrangian 2}
    \mathcal{L}(T,P_J,v,\lambda)=J(T,v)+\ell(P_J)-a(T,P_J,v)+\int_{\mathrm{\Gamma}_D}\lambda (T-T_D)~d\mathrm{\Gamma},
\end{equation} 
where $P_J$ and $\lambda$ are the Lagrange multipliers of the weak form as well as Dirichlet boundary condition as defined in \erefs{eq:equality constraint_1}-(\ref{eq:equality constraint_2}).
\par The optimality conditions of the minimization problem are derived as the stationary conditions of the Lagrangian. The stationary conditions with respect to $\lambda$ and $P_J$ give back the weak formulation of the boundary value problem as stated in \eref{eq:weak_form}, which can be satisfied by solving the matrix system given in \eref{eq:Linear matrix system} for the state variable $T$. The stationary condition with respect to $T$,
\begin{equation}
    \left<\dfrac{\partial \mathcal{L}(T,P,\varPhi,\lambda)}{\partial T}, \delta T \right>=0,
\end{equation} 
combined with the Dirichlet boundary condition, gives a well-posed adjoint problem as follows:
\begin{subequations}  \label{eq:Adjoint BVP}
\begin{align} 
\nabla \cdot \left( \boldsymbol{\kappa}\nabla P_J\right) &= \dfrac{\partial J(T,v)}{\partial T} \quad &&\textrm{in} \quad \mathrm{\Omega}, \label{eq:Adjoint Laplace equation}\\
 P_J &= 0 \quad &&\textrm{on} \quad \mathrm{\Gamma}_D, \label{eq:Adjoint Boundary conditions a}\\
  (\boldsymbol{\kappa}(v)\nabla P_J )\cdot \mathbf{n} &= 0 \quad &&\textrm{on} \quad \mathrm{\Gamma}_N, \label{eq:Adjoint Boundary conditions b}\\
  (\boldsymbol{\kappa}(v)\nabla P_J)\cdot \mathbf{n} &= -hP_J, \quad &&\textrm{on} \quad \mathrm{\Gamma}_R, \label{eq:Adjoint Boundary conditions c} \\
  \left\llbracket P_J\right\rrbracket &=0 \quad &&\textrm{on} \quad \mathrm{\Gamma}_I, \label{eq:Adjoint Interface conditions a}\\
  \mathbf{n}\cdot \left\llbracket \boldsymbol{\kappa}\nabla P_J\right\rrbracket &= 0 \quad &&\textrm{on} \quad \mathrm{\Gamma}_I, \label{eq:Adjoint Interface conditions b}
\end{align}
\end{subequations}
where $P_J$ is the adjoint temperature field. By employing the trial and test  function approximations, the adjoint problem can be written in the matrix form as:
\begin{equation} \label{eq:adjoint eq. matrix form}
\mathbf{K}^{\rm T} \mathbf{P}_J= \mathbf{F}_{J},
\end{equation}
where $\mathbf{P}_J$ is the vector of adjoint temperatures at control points, $\mathbf{F}_{J}$ is the global adjoint flux vector defined as:
\begin{equation}
\mathbf{F}_{J}= \int _{\Omega_{J}} \mathbf{N}^{\textrm{T}} \dfrac{\partial J_b}{\partial T}~d\Omega + \int _{\Gamma_{J}} \mathbf{N}^{\textrm{T}} \dfrac{\partial J_s}{\partial T}~d\Gamma,
\end{equation} 
\par At last, with the fulfillment of all three stationary conditions, the sensitivity of the objective function $\left({dJ}/{dv_i}\right)$ becomes equal to the total derivative of Lagrangian $\mathcal{L}$ with respect to a design variable $v_i$. Therefore:
\begin{equation} 
\dfrac{d J}{dv_i} = \int _{\Omega_{J}} \dfrac{\partial J_b}{\partial v}~\dfrac{dv}{dv_i}~d\Omega + \int _{\Gamma_{J}} \dfrac{\partial  J_s}{\partial v}~\dfrac{dv}{dv_i}~d\Gamma -(\mathbf{P}_J)^{\rm T}~\dfrac{d \mathbf{K}}{dv}~\dfrac{dv}{dv_i}~\mathbf{T}.
\end{equation}
By applying the density parameterization from \eref{eq:Density parameterization}, the sensitivity equation can be simplified as:
\begin{equation} \label{eq:objective fn sensitivity}
\dfrac{d J}{dv_i} = \int _{\Omega_{J}} \dfrac{\partial J_b}{\partial v}~R_i~d\Omega + \int _{\Gamma_{J}} \dfrac{\partial  J_s}{\partial v}~R_i~d\Gamma -(\mathbf{P}_J)^{\rm T}~\dfrac{d \mathbf{K}}{dv}~R_i~\mathbf{T}.
\end{equation}
A derivative of the global stiffness matrix $\mathbf{K}$ with respect to relative density $v$ is provided in \ref{sec:Appendix A}. The first and second terms in \eref{eq:objective fn sensitivity} are related to the explicit dependency of the objective function on relative density, while the last term is related to its dependency on relative density via the state variable, \textit{i.e.}, the temperature $T$.  
\par With the same procedure, the constraint sensitivities can be formulated too. In our case, we define a generalized equality constraint $h$ and a generalized inequality constraint $g$ in a similar manner to the objective function by the summation of a domain integral and a surface integral as follows: 
\begin{equation} \label{eq:constraint fn definition a}
h(T,v)= \int _{\Omega_{h}} h_b(T,v)~d\Omega + \int _{\Gamma_{h}} h_s(T,v)~d\Gamma,
\end{equation}
\begin{equation} \label{eq:constraint fn definition b}
g(T,v)= \int _{\Omega_{g}} g_b(T,v)~d\Omega + \int _{\Gamma_{g}} g_s(T,v)~d\Gamma,
\end{equation}
where $\Omega_{h}, \Gamma_{h}, h_b, h_s$ \& $\Omega_{g}, \Gamma_{g}, g_b, g_s$ are equality and inequality constraints related quantities following the nomenclature analogous to the objective function as described earlier in the section.
Accordingly, the constraint sensitivities are defined as:
\begin{equation} \label{eq:constraint fn sensitivity a}
\dfrac{dh}{dv_i} = \int _{\Omega_{h}} \dfrac{\partial h_{b}}{\partial v}~R_i~d\Omega + \int _{\Gamma_{h}} \dfrac{\partial  h_{s}}{\partial v}~R_i~d\Gamma -(\mathbf{P}_{h})^{\rm T}~\dfrac{d \mathbf{K}}{dv}~R_i~\mathbf{T},
\end{equation}
\begin{equation} \label{eq:constraint fn sensitivity b}
\dfrac{dg}{dv_i} = \int _{\Omega_{g}} \dfrac{\partial g_{b}}{\partial v}~R_i~d\Omega + \int _{\Gamma_{g}} \dfrac{\partial  g_{s}}{\partial v}~R_i~d\Gamma -(\mathbf{P}_{g})^{\rm T}~\dfrac{d \mathbf{K}}{dv}~R_i~\mathbf{T}.
\end{equation}
 \par Corresponding adjoint temperature vectors $\mathbf{P}_{h}$  and $\mathbf{P}_{g}$ are obtained by solving extra adjoint systems similar to \eref{eq:adjoint eq. matrix form} with the global adjoint flux vectors $\mathbf{F}_{h}$ \& $\mathbf{F}_{g}$ given as follows:
\begin{equation} \label{eq:adjoint eq. matrix form a}
\mathbf{F}_{h}= \int _{\Omega_{h}} \mathbf{N}^{\textrm{T}} \dfrac{\partial h_{b}}{\partial T}~d\Omega + \int _{\Gamma_{h}} \mathbf{N}^{\textrm{T}} \dfrac{\partial h_{s}}{\partial T}~d\Gamma,
\end{equation}
\begin{equation} \label{eq:adjoint eq. matrix form b}
\mathbf{F}_{g}= \int _{\Omega_{g}} \mathbf{N}^{\textrm{T}} \dfrac{\partial g_{b}}{\partial T}~d\Omega + \int _{\Gamma_{g}} \mathbf{N}^{\textrm{T}} \dfrac{\partial g_{s}}{\partial T}~d\Gamma.
\end{equation}

%
%
%
%
\section{Thermal cloak}
\label{sec:Thermal cloak}
\par In this section, both 2D and 3D thermal cloaks are designed with several different conditions using the proposed method. Since we are using `fmincon' optimization toolbox in MATLAB, there are several inbuilt stopping criteria such as, `OptimalityTolerance' (tolerance value in first-order optimality measures), `StepTolerance' (tolerance value of the change in design variables' values), `ObjectiveLimit' (tolerance value of the objective function), `MaxFunctionEvaluations' (maximum number of function evaluations), `MaxIterations' (maximum number of iterations). We define `ObjectiveLimit'=$1\times 10^{-10}$, `StepTolerance'=$1\times 10^{-10}$, `OptimalityTolerance'=$1\times 10^{-10}$. Unless otherwise stated, the stopping criteria and tolerance values are the same for all the examples solved, and in case of any discrepancies, they will be mentioned explicitly. 
We explore all 6 material laws presented in \sref{sec:Material models} with copper \& PDMS or (only) copper (and porosity as the other material) as their constituents. The initial distributions of relative density are defined based on $v_i=0.5(v_{\rm min}+v_{\rm max}), i=1,2,..,N_{\rm var}$ to give equal importance to both limits. The values of the objective function for each optimized solution are provided in the captions of the different figures.
\subsection{2D thermal cloak}
\label{sec:Chen2015case 2D cloak}
 \begin{figure}[!htbp]
    \centering
    \setlength\figureheight{1\textwidth}
    \setlength\figurewidth{1\textwidth}
    \begin{subfigure}[t]{0.31\textwidth}{\centering\includegraphics[width=1\textwidth]{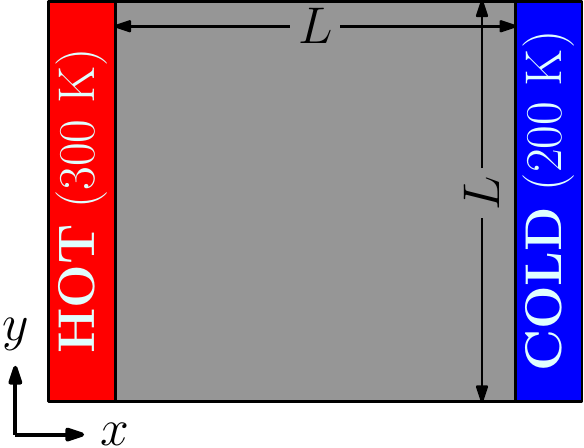}}
        \caption{A base material plate under constant heat flux.}
        \label{fig:Cloak problem schematics a}
    \end{subfigure}\quad
     \begin{subfigure}[t]{0.31\textwidth}{\centering\includegraphics[width=1\textwidth]{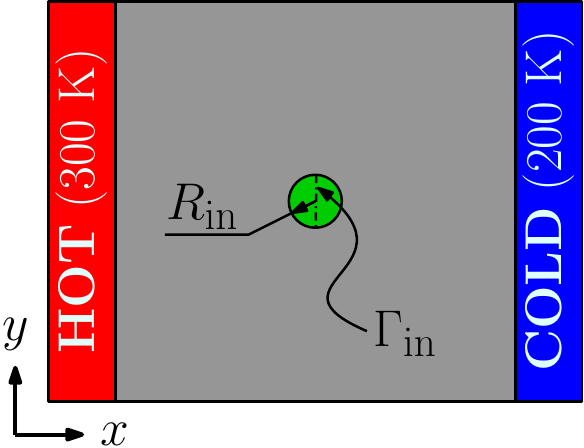}}
        \caption{An obstacle embedded in the plate.}
        \label{fig:Cloak problem schematics b}
    \end{subfigure}\quad
    \begin{subfigure}[t]{0.31\textwidth}{\centering\includegraphics[width=1\textwidth]{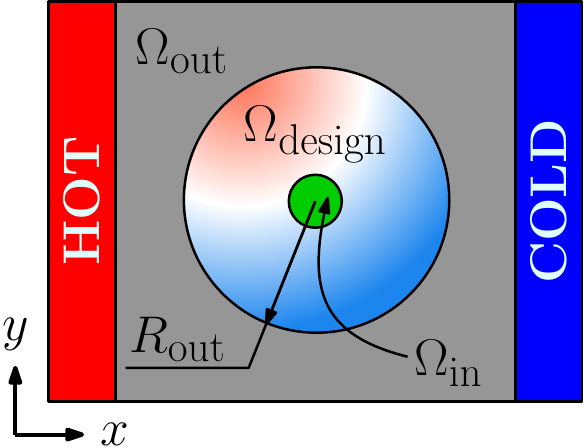}}
             \caption{A FGM-based cloak in the base material plate.}
             \label{fig:Cloak problem schematics c}
    \end{subfigure}
 \caption{Schematic design of (a) a base material plate  under constant heat flux applied by the high-temperature source on the left side and low-temperature sink on the right side; (b) a circular insulator embedded in the base material plate ($\mathrm{\Omega}_{\mathrm{in}}$ is the insulator); (c) the insulator and a surrounding FGM-based thermal cloak embedded in the base material plate; $\mathrm{\Omega}_{\mathrm{design}}$ is the domain of the cloak where the material distribution is optimized, $\mathrm{\Omega}_{\mathrm{out}}$ is the outside domain of remaining base material, where the temperature disturbance reduction is sought. $\mathrm{\Omega} = \mathrm{\Omega}_{\mathrm{in}} \cup \mathrm{\Omega}_{\mathrm{design}}\cup \mathrm{\Omega}_{\mathrm{out}}$.}
 \label{fig:chen2015case Schematics}
\end{figure}
\par We begin by designing 2D thermal cloaks, for which we consider an $L$ $\times$ $L$ square homogeneous base material plate. The plate is embedded by a circular insulator (with conductivity $\kappa_{\rm ins}=0.0001$~W/mK) of radius $R_{\rm in}$. Later, surrounding the insulator, an annular-shaped FGM-based thermal cloak with inner radius $R_{\rm in}$ and outer radius $R_{\rm out}$ is also added. Similar to the explanation in \sref{sec:Problem description}, by introduction of the insulator and thermal cloak, the plate domain $\Omega$ is divided into three parts: $\mathrm{\Omega}_{\rm in}$,~$\mathrm{\Omega}_{\rm design}$ and $\mathrm{\Omega}_{\rm out}$ that possess the material properties of the insulator, FGM and base material, respectively. Constant temperatures, 300~K on the left side and 200~K on the right side are applied. Adiabatic wall condition is imposed on the top and bottom edges. The schematics related to the geometry are given in \fref{fig:chen2015case Schematics}. This geometry will be consistently utilized in several of the examples throughout this article, potentially with different dimensions and/or material allocations. In this particular case, we take $L=140$~mm, $R_{\rm in}=10$~mm, $R_{\rm out}=50$~mm, and iron as the base material ($\kappa_{\rm base}=67$~W/mK).
\par The objective of a thermal cloak is to reduce the temperature disturbance in the outer region $\Omega_{\rm out}$ caused by the presence of an insulator. In that manner, a thermal cloak ensures that the insulator does not get detected in the two points/four points in-plane observation of the temperature profile. Keeping this in mind, we define the cloaking objective function as: 
\begin{equation} \label{eq:cloaking fn}
    J_{\mathrm{cloak}}=\dfrac{1}{\widetilde{J}_{\mathrm{cloak}}} \int_{\mathrm{\Omega}_{\mathrm{out}}}  (T - \overline{T})^2~d\mathrm{\Omega}, \quad \text{with} \quad  \widetilde{J}_{\mathrm{cloak}}= \int_{\mathrm{\Omega}_{\mathrm{out}}}  (\widetilde{T} - \overline{T} )^2~d\mathrm{\Omega}.
\end{equation}
where $\overline{T}$ is the temperature distribution of the reference case (a homogeneous base material plate without the presence of the insulator and thermal cloak), $\widetilde{T}$ is the temperature distribution when $\mathrm{\Omega}_{\rm design}$ is filled with the base material.
\par For NURBS parameterizations, the circumferential direction and radial direction are taken as the parametric directions. Also, second-order and first-order NURBS approximations are taken in circumferential direction and radial direction, respectively. A mesh model of a 2D NURBS patch is shown in \fref{fig:Chen2015case meshing strategy}. Most of the 2D thermal cloaks have symmetry (adiabatic symmetry) along $x$ and anti-symmetry (isothermal symmetry) along $y$-axes. Thus, we impose $x$ and $y$-axes symmetry for the design meshes to reduce the corresponding number of design variables. Following it, we perform a mesh sensitivity analysis to choose an appropriate solution mesh as presented in \ref{sec:2D cloak Mesh study}. Accordingly, we take a mesh with DOF=13167 as the solution mesh for $N_{\rm var}=25$ and $N_{\rm var}=81$.
\subsubsection{Design with various material models}
\label{sec:2D cloak Design with various material models}

\newcolumntype{M}[1]{>{\centering\arraybackslash}m{#1}}
\renewcommand{\arraystretch}{1.5}   
\begin{figure}[!htbp]
\centering
\scalebox{0.88}{
\begin{tabular}[c]{| M{5.4em} | M{5.45em} | M{5.45em} | M{5.45em}| M{5.45em} | M{7.7em} |}
\hline 
 \centering EMT \\ \vspace{0.1cm} $v_i\in[0,1]$\vspace{-0.2cm}& \centering Maxwell \\ \vspace{0.1cm} $v_i\in[0,1]$\vspace{-0.2cm} & \centering Porous Cu \\ \vspace{0.1cm} $v_i\in[0,0.7]$ \vspace{-0.2cm} & \centering Cu-Sn-Pb \\ \vspace{0.1cm} $v_i\in[0,0.3]$ \vspace{-0.2cm} & \centering TCOH \\ \vspace{0.1cm} ${v_i\in[0.2,0.8]}$\vspace{-0.2cm} & \begin{center}
    Gyroid\\ \vspace{0.1cm} $v_i\in[0.2,0.9]$\vspace{-0.2cm}
\end{center}
\\  
\hline
    \vspace{0.2cm}
    \begin{subfigure}[t]{0.15\textwidth}{\includegraphics[width=1\textwidth]{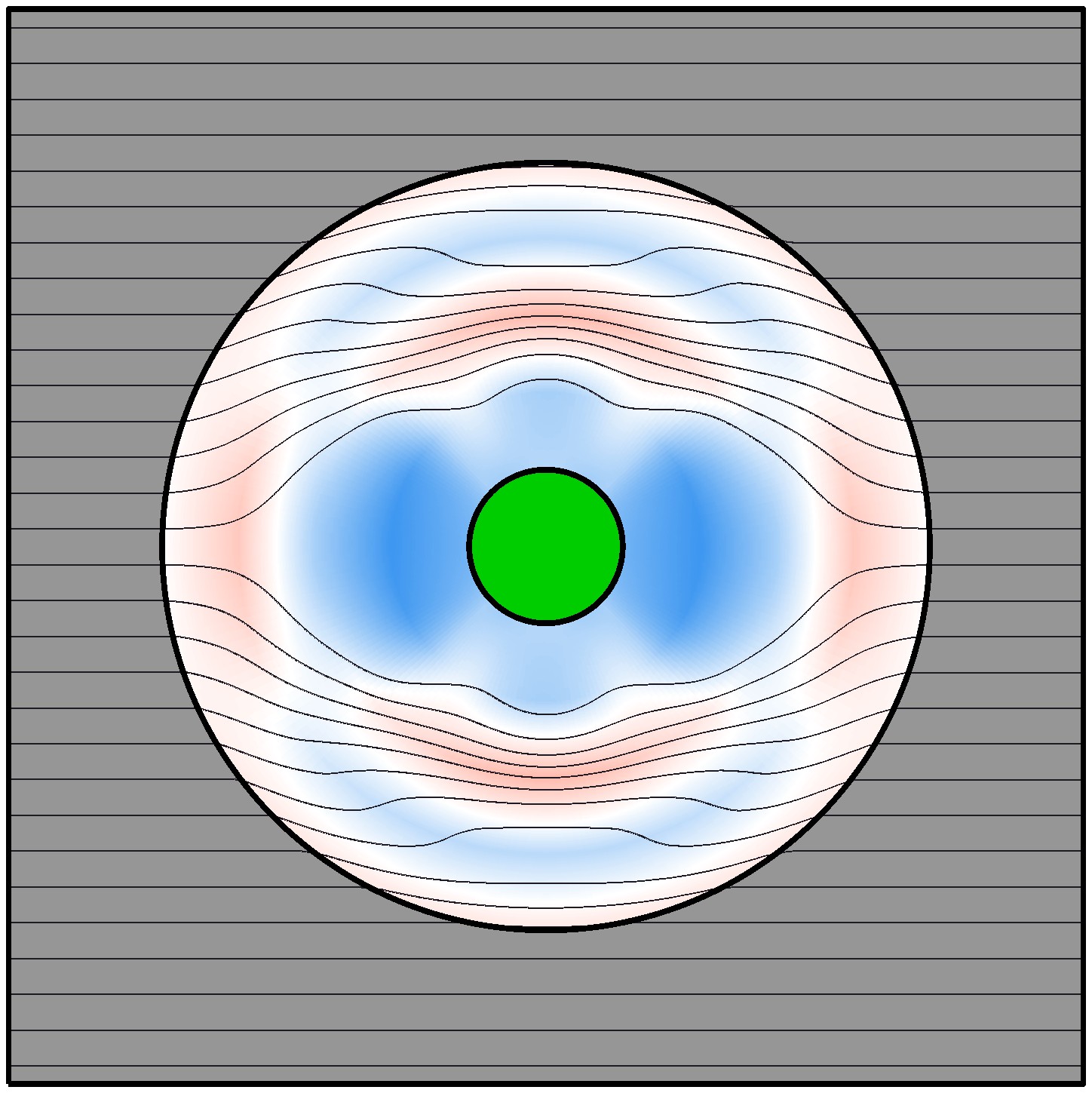}}
        \caption{\centering $N_{\rm var}=25$, $J=8.74\times 10^{-9}$}
    \end{subfigure}  & \vspace{0.2cm}
    \begin{subfigure}[t]{0.15\textwidth}{\includegraphics[width=1\textwidth]{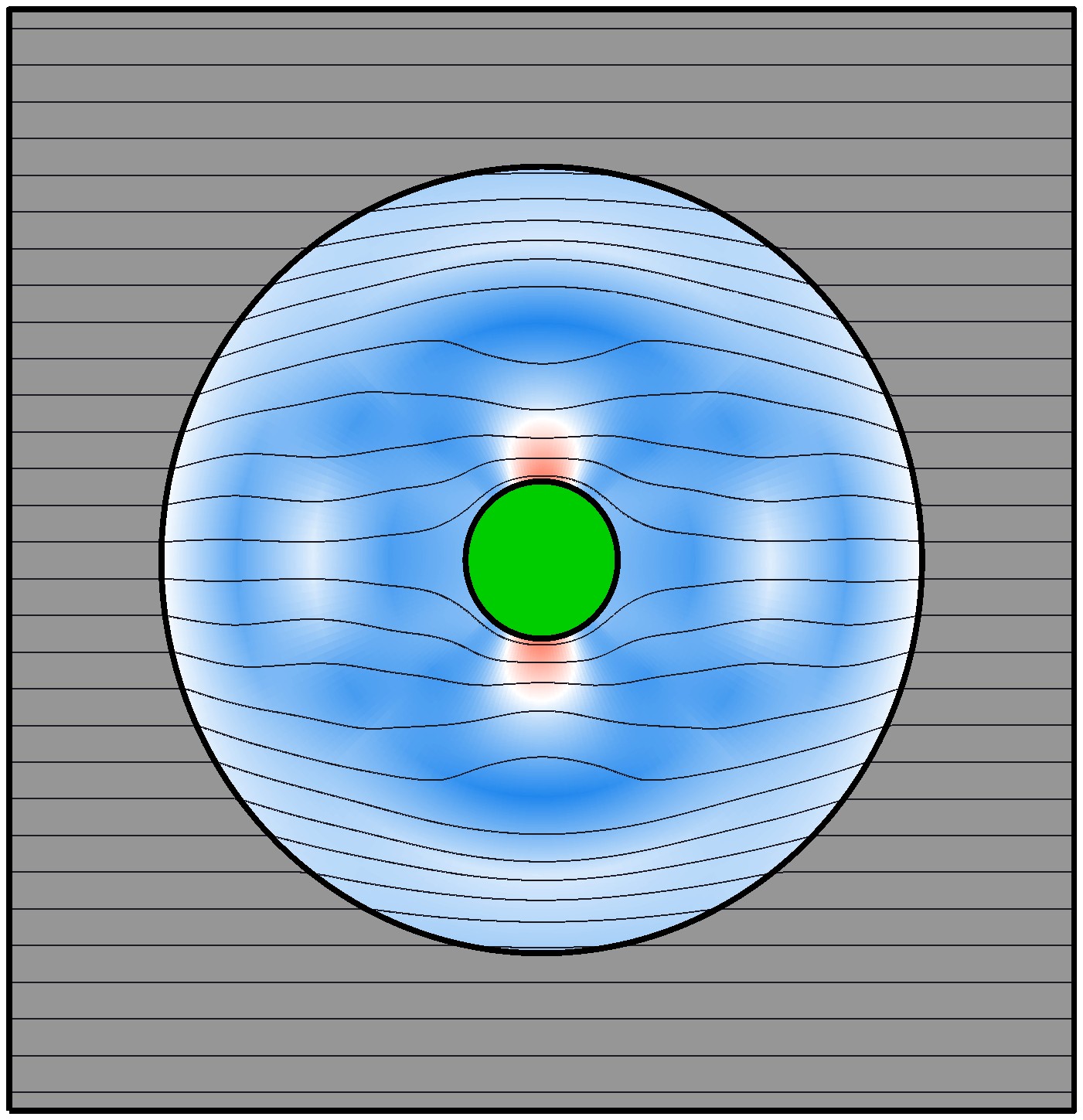}}
        \caption{\centering $N_{\rm var}=25$, $J=9.91\times 10^{-9}$}
    \end{subfigure} & \vspace{0.2cm}
    \begin{subfigure}[t]{0.15\textwidth}{\includegraphics[width=1\textwidth]{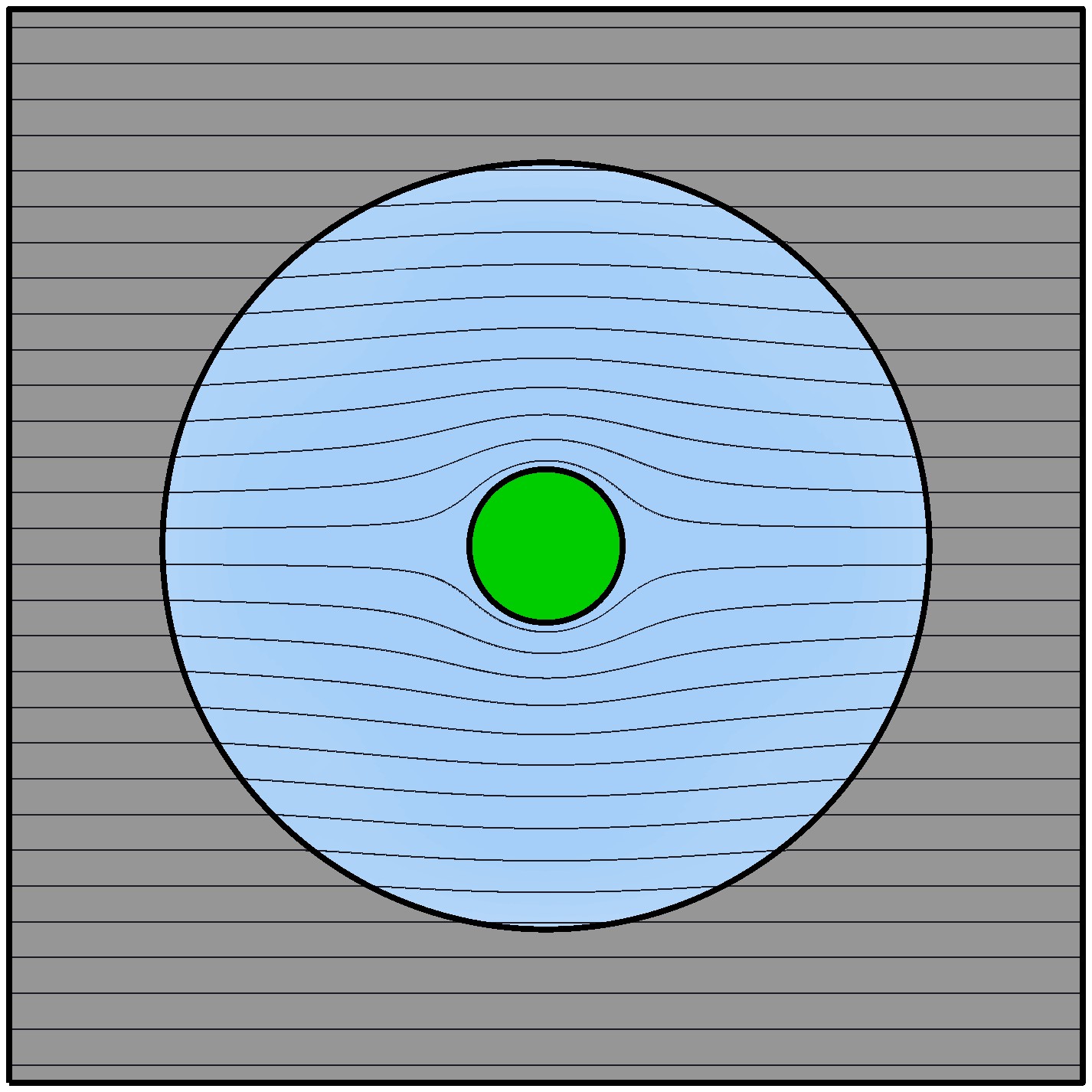}}
        \caption{\centering $N_{\rm var}=25$, $J= 2.01\times 10^{-8}$}
    \end{subfigure} & \vspace{0.2cm}
    \begin{subfigure}[t]{0.15\textwidth}{\includegraphics[width=1\textwidth]{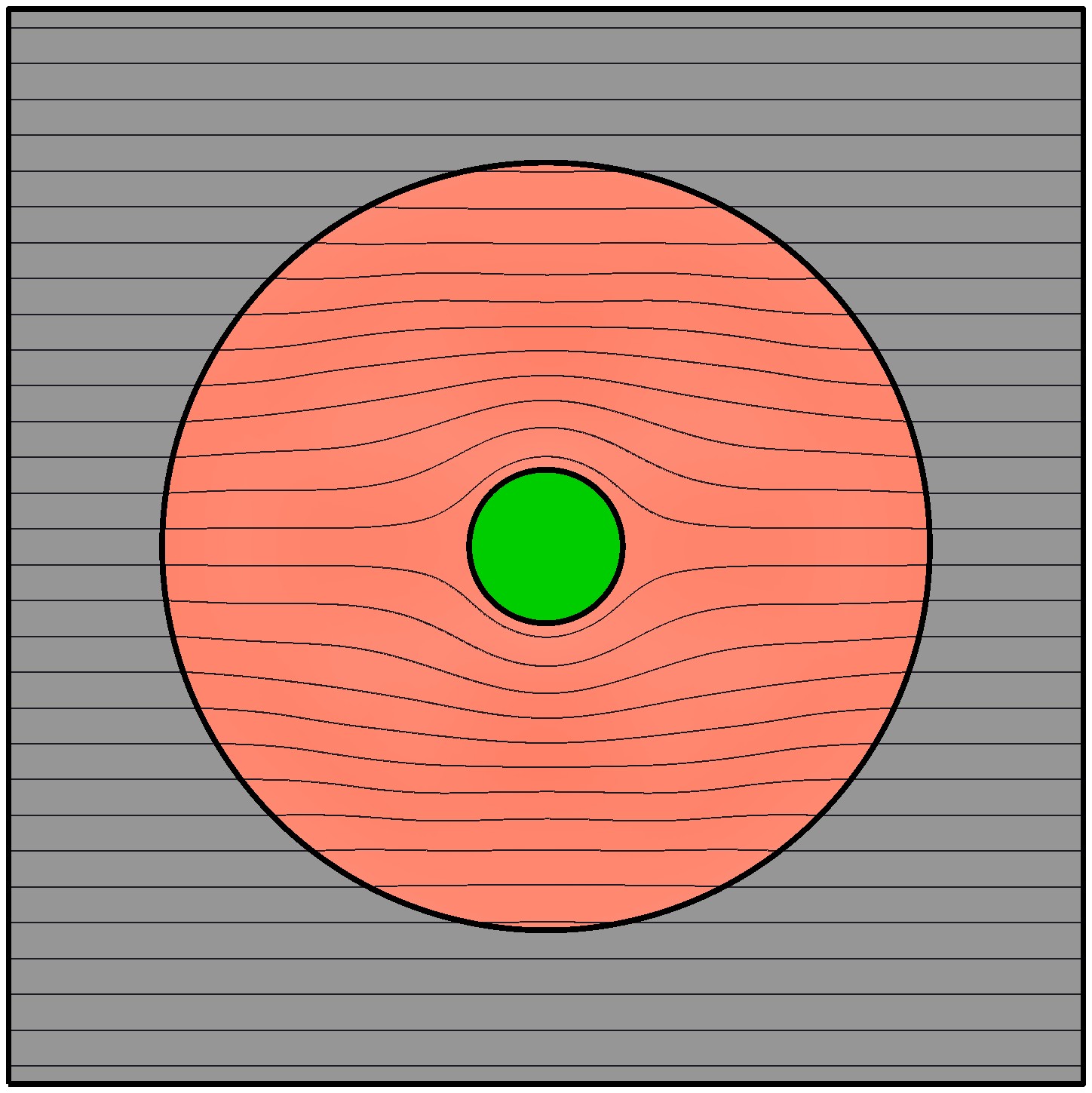}}
        \caption{\centering $N_{\rm var}=25$, $J=1.94\times 10^{-8}$}
    \end{subfigure}& \vspace{0.2cm}
    \begin{subfigure}[t]{0.15\textwidth}{\includegraphics[width=1\textwidth]{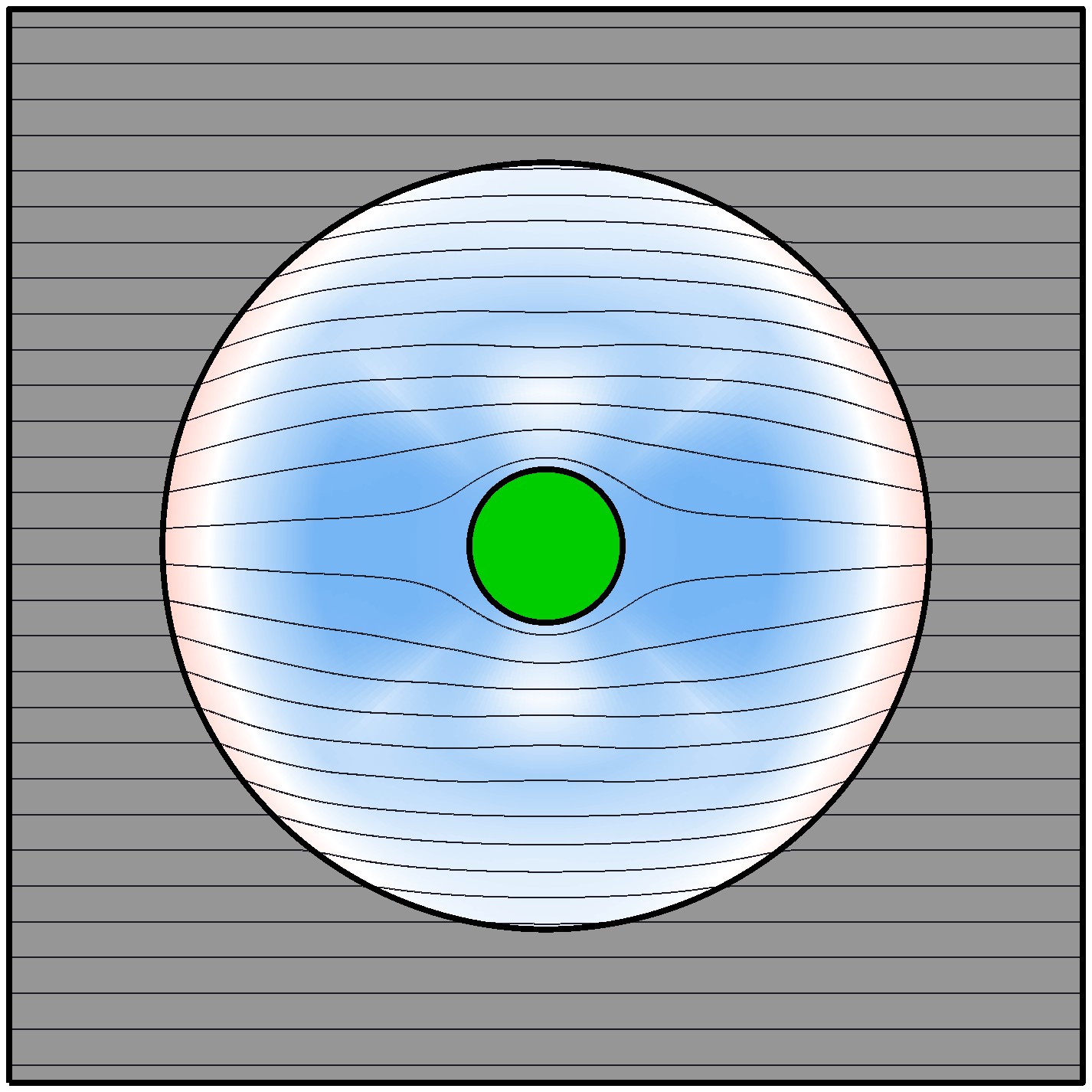}}
        \caption{\centering $N_{\rm var}=25$, $J=1.21\times 10^{-7}$}
    \end{subfigure}
    & \vspace{0.1cm}
    \begin{subfigure}[t]{0.15\textwidth}{\includegraphics[width=1\textwidth]{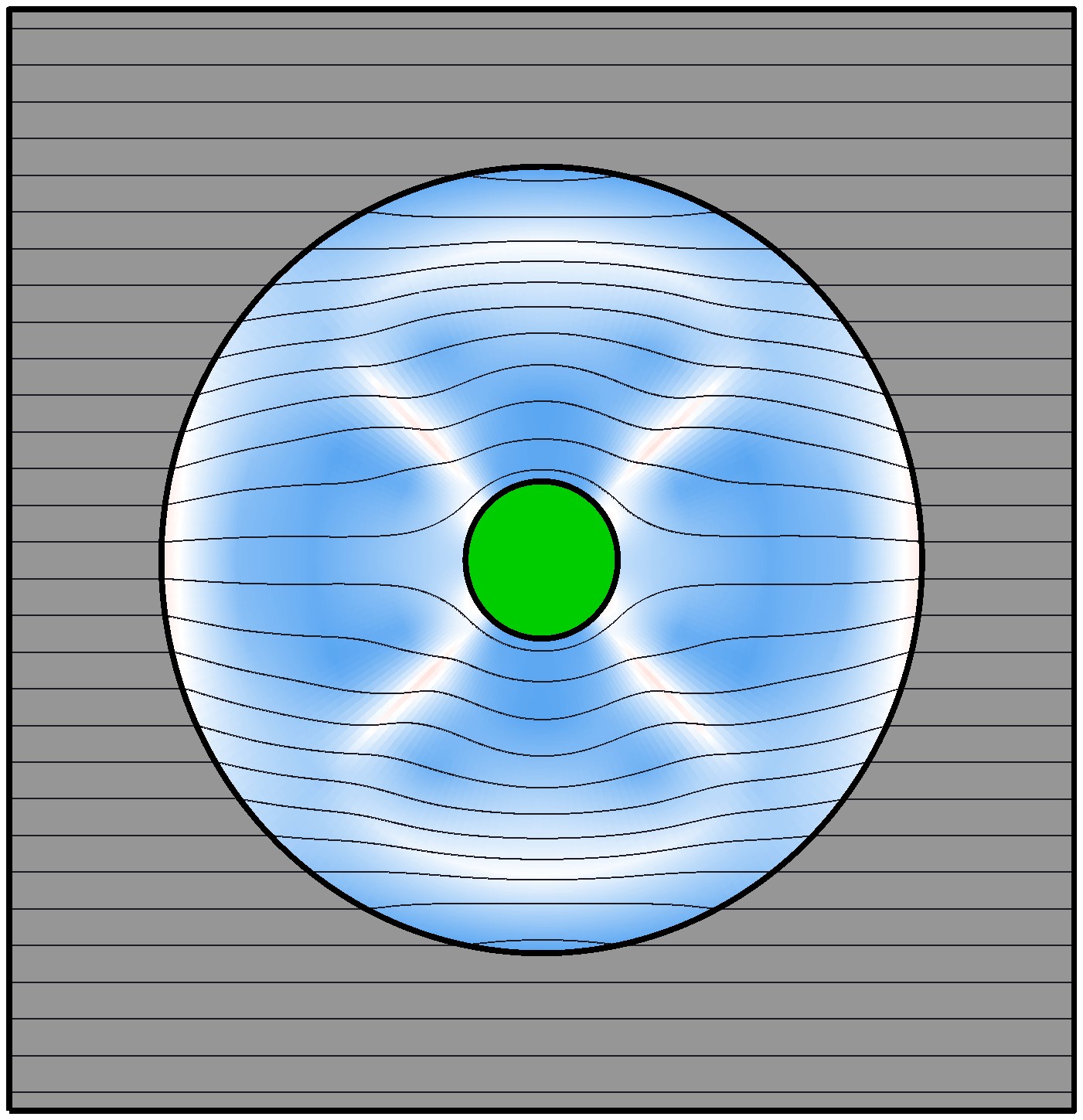}}
        \caption{\centering $N_{\rm var}=25$, $J=7.55\times 10^{-8}$}
    \end{subfigure}~\begin{subfigure}[b]{0.05\textwidth}{
\includegraphics[keepaspectratio=false,width=1.1\textwidth,height=2.45cm]{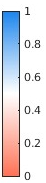}}  
\end{subfigure}\\   
\hline

    \end{tabular}
}

\caption{Optimized material distributions for the thermal cloak problem. Six material models and $N_{\rm var}=25$ are considered. Optimized objective function values are of order $10^{-7}$-$10^{-9}$. In optimized material distributions, almost the entire design domain filled with intermediate densities.}  
    \label{fig:chen2015case cloak}
\end{figure}

\par In \fref{fig:chen2015case cloak}, we show the optimized material distributions for thermal cloak using 6 different effective thermal conductivity models (given in \tref{table: effective thermal conductivity models}) for $N_{\rm var}=25$. As the solution of the thermal cloak problem is non-unique~\cite{calderon_inverse_1980,uhlmann_electrical_2009,Greenleaf2003nonuniqueness} and most topology optimization problems are non-convex, our optimization cases reach the nearest local minimizers. Therefore, as depicted in the figure, the optimized material distributions remain close to the initial material distributions with almost the entire domain filled with intermediate densities. All these distributions are valid solutions and exhibit objective function value of order $10^{-7}$-$10^{-9}$. 
\begin{figure}[!htbp]
    \centering
    \setlength\figureheight{1\textwidth}
    \setlength\figurewidth{1\textwidth}
\includegraphics[width=0.8\textwidth]{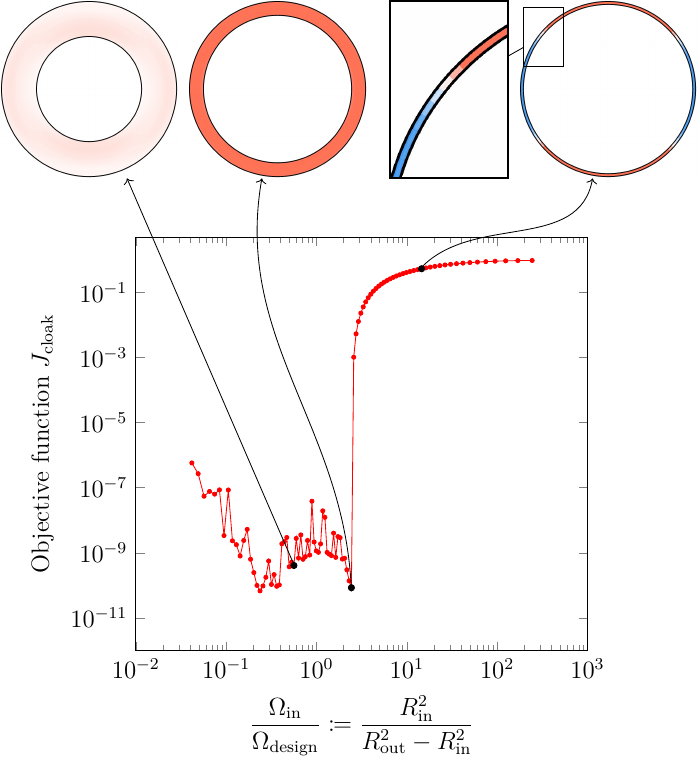}
 \caption{Trend of the objective function $J_{\rm cloak}$ vs. $\Omega_{\rm in}/\Omega_{\rm design}$. $R_{\rm in}$ is varied between $10$~mm to 49~mm, while keeping $R_{\rm out}=50$~mm constant. For $R_{\rm in}\lessapprox42.2$~mm, the designed cloaks have satisfactory cloaking function with $J_{\rm cloak}<10^{-6}$. As the insulator size increases more than 42.2~mm, $J_{\rm cloak}$ increases exponentially.}
 \label{fig:Chen2015case cloak riVar}
\end{figure}
\par Next, we want to study the effect of design domain size with respect to the insulator size on the overall optimization results. Therefore, we run optimization with different values of $R_{\rm in}$ (between $10$~mm and 50~mm) while keeping $R_{\rm out}=50$~mm constant. The plot of $J_{\rm cloak}$ vs ${\Omega_{\rm in}}/{\Omega_{\rm design}}$ for $N_{\rm var}=25$ and EMT model is presented in \fref{fig:Chen2015case cloak riVar}. From the figure, it is evident that the proposed method is capable of designing thermal cloaks of substantial quality for $R_{\rm in} \lessapprox 42.2$~mm (or ${\Omega_{\rm in}}/{\Omega_{\rm design}}\lessapprox 2.48$) with $J_{\rm cloak}<6\times 10^{-7}$. At $R_{\rm in} \approx 42.2$~mm, the optimized material distribution completely filled $\Omega_{\rm design}$ with copper. This optimized material distribution aligns with the concept of a bilayer cloak, as proposed by Han et al. \cite{Han2014} and verified in our previous study~\cite{jansari2024design}. A bilayer cloak consists of two annular layers: an outer layer made of a material with higher conductivity and an inner layer composed of an insulating material. The radius of the circular interface between these layers is uniquely determined by the conductivities of the materials involved. In our scenario, this interface radius would be $R_{\rm out}\sqrt{(\kappa_{\rm copper}-\kappa_{\rm base})/(\kappa_{\rm copper}+\kappa_{\rm base})}$=~42.18~mm. Thus, we can say that $R_{\rm in} \lessapprox 42.18$~mm represents a bilayer cloak without an inner layer. This also shows the minimum required design domain to attain a significant cloaking effect based on the given conductivities. Reducing the design domain below this threshold limits the available design space, consequently deteriorating optimization scope severely. Therefore, as $R_{\rm in}$ becomes greater than $42.2$~mm, $J_{\rm cloak}$ starts rising exponentially to approach 1. 
\par
\begin{figure}[!htbp]
\centering
\scalebox{0.98}{
\begin{tabular}[c]{|M{0.8em}|m{7.6em}|m{7.6em}|M{7.6em}|M{7.6em}|}
\hline	
   & \centering A square homogeneous base material plate
    & \centering A square base material plate embedded by a circular insulator ($J=1$)
    &  \multicolumn{2}{M{15.2em}|}{A square base material plate embedded by a circular insulator and surrounding thermal cloak}       
  \\
\cline{4-5}
   &  &  &  EMT ($J=1.97 \times 10^{-9}$)
&  Cu-Sn-Pb ($J=4.32 \times 10^{-9}$)\\ 
\hline 
  \vspace{0.05cm}  
  \rotatebox{90}{\centering \footnotesize Flux flow}  & \vspace{0.2cm} 
    \begin{subfigure}[t]{0.1785\textwidth}\begin{center}{
\includegraphics[width=1\textwidth]{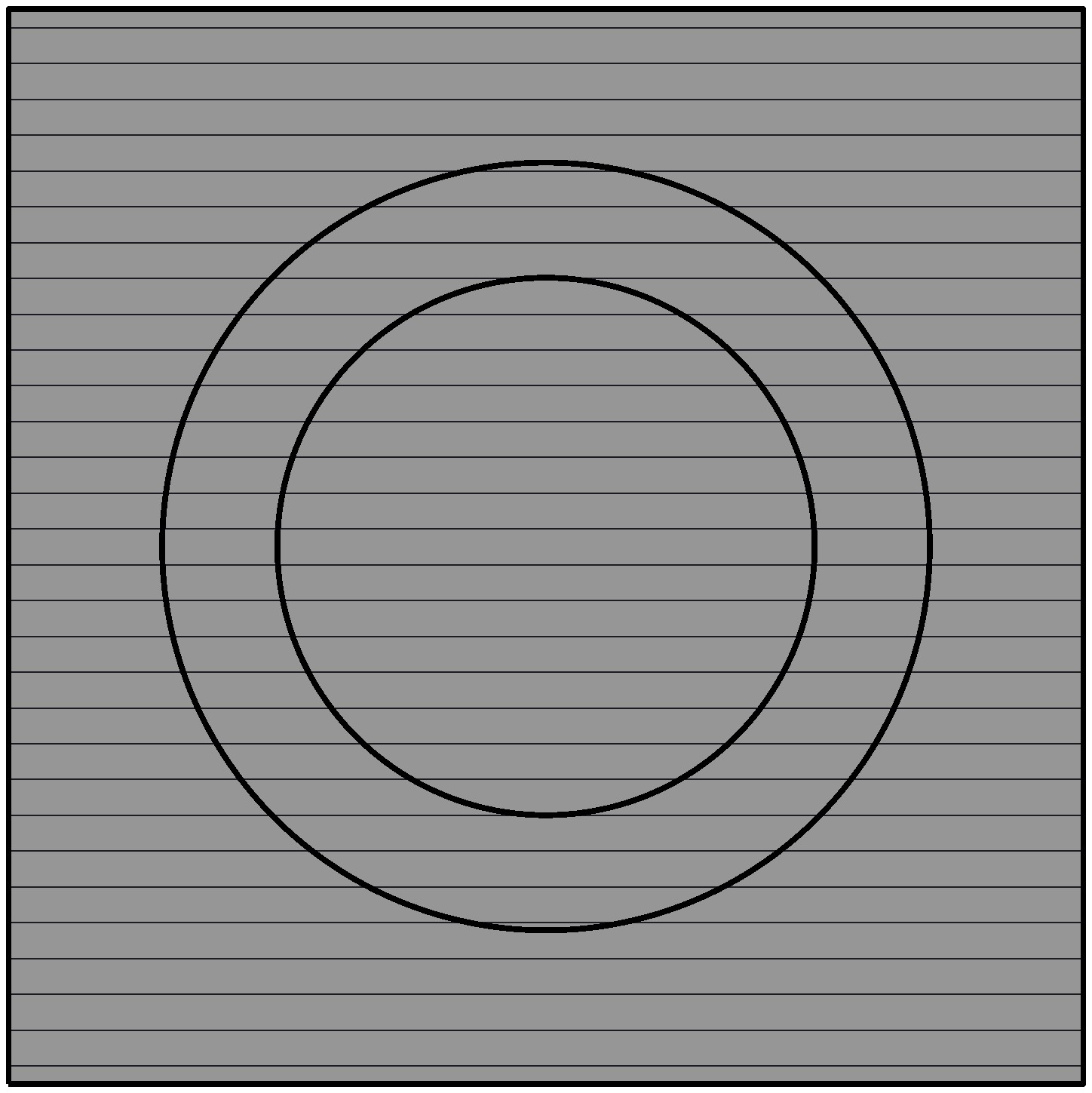}}      
    \end{center}
    \end{subfigure} & \vspace{0.2cm} 
    \begin{subfigure}[t]{0.1785\textwidth}\begin{center}{
\includegraphics[width=1\textwidth]{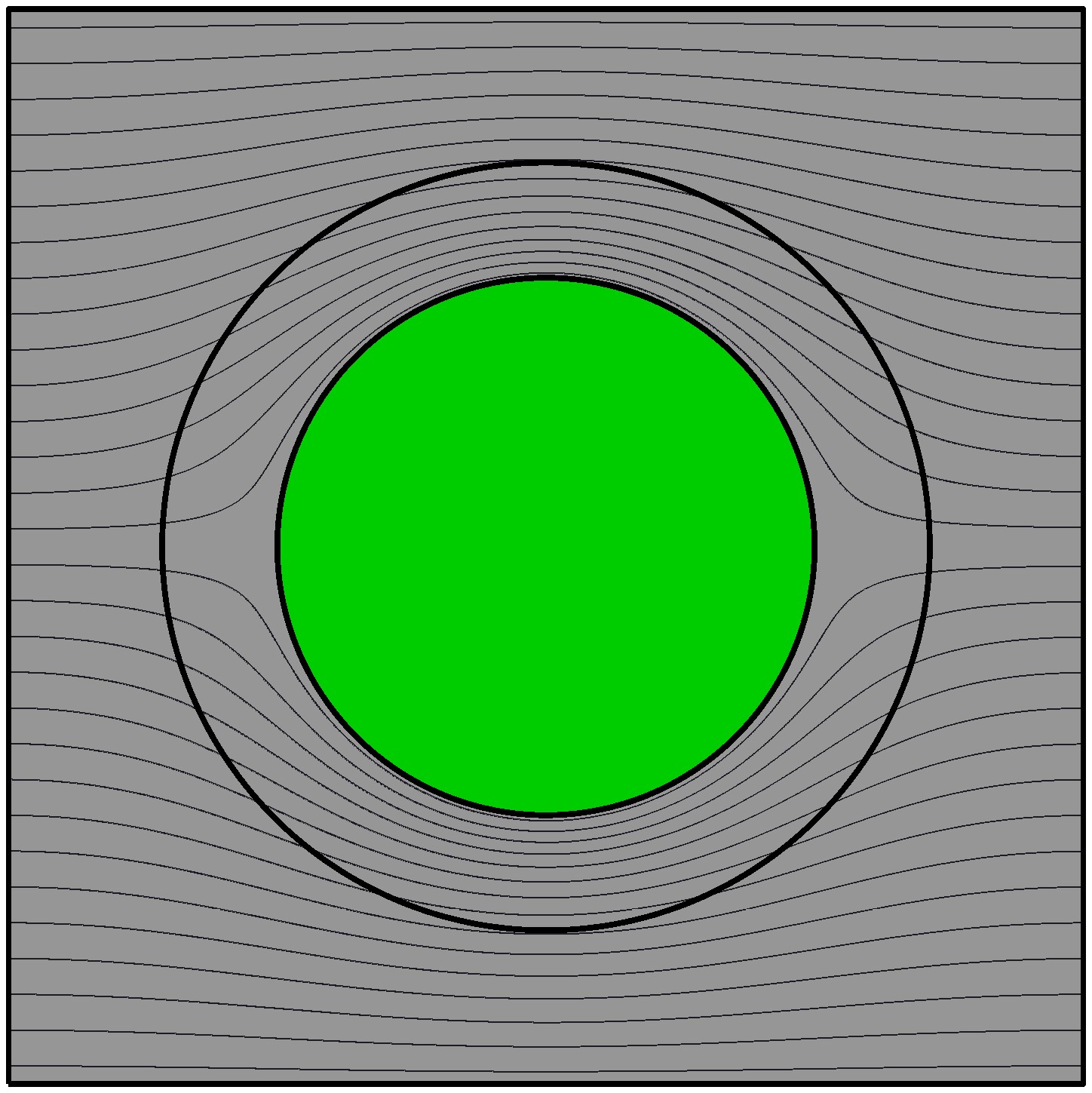}}       
    \end{center}
    \end{subfigure} & \vspace{0.2cm} 
    \begin{subfigure}[t]{0.2184\textwidth}\begin{center}{
\includegraphics[width=1\textwidth]{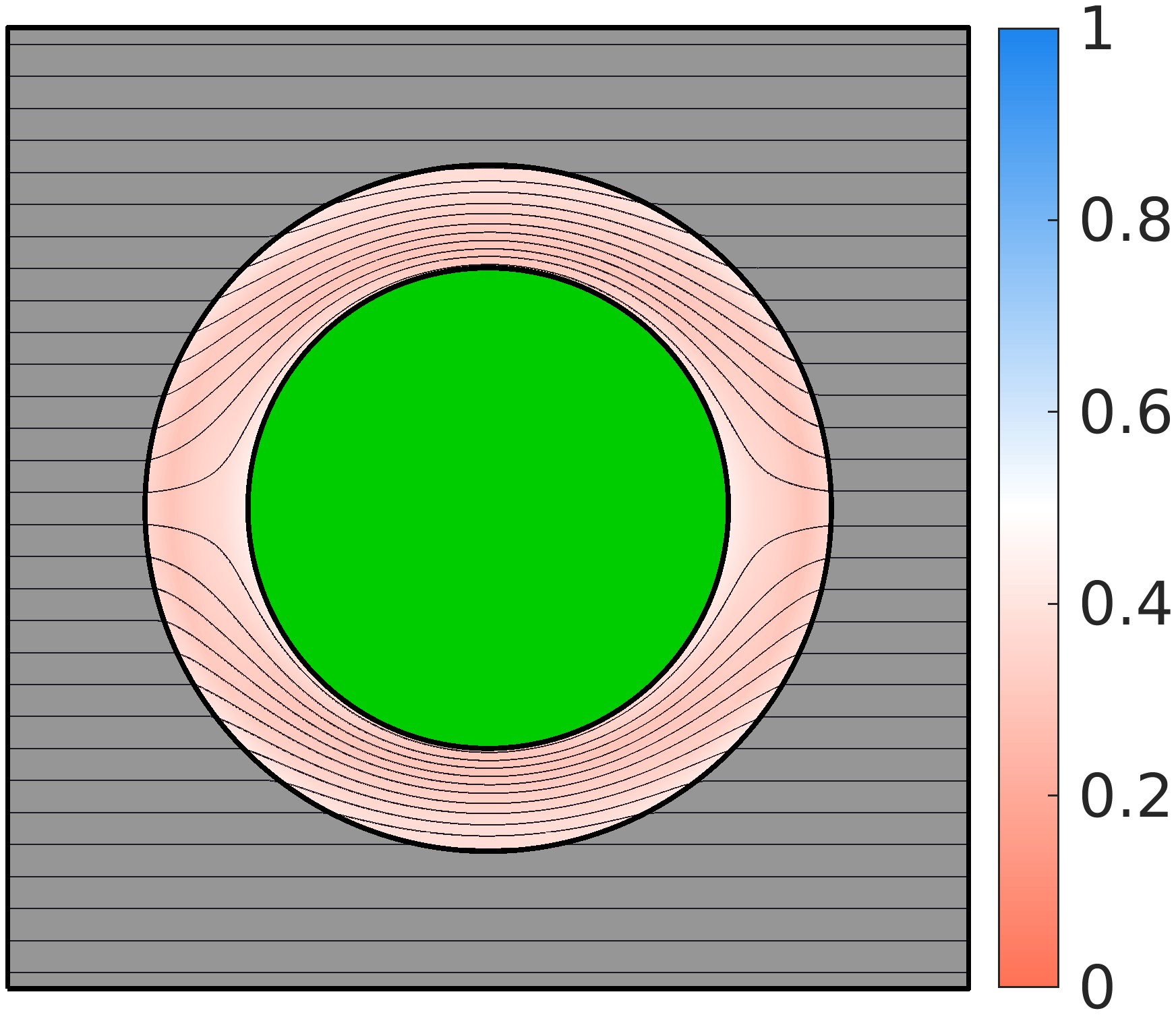}}   
    \end{center}
    \end{subfigure} 
    & \vspace{0.2cm} \begin{subfigure}[t]{0.2184\textwidth}\begin{center}{
\includegraphics[width=1\textwidth]{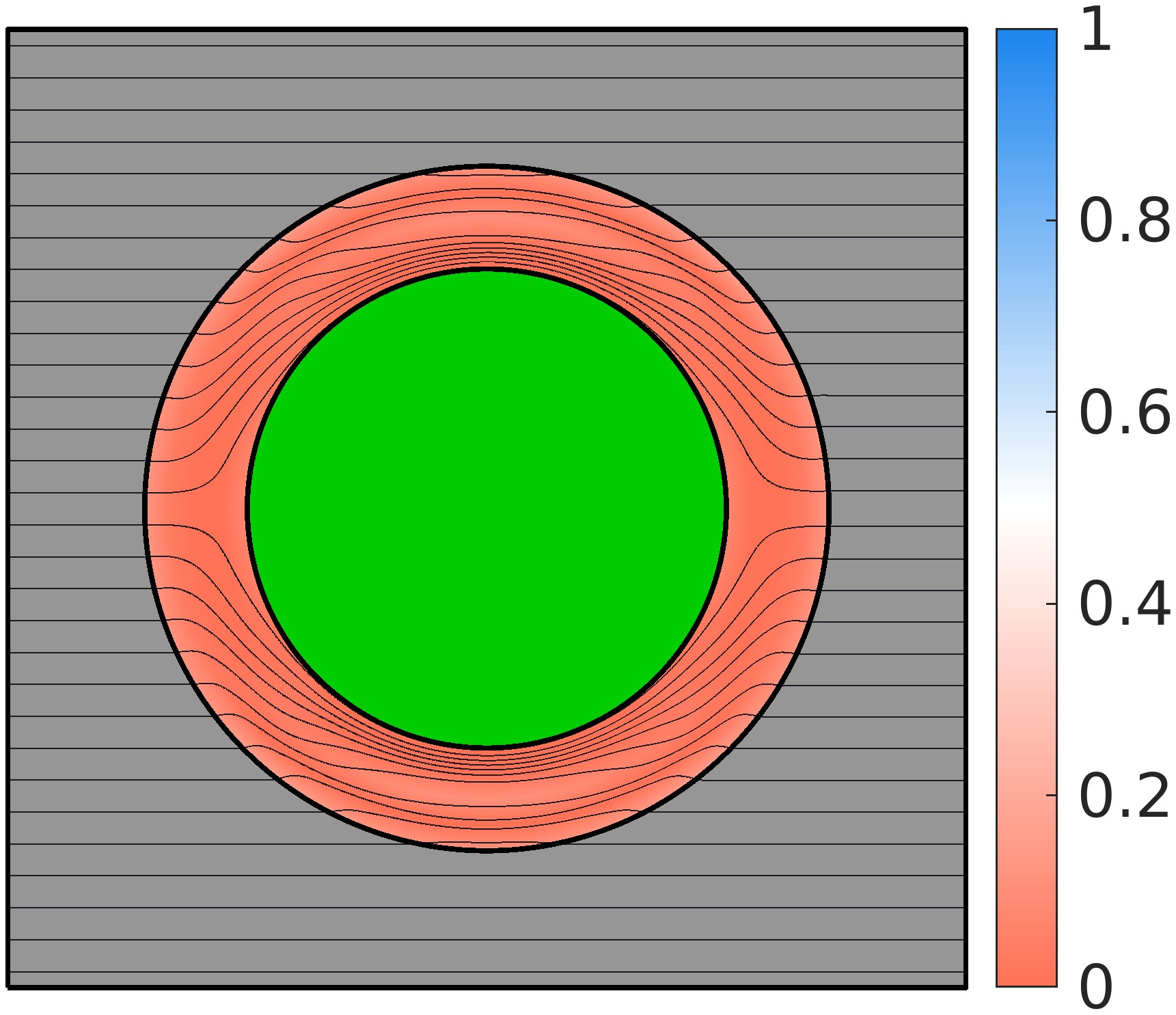}}    
    \end{center}
    \end{subfigure} 
    \\  
    \rotatebox{90}{\centering \footnotesize  Temp. $T$}  &
      \begin{subfigure}[t]{0.217\textwidth}\begin{center}{
\includegraphics[width=1\textwidth]{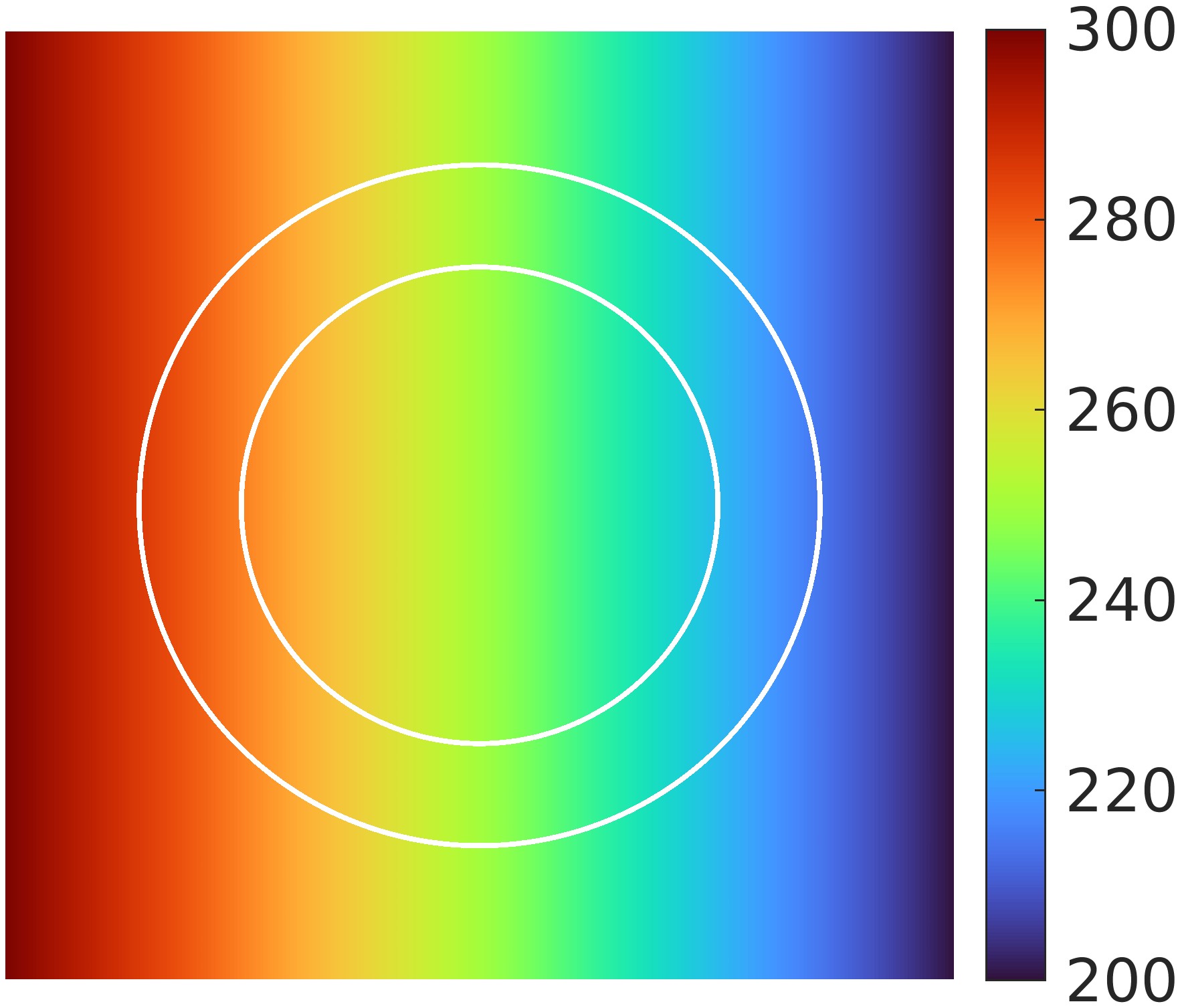}}     
    \end{center}
    \end{subfigure} 
    &
    \begin{subfigure}[t]{0.217\textwidth}\begin{center}{
\includegraphics[width=1\textwidth]{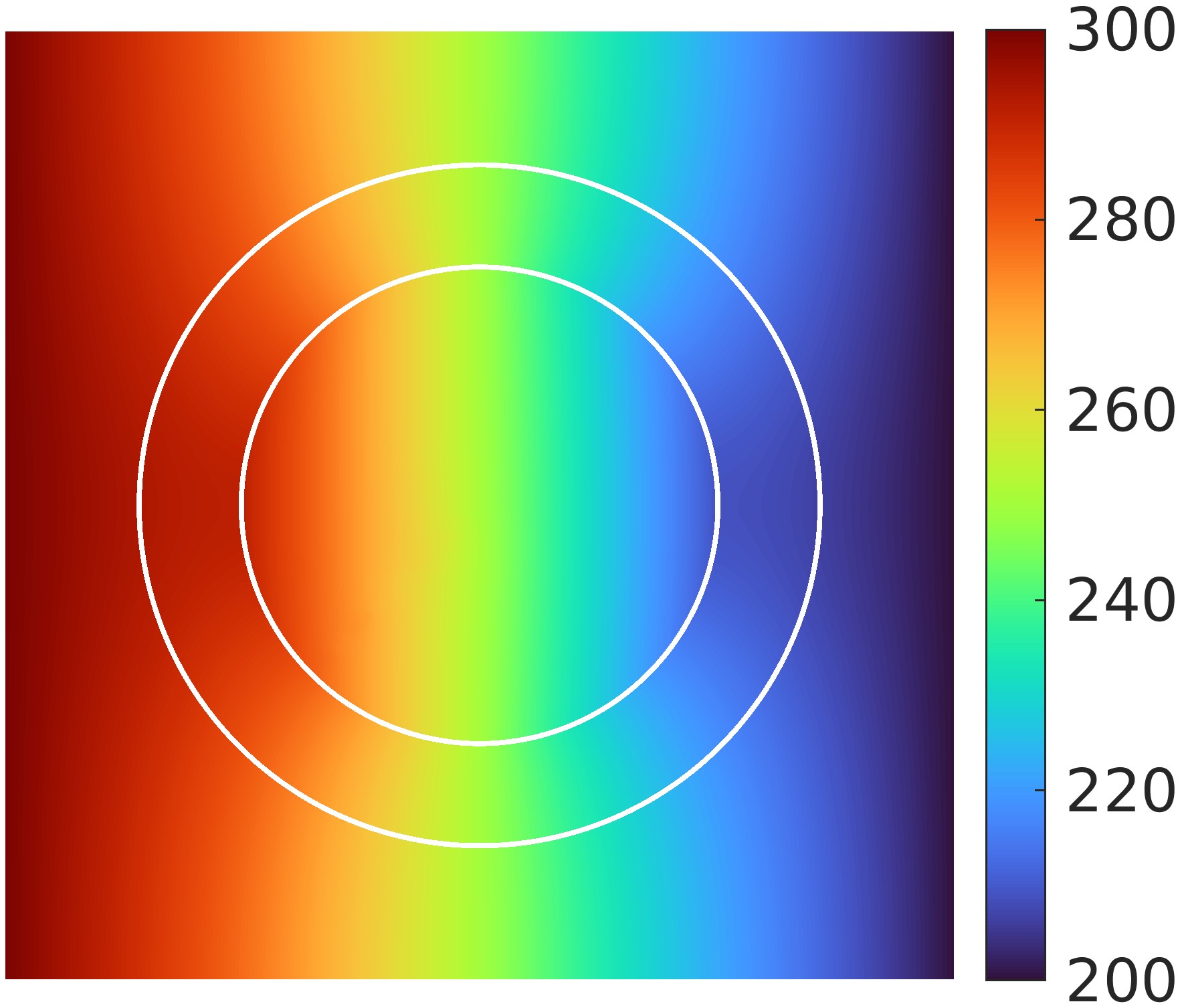}}    
    \end{center}
    \end{subfigure}
    &
    \begin{subfigure}[t]{0.217\textwidth}\begin{center}{
\includegraphics[width=1\textwidth]{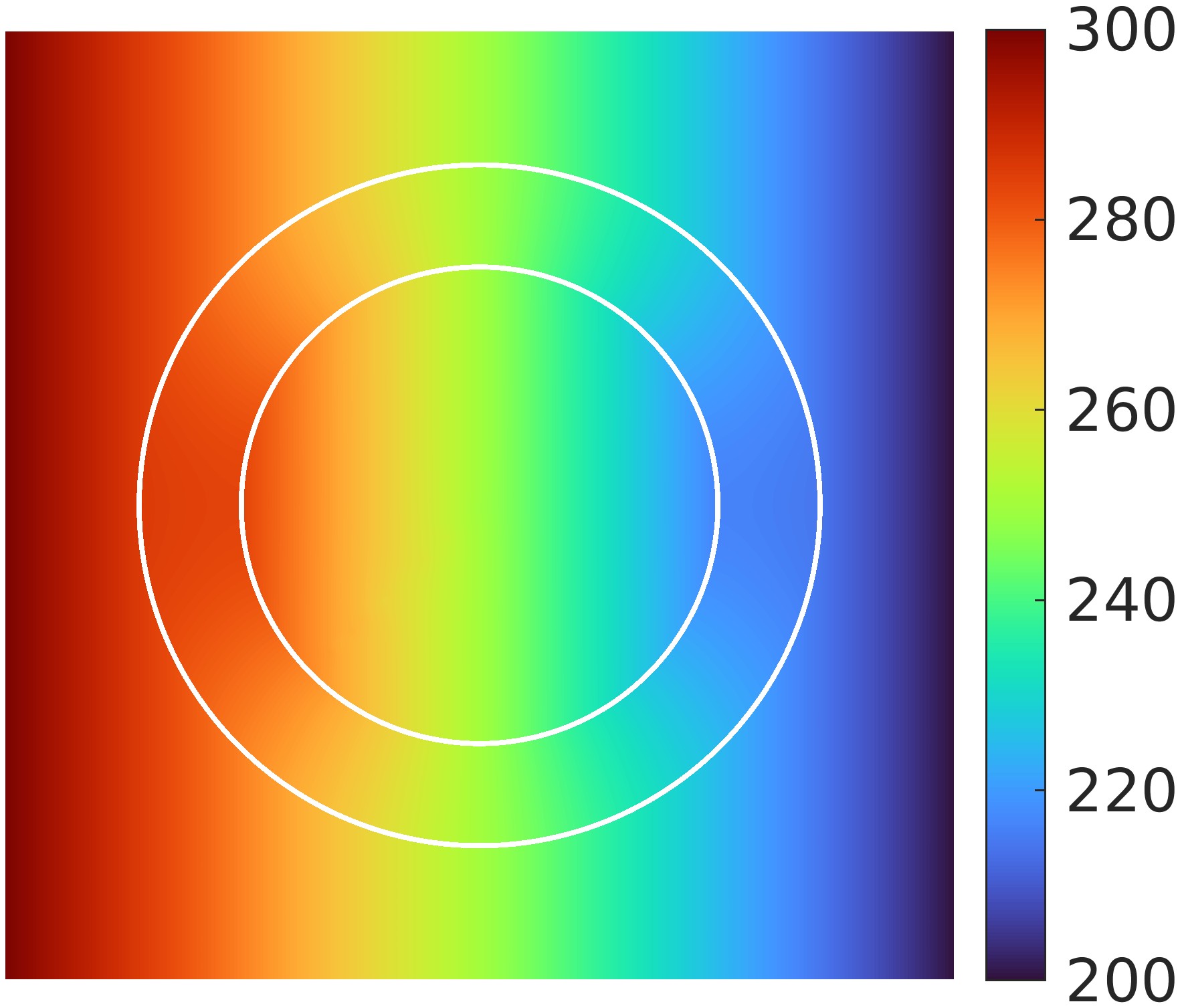}}      
    \end{center}
    \end{subfigure} 
    &
    \begin{subfigure}[t]{0.217\textwidth}\begin{center}{
\includegraphics[width=1\textwidth]{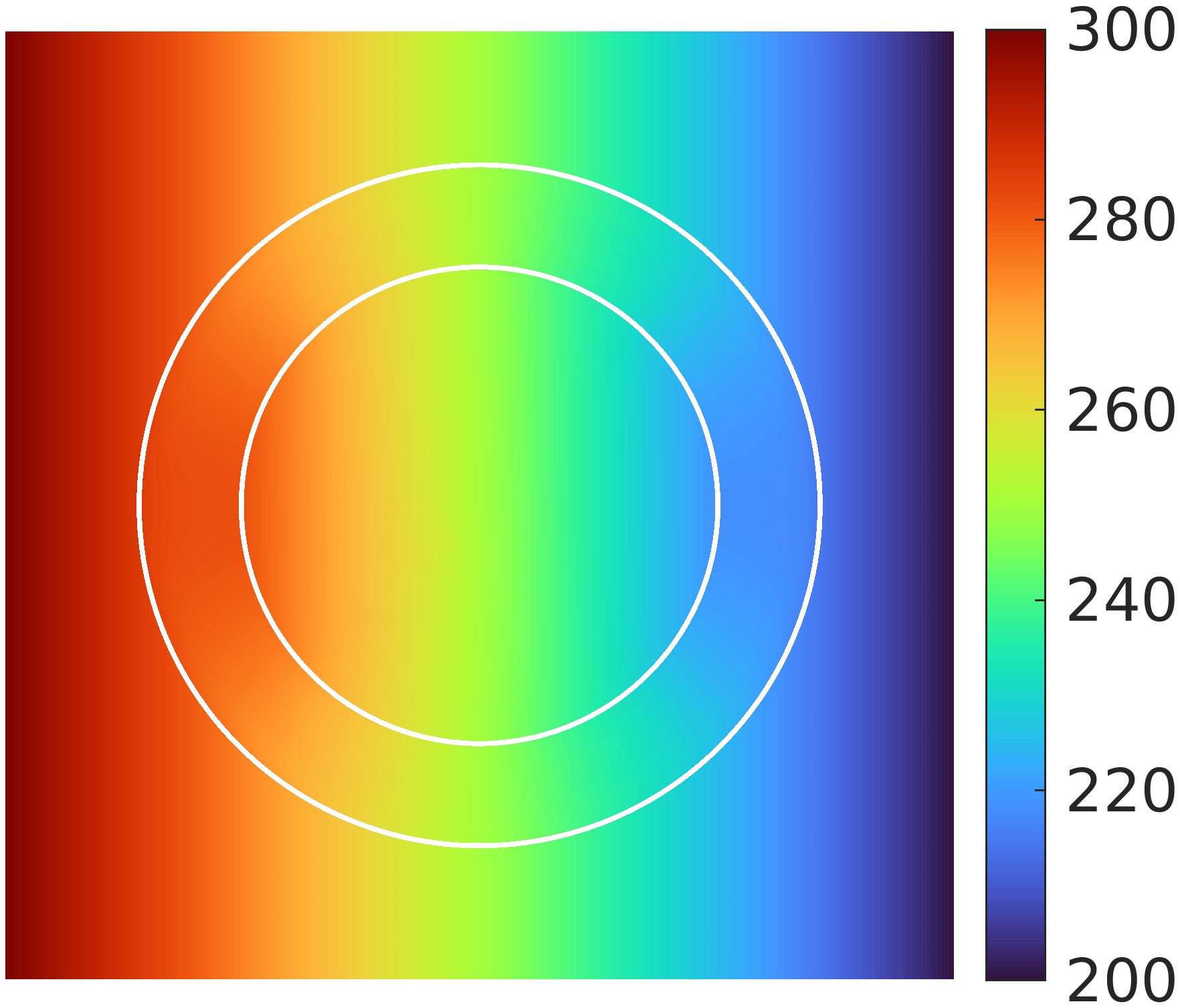}}      
    \end{center}
    \end{subfigure} 
    \\  
    \rotatebox{90}{\centering \footnotesize Temp. Diff. $T-\overline{T}$}  &
       &
    \begin{subfigure}[t]{0.2205\textwidth}\begin{center}{
\includegraphics[width=1\textwidth]{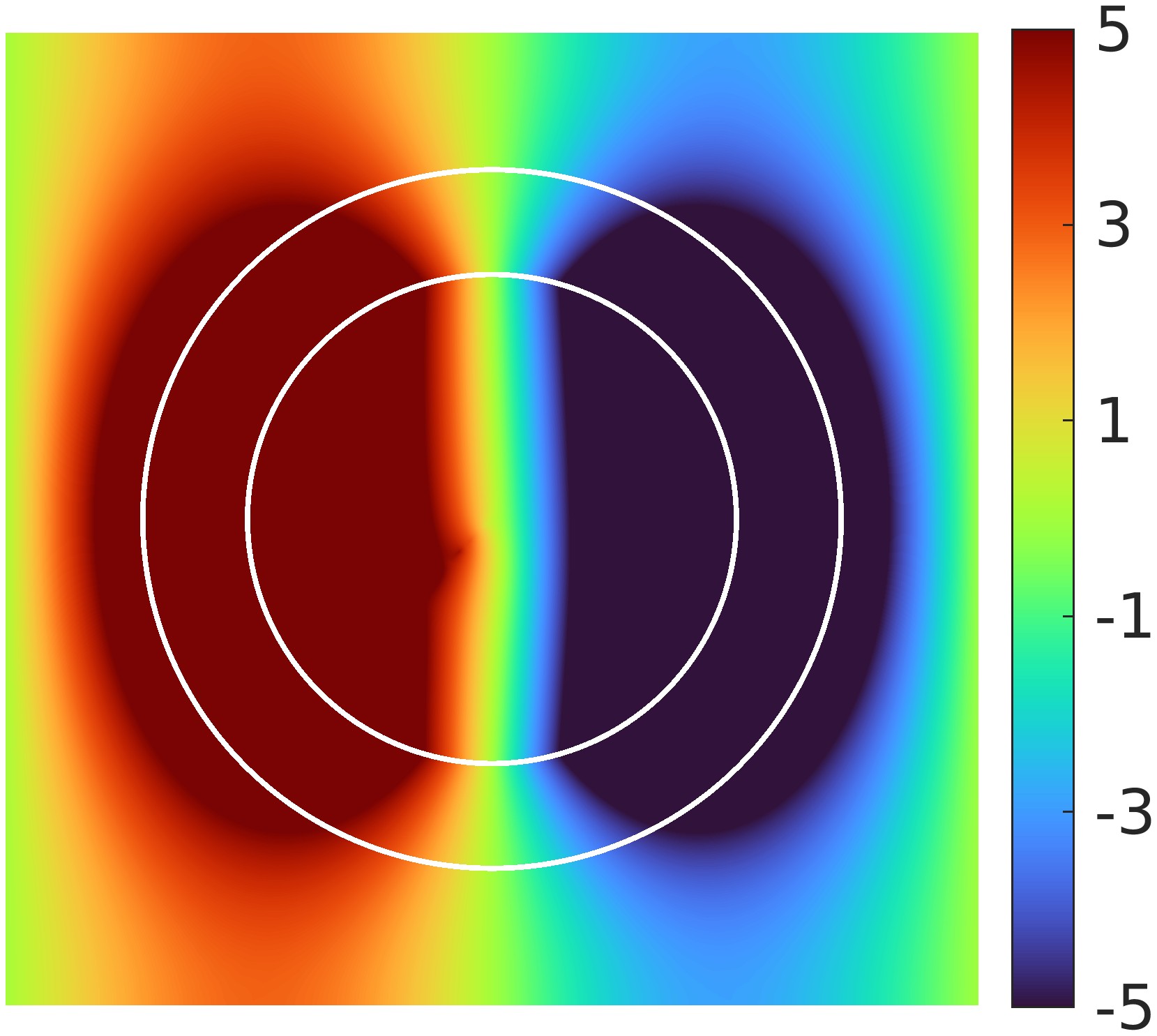}}   
    \end{center}
    \end{subfigure} 
    &
    \begin{subfigure}[t]{0.224
    \textwidth}\begin{center}{
\includegraphics[width=1\textwidth]{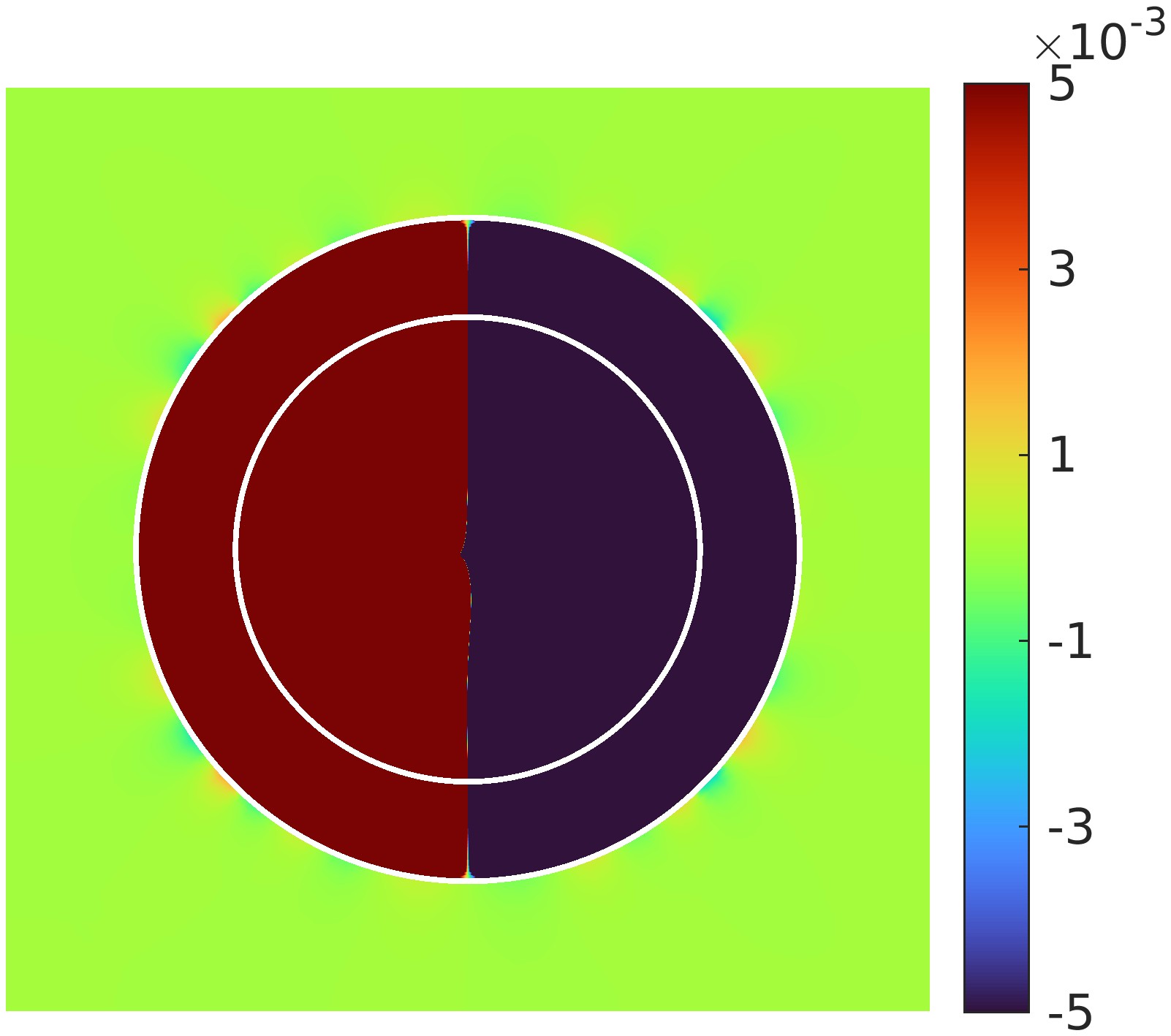}}      
    \end{center}
    \end{subfigure} 
    &
    \begin{subfigure}[t]{0.2204
    \textwidth}\begin{center}{
\includegraphics[width=1\textwidth]{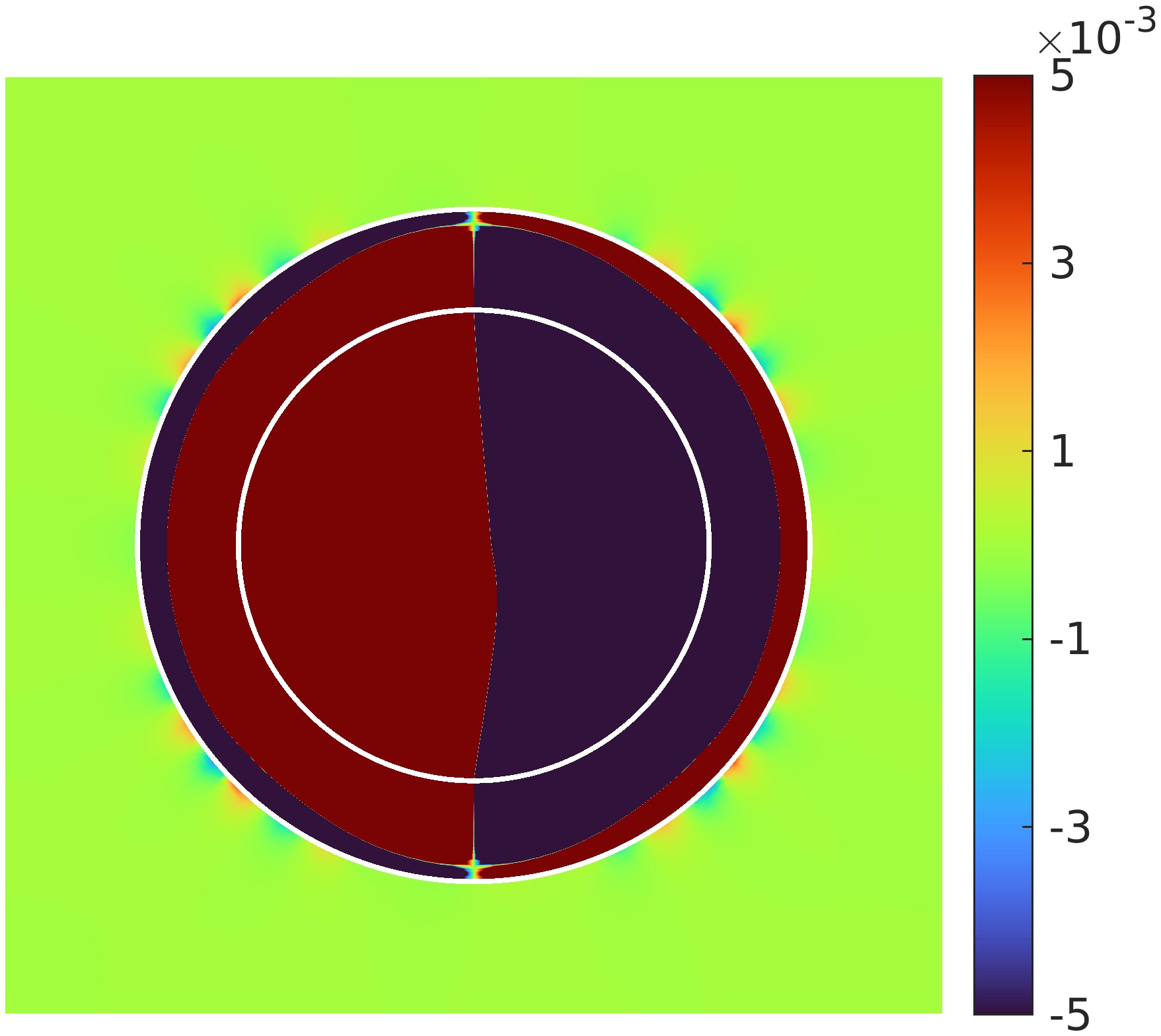}}       
    \end{center}
    \end{subfigure} 
    \\  
    \hline
    
\end{tabular}
}
\caption{Flux flow and temperature distribution $T$ for (first column) a homogeneous base material plate (reference case), (second column) a base material plate embedded with a circular insulator, and (third \& forth column) a base material plate embedded with a circular insulator and surrounding thermal cloak (optimized using EMT and Cu-Sn-Pb material models, $R_{\rm in}=35$~mm and $N_{\rm var}=25$). Temperature differences with respect to the reference case $T-\overline{T}$ are also presented. The thermal cloaks effectively diminish the temperature disturbance in $\mathrm{\Omega}_{\mathrm{out}}$. Temperature disturbances are almost nul in $\mathrm{\Omega}_{\mathrm{out}}$, even with the 3-4 order smaller temperature range.}
    \label{fig:chen2015case cloak tempDiff}
\end{figure}

 To showcase the cloaking performance of optimized cloaks, we present the flux flow, temperature profile and temperature difference with respect to a homogeneous base material plate in \fref{fig:chen2015case cloak tempDiff}. We exhibit the thermal cloaks obtained using EMT and Cu-Sn-Pb models for $N_{\rm var}=25$ and $R_{\rm in}=35$~mm. Here, we use a larger insulator to highlight the cloaking effect. We can witness substantial thermal cloaking in $\Omega_{\rm out}$, characterized by minimal temperature disruption, even within a temperature range that is 3-4  orders of magnitude smaller than the actual temperature range.  
\renewcommand{\arraystretch}{1.2}   
\par
\begin{figure}[!htbp]
\centering
\scalebox{1}{
\begin{tabular}[c]{| M{5.4em} | M{5.45em} | M{5.45em} | M{5.45em}| M{7.7em} |}
\hline 
 $v_i=0$ & $v_i=0.25$ & $v_i=0.5$ & $v_i=0.75$  & $v_i=1$ \\  
\hline
    \vspace{0.2cm}
    \begin{subfigure}[t]{0.15\textwidth}{\includegraphics[width=1\textwidth]{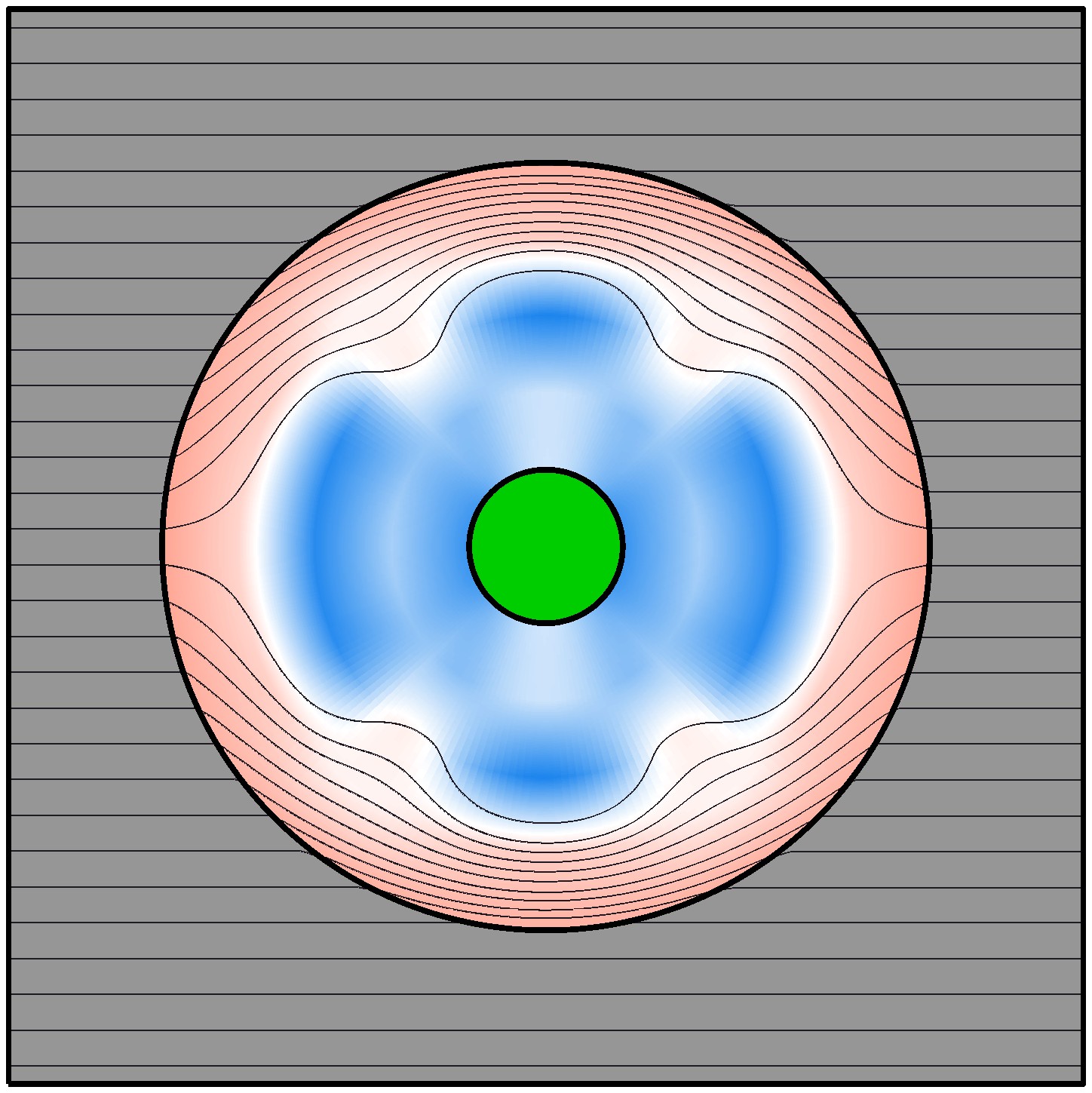}}
        \caption{\centering $N_{\rm var}=25$, $J=3.76\times 10^{-8}$}
    \end{subfigure}  & \vspace{0.2cm}
    \begin{subfigure}[t]{0.15\textwidth}{\includegraphics[width=1\textwidth]{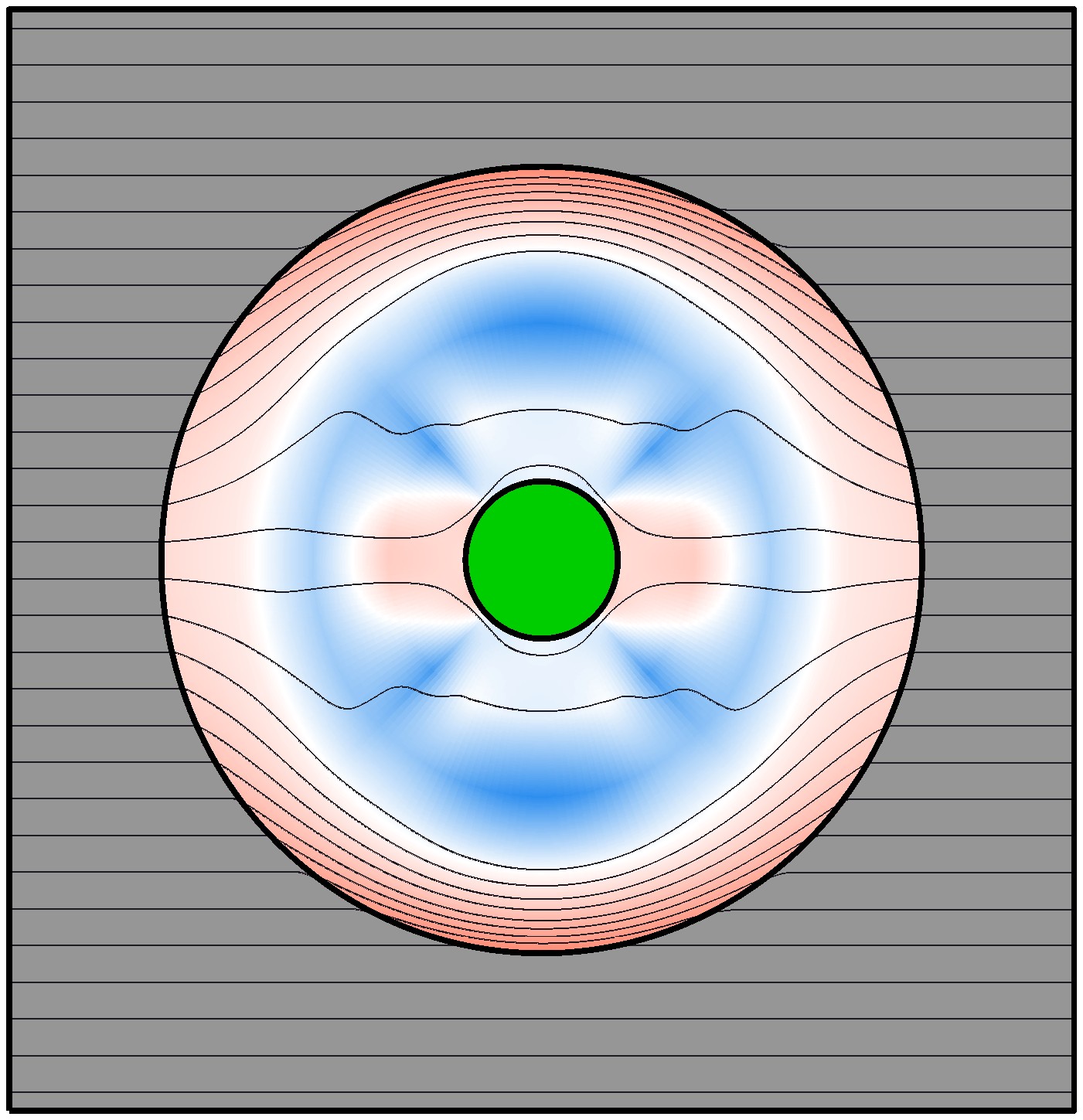}}
        \caption{\centering $N_{\rm var}=25$, $J=8.47\times 10^{-8}$}
    \end{subfigure} & \vspace{0.2cm}
    \begin{subfigure}[t]{0.15\textwidth}{\includegraphics[width=1\textwidth]{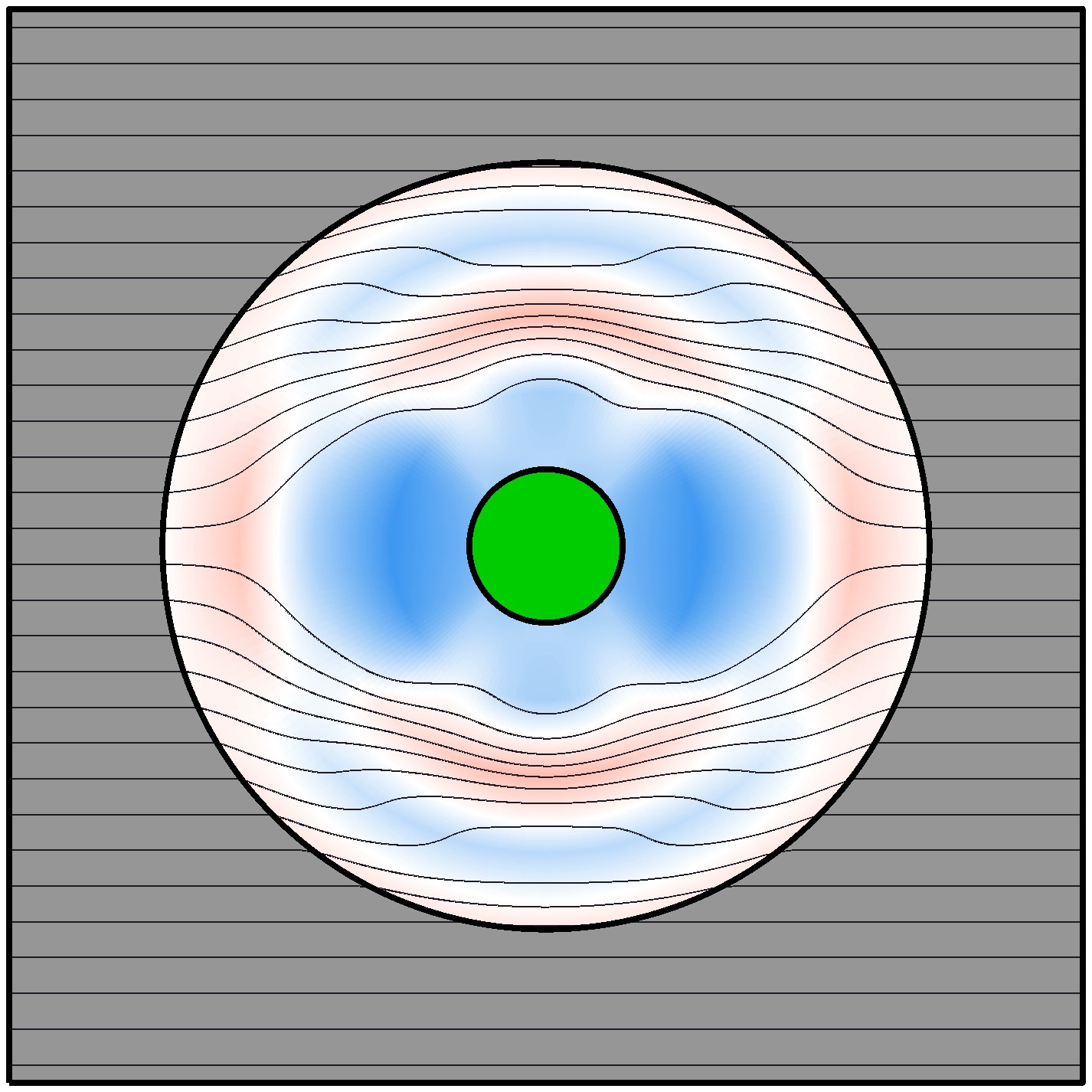}}
        \caption{\centering $N_{\rm var}=25$, $J= 8.74\times 10^{-9}$}
    \end{subfigure} & \vspace{0.2cm}
    \begin{subfigure}[t]{0.15\textwidth}{\includegraphics[width=1\textwidth]{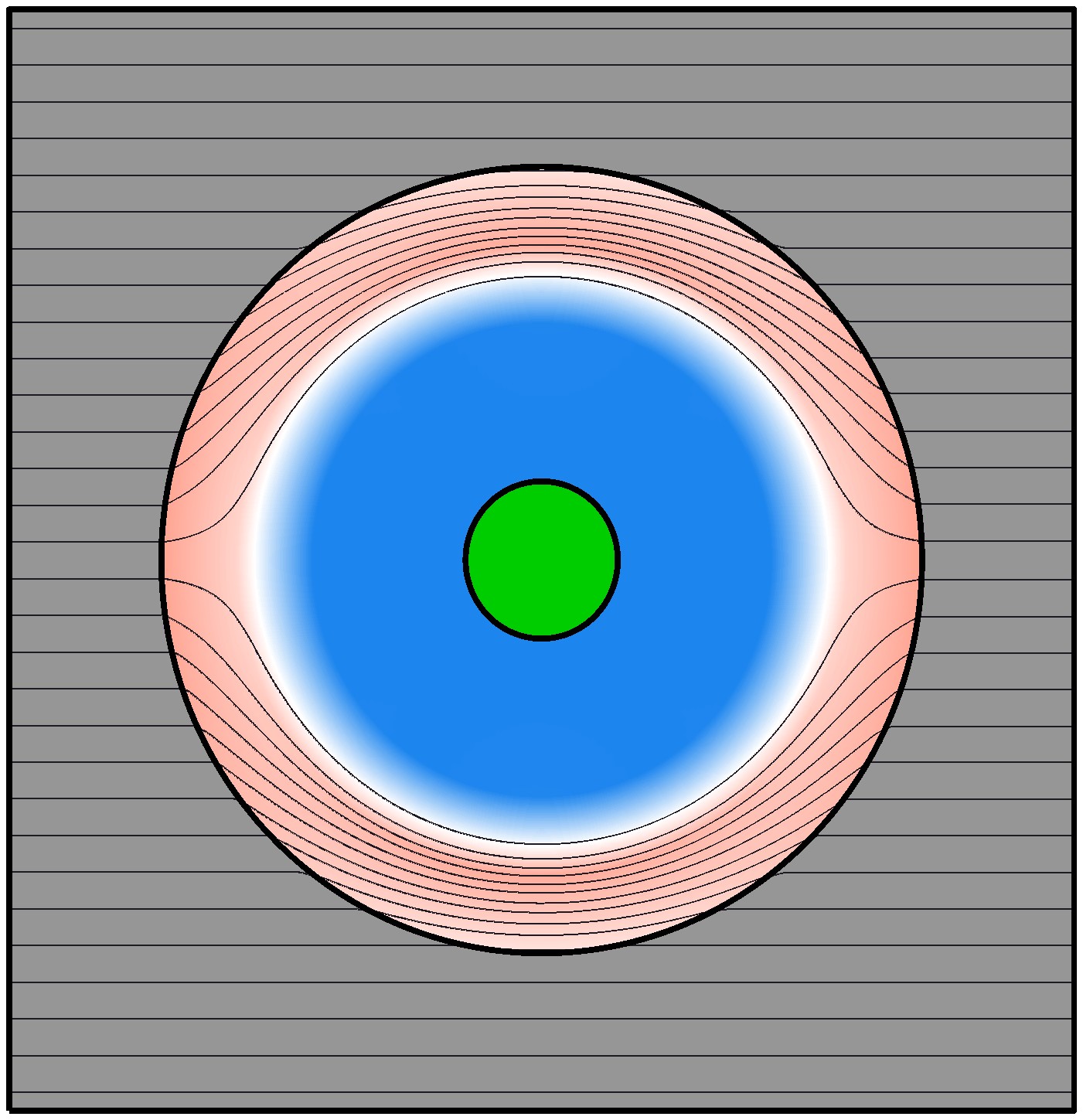}}
        \caption{\centering $N_{\rm var}=25$, $J=5.55\times 10^{-8}$}
    \end{subfigure}& \vspace{0.2cm}
    \begin{subfigure}[t]{0.15\textwidth}{\includegraphics[width=1\textwidth]{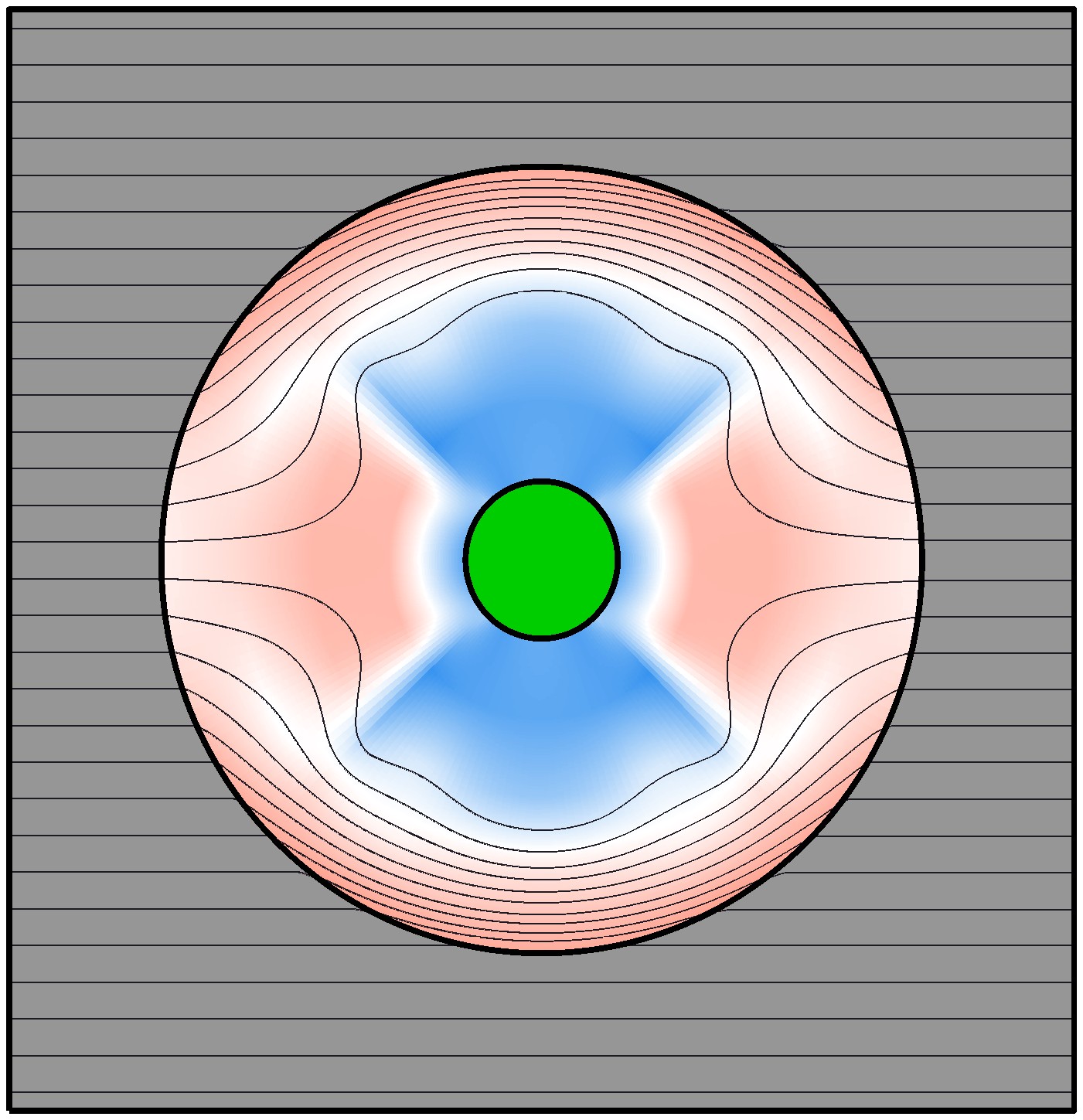}}
        \caption{\centering $N_{\rm var}=25$, $J=1.60\times 10^{-7}$}
    \end{subfigure}~\begin{subfigure}[b]{0.05\textwidth}{
\includegraphics[keepaspectratio=false,width=1.1\textwidth,height=2.45cm]{colorbar_VF2.jpg}}  
\end{subfigure}\\   
\hline
    \end{tabular}
}

\caption{Optimized material distributions for the thermal cloak problem with five different initial relative density distributions, $v_i=0,0.25,0.5,0.75,1$,~$i=1,2,...,N_{\rm var}$. EMT material model and $N_{\rm var}=25$ is considered. Optimized objective function values are of order $10^{-9}$-$10^{-11}$. Optimization results are dependent on the initial relative density distribution.}  
    \label{fig:chen2015case cloak initDistr}
\end{figure}

 As mentioned in the introduction, the thermal cloak problem falls under the Calder\'on tomography problem~\cite{calderon_inverse_1980,uhlmann_electrical_2009}, which has multiple solutions~\cite{Greenleaf2003nonuniqueness}. Due to this non-uniqueness, the optimization results are mesh as well as initial point-dependent. We tested five different values of $v_i$,~$i=1,2,...,N_{\rm var}$; $v_i=$ 0, 0.25, 0.5, 0.75, 1 for initial distributions for $N_{\rm var}=25$, $R_{\rm in}=10$~mm and EMT model. In \fref{fig:chen2015case cloak initDistr}, the optimization results clearly display the dependency on the initial distribution. This lack of well-posedness of the problem can be mitigated (but not avoided) by imposing regularizations/constraints.
\renewcommand{\arraystretch}{1.5}   
\subsubsection{Design with constraints}
\label{sec:2D cloak Design with constraints}
 \par A major limitation of conventional analytical methods is their inability to deal with free-form shapes, boundary conditions and design restrictions. This is one of the reasons why most thermal cloaks in the literature are designed for limited regular shapes (like circular~\cite{narayana2012heat,Han2014} and elliptical shapes~\cite{Han20181731}) and simple boundary conditions without any design constraints~\cite{narayana2012heat,Han2014,xu_ultrathin_2014,dai_transient_2018,xu_transformation_2020}. In the next few subsections, we explore the application of the proposed method to overcome above-mentioned limitations of the conventional methods. 
\par In order to make practical designs, the optimization often needs to follow design, manufacturing or material restrictions. This subsection focuses on the application of such restrictions on the thermal cloak design problem. We explore two schemes to apply restrictions, one via adding a regularization/penalty term in the primary objective function and another via including a constraint in the optimization problem. 
\par First, we consider the design requirement to have a large area covered by pure materials and only the most crucial areas occupied by the intermediate densities. To include this requirement, the objective function is augmented by an extra penalty term $J_{\rm Intpen}$ using the weight factor $\chi$. The penalty term $J_{\rm Intpen}$ penalizes the intermediate densities. The final objective function and the penalty term are presented as follows:
\begin{equation}
\label{eq:InterMediate Density penalization}
 J= J_{\rm cloak} + \chi J_{\mathrm{Intpen}} \quad \text{with}\quad
J_{\rm Intpen}= \int_{\Omega_{\rm design}} v^4(1-v)^4~d\Omega.
\end{equation}
By the weighting factor $\chi$, we delay the impact of the penalty term to secure a sufficient value of the primary objective function before the effect of penalization starts. Also, as mentioned in \sref{sec:Optimization problem description}, finding the derivative of $J_{\rm Intpen}$ with respect to design variables will be straightforward using \eref{eq:Density gradient}, owing to the NURBS parameterization for the density field. 

\renewcommand{\arraystretch}{1.5}   
\begin{figure}[!htbp]
\centering
\scalebox{1}{
\begin{tabular}[c]{| M{7.1em} | M{7.1em} | M{7.1em}| M{8.8em}|}
\hline 
  w/o reg.& $\chi=0.01$ 
  & $\chi =1$ 
  & $\chi =100$ \\  
\hline
    \vspace{0.2cm}
    \begin{subfigure}[t]{0.19\textwidth}{\includegraphics[width=1\textwidth]{FGM_objT_2_dVar25_HomMod_3_fluxPlot.jpg}}
        \caption{\centering $J=7.35\times 10^{-11}$, $J_{\rm cloak}=7.35\times 10^{-11}$.}
    \end{subfigure} & \vspace{0.2cm}
    \begin{subfigure}[t]{0.19\textwidth}{\includegraphics[width=1\textwidth]{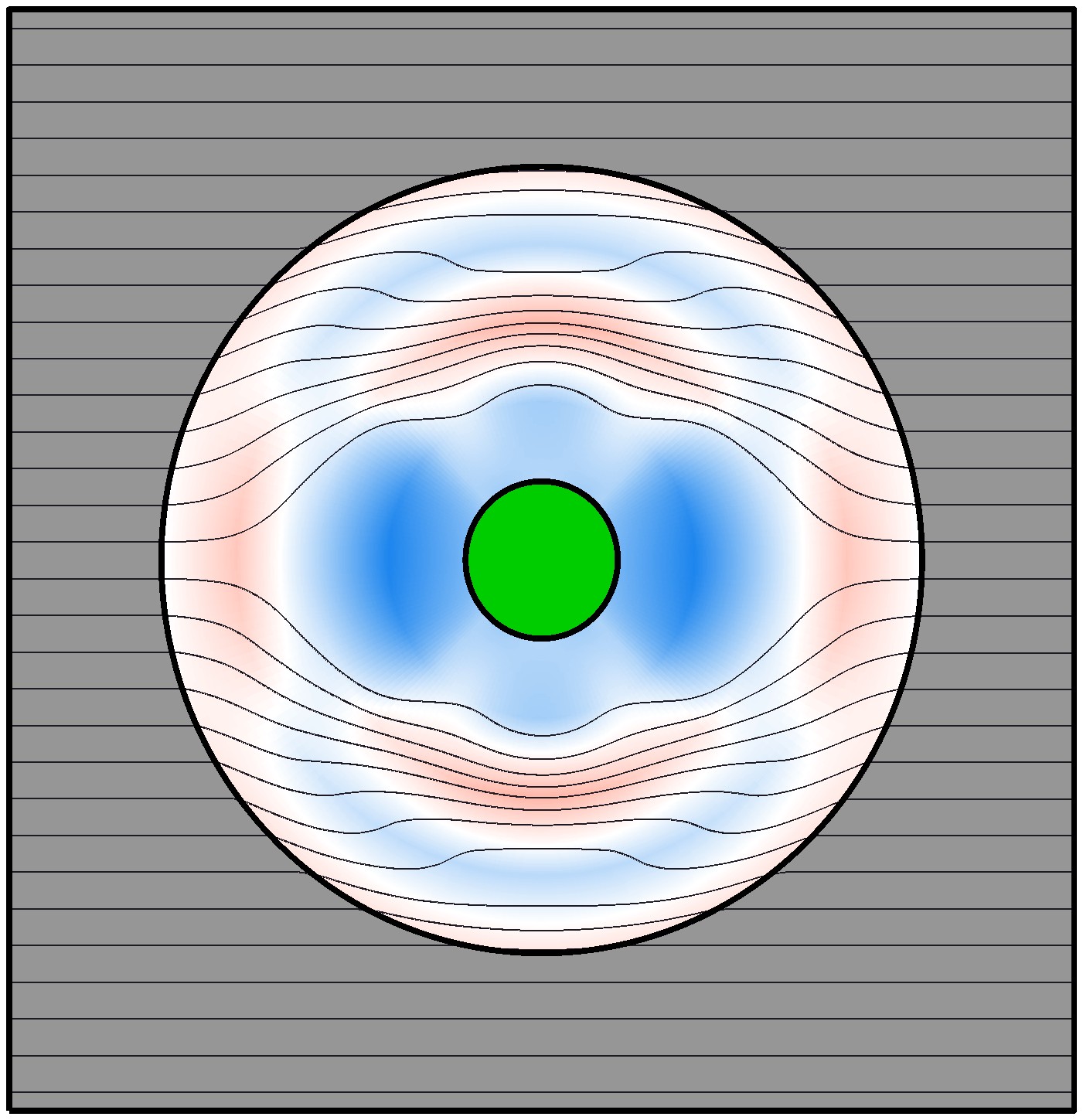}}
        \caption{\centering $J=2.94\times 10^{-7}$, $J_{\rm cloak}=6.05\times 10^{-8}$, $J_{\rm Intpen}=2.33\times 10^{-7}$.}
    \end{subfigure} & \vspace{0.2cm}
    \begin{subfigure}[t]{0.19\textwidth}{\includegraphics[width=1\textwidth]{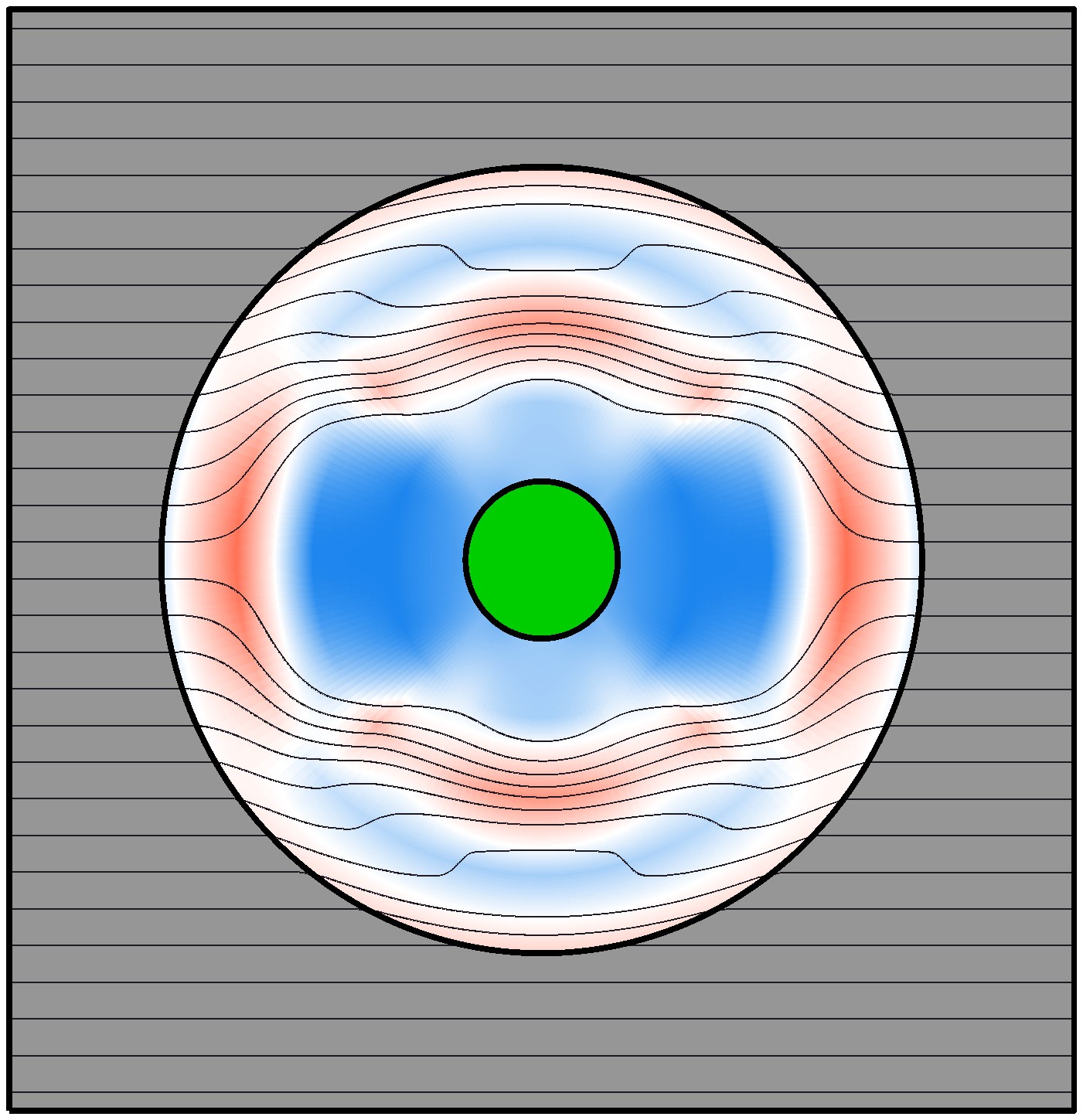}}
        \caption{\centering $J=1.97\times 10^{-5}$, $J_{\rm cloak}=2.56\times 10^{-7}$, $J_{\rm Intpen}=1.95\times 10^{-5}$.}
    \end{subfigure}& \vspace{0.2cm}
    \begin{subfigure}[t]{0.19\textwidth}{\includegraphics[width=1\textwidth]{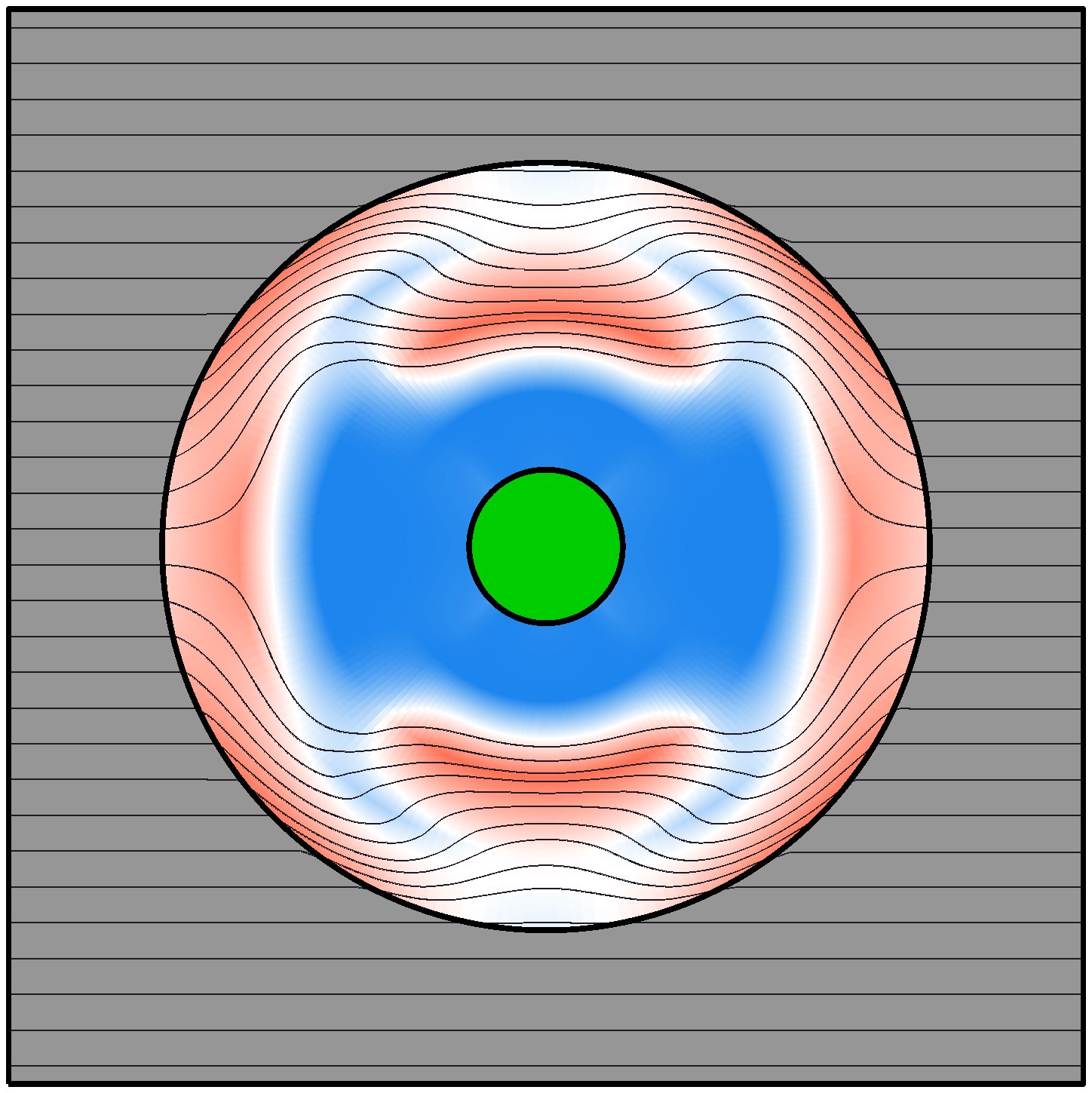}}
        \caption{\centering $J=1.24\times 10^{-3}$, $J_{\rm cloak}=1.27\times 10^{-5}$, $J_{\rm Intpen}=1.23\times 10^{-3}$.}
    \end{subfigure}~\begin{subfigure}[b]{0.05\textwidth}{
\includegraphics[keepaspectratio=false,width=1\textwidth,height=2.95cm]{colorbar_VF2.jpg}}  
\end{subfigure}\\   
\hline
\end{tabular}
}

\caption{(Columns 2-4) Optimized material distributions for the thermal cloak problem with the intermediate density penalization. EMT material model and $N_{\rm var}=25$ are considered. Three values of $\chi$, $\chi=~0.01$, $0.1$, $1$, are tested. Optimized objective function value are of order $10^{-4}-10^{-5}$ with $J_{\rm cloak}$ of order $10^{-7}$ and $J_{\rm Intpen}$ of order $10^{-4}-10^{-5}$. The left column represents the optimized material distribution of the cloak without penalization. The penalization effectively reduces the area with intermediate densities with a slight compromise in the main cloaking objective.}  
    \label{fig:chen2015case cloak wCnstr}
\end{figure}
\par In \fref{fig:chen2015case cloak wCnstr}, we have shown the results with penalization for the EMT model with $N_{\rm var}=25$. We consider three values of $\chi$, $\chi=10^{-2}$, $1$ and $100$. We can see that, by increasing the value of $\chi$, the optimized results move towards the material distributions with large areas of pure constituent materials. The penalization comes with a slight compromise in the main cloaking objective. Yet, all designs shown in \fref{fig:chen2015case cloak wCnstr} reach $J_{\rm cloak}$-value in the range of $10^{-5}$, showcasing satisfactory cloaking function. With a very large value of $\chi$, the method produces almost binary 0-1 type designs. This behaviour is similar to the SIMP interpolation scheme. The distinction, however, lies in the penalization, which comes from the extra objective function term in contrast to the material law as in the SIMP. A prediction of an appropriate value of $\chi$ is very difficult apriori and needs a trial-and-error study. Another point worth mentioning is one can use any power to the density terms in the definition of $J_{\rm Intpen}$. The power will regulate how sharply the effect of penalization changes with $\chi$.  
\par Secondly, we consider a localized maximum temperature constraint in the optimization problem. This constraint can be included when there is a requirement to maintain a specific maximum temperature in a local region of the domain. This requirement could attributed to the material or working environment limitations. To apply the maximum temperature numerically, we need a formula to approximate the maximum temperature, $\tau_{\rm max}$. Here, we exploit the approximation utilized in~\cite{Wen1989new,zhuang_temperature-constrained_2021} in a continuum framework, and the corresponding constraint is defined as:
\begin{equation}
    \tau_{\rm max}(\varphi)\leq T_{\rm max} \quad \text{with} \quad  \tau_{\rm max}=\dfrac{\int_{\Omega} T A^T H(\varphi) ~d\Omega}{\int_{\Omega}A^T H(\varphi) ~d\Omega},
\end{equation}
where $T_{\rm max}$ is the upper limit of the permissible temperature, $A$ is the constant that makes $\tau_{\rm max}$ the maximum approximate temperature as $A^T \rightarrow +\infty$ (In this example, we take $A=1.5$), $\varphi$ is the signed function to represent the temperature constrained region and $H(\varphi)$ is the Heaviside function.

 \begin{figure}[!htbp]
    \centering
    \setlength\figureheight{1\textwidth}
    \setlength\figurewidth{1\textwidth}
    \begin{subfigure}[b]{0.30\textwidth}{\centering\includegraphics[width=1\textwidth]{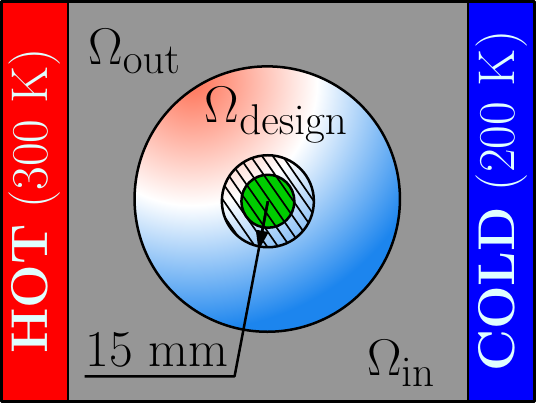}}
        \caption{\centering Problem domain}
        \label{fig:chen2015case cloak wCnstr 2 a}
    \end{subfigure} \\ 
    \begin{subfigure}[t]{0.28\textwidth}{\centering\includegraphics[width=1\textwidth]{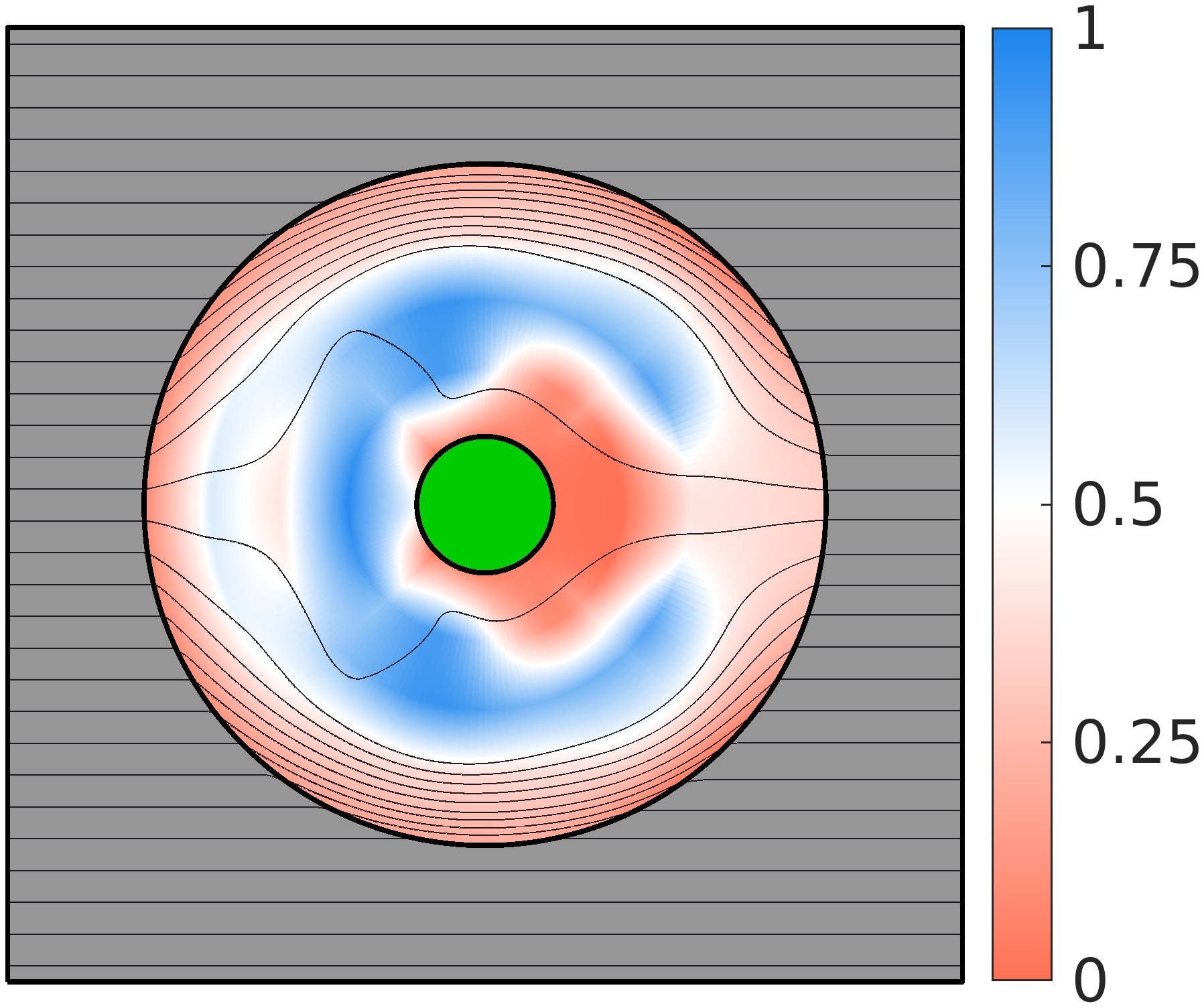}}
             \caption{\centering Optimized material distribution}
             \label{fig:chen2015case cloak wCnstr 2 b}
    \end{subfigure} \quad
    \begin{subfigure}[t]{0.28\textwidth}{\centering\includegraphics[width=1\textwidth]{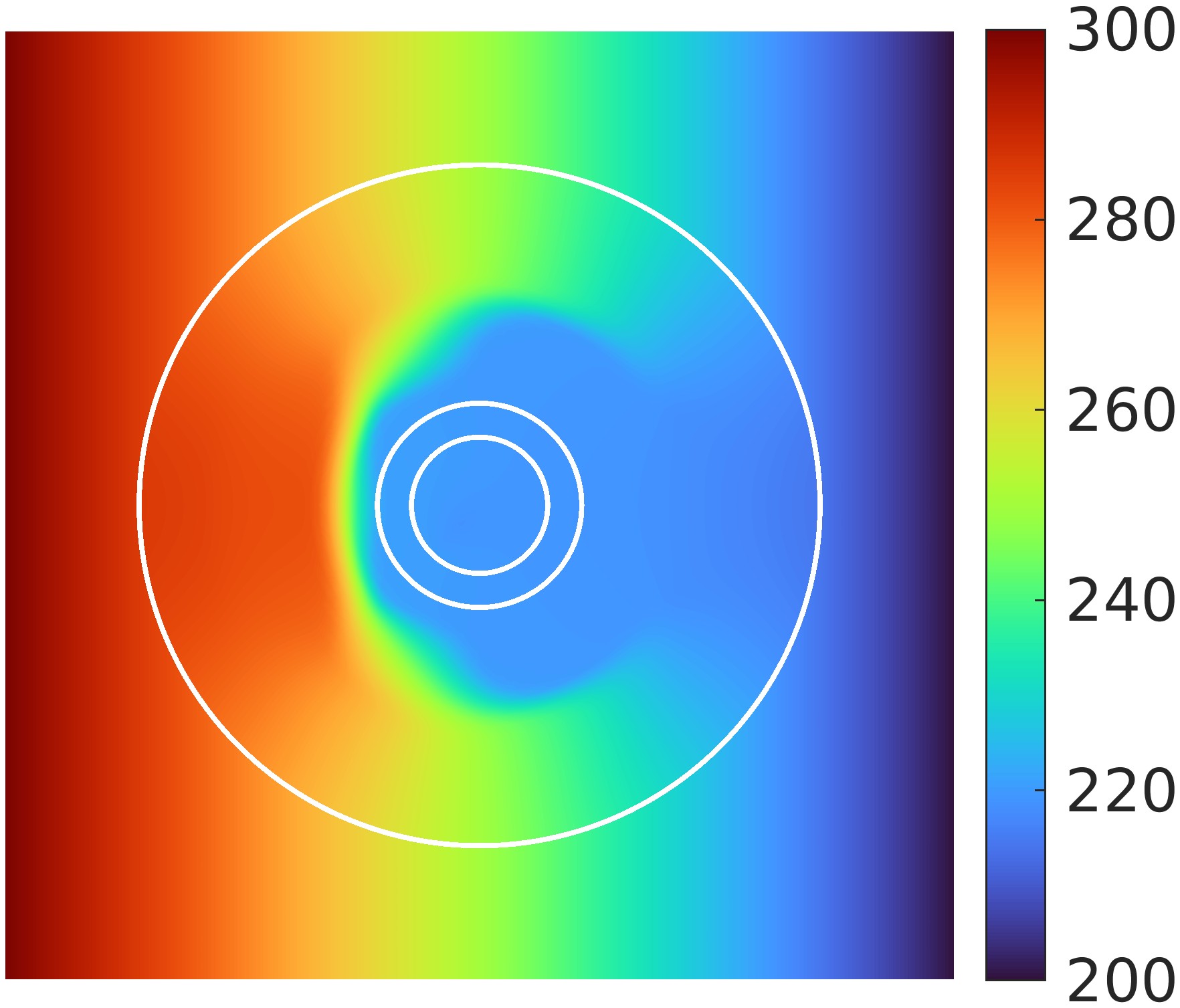}}
             \caption{\centering Temperature $T$ distribution}
             \label{fig:chen2015case cloak wCnstr 2 c}
    \end{subfigure} \quad
    \begin{subfigure}[t]{0.28\textwidth}{\centering\includegraphics[width=1\textwidth]{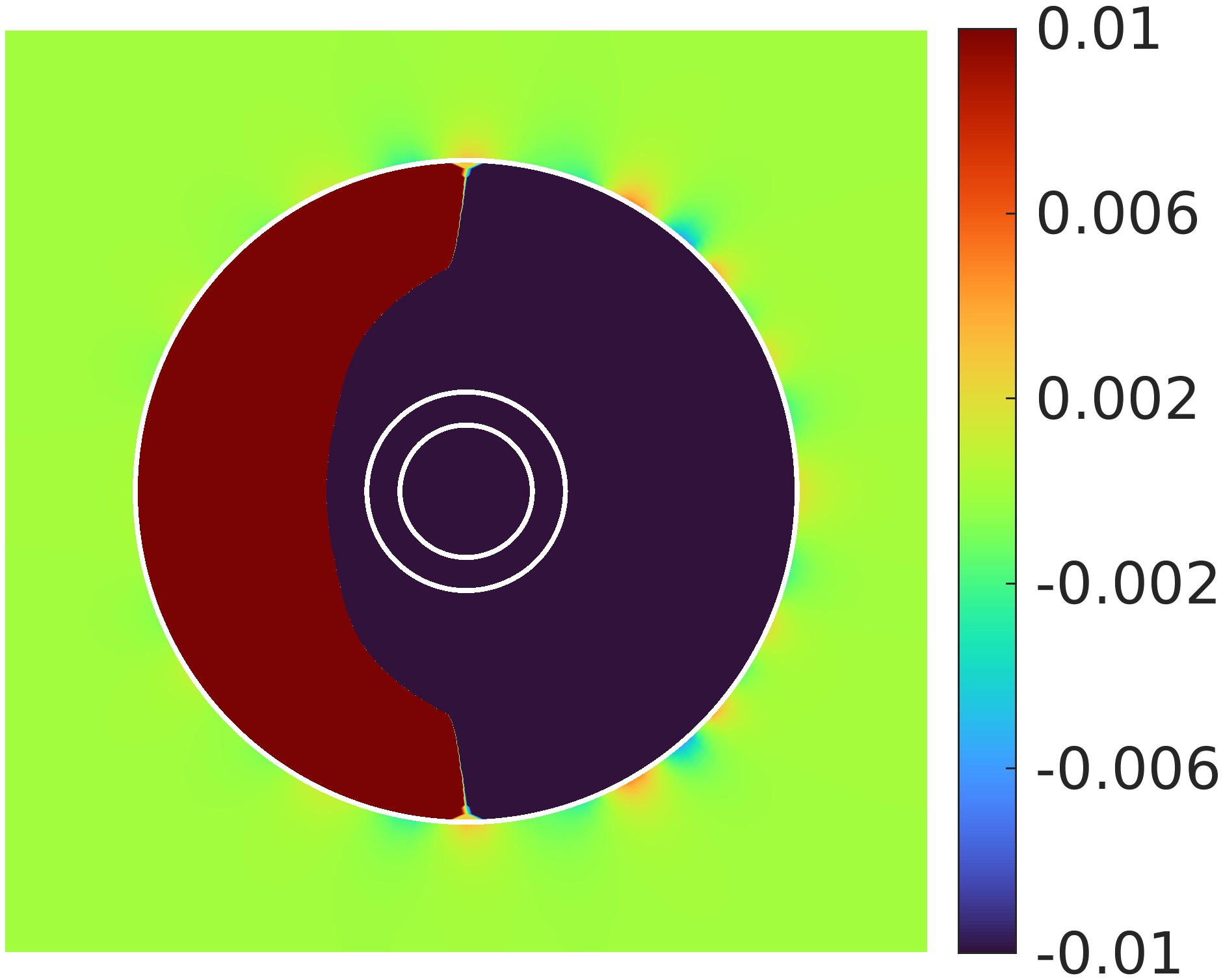}}
             \caption{\centering Temp difference $T-\overline{T}$}
             \label{fig:chen2015case cloak wCnstr 2 d}
    \end{subfigure}
 \caption{Problem domain, optimized material distribution, temperature $T$ distribution and temperature difference (with respect to the reference case) $T-\overline{T}$ for the thermal cloak problem with a localized maximum temperature constraint. EMT material model, $N_{\rm var}=50$ and $T_{\rm max}=220$~K are considered. The hatched area presents the region where the constraint is applied. Optimized material distribution maintains the temperature in the hatched area below $T_{\rm max}=220$~K while cloaking the insulator with the objective function value $J=1.68 \times 10^{-6}$.}
 \label{fig:chen2015case cloak wCnstr 2}
\end{figure}

\par For our design problem, we want to keep the temperature in a circular region with radius~15~mm at the center (covering the insulator and a small surrounding region) lower than or equal to $T_{\rm max}=220$~K. The local region under constraint is shown as the hatched area in \fref{fig:chen2015case cloak wCnstr 2 a}. The optimization problem with the constraint is not symmetric along $y$-axis, therefore we remove the symmetry condition along $y$-axis for the design variables. In order to solve the optimization problem, the optimization algorithm also necessitates the sensitivities of the constraint function with respect to the design variables. To calculate these constraint sensitivities, an extra adjoint problem is solved as presented in \erefs{eq:constraint fn sensitivity b}-(\ref{eq:adjoint eq. matrix form b}) at each optimization iteration. The optimization results with the constraint for the EMT model with $N_{\rm var}=50$ are presented in \frefs{fig:chen2015case cloak wCnstr 2 b}-\ref{fig:chen2015case cloak wCnstr 2 d}. From the figure, we can observe that the temperature in the hatched region is lower than or equal to $T_{\rm max}$. Also, from \fref{fig:chen2015case cloak wCnstr 2 d}, it is evident that the satisfactory cloaking function is achieved with $J=1.68 \times 10^{-6}$. 
\par Both examples demonstrate how effortlessly the proposed method includes constraints in the formulation, which could be regarded as the primary benefit of the the proposed method from the manufacturing point of view. Though two specific types of constraints are presented in the subsection, the method is rather general and other constraints can be applied with equal effectiveness.  
\subsubsection{Design with free-form geometries}
\label{sec:2D cloak Design with free-form geometries}
\begin{figure}[!htbp]
    \centering
    \includegraphics[width=1\linewidth]{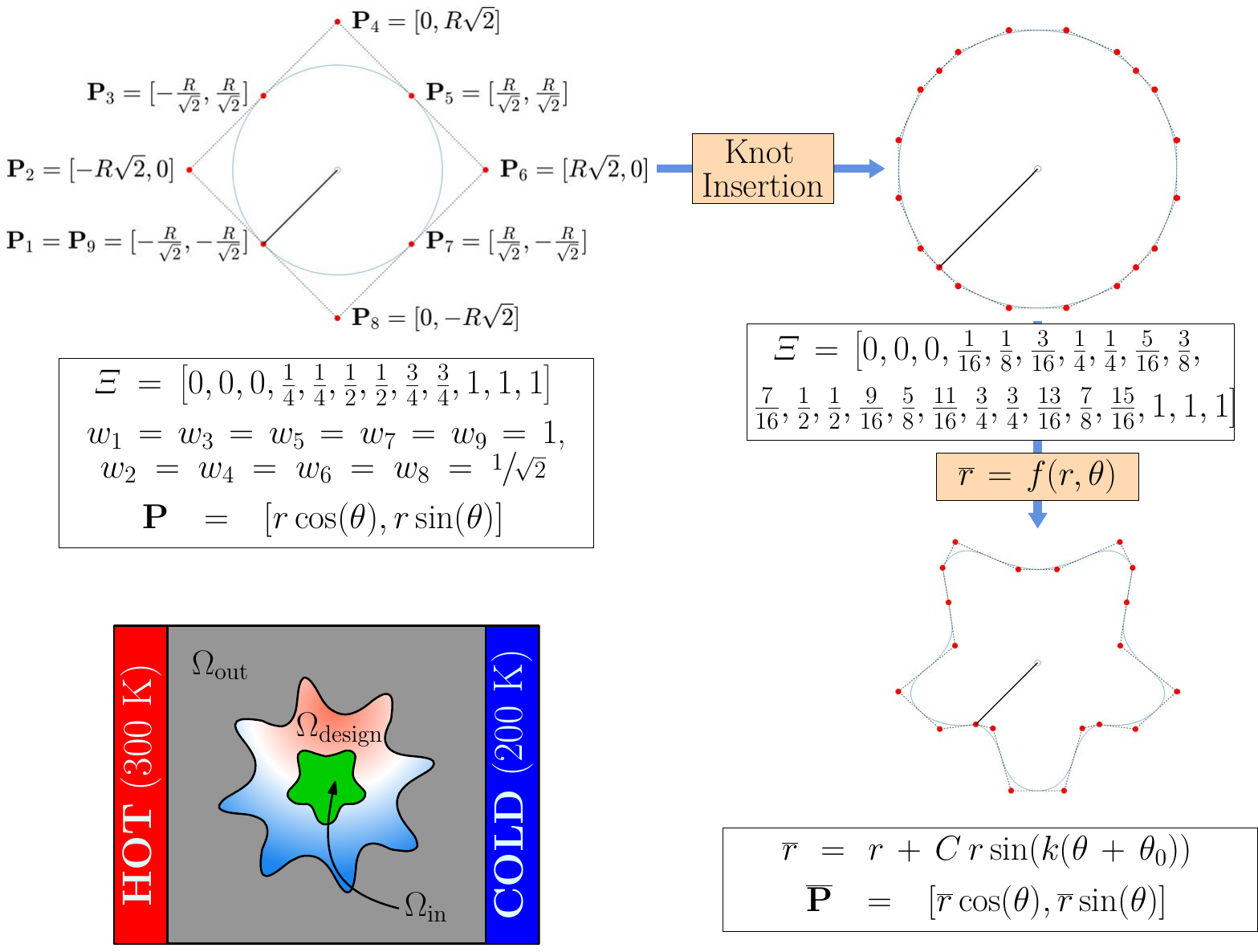}
 \caption{Steps to generate the star-shaped insulator \& thermal cloak and schematics of the final domain. The control points related to the circumferential parametric directions of the circles of radius $R_{\rm in}$ and $R_{\rm out}$ are perturbed. The steps for perturbation are as follows: (i) A NURBS-circle is created using a knot vector $\Xi=[0,0,0,\sfrac{1}{4},\sfrac{1}{4},\sfrac{1}{2},\sfrac{1}{2},\sfrac{3}{4},\sfrac{3}{4},1,1,1]$ with nine control points $\mathbf{P}_i$, $i=1,2,...,9$. The Cartesian coordinates and weights of the control points are shown in the first figure. (ii) The NURBS-circle is refined by adding knots $\sfrac{1}{16},\sfrac{1}{8},\sfrac{3}{16},\sfrac{5}{16},\sfrac{3}{8},\sfrac{7}{16},\sfrac{9}{16},\sfrac{5}{8},\sfrac{11}{16},\sfrac{13}{16},\sfrac{7}{8},\sfrac{15}{16}$ through knot insertion procedure. (iii) The control points of the refined circle are transformed into polar coordinates $(r,\theta)$ from Cartesian coordinates $\mathbf{P}=(x,y)$; the radial coordinates $r$ are perturbed by the function, $\overline{r}=r+Cr \sin{(k(\theta+\theta_0))}$, while keeping $\theta$ coordinates unchanged; the modified polar coordinates $(\overline{r},\theta)$ are transformed back into Cartesian coordinates $\overline{\mathbf{P}}=(\overline{x},\overline{y})$. For $R_{\rm in}=15$~mm, $C=0.3$, $k=5$ and $\theta_0=\pi$, while for $R_{\rm out}=40$~mm,  $C=0.4$, $k=8$ and $\theta_0=-\sfrac{\pi}{2}$.}
 \label{fig:Chen2015case_geo_creation}
\end{figure}
\par In this subsection, we study the effectiveness of our method for the star-shaped insulator and thermal cloak. To create both star-shaped geometries, we perturb the control points of NURBS-based circles of radius $R_{\rm in}$ and $R_{\rm out}$. For the perturbation, we perform knot insertion, coordinate-system transformations (from Cartesian to polar \& from polar to Cartesian) and functional transformation of radial coordinates. The detailed procedure and the final domain are presented in \fref{fig:Chen2015case_geo_creation}. For ${R_{\rm in}=15}$~mm, we use $C=0.3$, $k=5$ and $\theta_0=\pi$, while for $R_{\rm out}=40$~mm, we use $C=0.4$, $k=8$ and $\theta_0=\sfrac{-\pi}{2}$. Other details are kept the same as the original cloak problem. 
\par
\begin{figure}[!htbp]
\centering
\scalebox{1}{
\begin{tabular}[c]{| M{8.8em} | M{8.8em} | M{8.8em} |}
\hline
\multicolumn{1}{|c|}{\centering EMT}  & \multicolumn{1}{c|}{\centering Porous Cu} & \multicolumn{1}{c|}{\centering Gyroid} \\
\hline
\vspace{0.2cm}
    \begin{subfigure}[t]{0.25\textwidth}{\centering\includegraphics[width=1\textwidth]{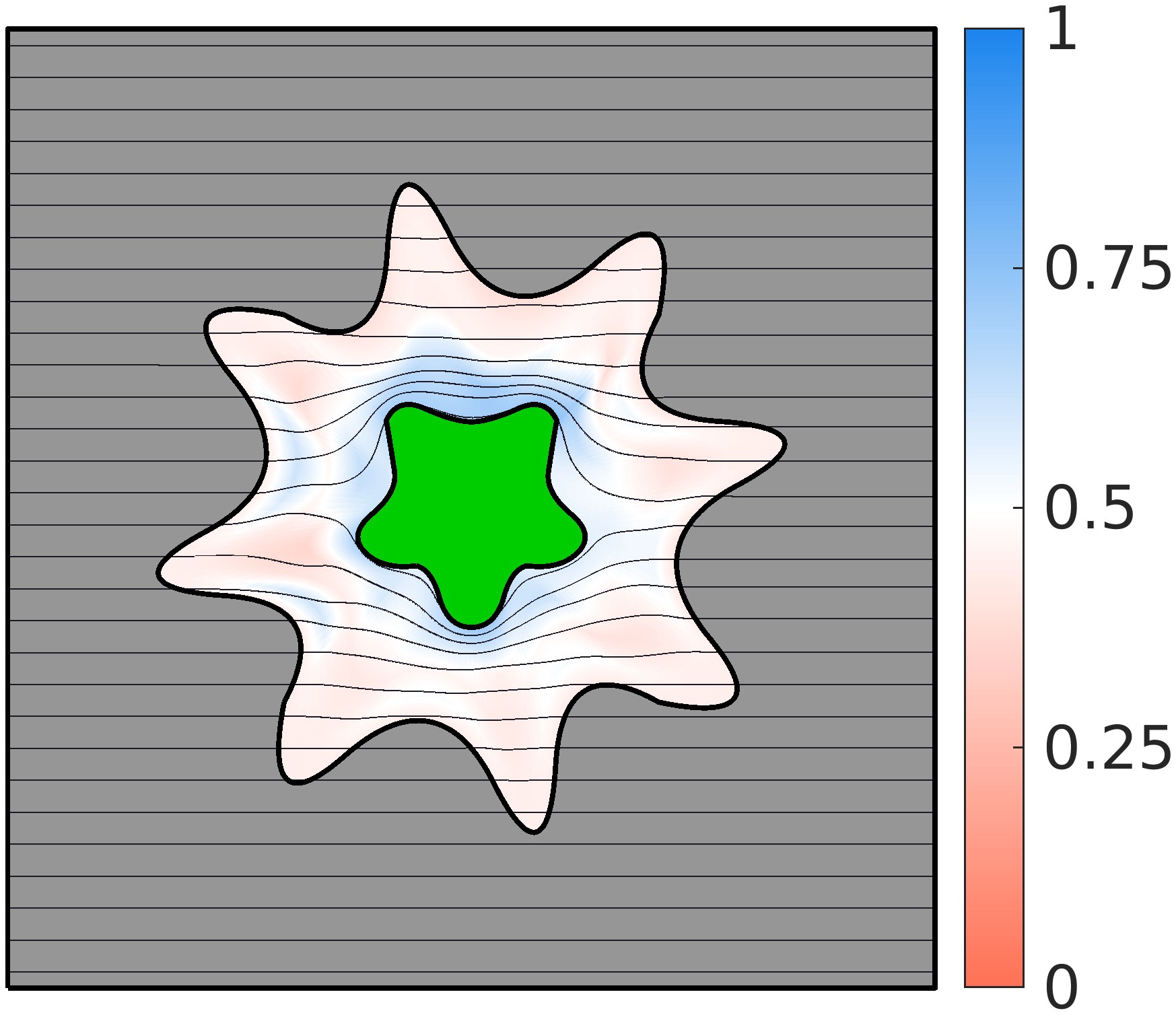}}
        \caption{\centering Optimized material distribution}
    \end{subfigure}& \vspace{0.2cm}
    \begin{subfigure}[t]{0.25\textwidth}{\centering\includegraphics[width=1\textwidth]{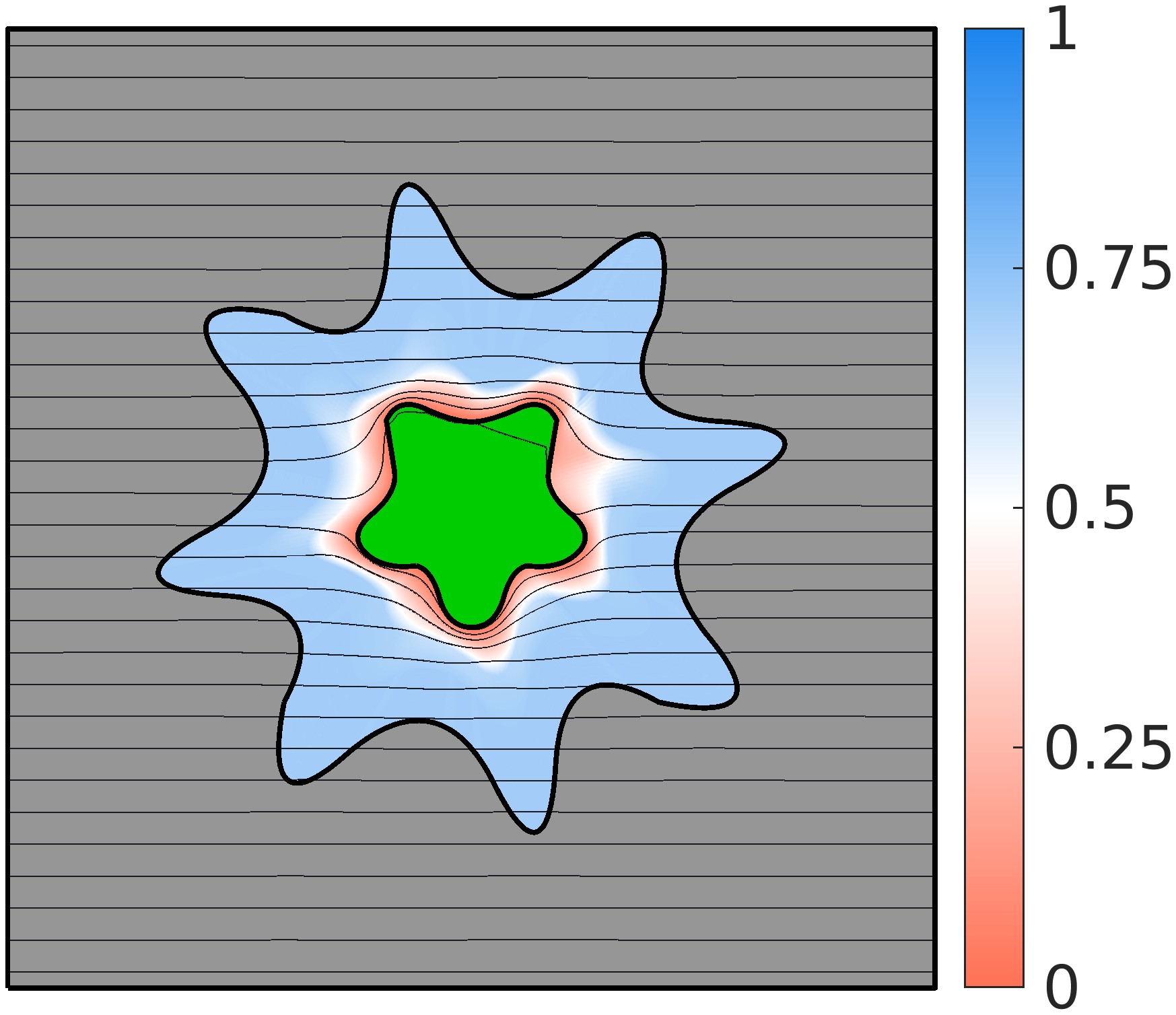}}
         \caption{\centering Optimized material distribution}
    \end{subfigure}
    & \vspace{0.2cm}
    \begin{subfigure}[t]{0.25\textwidth}{\centering\includegraphics[width=1\textwidth]{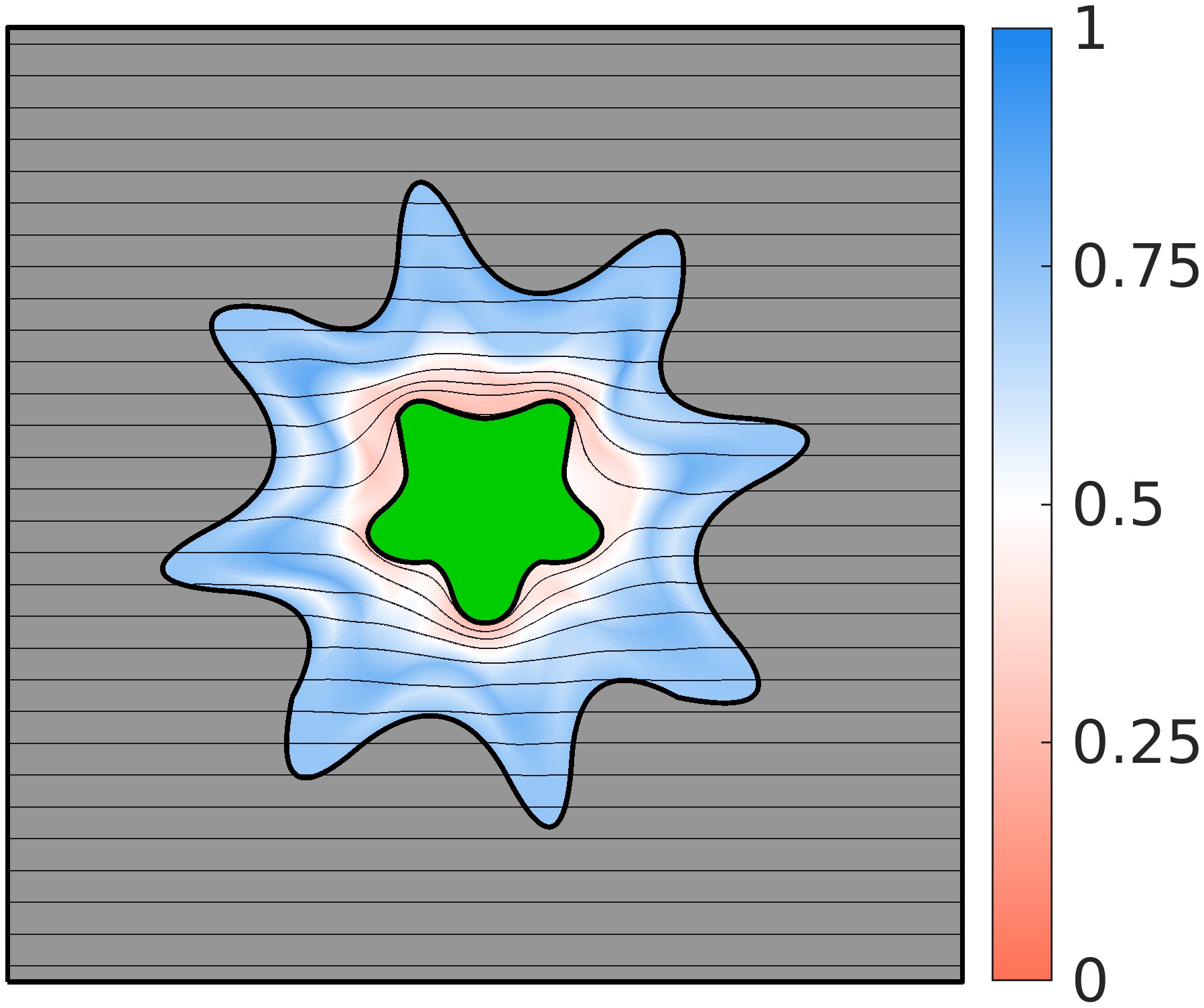}}
        \caption{\centering Optimized material distribution}
    \end{subfigure}\\ \vspace{0.2cm}
    \begin{subfigure}[t]{0.25\textwidth}{\centering\includegraphics[width=1\textwidth]{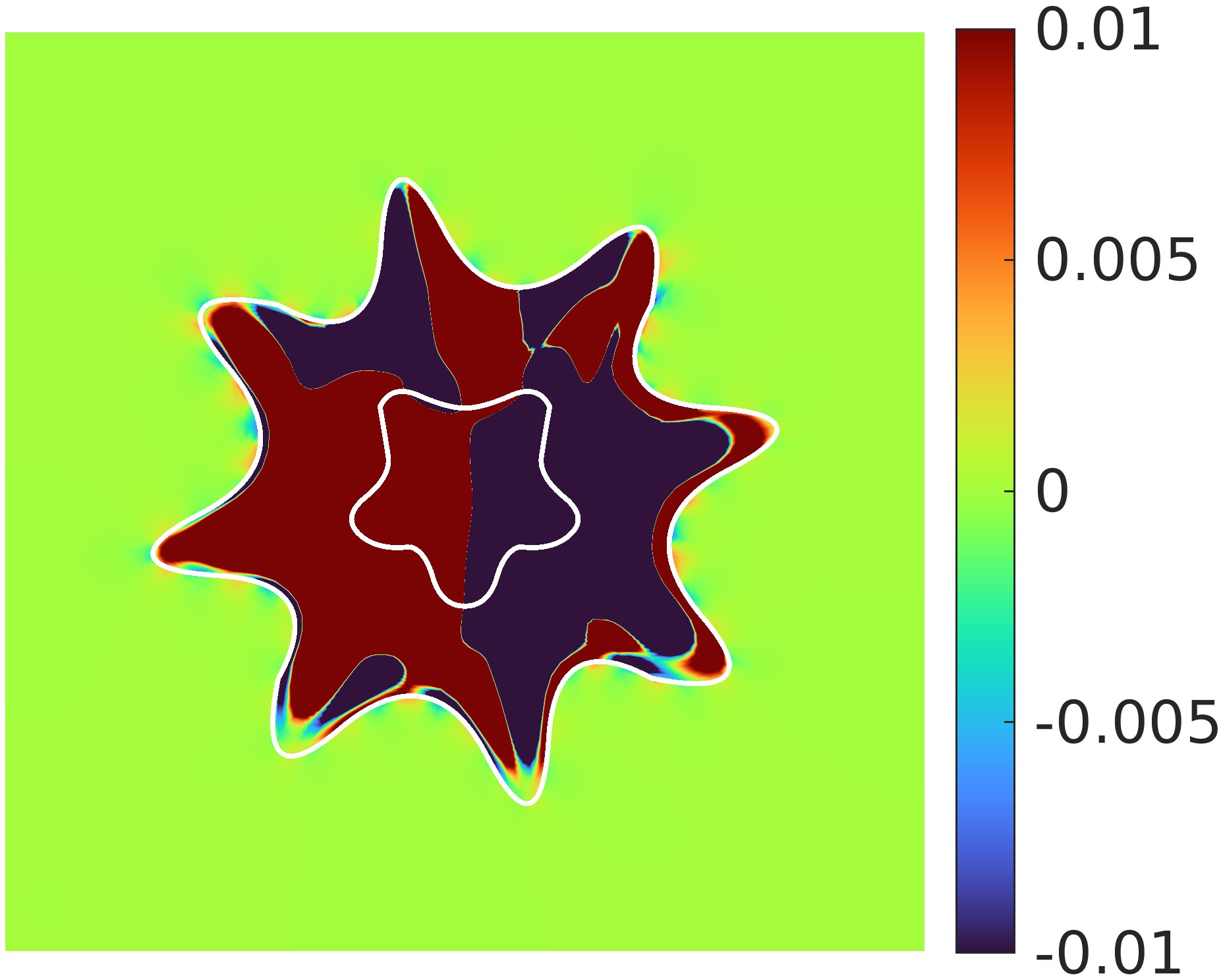}}
        \caption{\centering $T-\overline{T}$,\linebreak $J_{\rm cloak}=1.26\times 10^{-7}$}
    \end{subfigure}& \vspace{0.2cm}
    \begin{subfigure}[t]{0.25\textwidth}{\centering\includegraphics[width=1\textwidth]{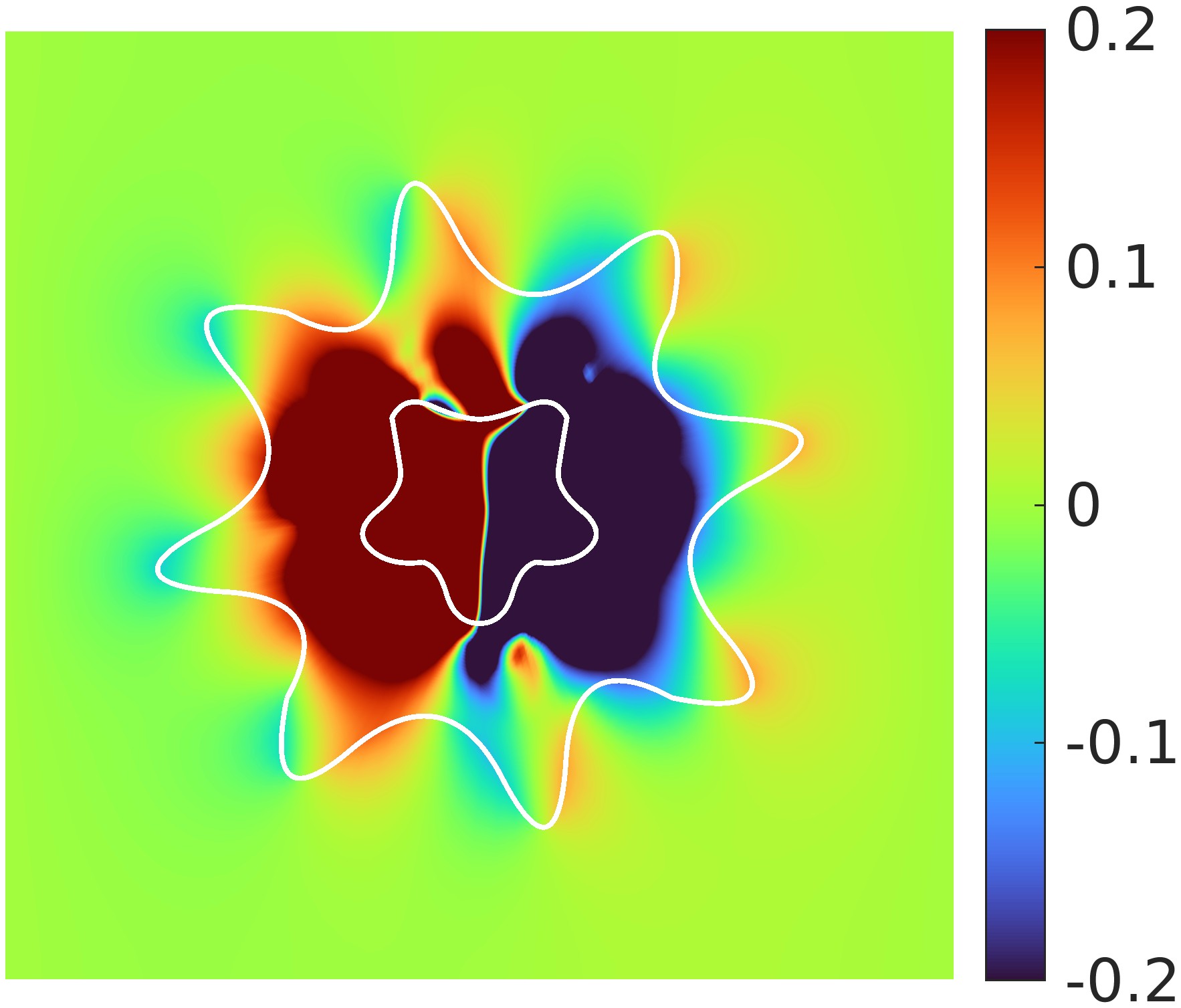}}
       \caption{\centering $T-\overline{T}$,\linebreak $J_{\rm cloak}=3.49\times 10^{-4}$}
    \end{subfigure}
    & \vspace{0.2cm}
    \begin{subfigure}[t]{0.25\textwidth}{\centering\includegraphics[width=1\textwidth]{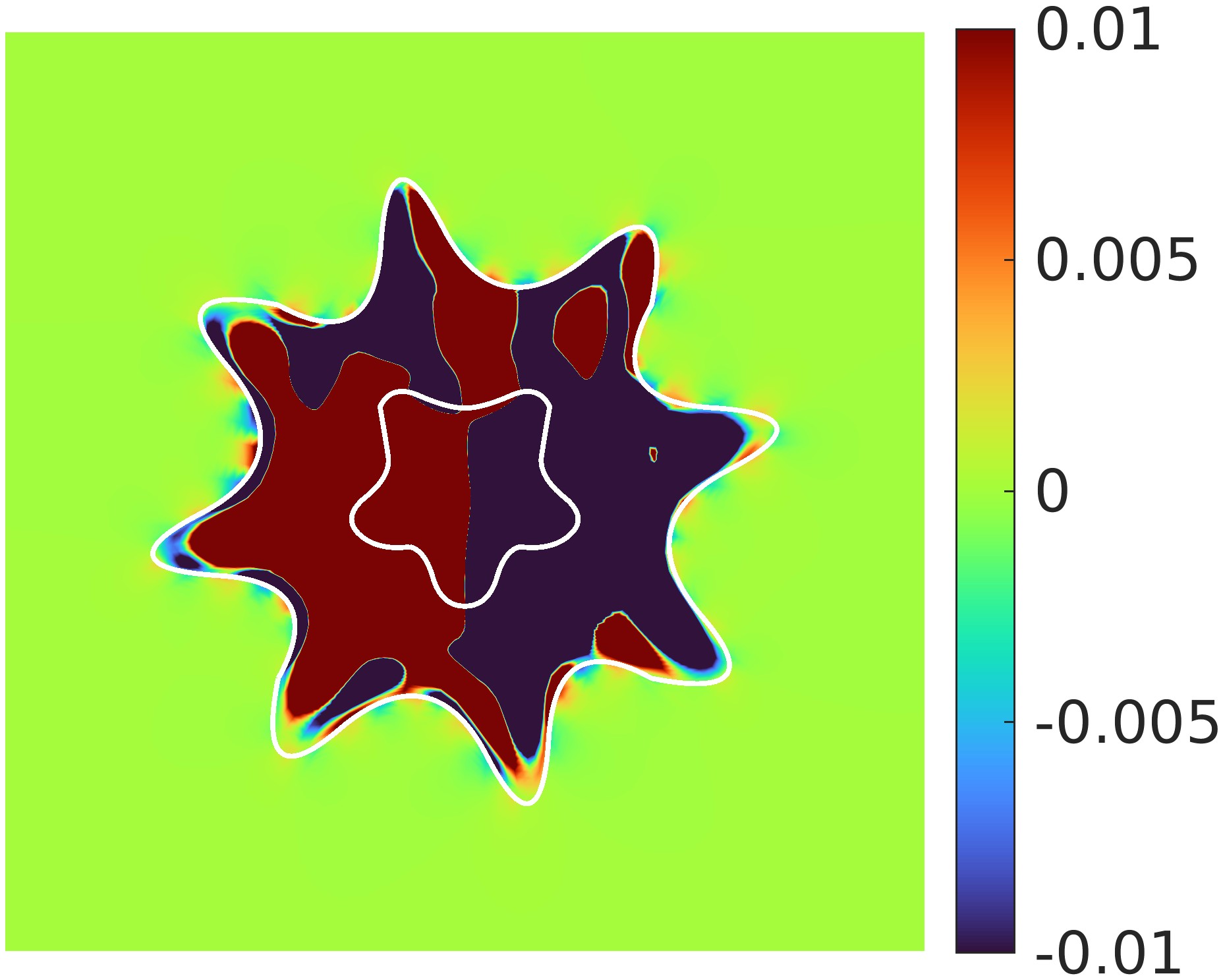}}
        \caption{\centering $T-\overline{T}$, \linebreak $J_{\rm cloak}=2.62\times 10^{-7}$}
    \end{subfigure}\\
\hline
\end{tabular}

 }
\caption{Optimized material distributions and temperature differences $T-\overline{T}$ for the thermal cloak problem with the star-shaped insulator and thermal cloak. Three material models (EMT, Porous Cu and Gyroid) and $N_{\rm var}=25$ are considered. Optimized objective function values are of order $10^{-4}$-$10^{-7}$.} 
\label{fig:chen2015case cloak Geo}
\end{figure}

 We consider three material models (EMT, Porous Cu and Gyroid) with ${N_{\rm var}=25}$. The optimized material distributions and temperature differences are shown in \fref{fig:chen2015case cloak Geo}. From the figure, we can see that the proposed method can effectively design the star-shaped thermal cloak around a free-form-shaped insulator without any issues. For EMT and Gyroid models, the optimization could achieve the objective function values of order $10^{-7}$. However, for Porous Cu, the optimization could only reach up to the order $10^{-4}$. This is due to the fact that the $\kappa_{\rm min}$ in the Porous Cu model, approximately 70.24~W/mK, is higher than the other two models as well as the conductivity of base material. Also, $\Omega_{\rm design}$ is relatively smaller than earlier circular geometry cases, which poses a limitation on creating an overall anisotropic effect required for cloaking. The results could be improved by increasing design freedom using a finer design mesh or by taking larger $\Omega_{\rm design}$. 

\renewcommand{\arraystretch}{1.5}   

\begin{figure}[!htbp]
    \centering
    \setlength\figureheight{1\textwidth}
    \setlength\figurewidth{1\textwidth}
    \begin{subfigure}[c]{0.38\textwidth}{\vspace{0.5cm}\centering\includegraphics[width=1\textwidth]{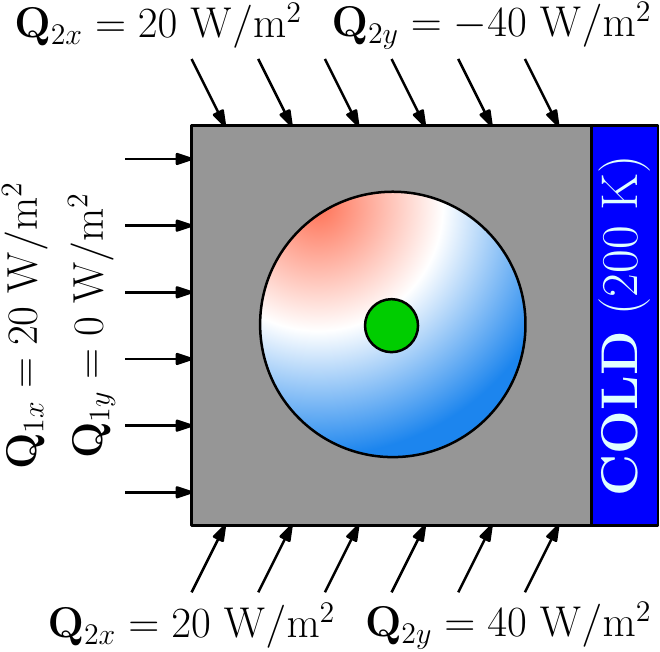}}
        \caption{Schematic of the thermal cloak problem under Neumann boundary condition.}
        \label{fig:chen2015case Schematics a}
    \end{subfigure}\quad \quad
     \begin{subfigure}[c]{0.31\textwidth}{\centering\includegraphics[width=1\textwidth]{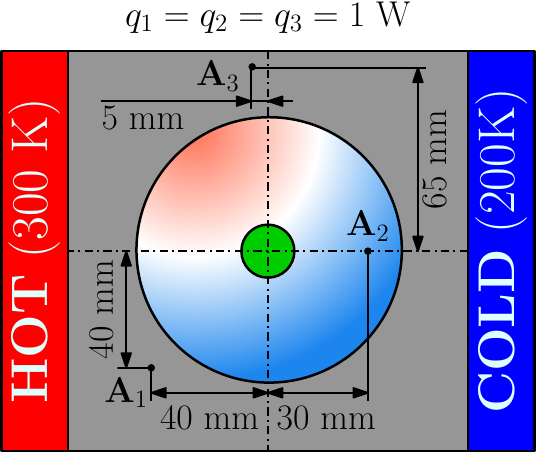}}
        \caption{Schematic of the thermal cloak problem with point heat sources in the domain.}
        \label{fig:chen2015case Schematics b}
    \end{subfigure}
 \caption{Schematics of thermal cloak problems with various boundary conditions. Two cases (a) one with the Neumann boundary conditions, and (b) another with the point heat sources are considered. The specifics of boundary conditions are described in the schematics.}
 \label{fig:Chen2015case schematics DiffBC}
\end{figure}
\subsubsection{Design with various boundary conditions}
\label{sec:2D cloak Design with intricate boundary conditions}
\par In the next few paragraphs, we design the thermal cloaks under various boundary conditions. We solve two cases, one with Neumann boundary conditions, and another with point heat sources as shown in \fref{fig:Chen2015case schematics DiffBC}. For the first problem, heat fluxes $\mathbf{Q}_1=[20,0]$~W/m$^{2}$, $\mathbf{Q}_2=[20,-40]$~W/m$^{2}$, and $\mathbf{Q}_3=[20,40]$~W/m$^{2}$ are applied on the left, top and bottom side, respectively. The right side is kept at $200$~K constant temperature. For the second problem, point heat sources $q_i=1$~W, $i=1,2,3$, are provided at the locations, $\mathbf{A}_i$, $i=1,2,3$. The right and left sides are kept at constant temperature $300$~K \& $200$~K, respectively. For numerical analysis, the point heat sources are modelled as the domain heat source using approximate Dirac delta function $\widetilde{\delta}$. The total heat source $q_b$ can be can be written as follows:
\begin{equation}
q_b(\mathbf{x})=\sum_{i=1}^{3}q_i \widetilde{\delta}(||\mathbf{x}-\mathbf{A}_i||_2) \quad \text{with} \quad  \widetilde{\delta}(\phi)= \begin{cases}
  \dfrac{3}{4\Delta}\left(1-\dfrac{\phi^2}{\Delta^2}\right) \quad &\text{if} \quad  \phi \leq \Delta \\
  0 \quad &\text{if} \quad \phi > \Delta
\end{cases}
\end{equation}
where $\Delta$ is a support bandwidth. In this example, we take $\Delta=0.005$. For both problems, we consider three material models (EMT, Porous Cu and Gyroid) same as the last problem.

\renewcommand{\arraystretch}{1.5}   
\begin{figure}[!]
\centering
\scalebox{1}{
\begin{tabular}[c]{| M{7.75em} | M{7.75em} | M{7.75em} | M{7.75em} |}
\hline
\multicolumn{1}{|c|}{\centering Reference case}  & 
\multicolumn{1}{c|}{\centering EMT}  & \multicolumn{1}{c|}{\centering Porous Cu} & \multicolumn{1}{c|}{\centering Gyroid} \\
\hline
\vspace{0.2cm}
    \begin{subfigure}[t]{0.22\textwidth}{\centering\includegraphics[width=1\textwidth]{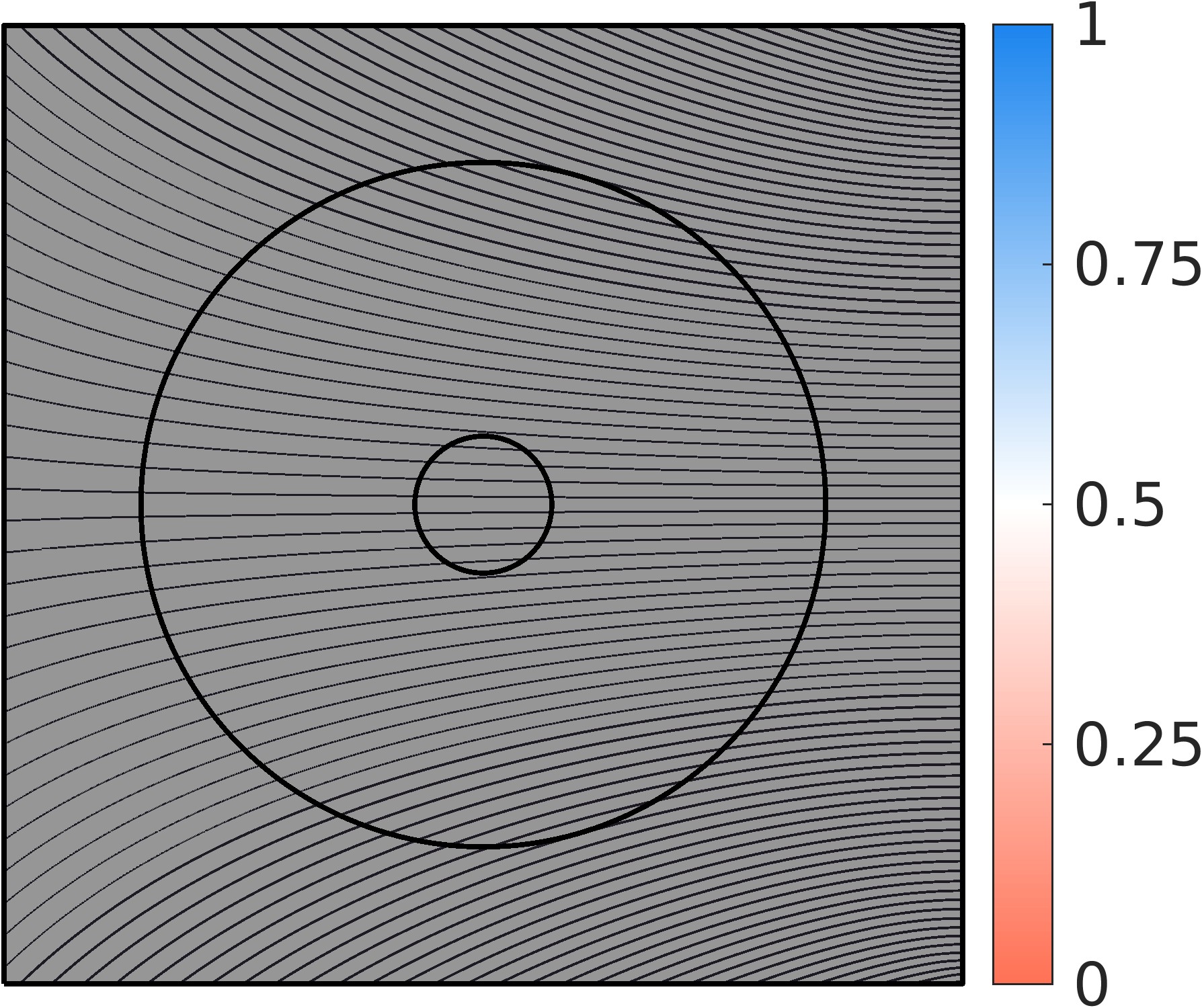}}
        \caption{\centering Reference case}
    \end{subfigure}&
\vspace{0.2cm}
    \begin{subfigure}[t]{0.22\textwidth}{\centering\includegraphics[width=1\textwidth]{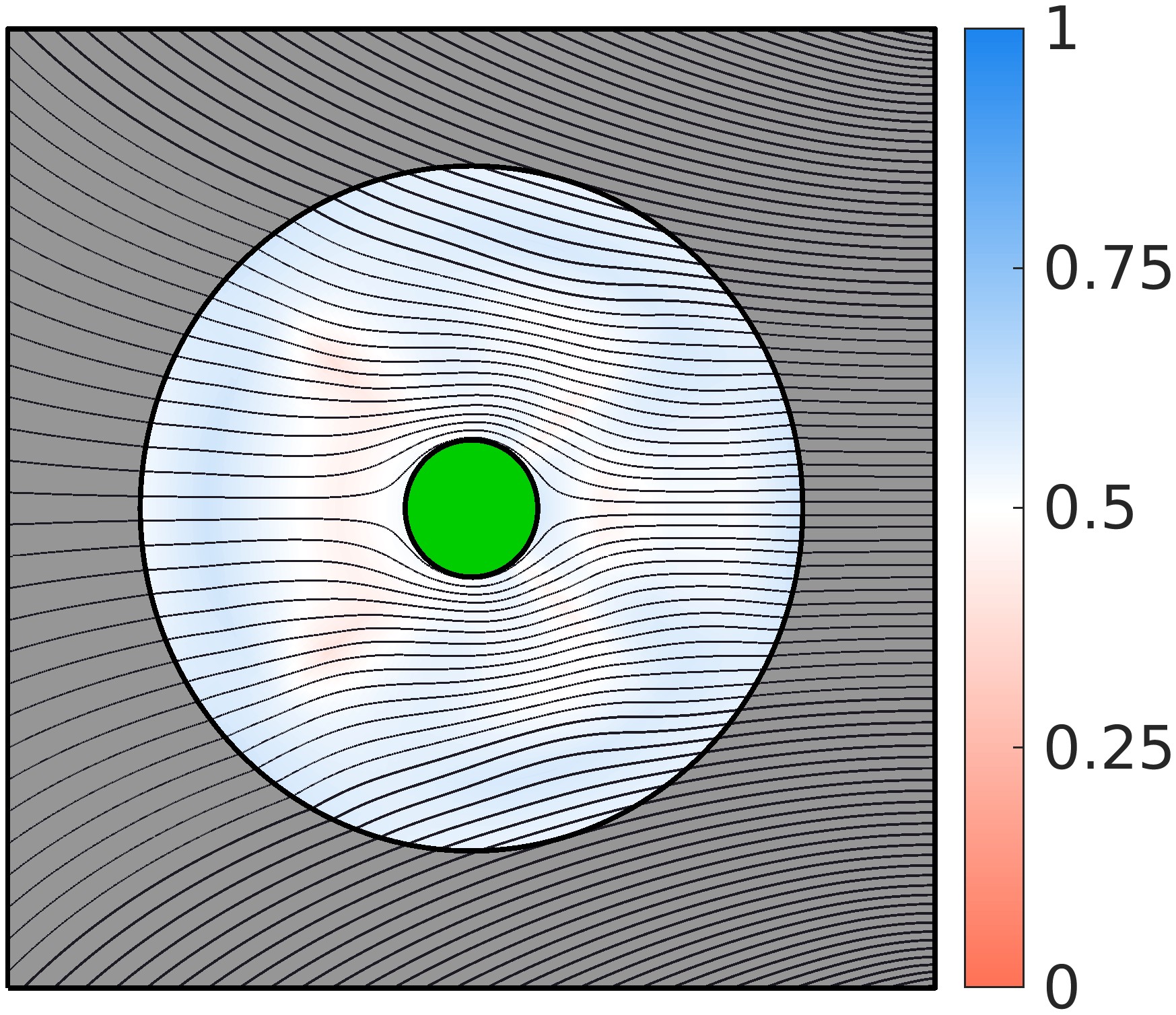}}
        \caption{\centering Optimized material distribution}
    \end{subfigure}& \vspace{0.2cm}
    \begin{subfigure}[t]{0.22\textwidth}{\centering\includegraphics[width=1\textwidth]{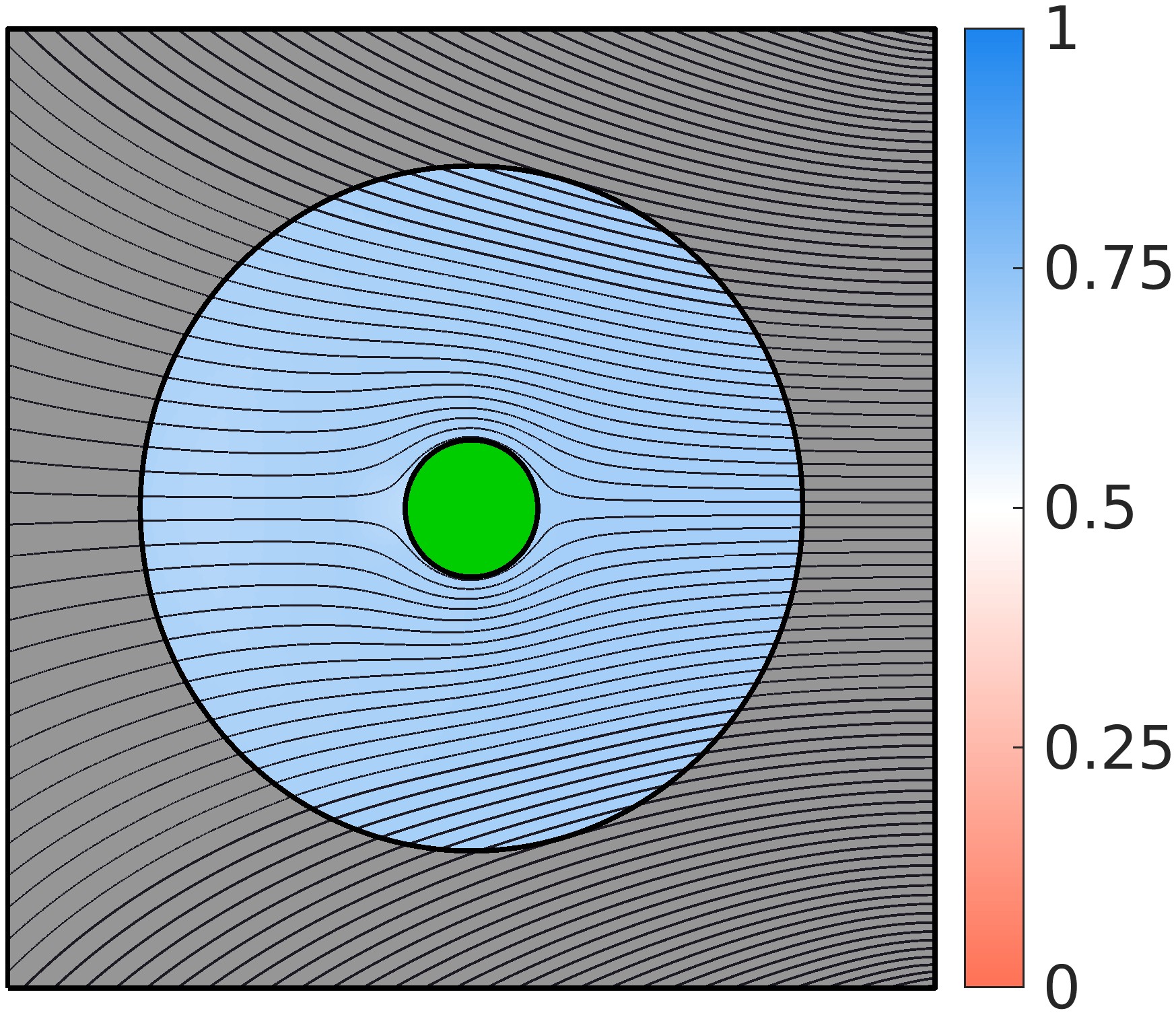}}
         \caption{\centering Optimized material distribution}
    \end{subfigure}
    & \vspace{0.2cm}
    \begin{subfigure}[t]{0.22\textwidth}{\centering\includegraphics[width=1\textwidth]{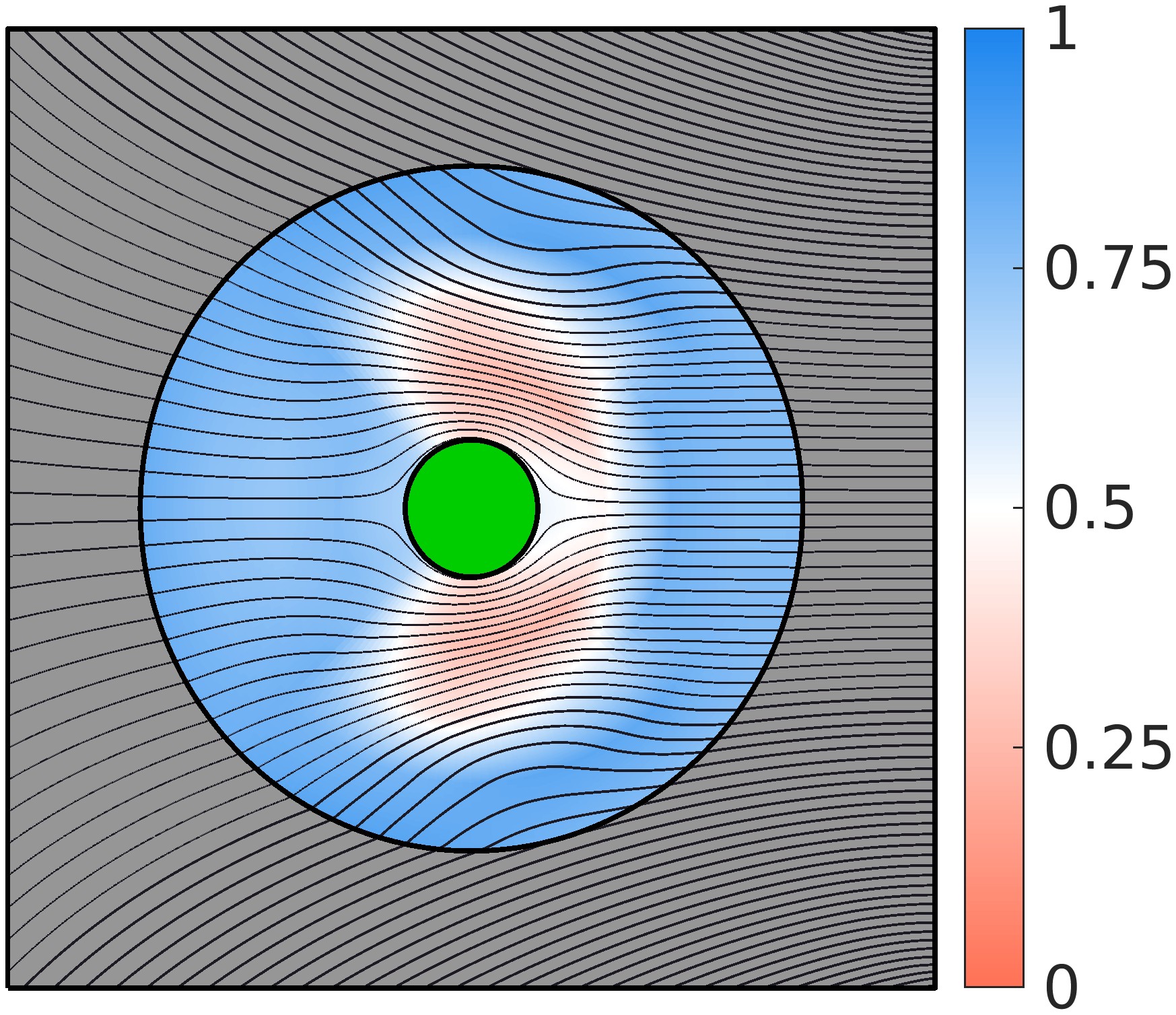}}
        \caption{\centering Optimized material distribution}
    \end{subfigure}\\ \vspace{0.2cm}
    \begin{subfigure}[t]{0.22\textwidth}{\centering\includegraphics[width=1\textwidth]{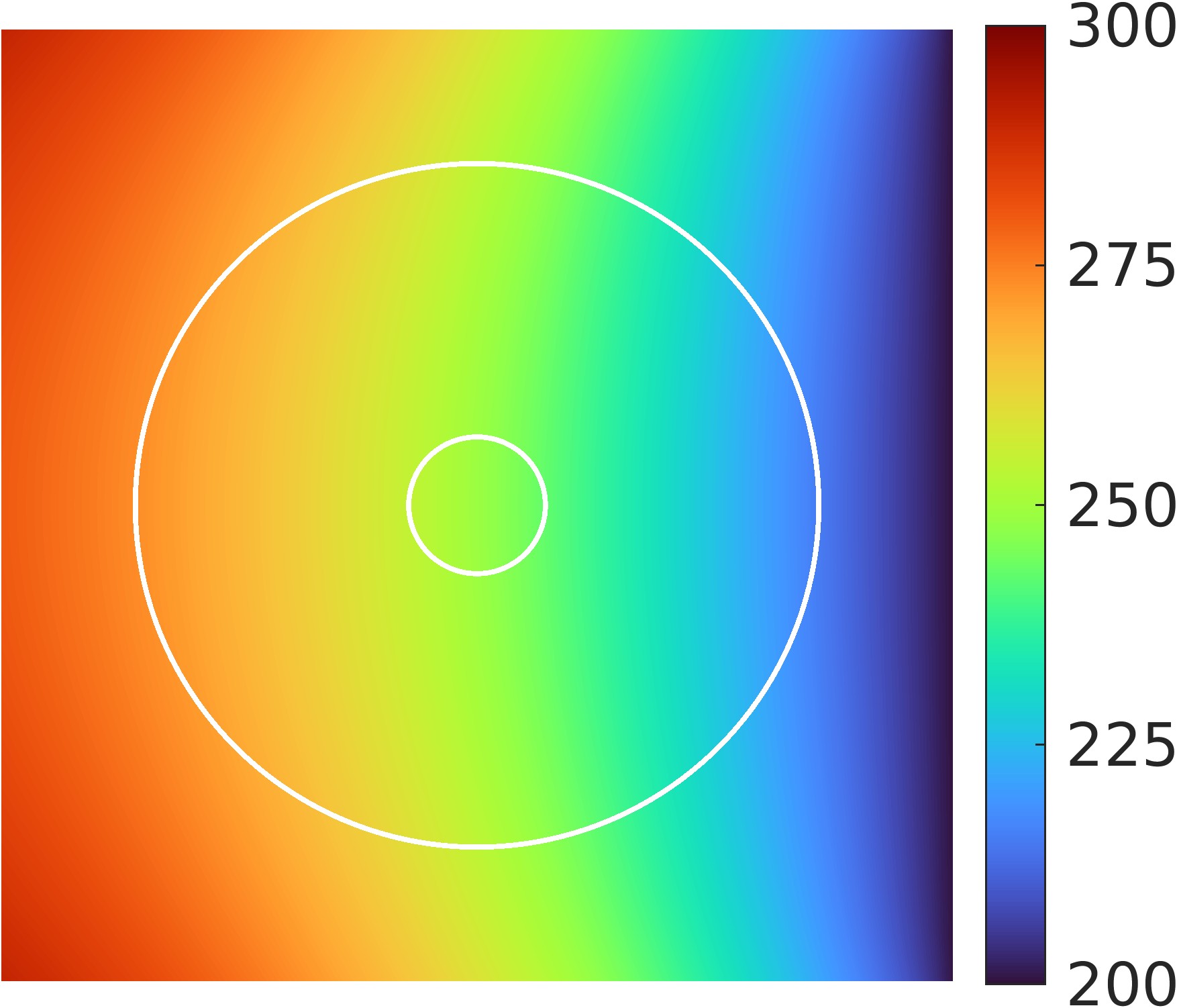}}
        \caption{\centering Reference  temp. distribution $\overline{T}$}
    \end{subfigure}&\vspace{0.2cm}
    \begin{subfigure}[t]{0.22\textwidth}{\centering\includegraphics[width=1\textwidth]{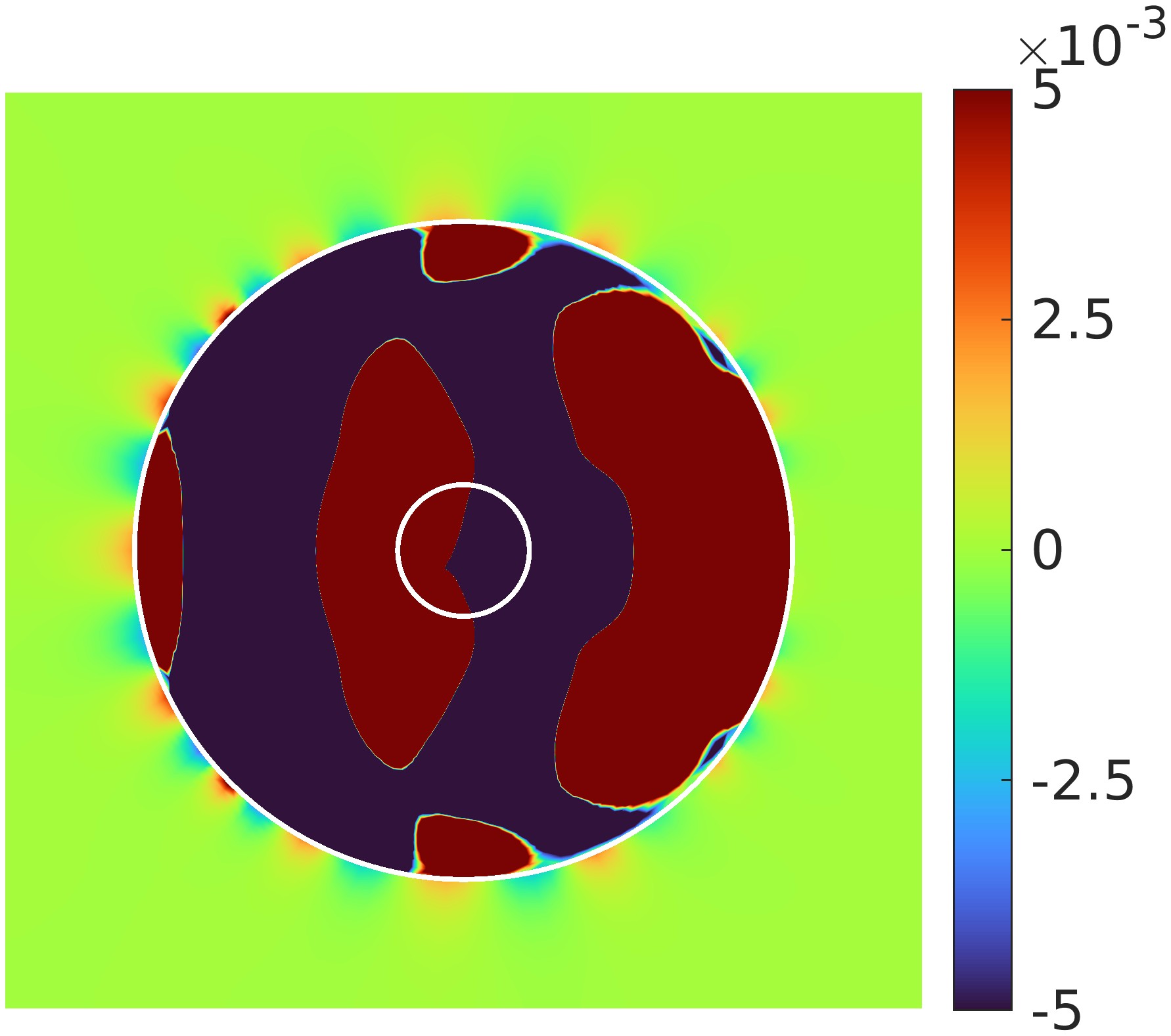}}
        \caption{\centering $T-\overline{T}$,\linebreak $J=4.93\times 10^{-8}$}
    \end{subfigure}& \vspace{0.2cm}
    \begin{subfigure}[t]{0.22\textwidth}{\centering\includegraphics[width=1\textwidth]{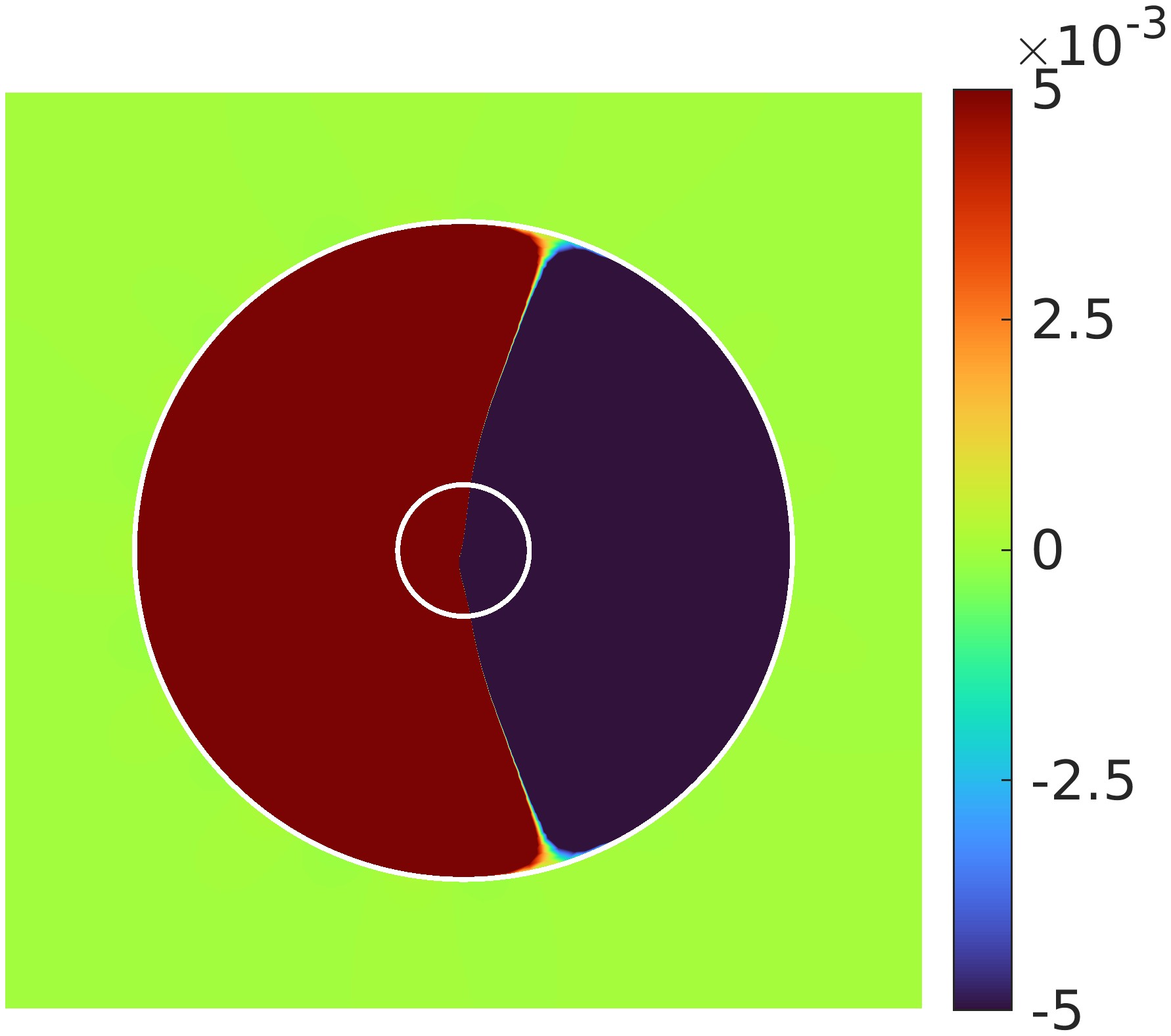}}
       \caption{\centering $T-\overline{T}$,\linebreak $J=9.51\times 10^{-11}$}
    \end{subfigure}
    & \vspace{0.2cm}
    \begin{subfigure}[t]{0.22\textwidth}{\centering\includegraphics[width=1\textwidth]{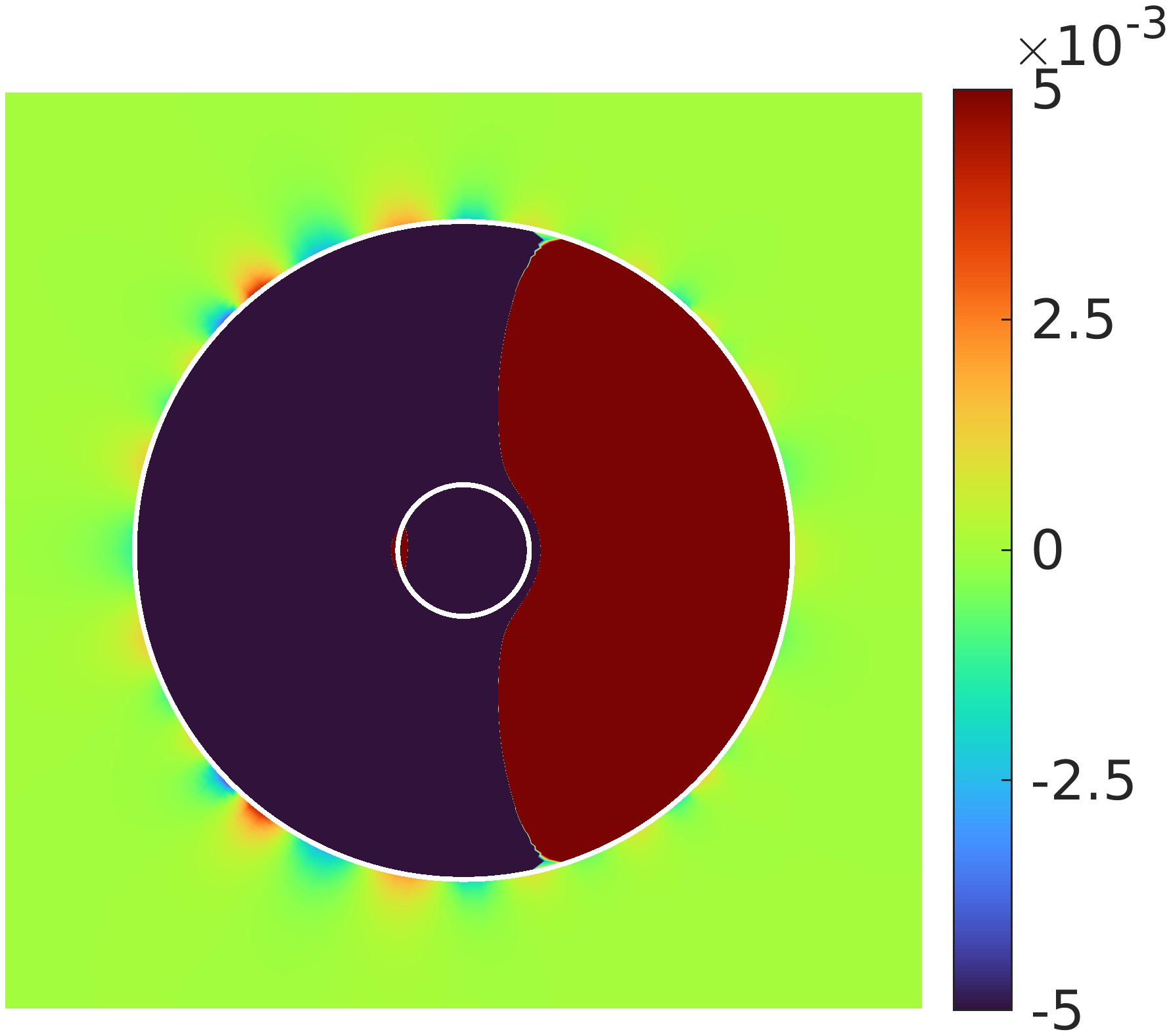}}
        \caption{\centering $T-\overline{T}$, \linebreak $J=3.08\times 10^{-8}$}
    \end{subfigure}\\
\hline
 \end{tabular}
}
\caption{(Columns 2-4) Optimized material distributions and temperature differences $T-\overline{T}$ for the thermal cloak problem with Neumann boundary conditions. Three material models (EMT, Porous Cu and Gyroid) and $N_{\rm var}=25$ are considered. Optimized objective function values are of order $10^{-8}$-$10^{-11}$. Column 1 represents the reference case under applied boundary conditions.}  
    \label{fig:chen2015case DiffBC2}
\end{figure}

\par
\begin{figure}[!htbp]
\centering
\scalebox{1}{
\begin{tabular}[c]{| M{7.75em} | M{7.75em} | M{7.75em} | M{7.75em} |}
\hline
\multicolumn{1}{|c|}{\centering Reference case}  & 
\multicolumn{1}{c|}{\centering EMT}  & \multicolumn{1}{c|}{\centering Porous Cu} & \multicolumn{1}{c|}{\centering Gyroid} \\
\hline
\vspace{0.2cm}
    \begin{subfigure}[t]{0.22\textwidth}{\centering\includegraphics[width=1\textwidth]{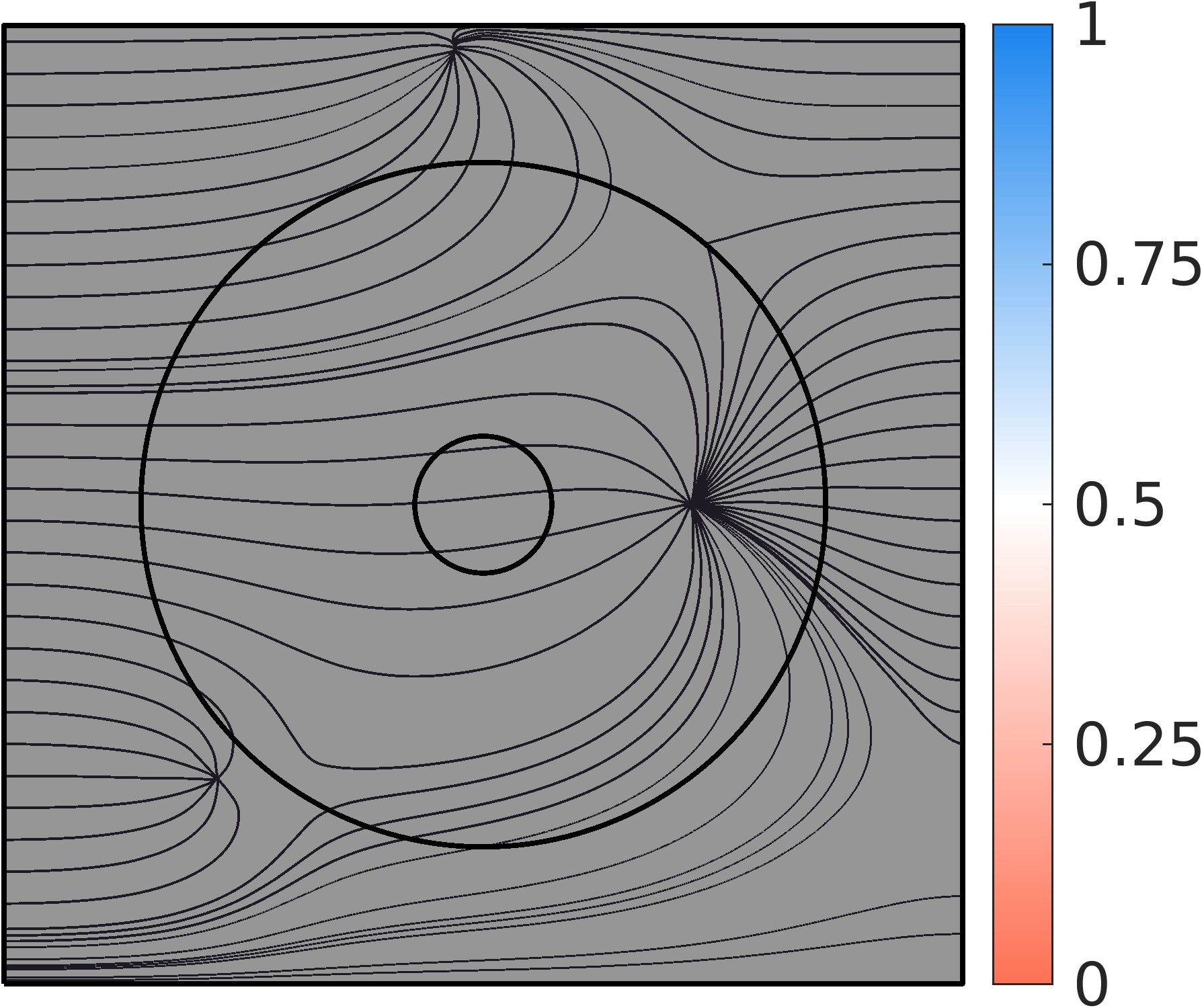}}
        \caption{\centering Reference case}
    \end{subfigure}&
\vspace{0.2cm}
    \begin{subfigure}[t]{0.22\textwidth}{\centering\includegraphics[width=1\textwidth]{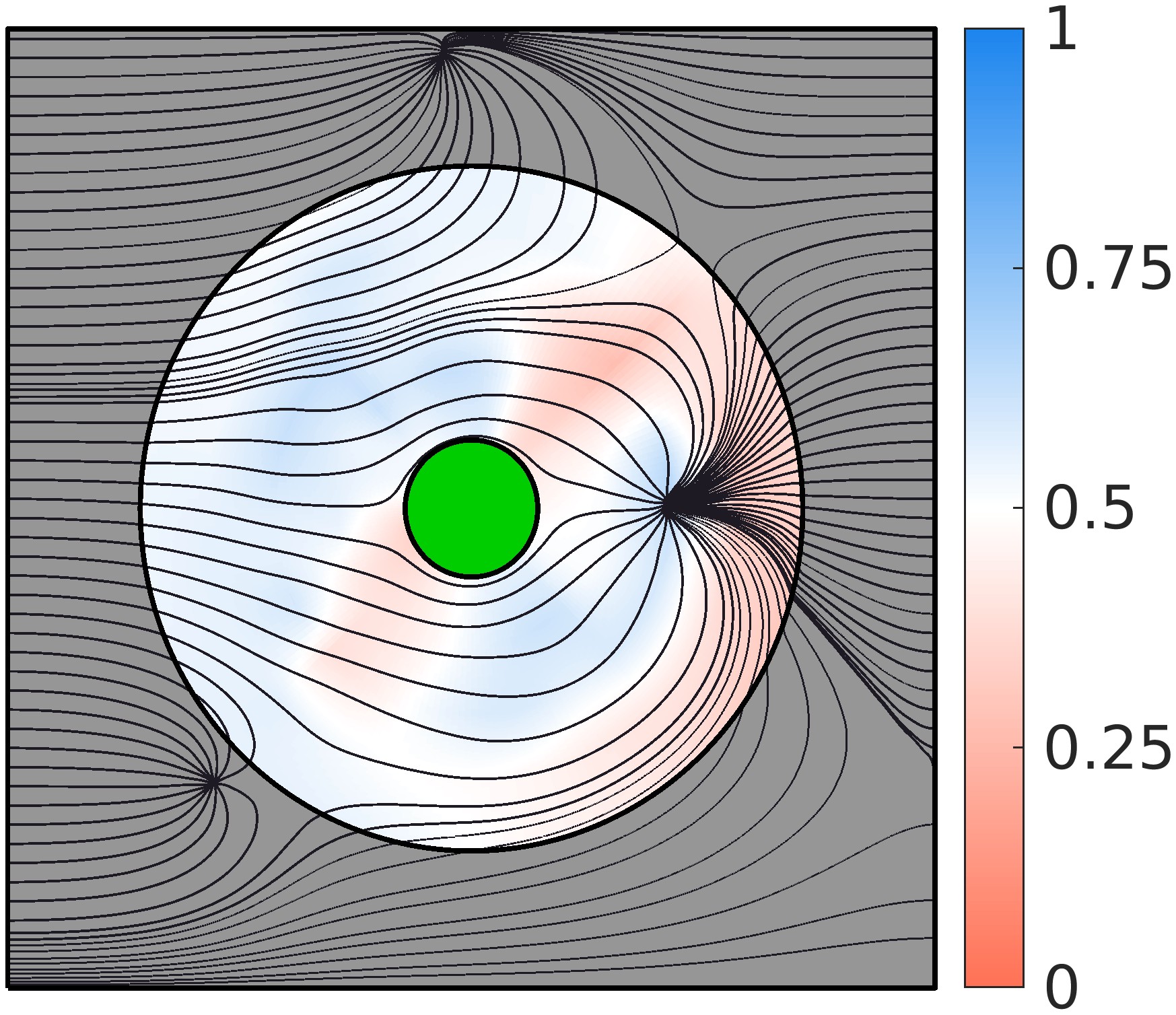}}
        \caption{\centering Optimized material distribution}
    \end{subfigure}& \vspace{0.2cm}
    \begin{subfigure}[t]{0.22\textwidth}{\centering\includegraphics[width=1\textwidth]{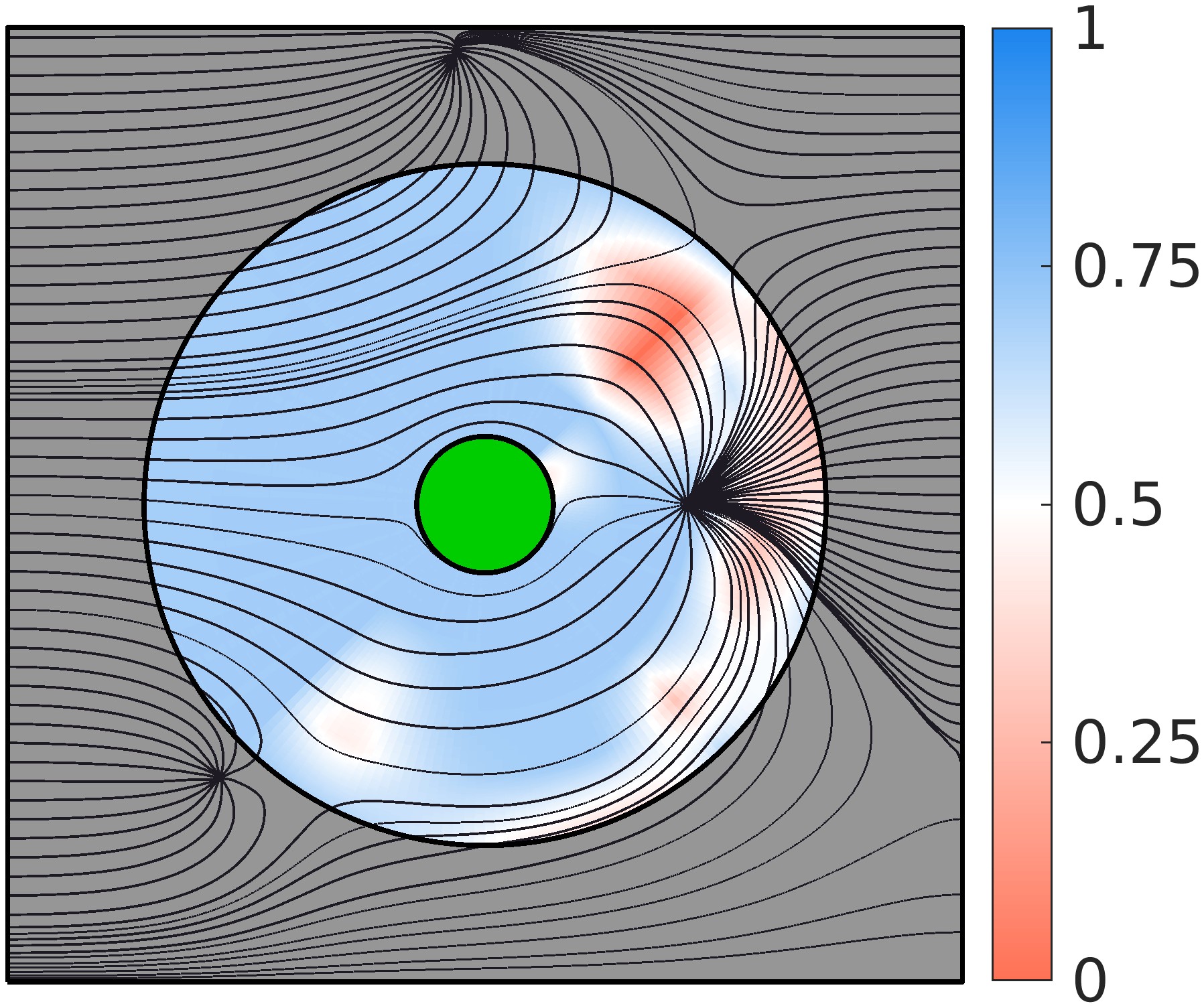}}
         \caption{\centering Optimized material distribution}
    \end{subfigure}
    & \vspace{0.2cm}
    \begin{subfigure}[t]{0.22\textwidth}{\centering\includegraphics[width=1\textwidth]{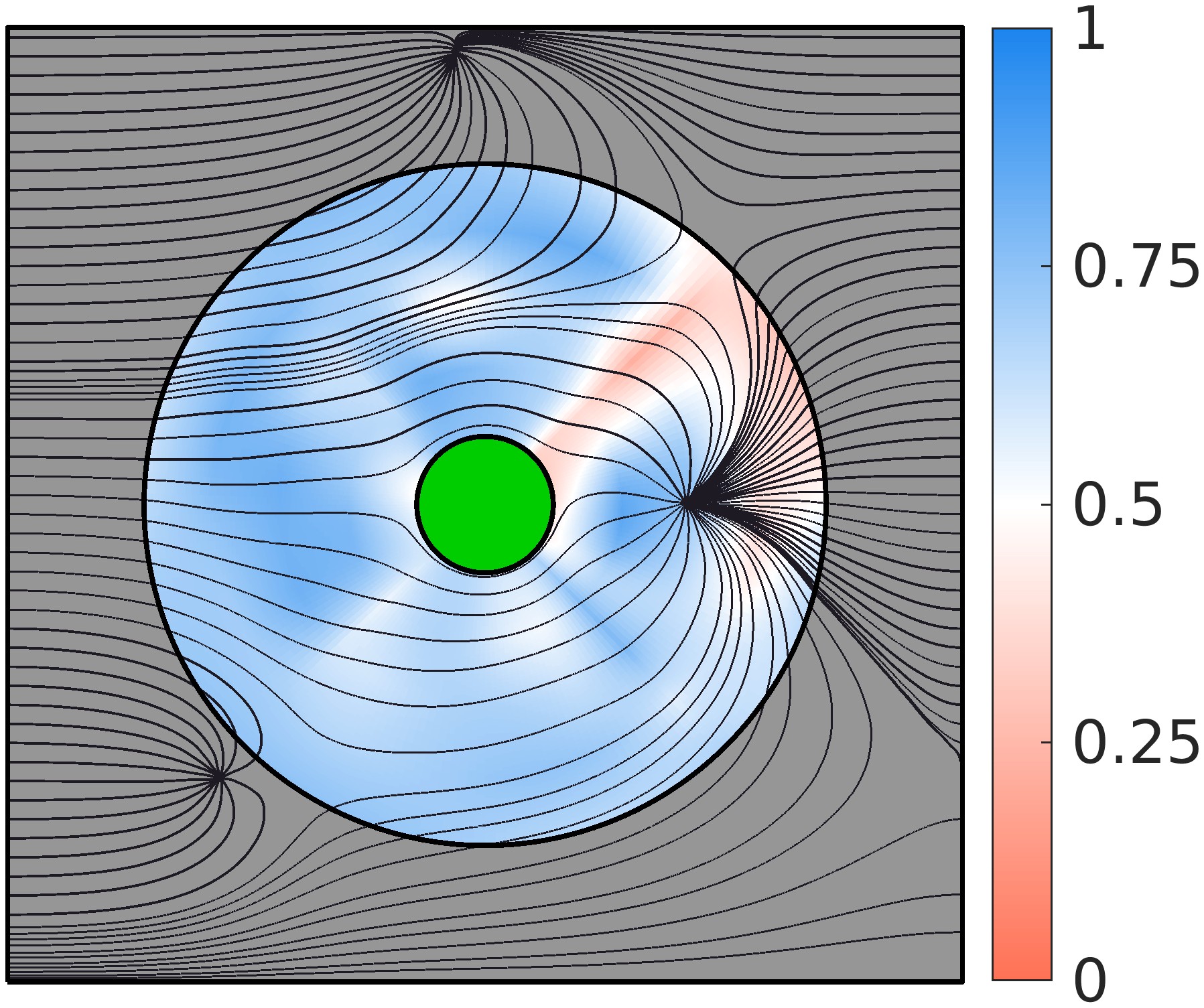}}
        \caption{\centering Optimized material distribution}
    \end{subfigure}\\ 
    \vspace{0.2cm}
    \begin{subfigure}[t]{0.22\textwidth}{\centering\includegraphics[width=1\textwidth]{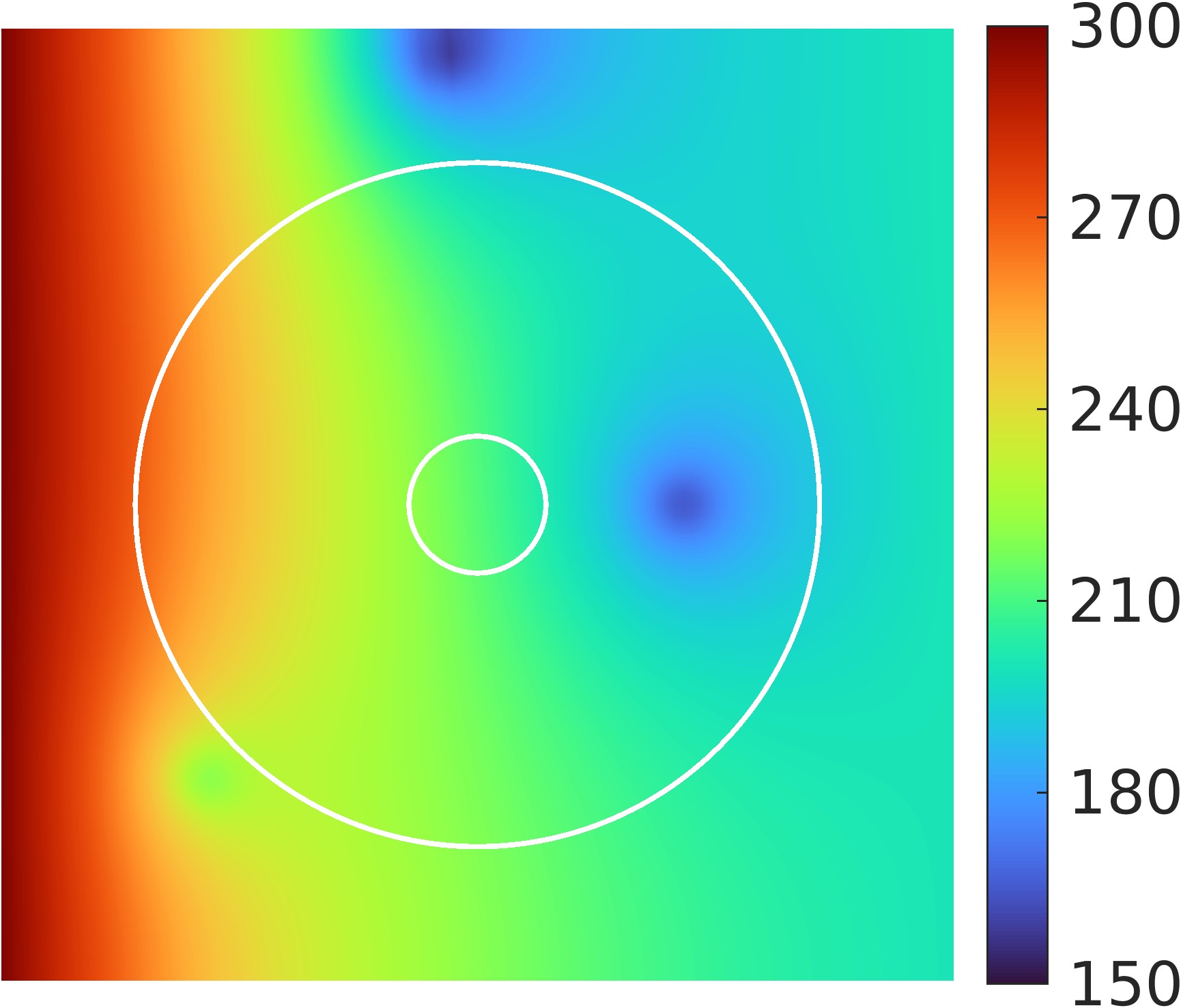}}
        \caption{\centering Reference temp. distribution $\overline{T}$}
    \end{subfigure}&\vspace{0.2cm}
    \begin{subfigure}[t]{0.22\textwidth}{\centering\includegraphics[width=1\textwidth]{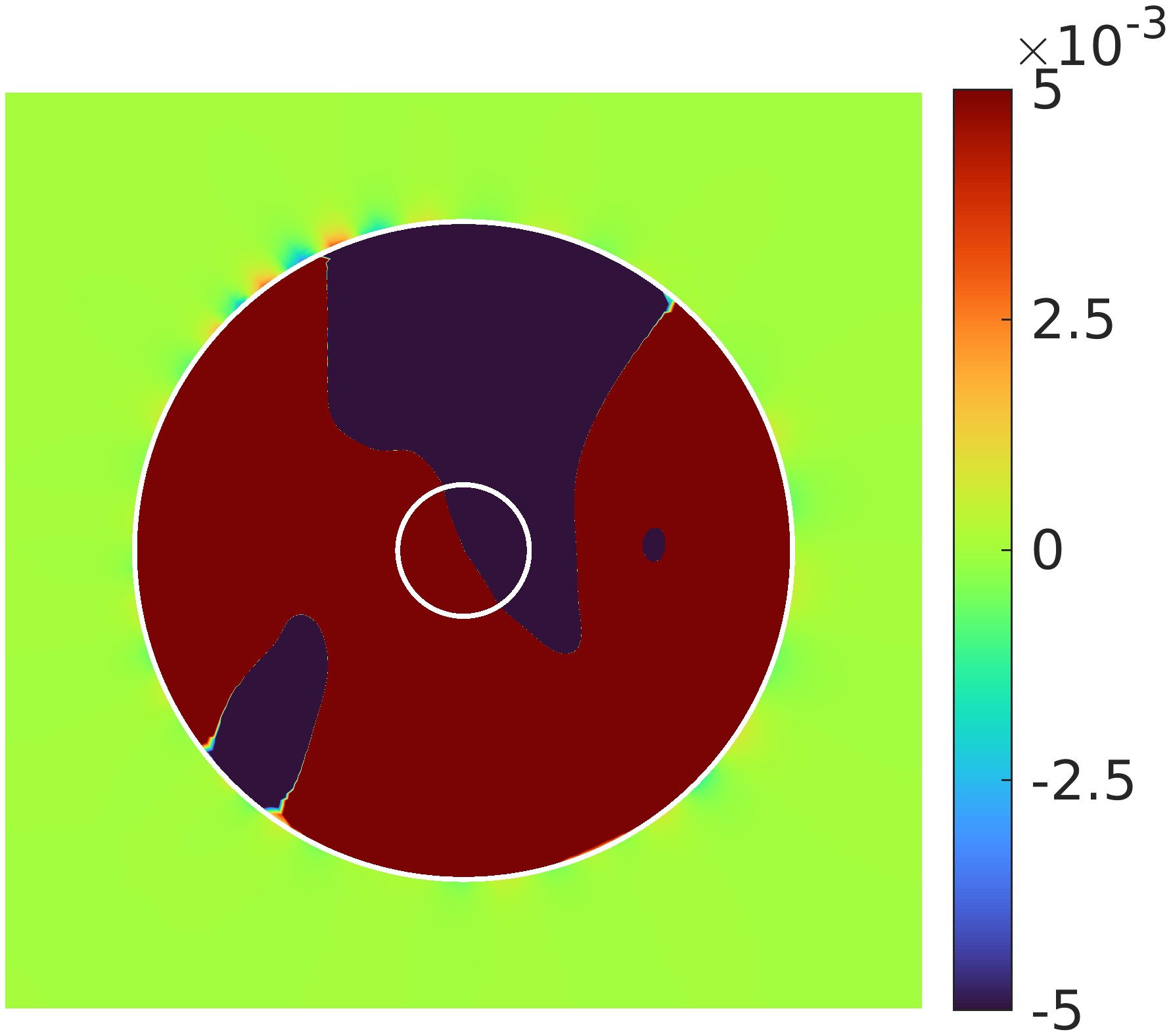}}
        \caption{\centering $T-\overline{T}$,\linebreak $J=1.05\times 10^{-7}$}
    \end{subfigure}& \vspace{0.2cm}
    \begin{subfigure}[t]{0.22\textwidth}{\centering\includegraphics[width=1\textwidth]{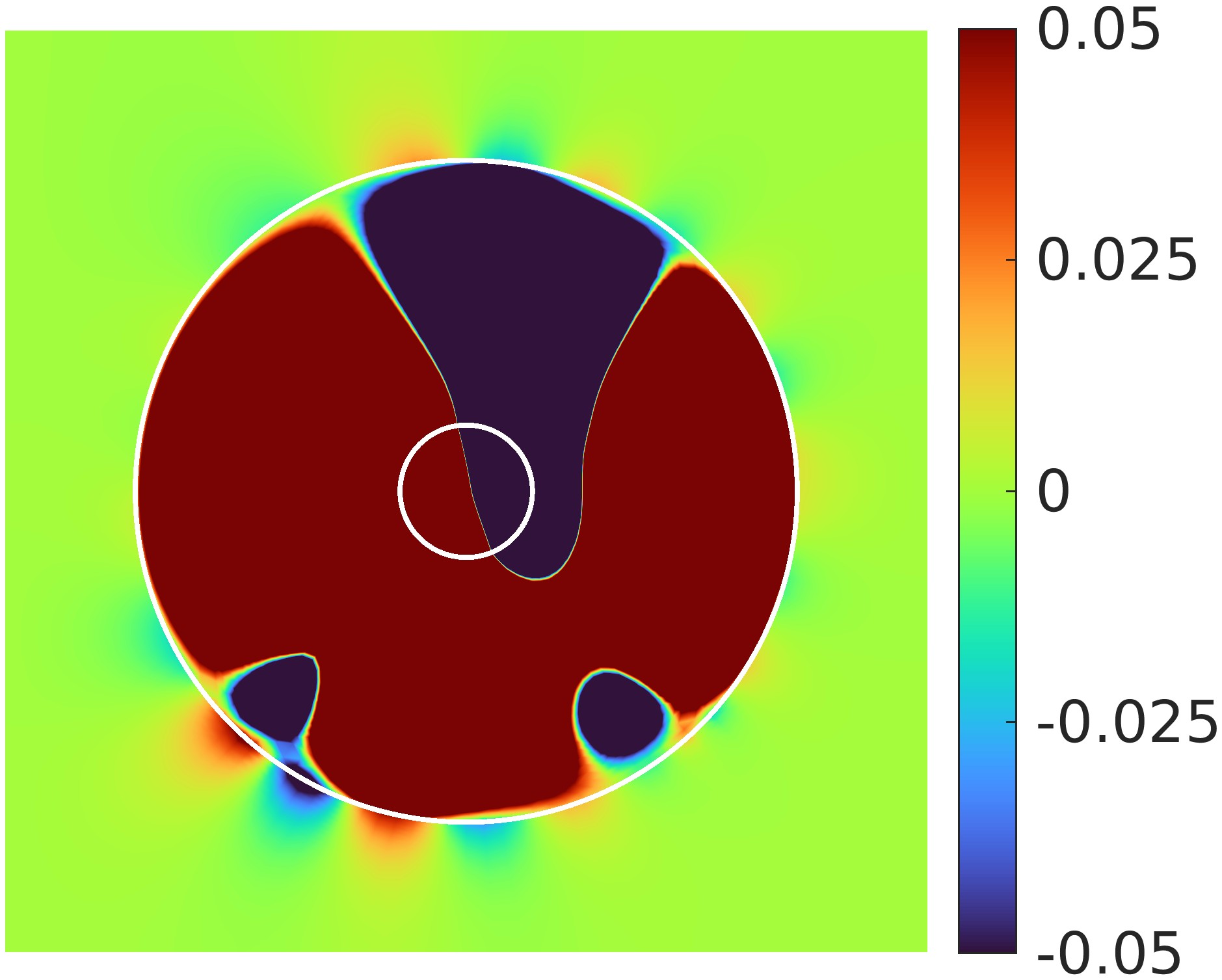}}
       \caption{\centering $T-\overline{T}$,\linebreak $J=1.19\times 10^{-4}$}
    \end{subfigure}
    & \vspace{0.2cm}
    \begin{subfigure}[t]{0.22\textwidth}{\centering\includegraphics[width=1\textwidth]{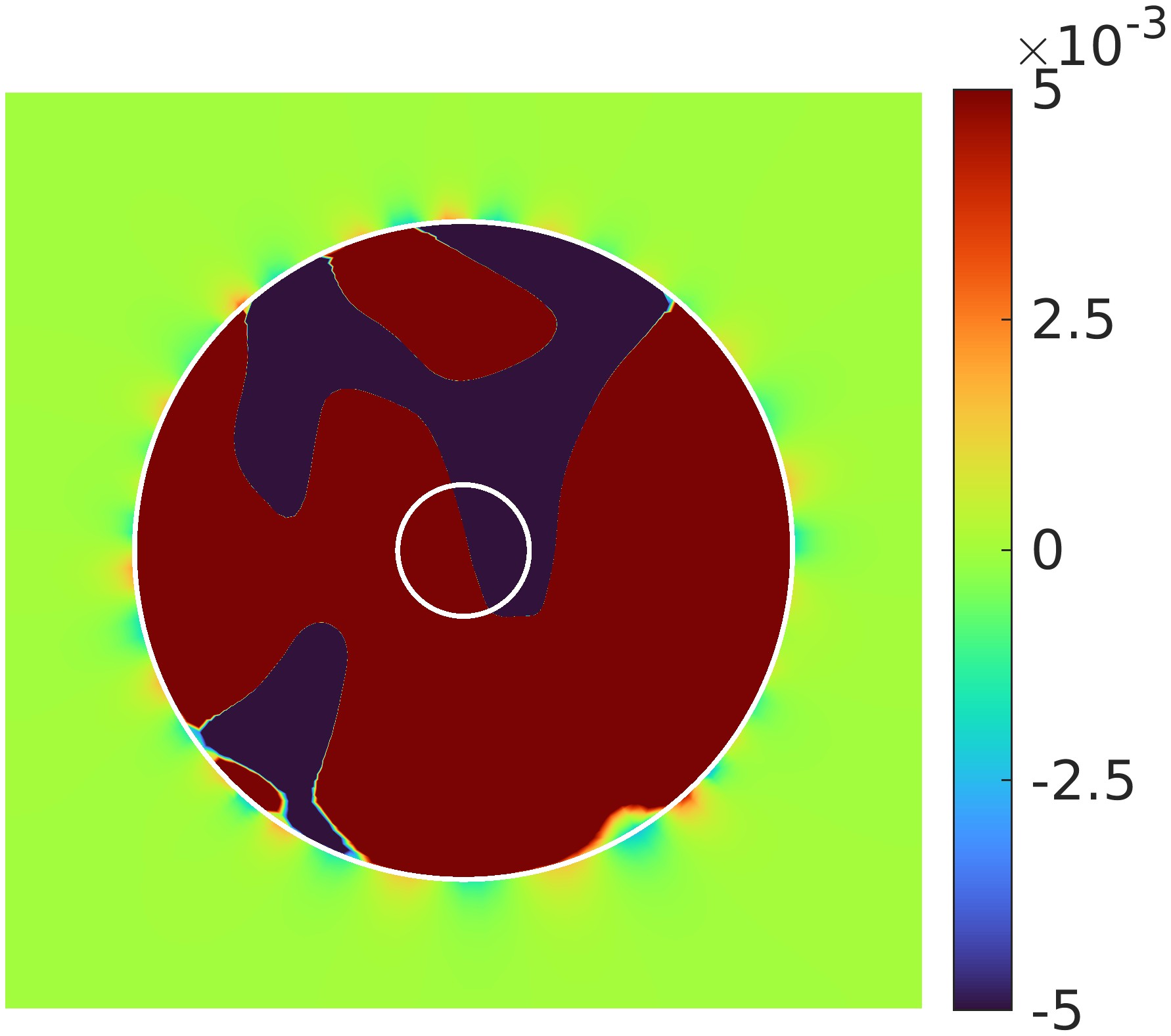}}
        \caption{\centering $T-\overline{T}$, \linebreak $J=2.98\times 10^{-7}$}
    \end{subfigure}\\
\hline
 \end{tabular}
}

\caption{(Columns 2-4) Optimized material distributions and temperature differences $T-\overline{T}$ for the thermal cloak problem with the point heat sources inside the domain. Three material models (EMT, Porous Cu and Gyroid) and $N_{\rm var}=25$ are considered. Optimized objective function values are of order $10^{-4}$-$10^{-7}$. Column 1 represents the reference case under applied boundary conditions.}  
    \label{fig:chen2015case DiffBC3}
\end{figure}

 The optimization results are shown in \frefs{fig:chen2015case DiffBC2}-\ref{fig:chen2015case DiffBC3}. The proposed method can effectively design thermal cloaks for both problems with substantial cloaking function ($J_{\rm cloak}<10^{-6}$), except for Porous Cu model with point heat sources. As explained in the previous subsection, this is primarily due to a smaller $\kappa$-range for Porous Cu model. Improvements in $J_{\rm cloak}$ can be achieved by increasing design freedom or expanding the size of $\Omega_{\rm design}$.

\renewcommand{\arraystretch}{1.5}   
\subsection{3D thermal cloak}
\label{sec:Chen2015case 3D cloak}
\begin{figure}[!htbp]
\centering
\centering\includegraphics[width=0.7\textwidth]{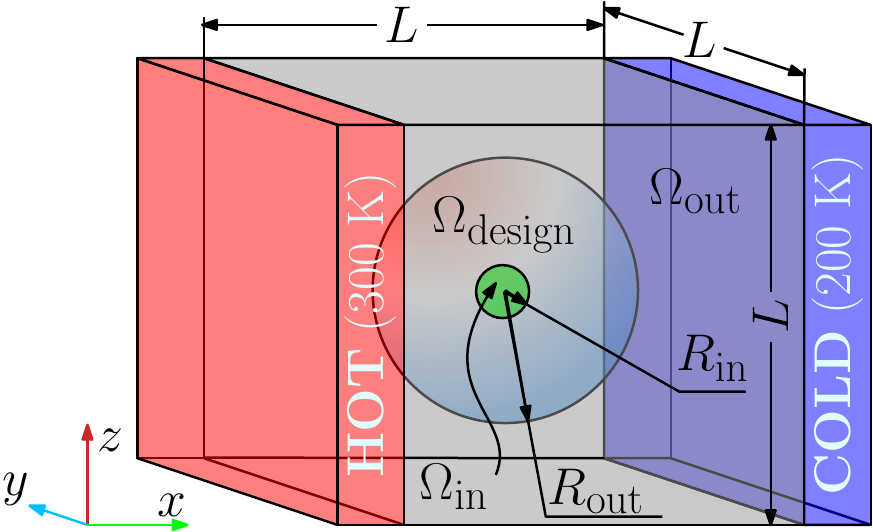} \caption{Schematics of a spherical insulator ($\mathrm{\Omega}_{\mathrm{in}}$) and a surrounding spherical FGM-based thermal cloak embedded ($\mathrm{\Omega}_{\mathrm{design}}$) in the base material cube $\mathrm{\Omega}$; $\mathrm{\Omega}_{\mathrm{design}}$ is the domain of the cloak where the material distribution is optimized, $\mathrm{\Omega}_{\mathrm{out}}$ is the outside domain of remaining base material, where the temperature disturbance is sought to be reduced. $\mathrm{\Omega} = \mathrm{\Omega}_{\mathrm{in}} \cup \mathrm{\Omega}_{\mathrm{design}}\cup \mathrm{\Omega}_{\mathrm{out}}$.}
 \label{fig:Chen2015case schematics 3D cloak}
\end{figure}

In this subsection, we design 3D thermal cloaks using the proposed method. The geometry, boundary conditions and NURBS parameterizations are simple extensions of 2D problems as mentioned \sref{sec:Chen2015case 2D cloak}. The 3D cube of base material with embedded spherical insulator and thermal cloak is shown in \fref{fig:Chen2015case schematics 3D cloak}. All dimensions, conductivities, objective function and material distributions are assumed to be the same as those described in \sref{sec:Chen2015case 2D cloak}. For 3D thermal cloaks, we provide $x$, $y$ and $z$-plane symmetry for design meshes. Also, 3D problems are solved only on $\sfrac{1}{4^{th}}$ domain (by giving symmetry conditions along $y$ \& $z$-planes) to avoid the extra computational burden.
\subsubsection{Design with various material models} 
\label{sec:3D cloak Design with various material models}

\renewcommand{\arraystretch}{1.5}   
\begin{figure}[!htbp]
\centering
\scalebox{1}{
\begin{tabular}[c]{| M{16em} | M{16em} |}
\hline
\multicolumn{1}{|c|}{\centering EMT}  
& \multicolumn{1}{c|}{\centering Gyroid} \\
\hline
\vspace{0.2cm}
    \begin{subfigure}[t]{0.43\textwidth}{\centering\includegraphics[width=0.31\textwidth]{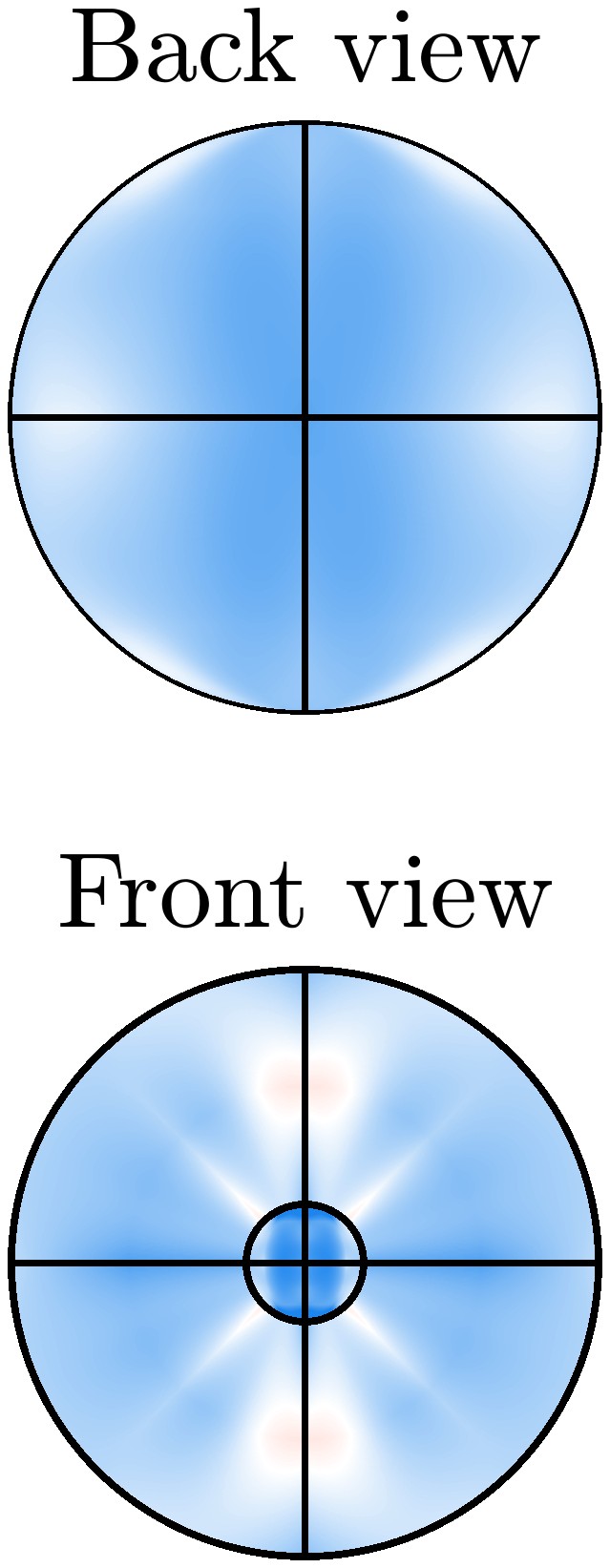}\includegraphics[width=0.69\textwidth]{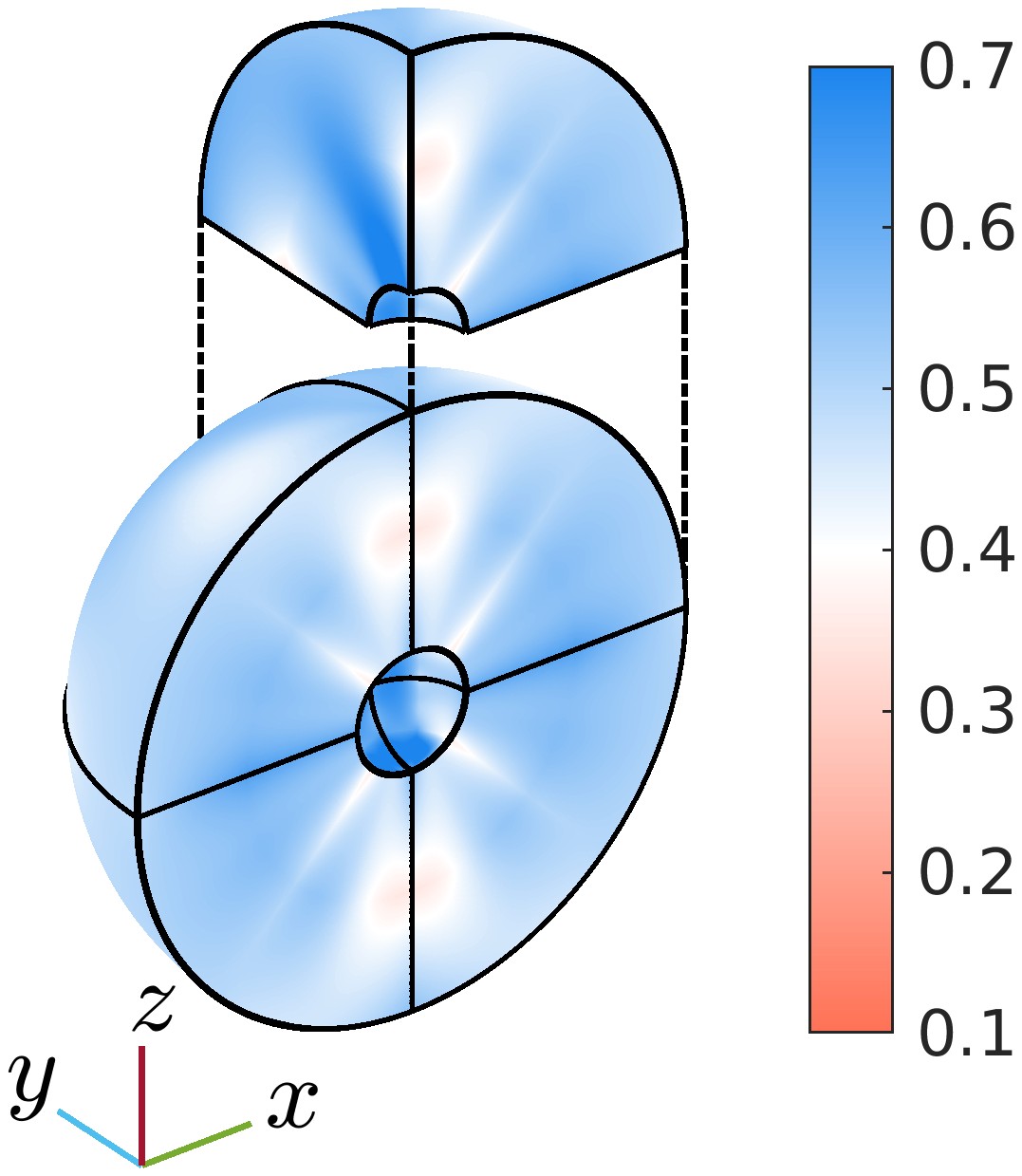}}
        \caption{\centering $J_{\rm cloak}=1.23 \times 10^{-3}$}
    \end{subfigure}
    & \vspace{0.2cm}
    \begin{subfigure}[t]{0.43\textwidth}{\centering\includegraphics[width=0.31\textwidth]{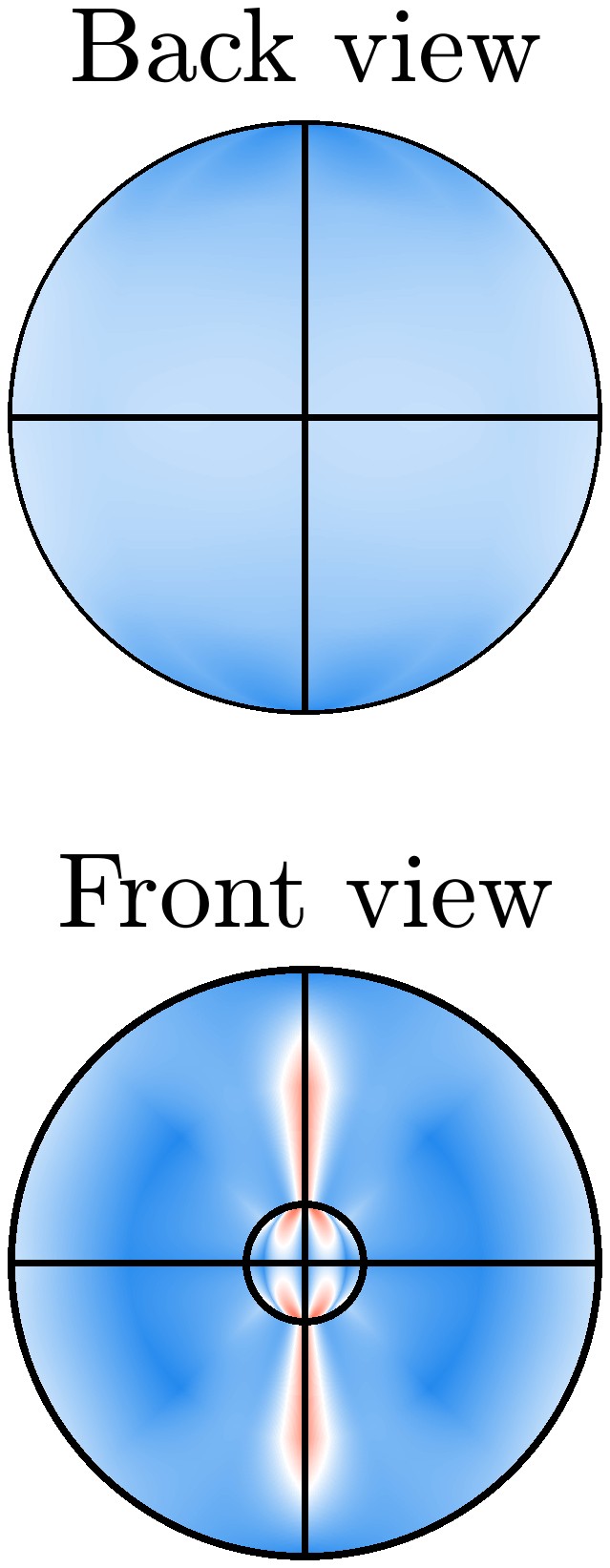}\includegraphics[width=0.69\textwidth]{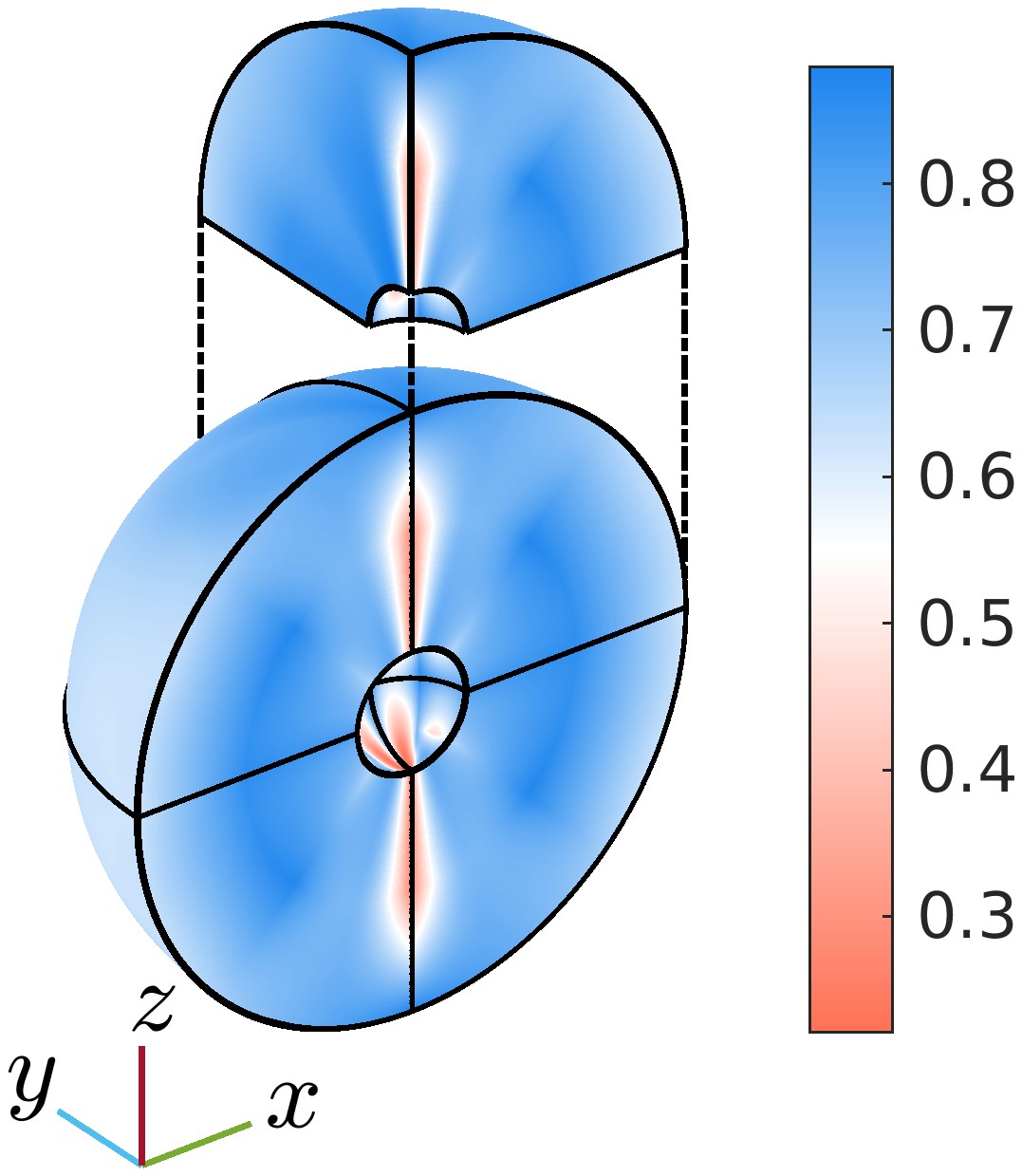}}
        \caption{\centering $J_{\rm cloak}=6.68 \times 10^{-4}$}
    \end{subfigure} \\
\hline
 \end{tabular}
}

\caption{Optimized material distribution for the 3D thermal cloaks. Two material models (EMT and Gyroid) and $N_{\rm var}=129$ are considered. Optimized objective function values are of order  $10^{-3}-10^{-4}$. Optimized material distributions remain close to the initial material distributions with almost the entire domain filled with intermediate densities.}  
    \label{fig:Chen2015case 3D cloak}
\end{figure}

\begin{figure}[!htbp]
    \centering
    \begin{subfigure}[b]{0.46\textwidth}{\centering\includegraphics[width=1\textwidth]{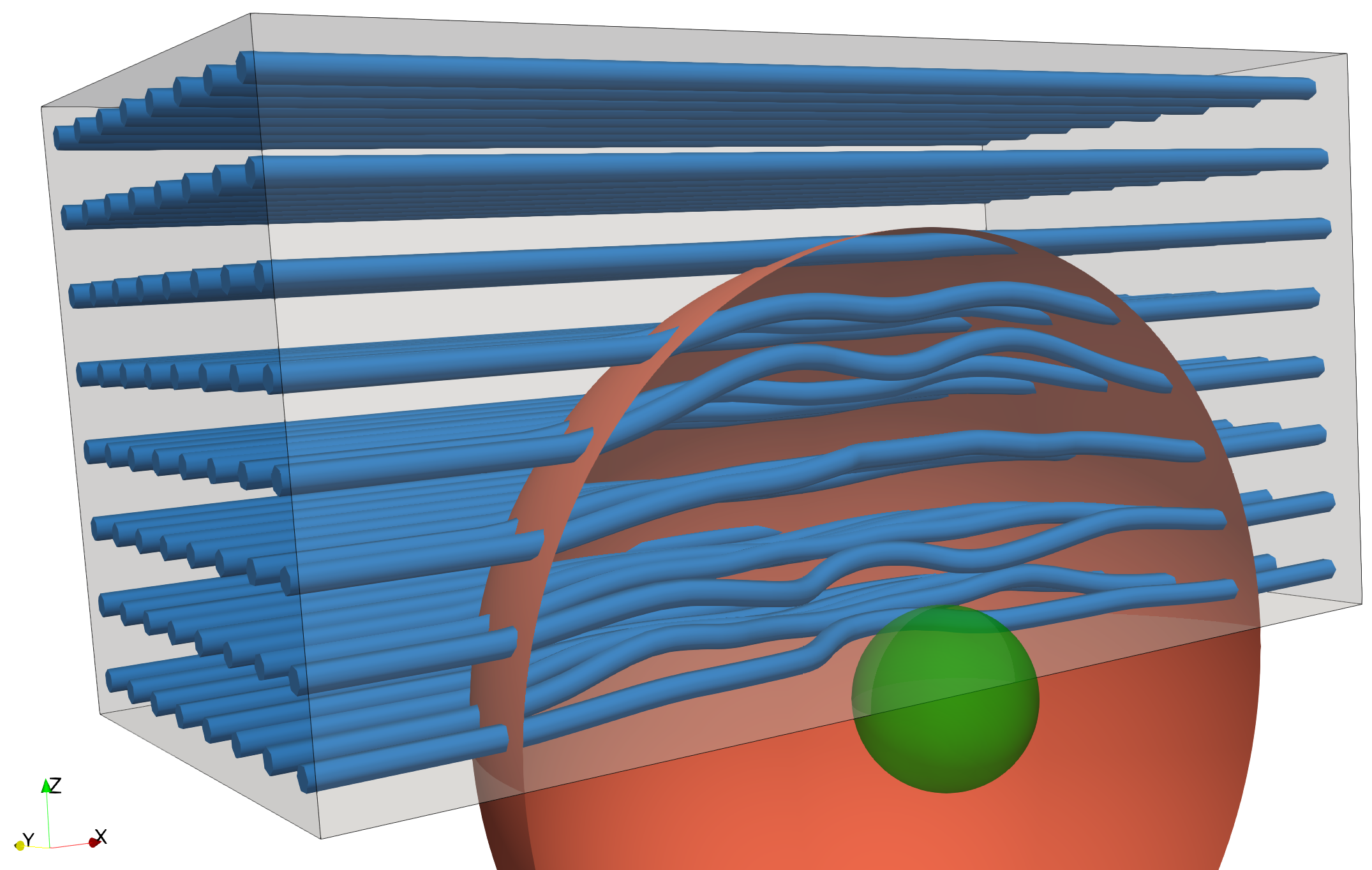}}
        \caption{Flux flow}
    \end{subfigure}\quad
    \begin{subfigure}[b]{0.5\textwidth}{\centering\includegraphics[width=1\textwidth]{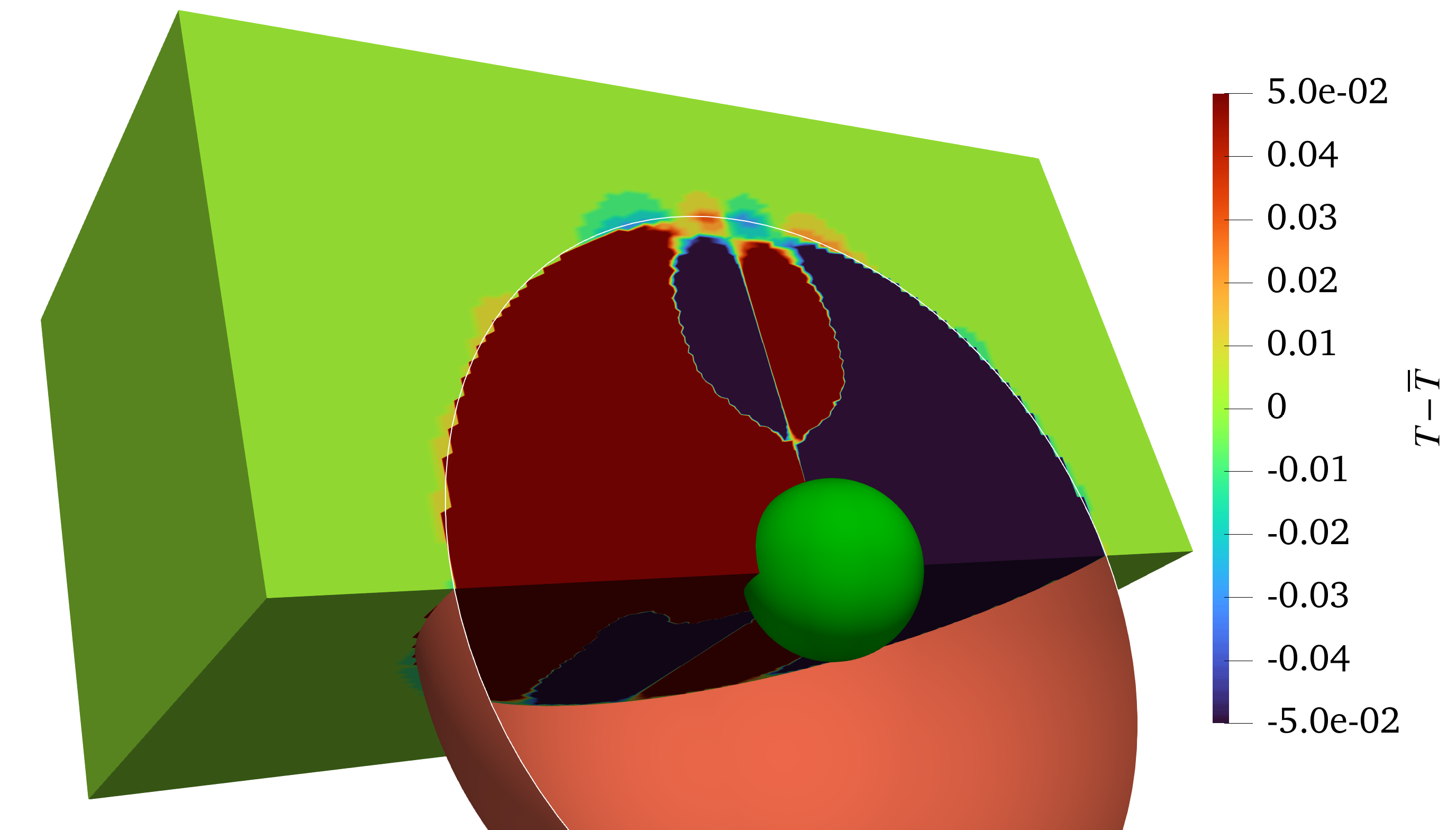}}
             \caption{Temperature difference $T-\overline{T}$}
    \end{subfigure} 
 \caption{Flux flow and temperature difference (with respect to the reference case) $T-\overline{T}$ for the 3D thermal cloak. EMT model and $N_{\rm var}=129$ are considered. Optimized objective function value $J_{\rm cloak}=1.23 \times 10^{-3}$. The 3D thermal cloak effectively diminishes the temperature disturbance in $\mathrm{\Omega}_{\mathrm{out}}$. The thermal cloak keeps the flux streamlines undisturbed and diminishes the temperature disturbance in $\mathrm{\Omega}_{\mathrm{out}}$. Temperature disturbances are almost nul in $\mathrm{\Omega}_{\mathrm{out}}$, even with the 3-order smaller temperature range.}
 \label{fig:Chen2015case 3D cloak tempDiff}
\end{figure}
\par The optimized material distributions of thermal cloaks design with two material models (EMT and Gyroid) are shown in \fref{fig:Chen2015case 3D cloak}. We use a design mesh with $N_{\rm var}=129$ and a solution mesh with DOF=124950. Similar to 2D thermal cloaks, the optimizations settle to the nearest local solutions for 3D thermal cloaks. Optimized material distributions reach the objective function value of order $10^{-3}-10^{-4}$. The material distribution varies slightly according to each material law and its limits on relative density, $v_{\rm min}$ \& $v_{\rm max}$. We exhibited the cloaking performance of the thermal cloak obtained using EMT in \fref{fig:Chen2015case 3D cloak tempDiff}. The figure presents the flux flow and temperature difference with respect to a homogeneous base material plate. From the figure, it is observed that the optimized thermal cloaks diminish the temperature disturbance in $\Omega_{\rm out}$. Temperature difference $T-\overline{T}$ is almost negligible in $\Omega_{\rm out}$ in the temperature range which is 3 orders smaller than actual temperature values.  
\subsubsection{Design with free-form geometries}
\label{sec:3D cloak Design with free-form geometries}
\begin{figure}[!htbp]
    \centering
    \includegraphics[width=1\linewidth]{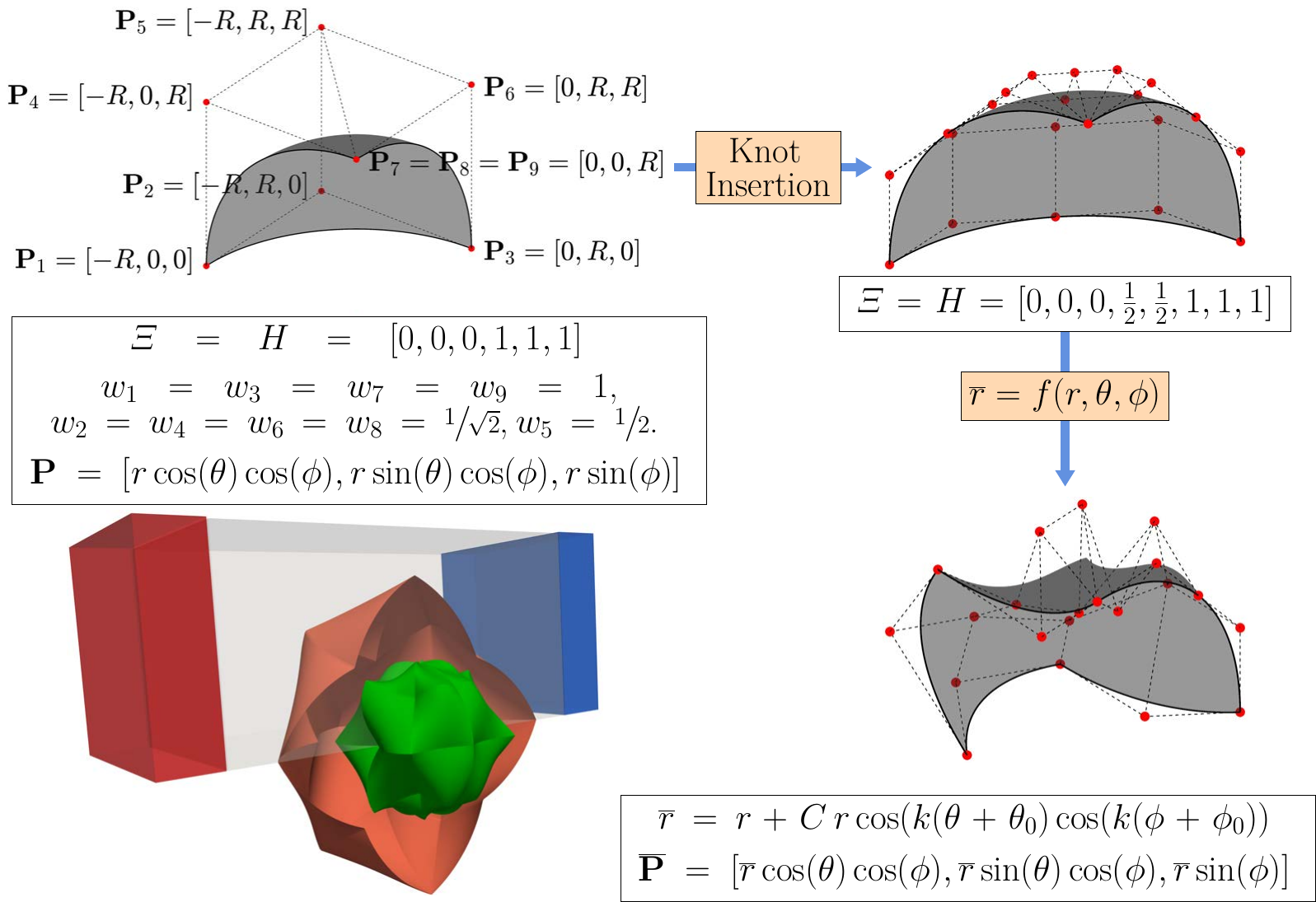}
 \caption{Steps to generate the star-shaped 3D insulator \& 3D thermal cloak and schematics of the final domain. The control points related to the circumferential parametric directions of the sphere of radius $R_{\rm in}$ and $R_{\rm out}$ are perturbed. The steps for perturbation are as follows: (i) $\sfrac{1}{4^{th}}$ of a NURBS-sphere is created using the knot vectors $\Xi=H=[0,0,0,1,1,1]$ with nine control points $\mathbf{P}_i$, $i=1,2,...,9$. The Cartesian coordinates and weights of the control points are shown in the first figure. (ii) The NURBS-geometry is refined by adding knots $\sfrac{1}{2},\sfrac{1}{2}$ through the knot insertion procedure in both $\Xi$ and $H$. (iii) The control points of the refined geometry are transformed into spherical coordinates $(r,\theta,\phi)$ from Cartesian coordinates $\mathbf{P}=(x,y,z)$; the radial coordinates $r$ are perturbed by the function, $\overline{r}=r+Cr \cos{(k(\theta+\theta_0))} \cos{(k(\phi+\phi_0))}$, while keeping $\theta$ and $\phi$ coordinates unchanged; the modified spherical coordinates $(\overline{r},\theta,\phi)$ are transformed back into Cartesian coordinates $\overline{\mathbf{P}}=(\overline{x},\overline{y},\overline{z})$. For $R_{\rm in}=25$~mm,  $C=0.3$, $k=5$, $\theta_0=\pi$ and $\phi_0=0$, while for $R_{\rm out}=50$~mm,  $C=0.2$, $k=4$, $\theta_0=-\pi$ and $\phi_0=\pi$. The entire geometry is created then using symmetry conditions along $y$-and $z$-planes.}
 \label{fig:Chen2015case_3D_geo_creation}
\end{figure}
\par
\begin{figure}[!htbp]
    \centering
    \begin{subfigure}[b]{0.43\textwidth}{\centering\includegraphics[width=0.31\textwidth]{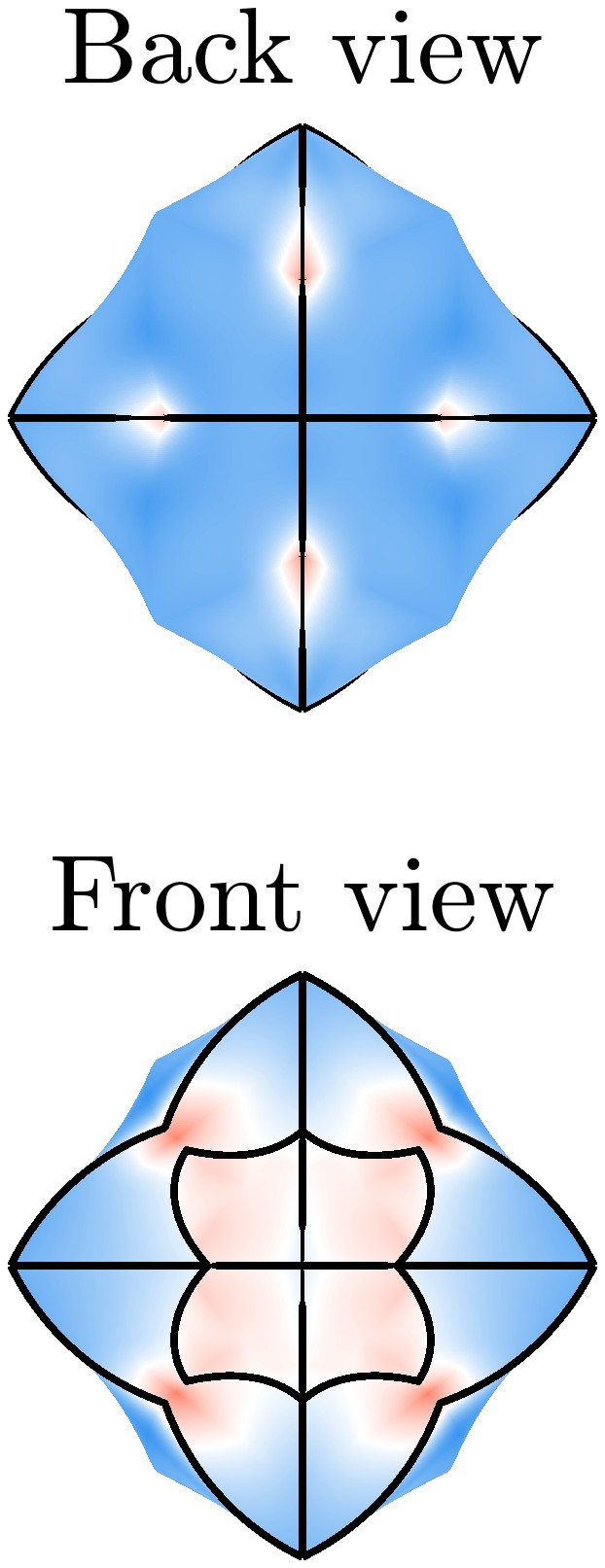}\includegraphics[width=0.69\textwidth]{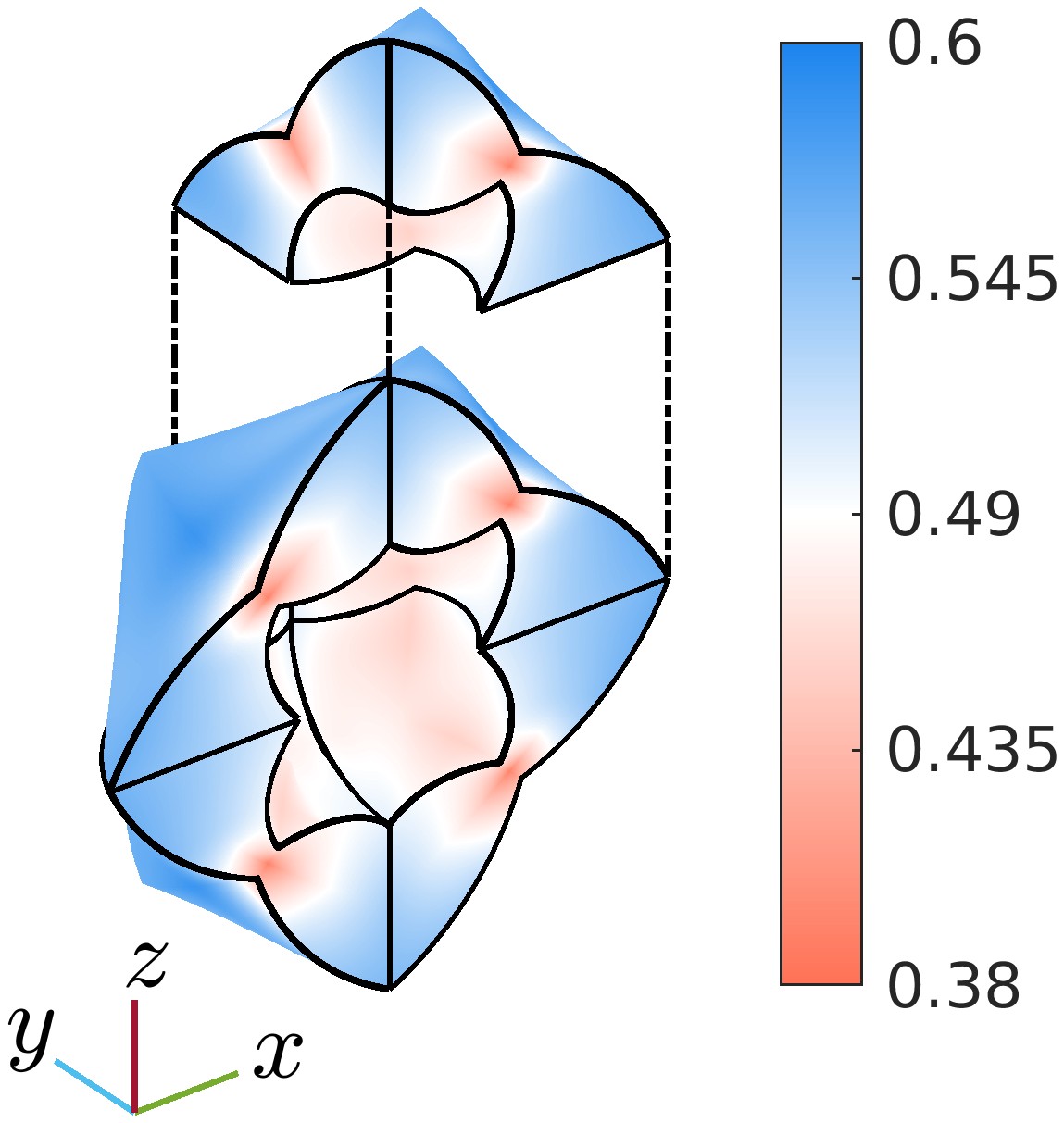}}
           \caption{Optimized material distribution}
            \label{fig:Chen2015case 3D cloak DiffGeo a}
    \end{subfigure} \\
    \begin{subfigure}[b]{0.445\textwidth}{\centering\includegraphics[width=1\textwidth]{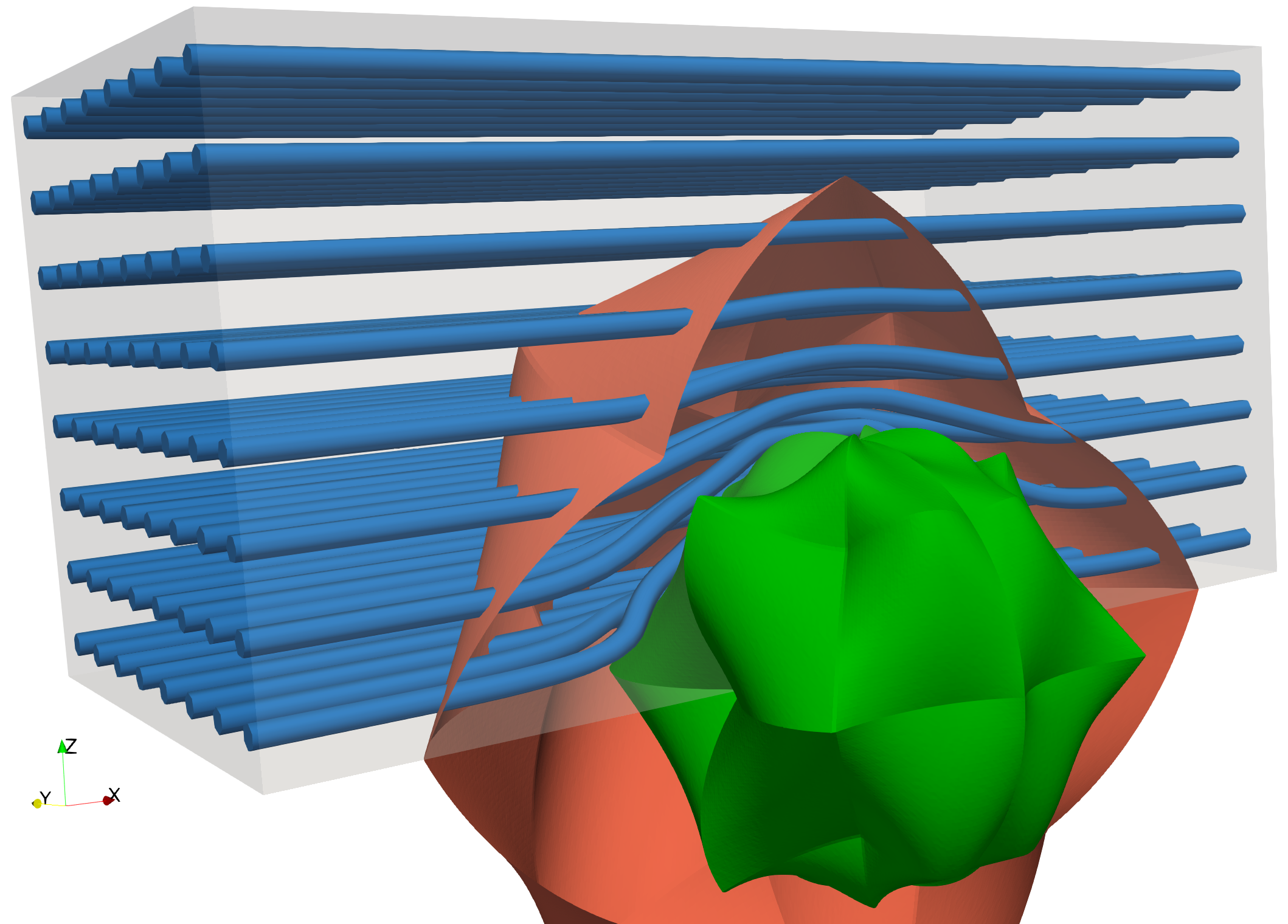}}
        \caption{Flux flow}
         \label{fig:Chen2015case 3D cloak DiffGeo b}
    \end{subfigure}
    \quad
    \begin{subfigure}[b]{0.51\textwidth}{\centering\includegraphics[width=1\textwidth]{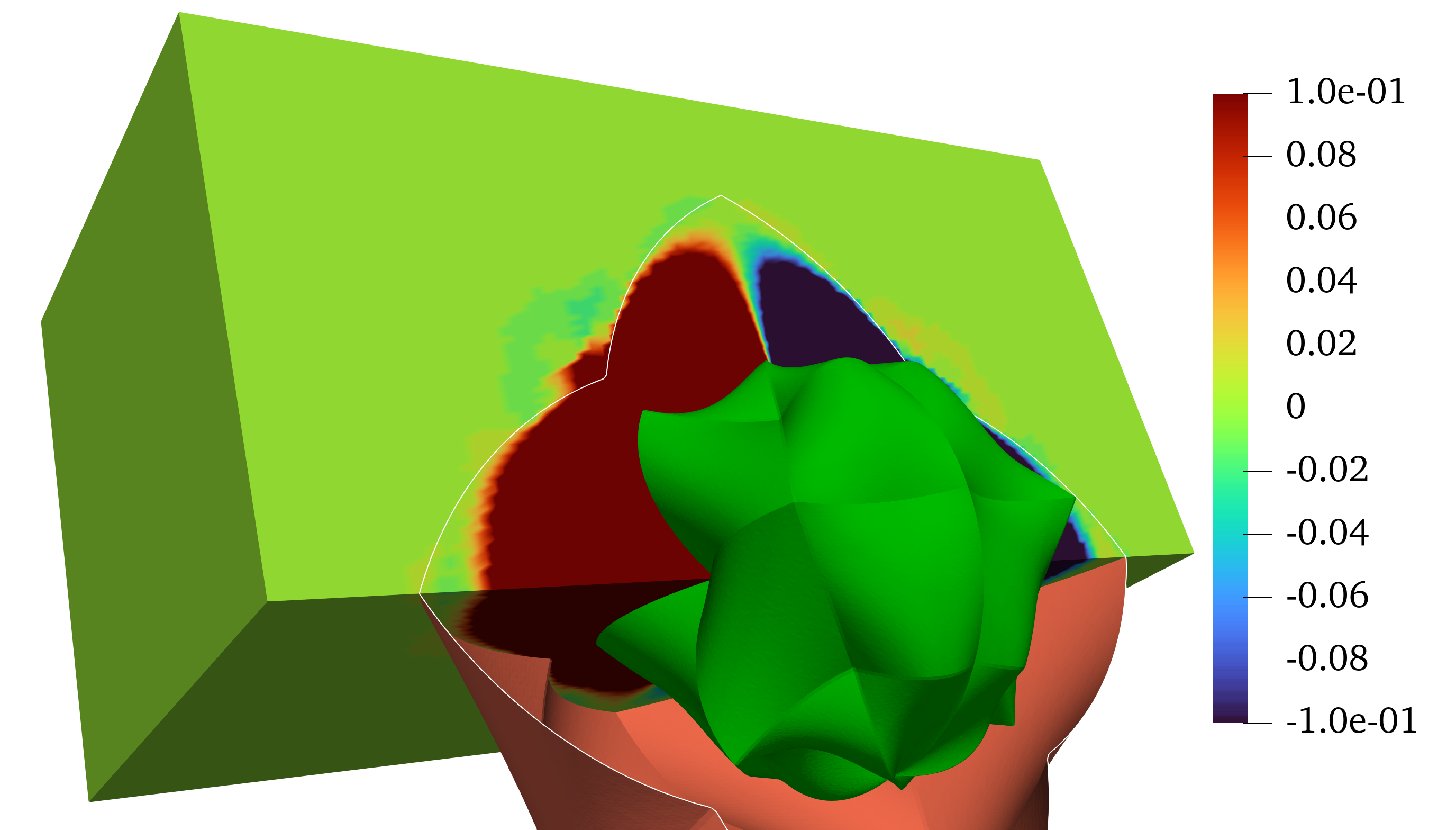}}
             \caption{Temperature difference $T-\overline{T}$}
              \label{fig:Chen2015case 3D cloak DiffGeo c}
    \end{subfigure} 
 \caption{Optimized material distribution, flux flow and temperature difference (with respect to the reference case) $T-\overline{T}$ for the 3D thermal cloak problem with complex shaped insulator and thermal cloak. EMT model and $N_{\rm var}=129$ are considered. Optimized objective function value $J_{\rm cloak}=2.35 \times 10^{-5}$. The 3D thermal cloak keeps the flux streamlines undisturbed and diminishes the temperature disturbance in $\mathrm{\Omega}_{\mathrm{out}}$. Temperature disturbances are almost nil in $\mathrm{\Omega}_{\mathrm{out}}$, even with the 3-order smaller temperature range.}
 \label{fig:Chen2015case 3D cloak DiffGeo}
\end{figure}

 In this subsection, we check a free-form geometry for a 3D thermal cloak and insulator. Similar to 2D free-form geometries as in \sref{sec:2D cloak Design with free-form geometries}, we define the free-form-shaped 3D insulator and thermal cloak by the method of perturbation. We consider the geometry to be symmetric along all three axes, and only $\sfrac{1}{8^{th}}$ part of spheres of radius $R_{\rm in}=25$~mm and $R_{\rm out}=50$~mm are taken into account for perturbation. Later, the knot insertion, coordinate-system transformations (from Cartesian to spherical \& from spherical to Cartesian) and functional transformation of radial coordinates are performed to perturb their control points. The detailed procedure is presented in \fref{fig:Chen2015case_3D_geo_creation}. We use $C=0.3$, $k=5$ $\theta_0=\pi$ and $\phi_0=0$ for $R_{\rm in}=25$~mm, while $C=0.2$, $k=4$ $\theta_0=\pi$ and $\phi_0=\pi$ for $R_{\rm in}=50$~mm. To create overall $\sfrac{1}{4^{th}}$ of the entire geometry for numerical analysis, the mirror images of both spheres are taken along $x$-plane. The final domain created for the problem is also shown in \fref{fig:Chen2015case_3D_geo_creation}. The results of the optimization are presented in \fref{fig:Chen2015case 3D cloak DiffGeo}. From the figure, we can observe that the material distribution presented in \fref{fig:Chen2015case 3D cloak DiffGeo a} diminishes the temperature disturbance created by the insulator with the objective function value $J=2.35 \times 10^{-5}$. Also, the flux flow shows undisturbed, parallel to $x$-axes streamlines in $\Omega_{\rm out}$ region. 

%
%
%
%
\section{Other thermal meta-structures}
\label{sec:Other thermal metamaterials}
\par In this section, we investigate various other thermal meta-structures such as thermal concentrators, thermal rotators, thermal cloaking sensors, thermal cloaking concentrators and thermal bidirectional cloak-concentrators. Each of these manipulators is associated with a distinct objective function. For all manipulators, we take an geometry identical to that of the thermal cloak problem as depicted in \fref{fig:chen2015case Schematics} for 2D cases and in \fref{fig:Chen2015case schematics 3D cloak} for 3D cases. The dimensions and material allocations can vary example-wise, which will be mentioned in their respective descriptions.
\subsection{Thermal concentrator}
\label{sec:thermal concentrator} 
\par In this example, we design both 2D and 3D thermal concentrators. For the concentrator, we consider the same dimensions and boundary conditions as the thermal cloak problem. Even the materials involved are the same with only differences in their allocation. In this thermal meta-structure, $\Omega_{\rm in}$ is filled with the base material iron instead of the insulator. 
\par The objective of a thermal concentrator is to concentrate the flux inside region $\Omega_{\rm in}$. Thus, we define the concentration function in a mathematical sense as follows:
\begin{equation} \label{eq:concentrating fn}
    \varPsi_{\mathrm{cntr}}=\dfrac{1}{\widetilde{\varPsi}_{\mathrm{cntr}}} \int_{\Gamma_{\mathrm{in}}} - \kappa_{\mathrm{in}} \nabla T \cdot \mathbf{n}~d\Gamma, \quad \text{with} \quad
    \widetilde{\varPsi}_{\mathrm{cntr}}= \int_{\Gamma_{\mathrm{in}}} - \overline{\kappa}_{\mathrm{in}} \nabla \overline{T} \cdot \mathbf{n}~d\Gamma, 
\end{equation}
where $\overline{T}$ is the temperature when entire $\mathrm{\Omega}$ is filled with the base material. 
From a physical perspective, the concentration function value represents the concentrated flux as a multiple of the flux in the homogeneous base plate case. Since we solve the minimization problem in optimization, the original objective function is defined as the inverse of the absolute value of a concentration function, $J_{\rm cntr}= \dfrac{1}{ \varPsi_{\mathrm{cntr}}}$. Before running the optimizations, we also perform a mesh sensitivity analysis and take a mesh with DOF=13167 as the solution mesh, similar to \sref{sec:Chen2015case 2D cloak} . The results of the mesh sensitivity analysis are also presented in \ref{sec:2D cloak Mesh study}.

\renewcommand{\arraystretch}{1.5}   
\begin{figure}[!htbp]
\centering
\scalebox{0.88}{
\begin{tabular}[c]{| M{5.4em} | M{5.45em} | M{5.45em} | M{5.45em}| M{5.45em} | M{7.7em} |}
\hline 
 \centering EMT \\ \vspace{0.1cm} $v_i\in[0,1]$\vspace{-0.2cm}& \centering Maxwell \\ \vspace{0.1cm} $v_i\in[0,1]$\vspace{-0.2cm} & \centering Porous Cu \\ \vspace{0.1cm} $v_i\in[0,0.7]$ \vspace{-0.2cm} & \centering Cu-Sn-Pb \\ \vspace{0.1cm} $v_i\in[0,0.3]$ \vspace{-0.2cm} & \centering TCOH \\ \vspace{0.1cm} ${v_i\in[0.2,0.8]}$\vspace{-0.2cm} & \begin{center}
    Gyroid\\ \vspace{0.1cm} $v_i\in[0.2,0.9]$\vspace{-0.2cm}
\end{center}
\\  
\hline
   \vspace{0.2cm}
    \begin{subfigure}[t]{0.15\textwidth}{\includegraphics[width=1\textwidth]{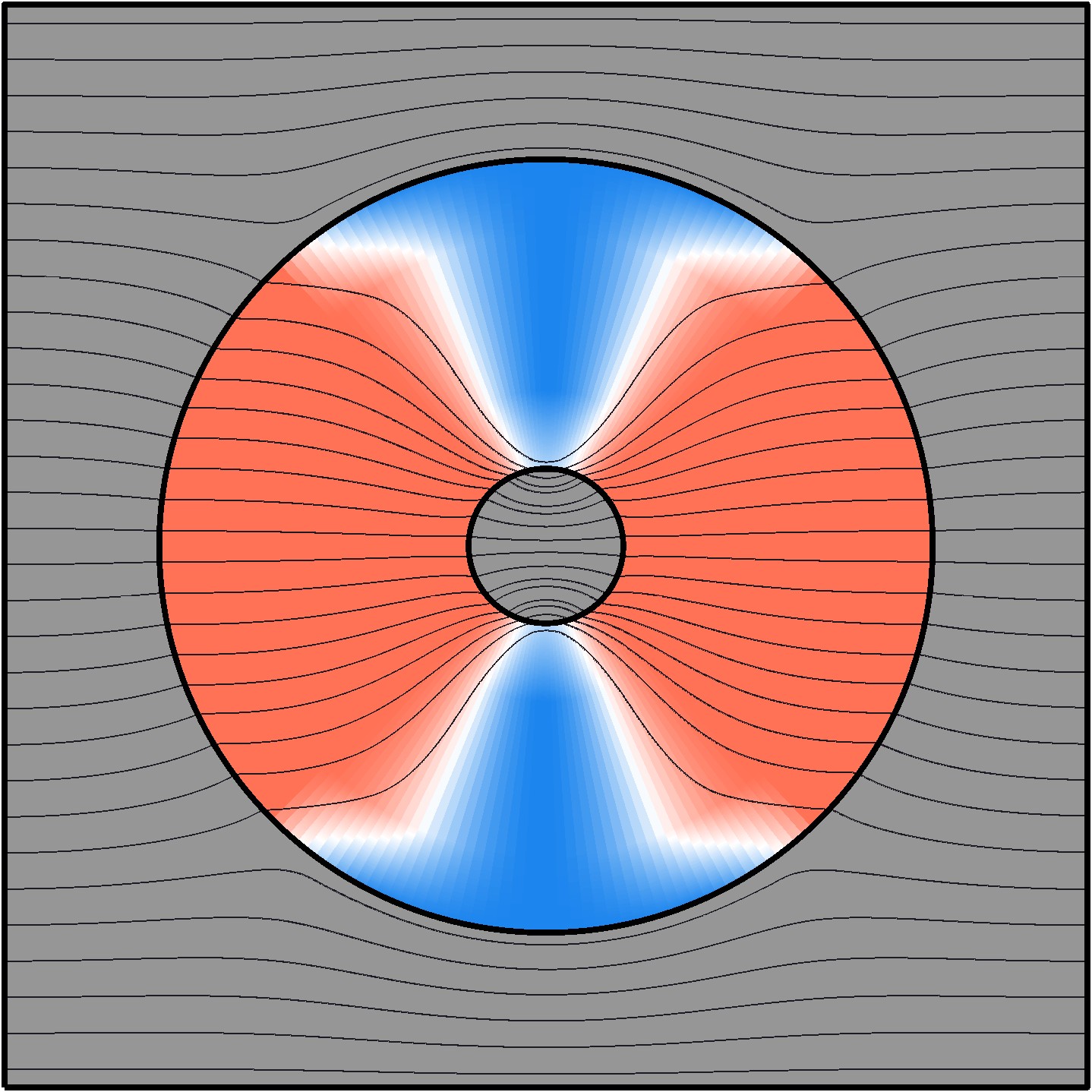}}
        \caption{\centering $N_{\rm var}=25$, $J=2.19\times 10^{-1}$, $\varPsi_{\rm cntr}=4.57$}
    \end{subfigure}  & \vspace{0.2cm}
    \begin{subfigure}[t]{0.15\textwidth}{\includegraphics[width=1\textwidth]{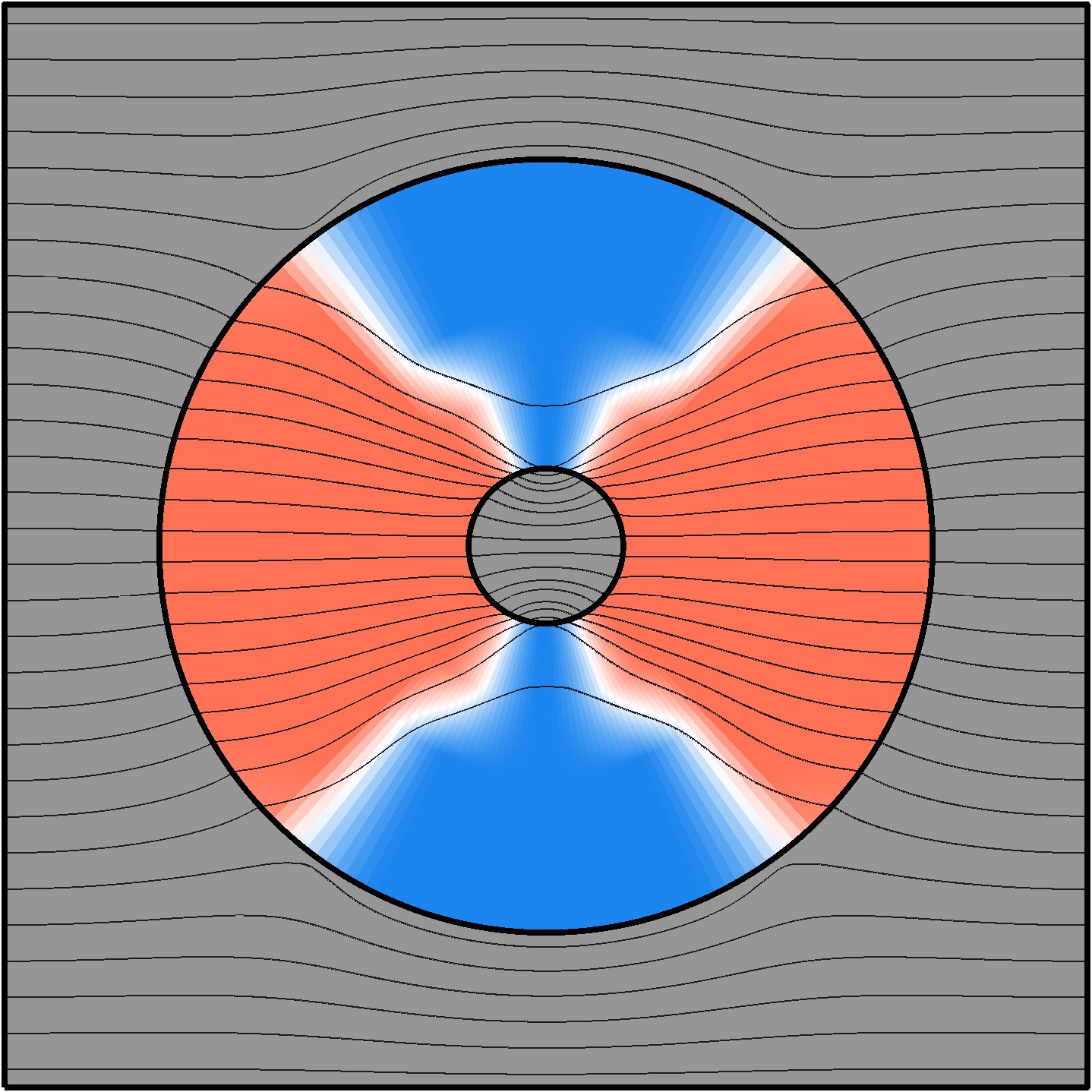}}
        \caption{\centering $N_{\rm var}=25$, $J=2.27\times 10^{-1}$, $\varPsi_{\rm cntr}=4.41$}
    \end{subfigure} & \vspace{0.2cm}
    \begin{subfigure}[t]{0.15\textwidth}{\includegraphics[width=1\textwidth]{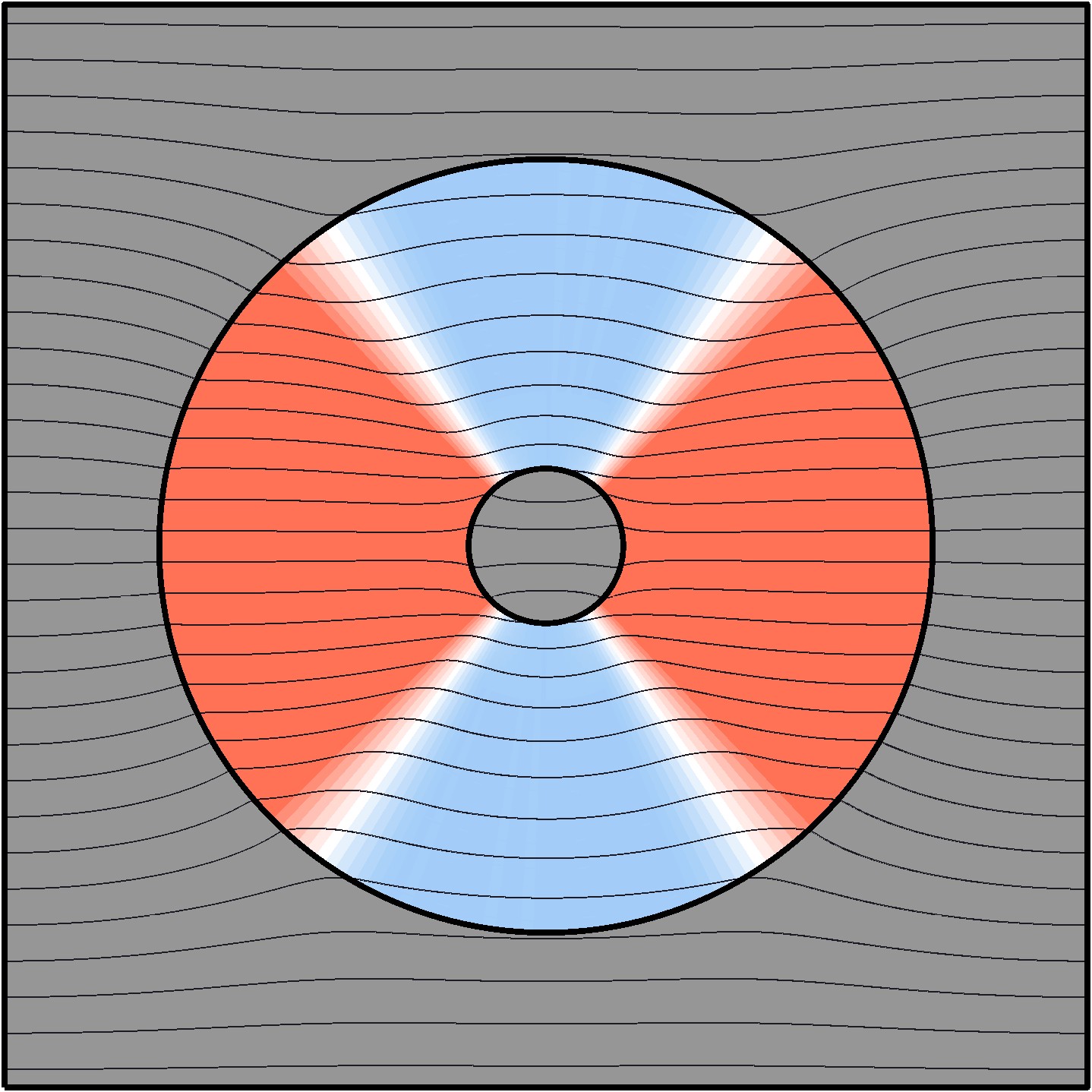}}
        \caption{\centering $N_{\rm var}=25$, $J=4.72\times 10^{-1}$, $\varPsi_{\rm cntr}=2.12$}
    \end{subfigure} & \vspace{0.2cm}
    \begin{subfigure}[t]{0.15\textwidth}{\includegraphics[width=1\textwidth]{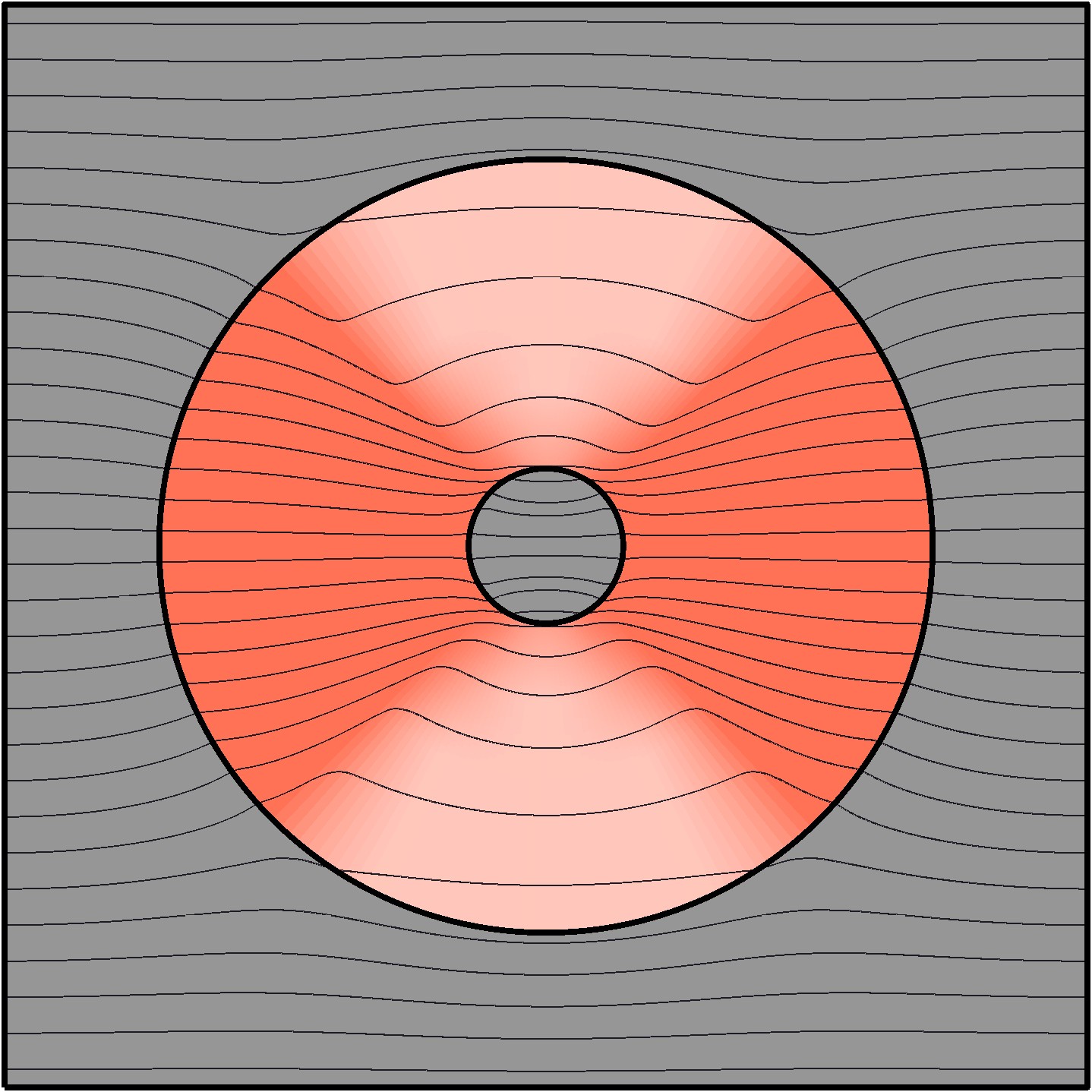}}
        \caption{\centering $N_{\rm var}=25$, $J=3.35\times 10^{-1}$, $\varPsi_{\rm cntr}=2.98$}
    \end{subfigure}& \vspace{0.2cm}
    \begin{subfigure}[t]{0.15\textwidth}{\includegraphics[width=1\textwidth]{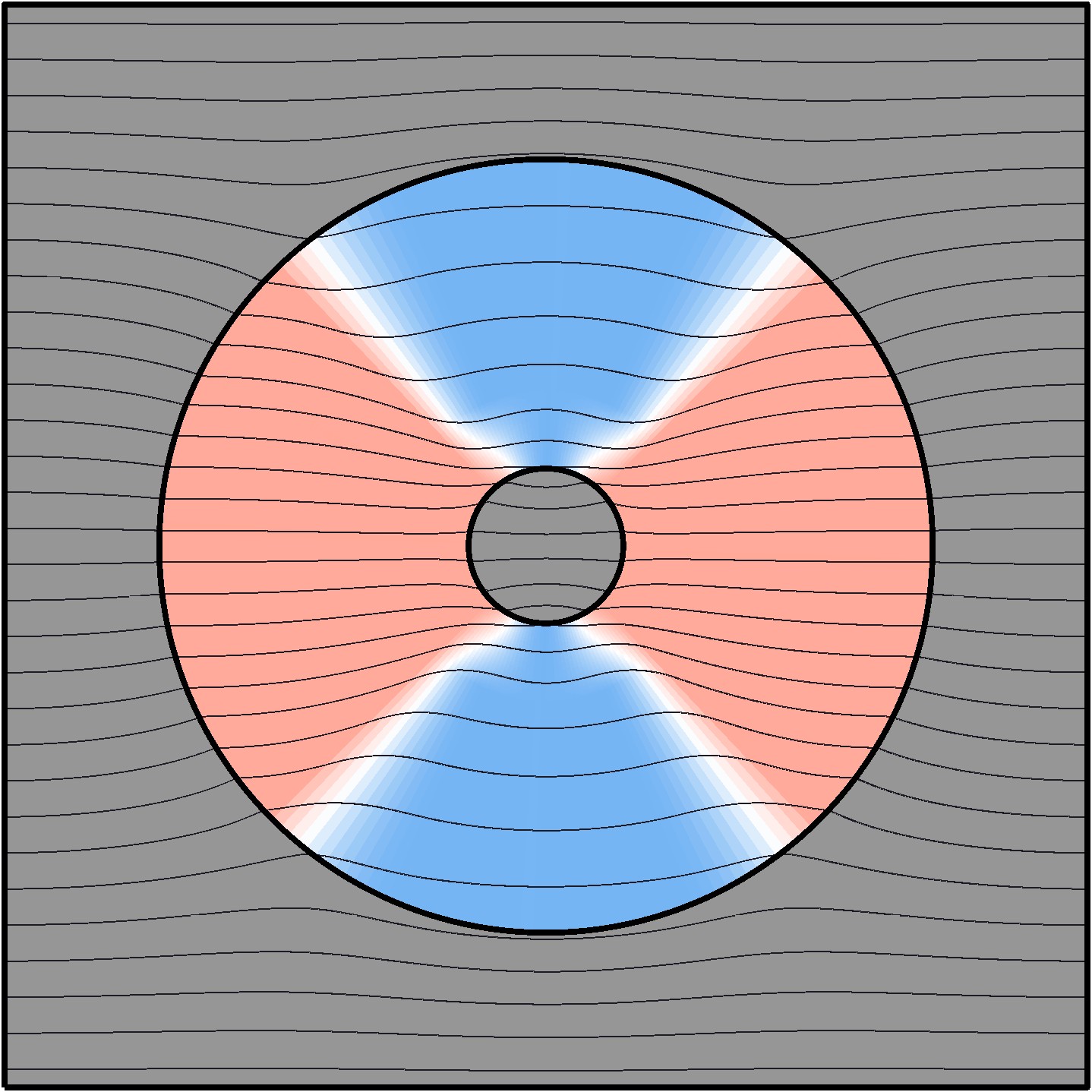}}
        \caption{\centering $N_{\rm var}=25$, $J=4.34\times 10^{-1}$, $\varPsi_{\rm cntr}=2.31$}
    \end{subfigure}
    & \vspace{0.1cm}
    \begin{subfigure}[t]{0.15\textwidth}{\includegraphics[width=1\textwidth]{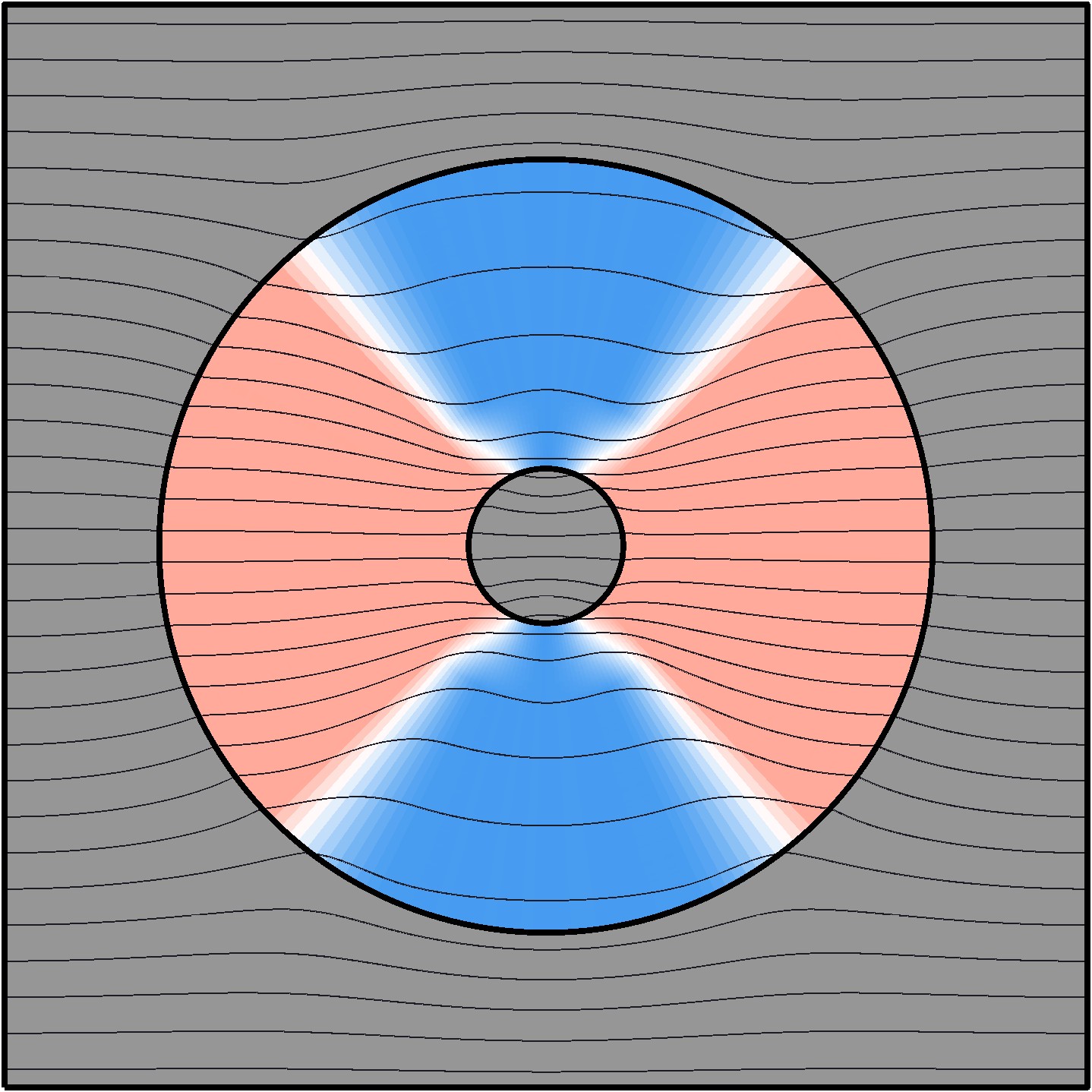}}
        \caption{\centering $N_{\rm var}=25$, $J=3.72\times 10^{-1}$, $\varPsi_{\rm cntr}=2.69$}
    \end{subfigure}~\begin{subfigure}[b]{0.05\textwidth}{
\includegraphics[keepaspectratio=false,width=1.1\textwidth,height=2.45cm]{colorbar_VF2.jpg}} 
\end{subfigure} \\   
\hline
   \vspace{0.2cm}
    \begin{subfigure}[t]{0.15\textwidth}{\includegraphics[width=1\textwidth]{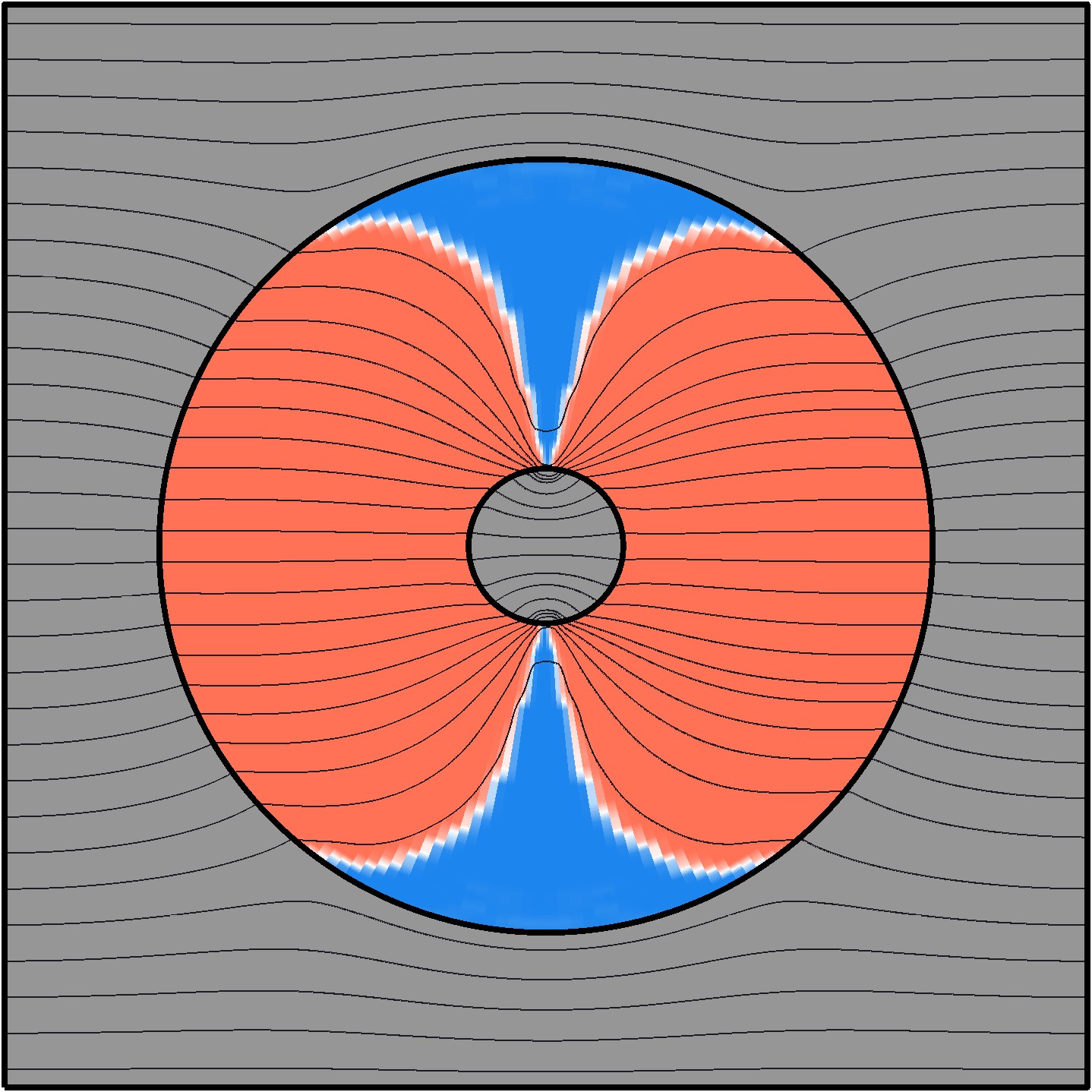}}
        \caption{\centering $N_{\rm var}=1089$, $J=1.73\times 10^{-1}$, $\varPsi_{\rm cntr}=5.77$}
    \end{subfigure}  & \vspace{0.2cm}
    \begin{subfigure}[t]{0.15\textwidth}{\includegraphics[width=1\textwidth]{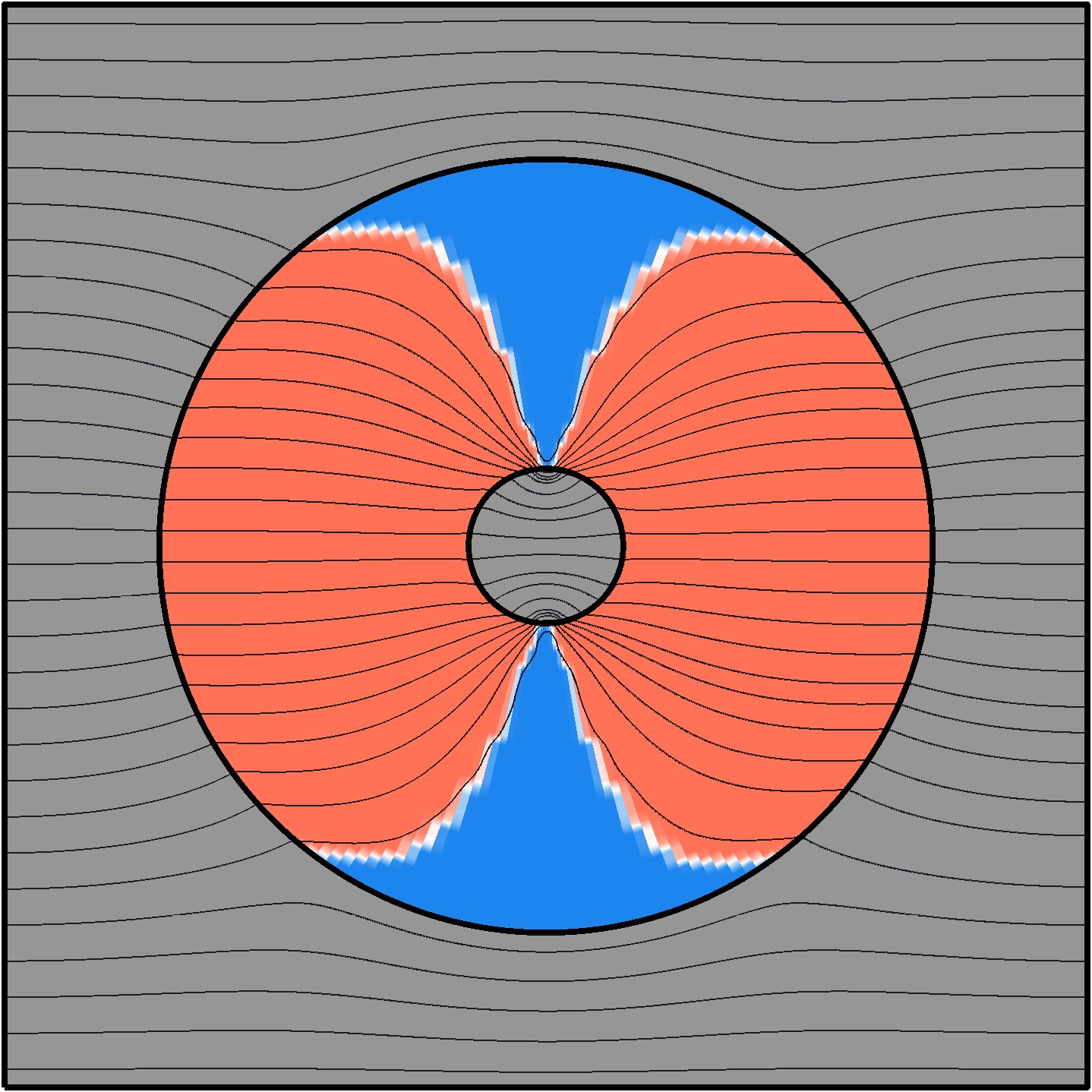}}
        \caption{\centering $N_{\rm var}=1089$, $J=1.86\times 10^{-1}$, $\varPsi_{\rm cntr}=5.39$}
    \end{subfigure} & \vspace{0.2cm}
    \begin{subfigure}[t]{0.15\textwidth}{\includegraphics[width=1\textwidth]{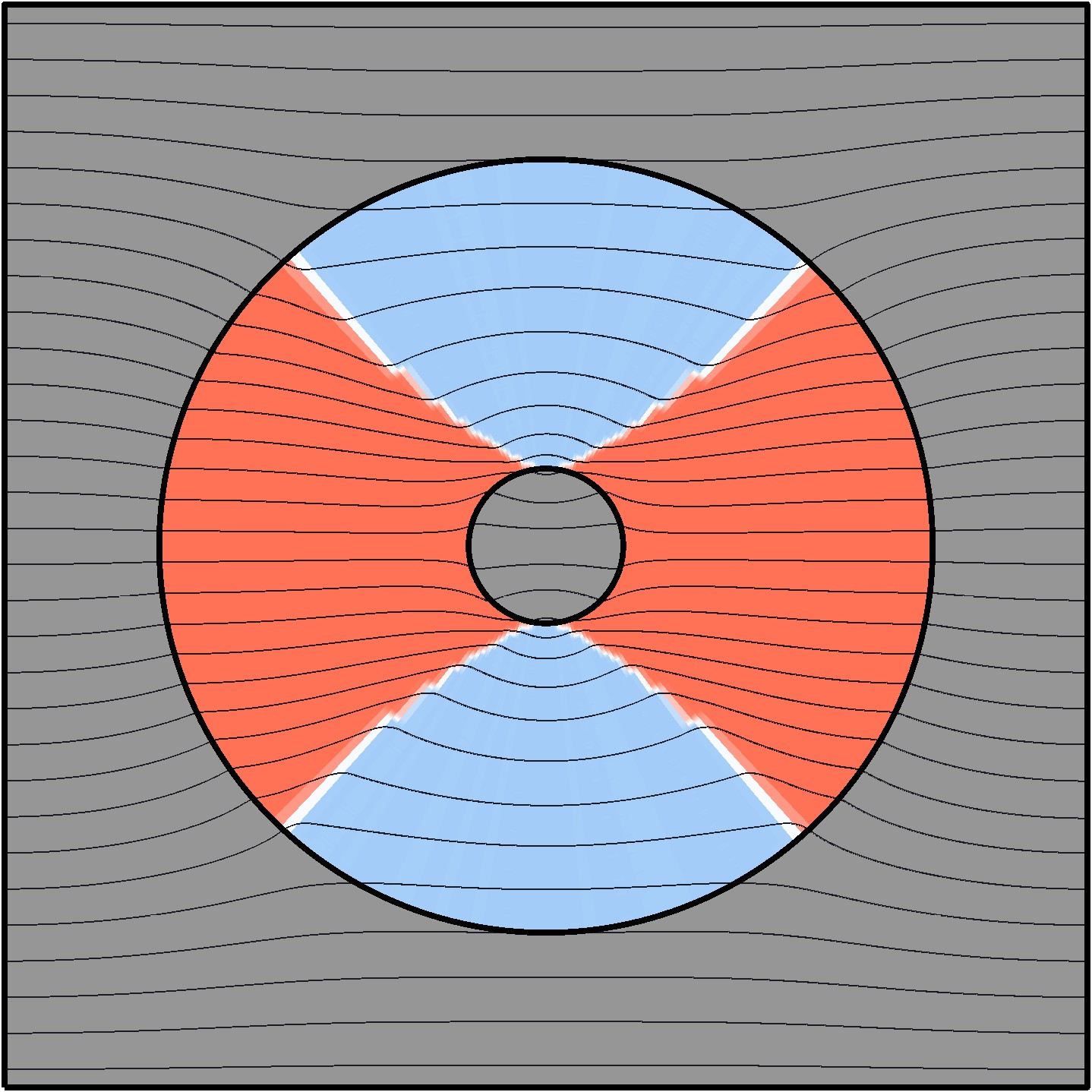}}
        \caption{\centering $N_{\rm var}=1089$, $J=4.38\times 10^{-1}$, $\varPsi_{\rm cntr}=2.28$}
    \end{subfigure} & \vspace{0.2cm}
    \begin{subfigure}[t]{0.15\textwidth}{\includegraphics[width=1\textwidth]{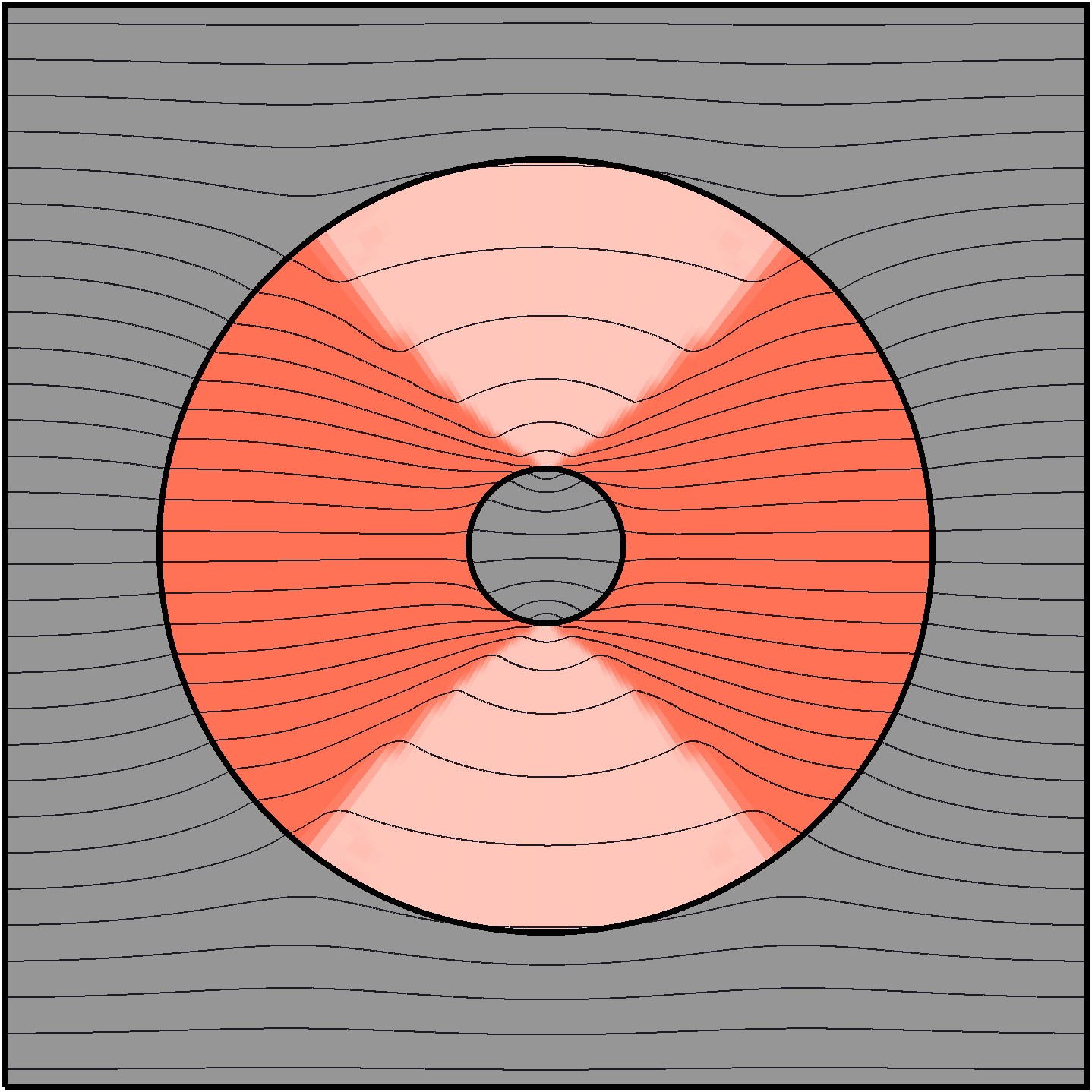}}
        \caption{\centering $N_{\rm var}=1089$, $J=3.01\times 10^{-1}$, $\varPsi_{\rm cntr}=3.32$}
    \end{subfigure}& \vspace{0.2cm}
    \begin{subfigure}[t]{0.15\textwidth}{\includegraphics[width=1\textwidth]{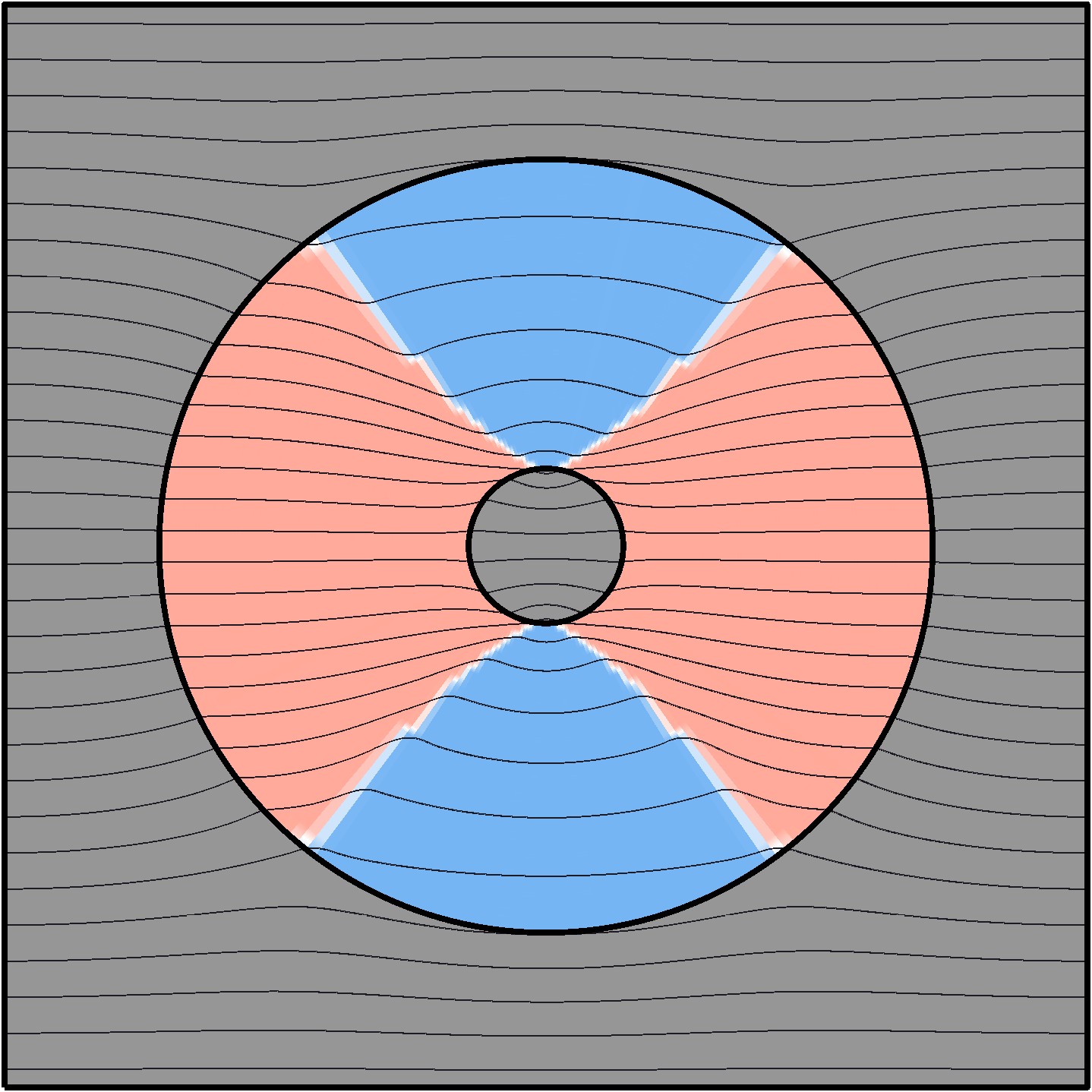}}
        \caption{\centering $N_{\rm var}=1089$, $J=4.02\times 10^{-1}$, $\varPsi_{\rm cntr}=2.49$}
    \end{subfigure}
    & \vspace{0.1cm}
    \begin{subfigure}[t]{0.15\textwidth}{\includegraphics[width=1\textwidth]{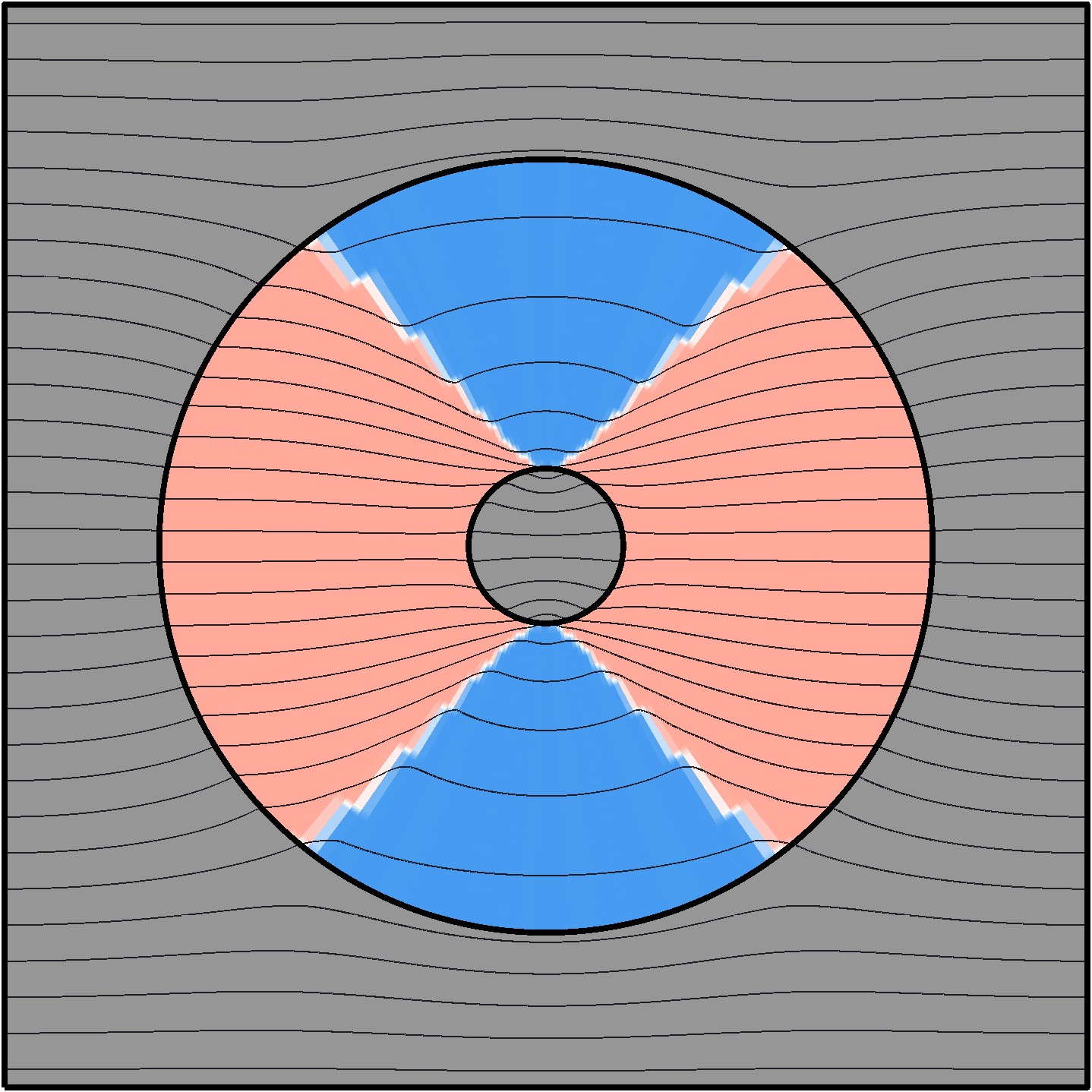}}
        \caption{\centering $N_{\rm var}=1089$, $J=3.36\times 10^{-1}$, $\varPsi_{\rm cntr}=2.97$}
    \end{subfigure}~\begin{subfigure}[b]{0.05\textwidth}{
\includegraphics[keepaspectratio=false,width=1.1\textwidth,height=2.45cm]{colorbar_VF2.jpg}}  
\end{subfigure}\\ 
    \hline 
    \end{tabular}
}

\caption{Optimized material distributions for the thermal concentrator problem. Six material models and $N_{\rm var}=25$,  $1089$ are considered. Optimized concetrator function $\varPsi_{\rm cntr}$ values are from $2.12$ to $5.77$. All optimized material distributions are close to the sector-type geometry with alternative sectors of $\kappa_{\rm max}$-material
and $\kappa_{\rm min}$-material. The intermediate densities/conductivities do not play much role
in the design, and they are only present as the thinnest possible transition from $\kappa_{\rm min}$ to $\kappa_{\rm max}$ with the given design mesh.}  
    \label{fig:chen2015case cntr}
\end{figure}

\begin{figure}[!htbp]
    \centering
    \setlength\figureheight{1\textwidth}
    \setlength\figurewidth{1\textwidth}
\includegraphics[width=0.8\textwidth]{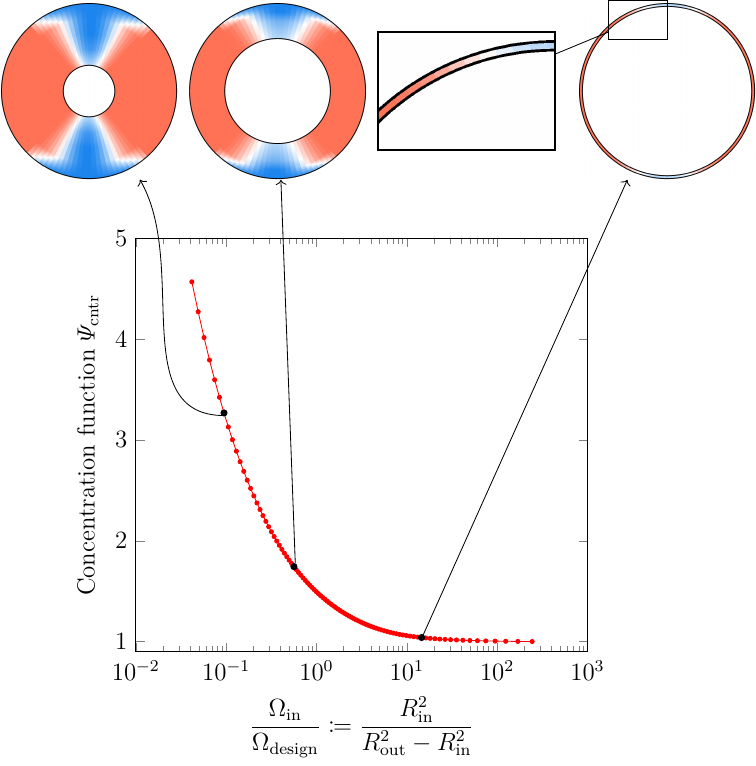}
 \caption{Change in the objective function $J_{\rm cntr}$ with respect to $\Omega_{\rm in}/\Omega_{\rm design}$. $R_{\rm in}$ is varied between $10$~mm to 49~mm, while keeping $R_{\rm out}=50$~mm constant. The optimized material distribution does not show any significant effect related to $R_{\rm in}$ increment, and maintains sector-type structure throughout.}
 \label{fig:Chen2015case cntr riVar}
\end{figure}
\par Next, we design 2D thermal concentrators using all 6 material models and two design meshes with $N_{\rm var}=25$ and $1089$. The results are shown in \fref{fig:chen2015case cntr}. We observe that the designed thermal concentrators can concentrate from 2.12 to 5.77 times more flux compared to a homogeneous plate. All optimized material distributions are close to the sector-type geometry with alternative sectors of $\kappa_{\rm max}$-material and $\kappa_{\rm min}$-material. The intermediate densities/conductivities do not play much role in the design, and they are only present as the thinnest possible transition from $\kappa_{\rm min}$ to $\kappa_{\rm max}$ with the given design mesh. Therefore, for a larger value of $N_{\rm var}$ such as $N_{\rm var}=1089$, the intermediate densities almost vanish before making the material distribution close to $\kappa_{\rm min}$-$\kappa_{\rm max}$ design.
\par Thermal concentrators made of sector-type geometries are already studied and experimentally demonstrated in Chen~\textit{et.al.}~\cite{chenExperimentalRealizationExtreme2015}.  Our current results align well with the earlier results from Jansari \textit{et.al.}~\cite{jansari2022design}, showcasing that the geometry with 4 sectors gives the best results. From both results, we can say that to concentrate more flux in $\Omega_{\rm in}$, the $\kappa_{\rm max}$ sectors need to widen their arc at $R_{\rm in}$ as well as $R_{\rm out}$. However, how much widening is feasible with the given configuration is dependent on the available thermal conductivities and the design freedom. 
\par For all models with $N_{\rm var}=1089$, the arcs of $\kappa_{\rm max}$ sectors at $R_{\rm in}$ cover almost the entire perimeter, only leaving a very small length for $\kappa_{\rm min}$ sectors. One the other hand, the arcs of $\kappa_{\rm max}$ at $R_{\rm out}$ are almost similar for $N_{\rm var}=25$ \& $N_{\rm var}=1089$. The curvature between $\kappa_{\rm min}$ \& $\kappa_{\rm max}$ sectors also play an important part in guiding the extra flux towards $\Omega_{\rm in}$. For EMT and Maxwell models, the curvature is bigger compared to other models for $N_{\rm var}=1089$. This can mainly be attributed to the fact that the larger difference between $\kappa_{\rm min}$-$\kappa_{\rm max}$ allows to generate bigger curvature while maintaining the interface conditions. For other models, the curvature is constrained by a small difference between $\kappa_{\rm min}$ \& $\kappa_{\rm max}$. 
\renewcommand{\arraystretch}{1.5}   
\begin{figure}[!htbp]
\centering
\scalebox{1}{
\begin{tabular}[c]{| M{5.4em} | M{5.45em} | M{5.45em} | M{5.45em}| M{7.7em} |}
\hline 
 $v_i=0$ & $v_i=0.25$ & $v_i=0.5$ & $v_i=0.75$  & $v_i=1$ \\  
\hline
    \vspace{0.2cm}
    \begin{subfigure}[t]{0.15\textwidth}{\includegraphics[width=1\textwidth]{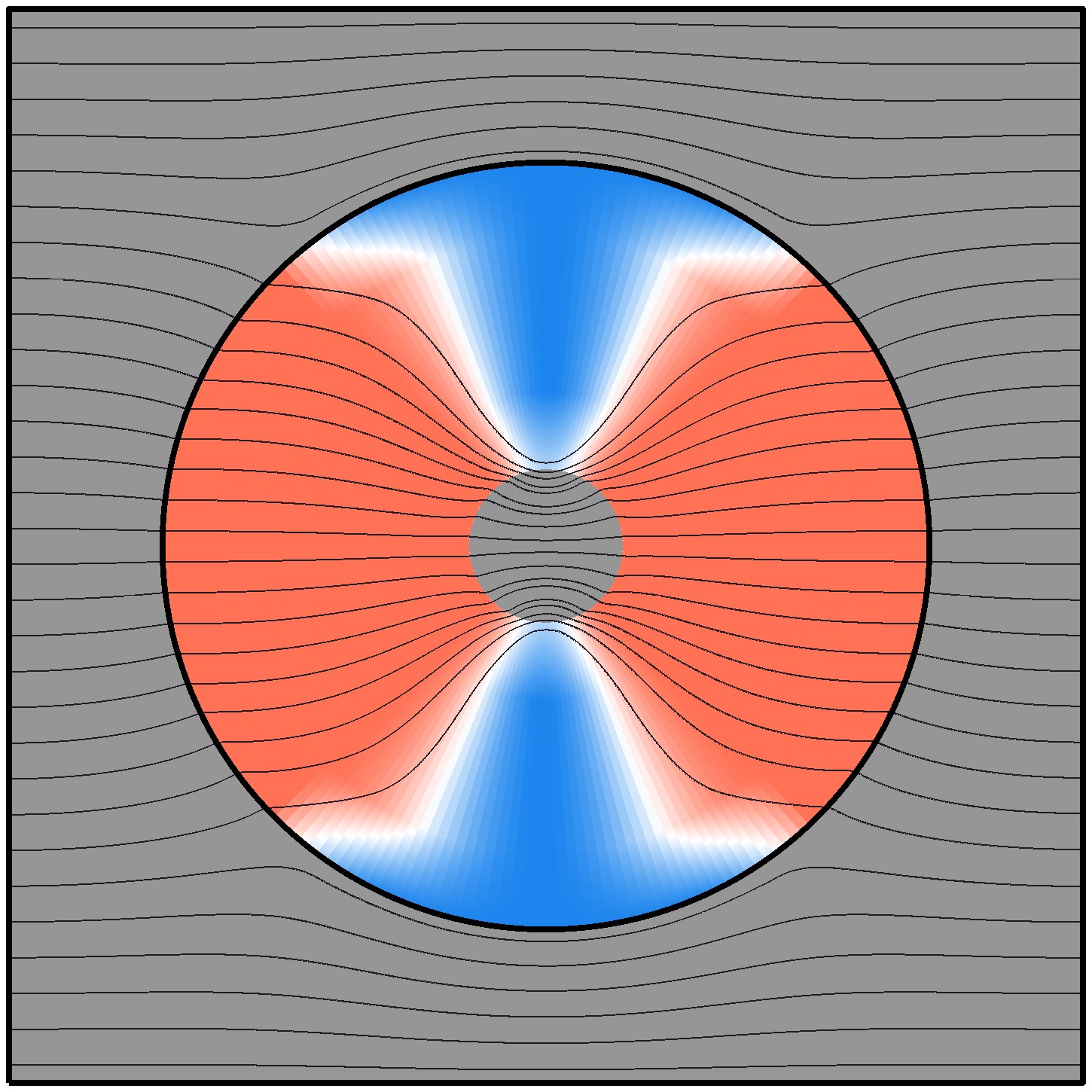}}
        \caption{\centering $N_{\rm var}=25$, $J=2.19\times 10^{-1}$}
    \end{subfigure}  & \vspace{0.2cm}
    \begin{subfigure}[t]{0.15\textwidth}{\includegraphics[width=1\textwidth]{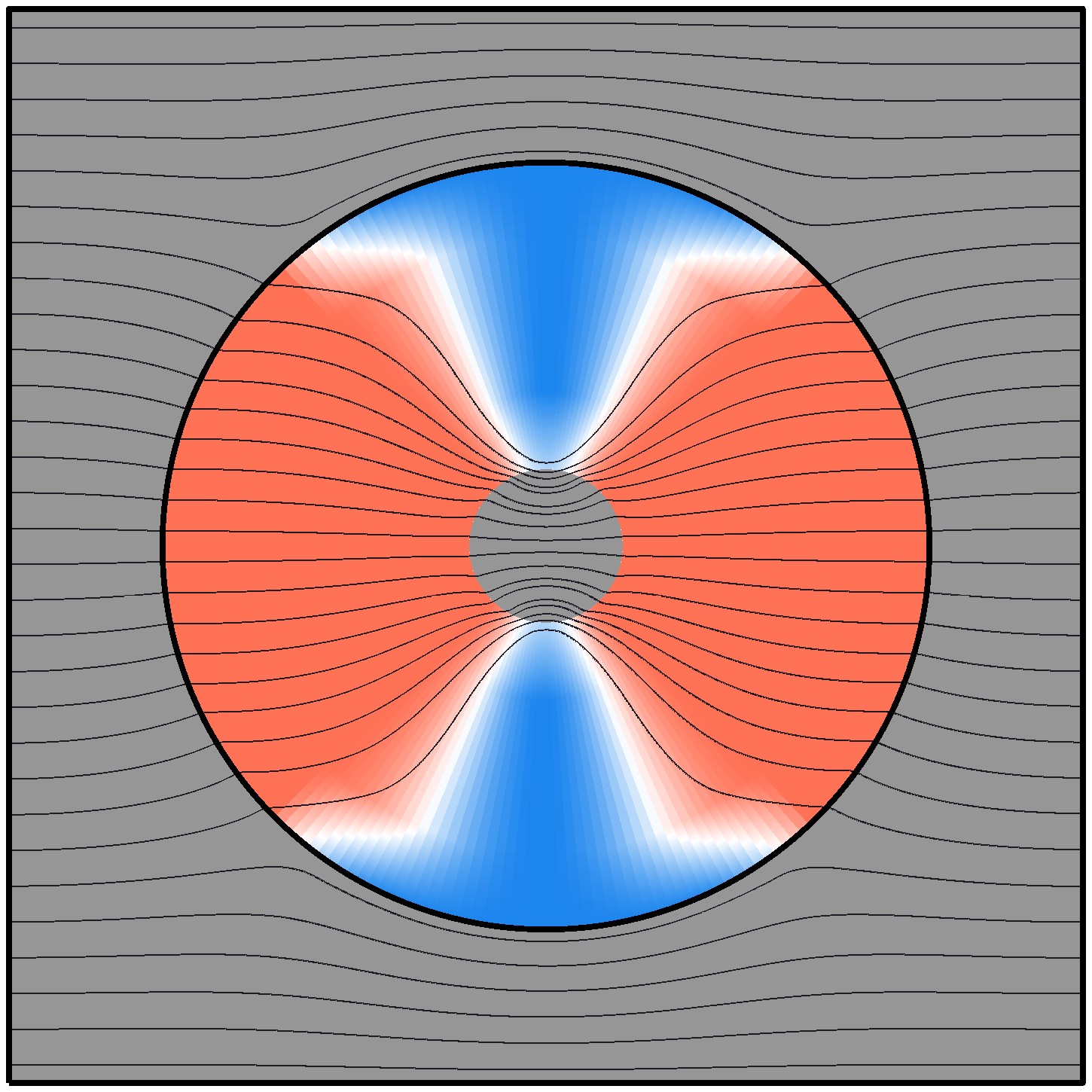}}
        \caption{\centering $N_{\rm var}=25$, $J=2.19\times 10^{-1}$}
    \end{subfigure} & \vspace{0.2cm}
    \begin{subfigure}[t]{0.15\textwidth}{\includegraphics[width=1\textwidth]{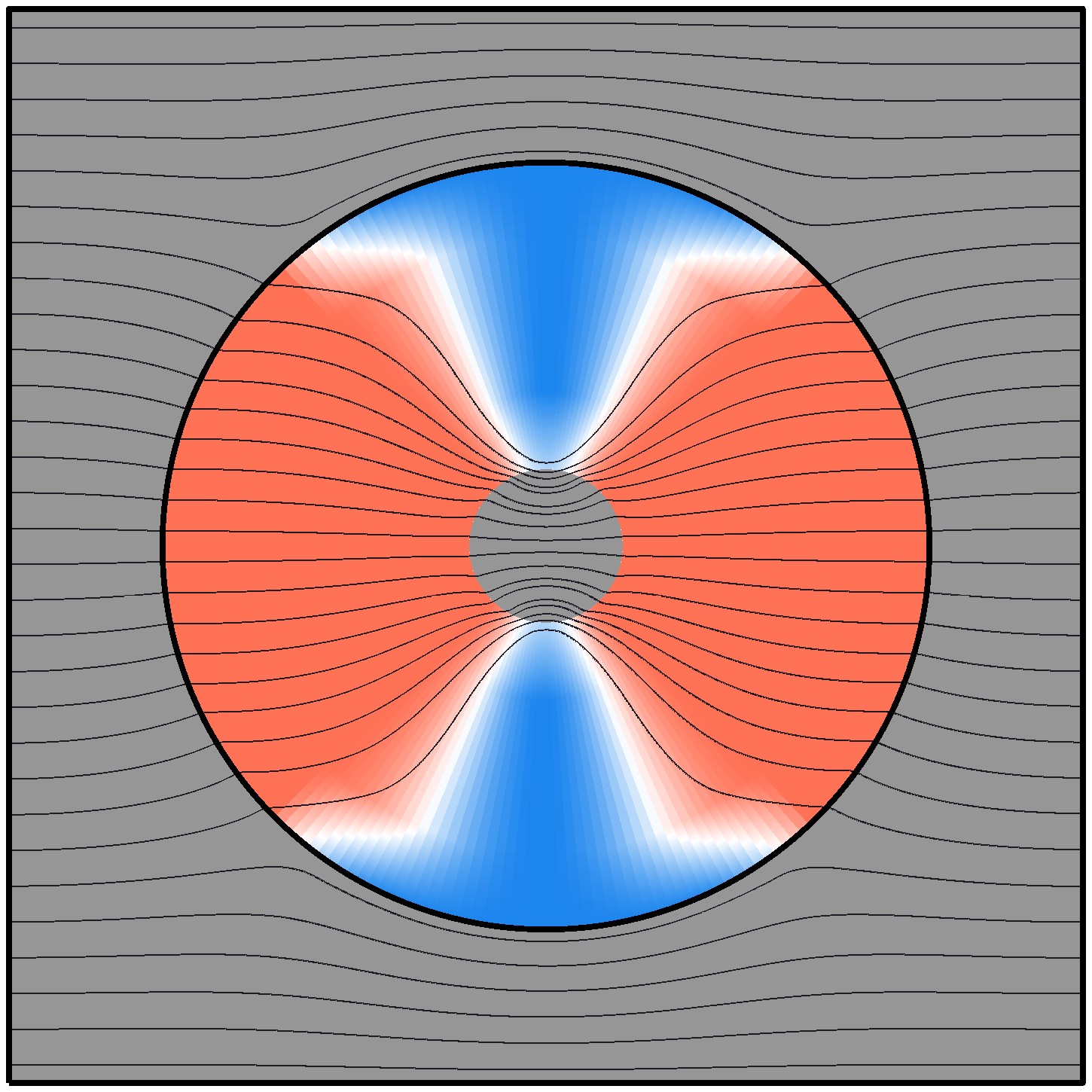}}
        \caption{\centering $N_{\rm var}=25$, $J=2.19\times 10^{-1}$}
    \end{subfigure} & \vspace{0.2cm}
    \begin{subfigure}[t]{0.15\textwidth}{\includegraphics[width=1\textwidth]{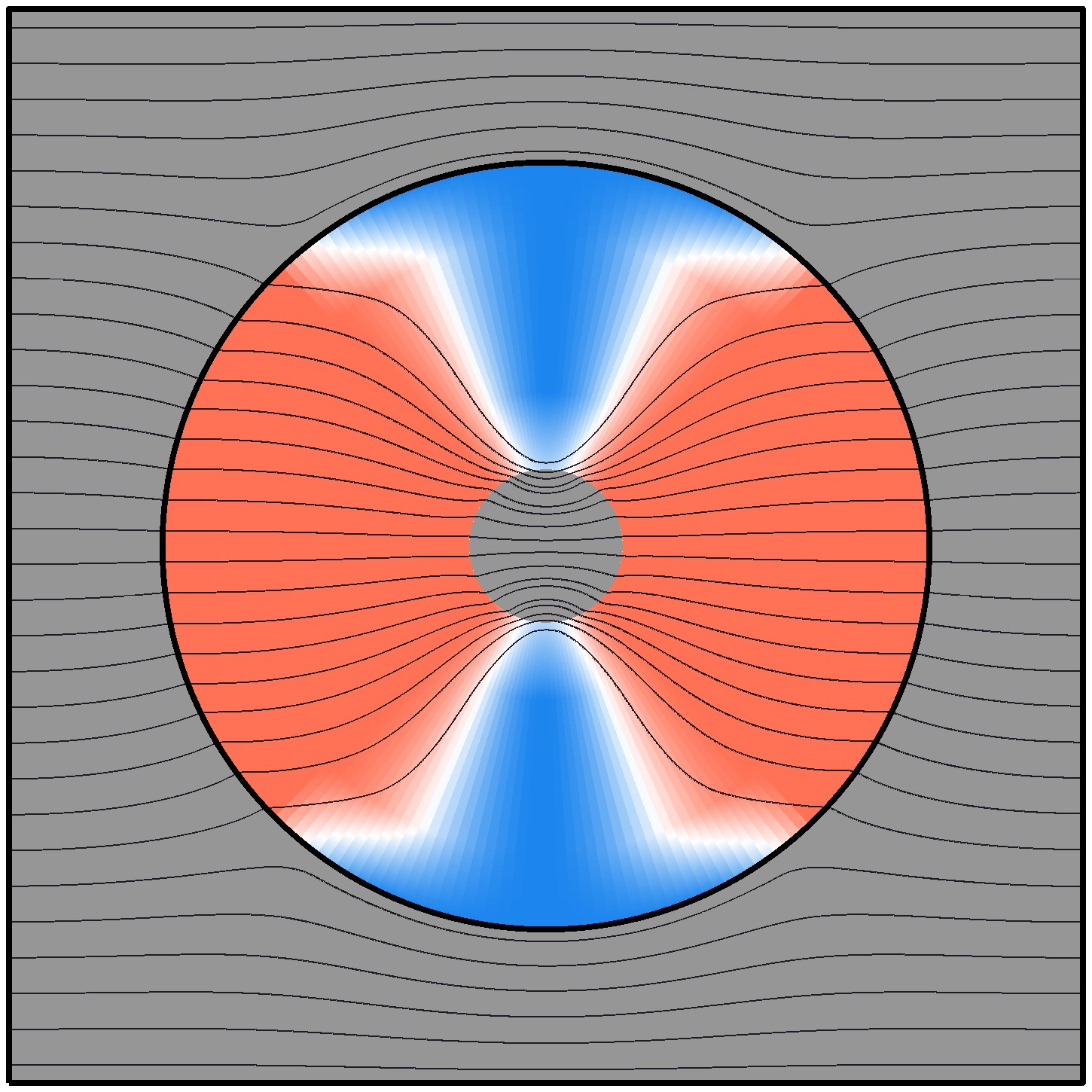}}
        \caption{\centering $N_{\rm var}=25$, $J=2.19\times 10^{-1}$}
    \end{subfigure}& \vspace{0.2cm}
    \begin{subfigure}[t]{0.15\textwidth}{\includegraphics[width=1\textwidth]{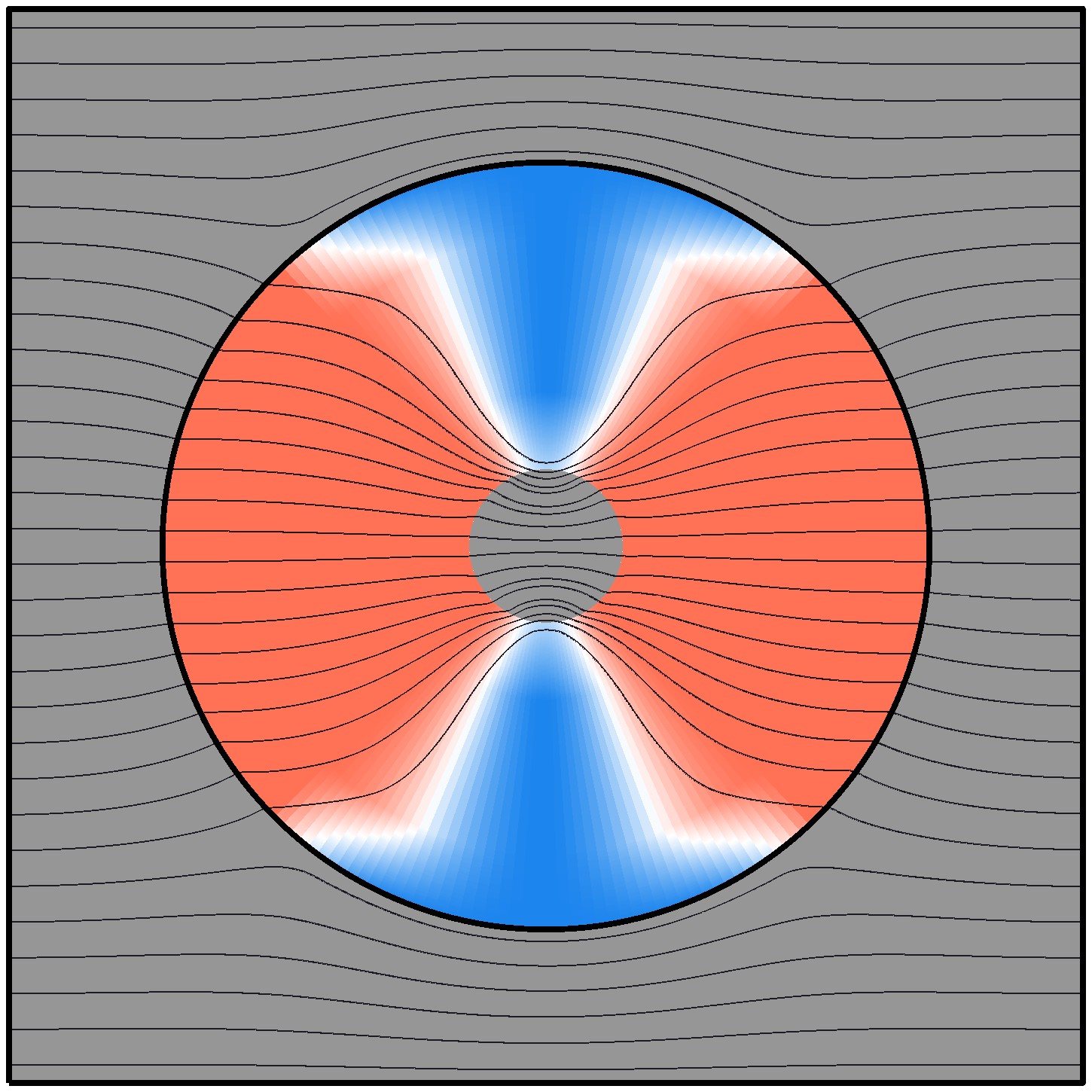}}
        \caption{\centering $N_{\rm var}=25$, $J=2.19\times 10^{-1}$}
    \end{subfigure}~\begin{subfigure}[b]{0.05\textwidth}{
\includegraphics[keepaspectratio=false,width=1.1\textwidth,height=2.45cm]{colorbar_VF2.jpg}}  
\end{subfigure}\\   
\hline
    \end{tabular}
}

\caption{Optimized material distributions for the thermal concentrator problem for $N_{\rm var}=25$ and EMT material model. Five different initial relative density distributions are considered with $v_i=0,0.25,0.5,0.75,1$,~$i=1,2,...,N_{\rm var}$. All initial distributions reach the same solution with the objective function value $J=2.19\times 10^{-1}$.}  
    \label{fig:chen2015case cntr inB}
\end{figure}
\par Similar to the thermal cloak problem, we also study the effect of $\Omega_{\rm in}$/$\Omega_{\rm design}$ as well as different initial material distribution on optimized material distribution. For both studies, we consider $N_{\rm var}=25$ and EMT model. For the first study, we vary $R_{\rm in}$ between $10$~mm and 50~mm while maintaining $R_{\rm out}=50$~mm. Corresponding, optimized $\varPsi_{\rm cntr}$ are plotted in \fref{fig:Chen2015case cntr riVar}. For the figure, we can say that the optimized material distribution does not show any significant effect related to $R_{\rm in}$ increment, and maintains sector-type structure throughout. For the second study, we take five initial distributions with $v_i=$ 0, 0.25, 0.5, 0.75, 1,~$i=1,2,...,N_{\rm var}$. The optimization results are shown in \fref{fig:chen2015case cntr inB}. All cases reach the same solution with $J=2.19 \times 10^{-1}$ irrespective of their starting points. Results from both \fref{fig:chen2015case cntr} and \fref{fig:chen2015case cntr inB} exhibits the thermal concentrator problem has a lower degree of non-convexity than the thermal cloak problem. 
\par
\begin{figure}[!htbp]
\centering
\scalebox{1}{
\begin{tabular}[c]{| M{16em} | M{16em} |}
\hline
\multicolumn{1}{|c|}{\centering EMT}  
& \multicolumn{1}{c|}{\centering Gyroid} \\
\hline
\vspace{0.2cm}
    \begin{subfigure}[t]{0.43\textwidth}{\centering\includegraphics[width=0.31\textwidth]{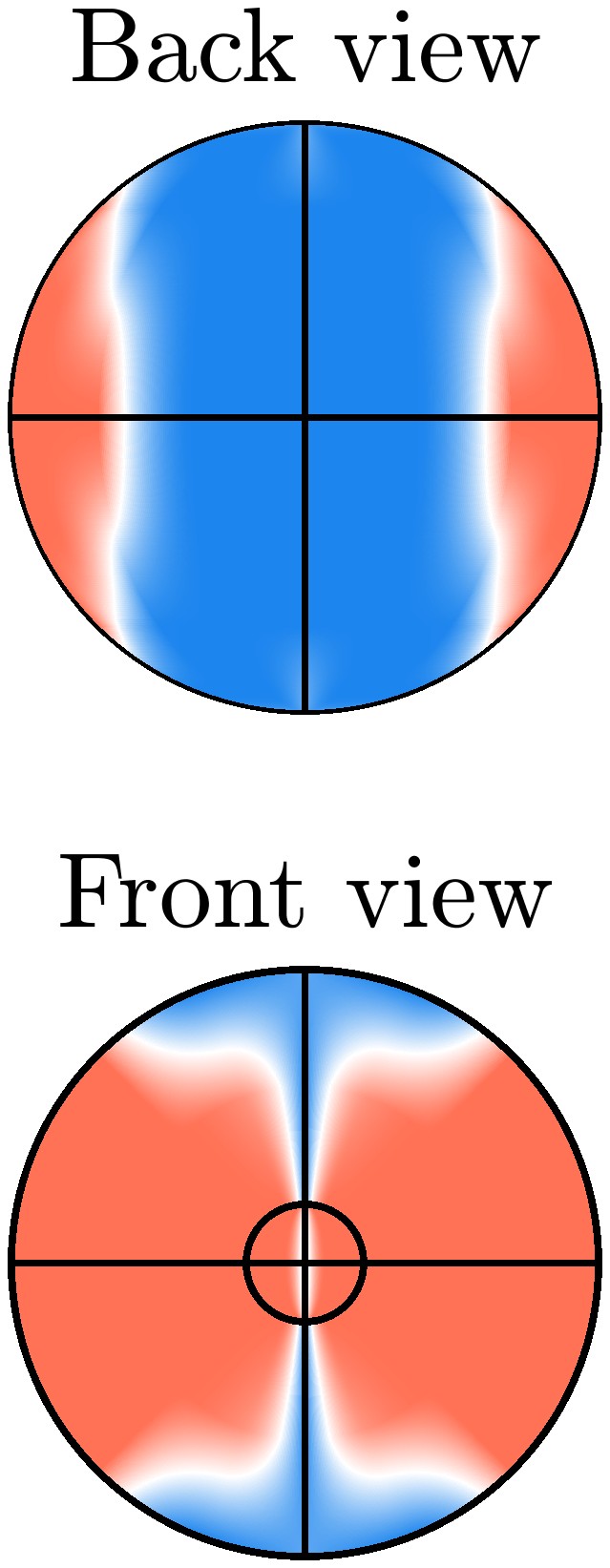}\includegraphics[width=0.69\textwidth]{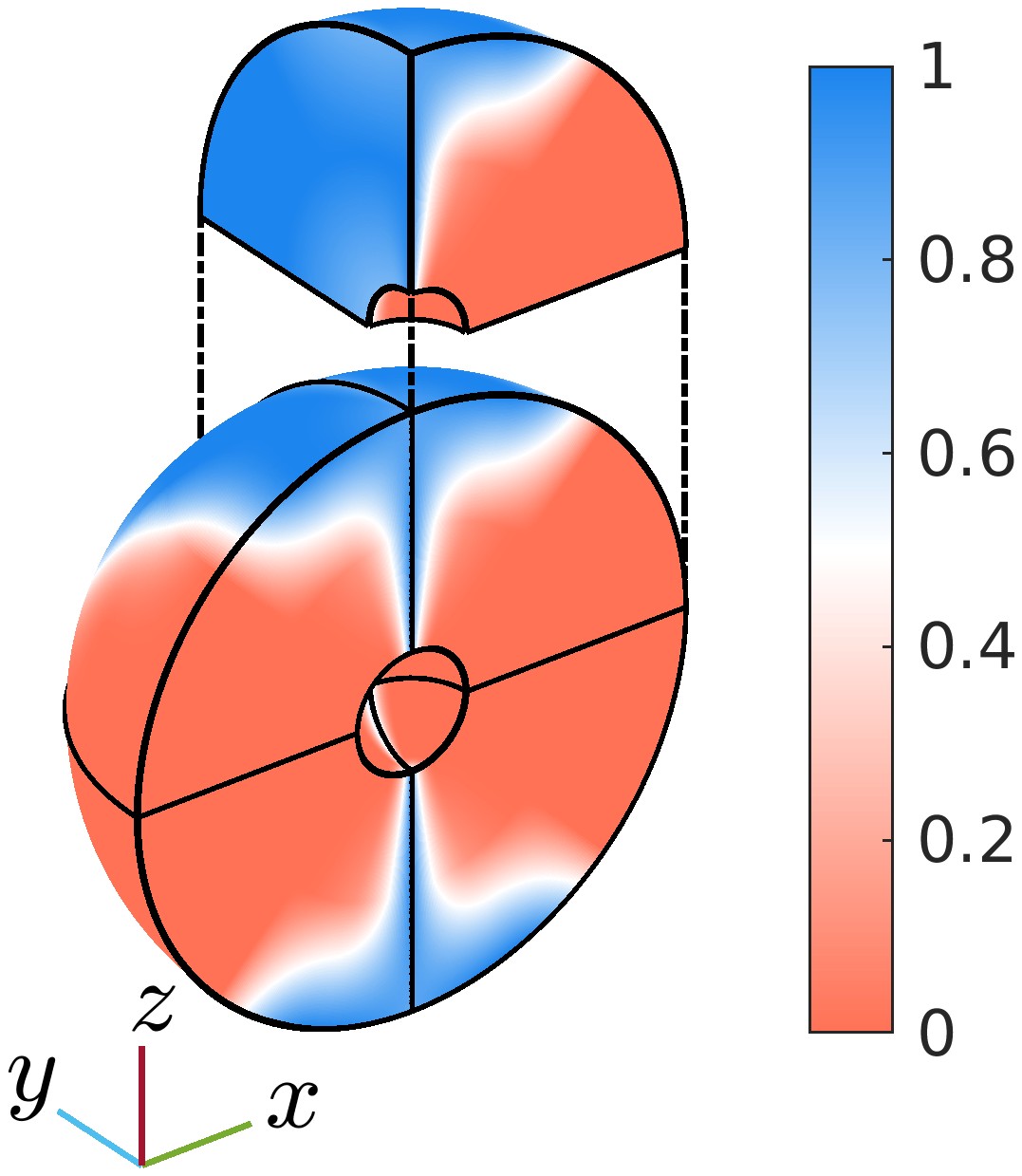}}
        \caption{\centering $J_{\rm cntr}=8.81\times 10^{-2}$, $\varPsi_{\rm cntr}=11.36$}
    \end{subfigure}
    & \vspace{0.2cm}
    \begin{subfigure}[t]{0.43\textwidth}{\centering\includegraphics[width=0.31\textwidth]{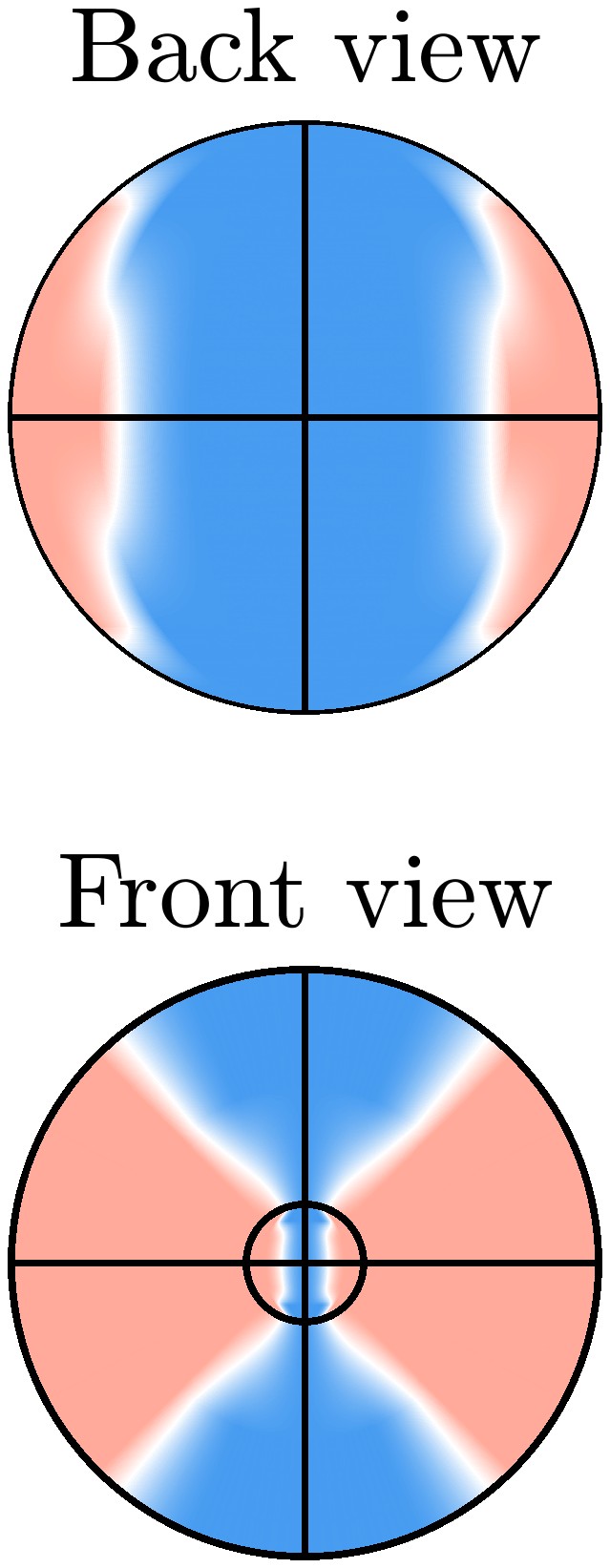}\includegraphics[width=0.69\textwidth]{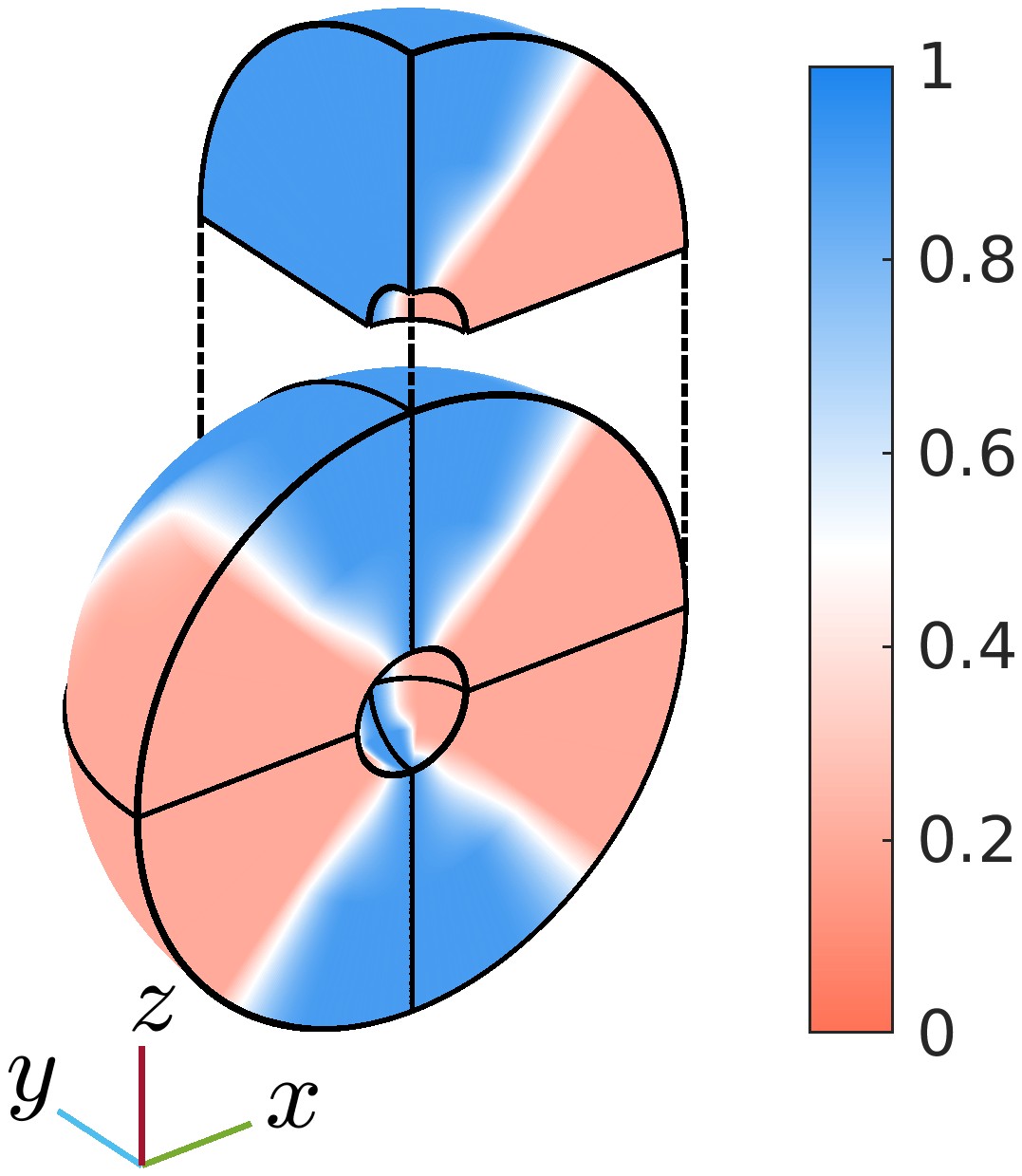}}
        \caption{\centering $J_{\rm cntr}=2.88 \times 10^{-1}$, $\varPsi_{\rm cntr}=3.47$}
    \end{subfigure} \\
\hline
 \end{tabular}
}

\caption{Optimized material distribution for the 3D thermal concentrators. Two material models (EMT and Gyroid) and $N_{\rm var}=129$ are explored. The proposed method could effectively design thermal concentrators concentrating 3 to 11 times more flux than a homogeneous plate. Analogous to the results of the 2D
thermal concentrators, two spherical cones made of $\kappa_{\rm max}$-the material is found along the x-axis. The remaining design domain is filled with $\kappa_{\rm min}$, leaving a very small transition region for intermediate densities.}  
\label{fig:Chen2015case 3D cntr}
\end{figure}

\begin{figure}[!htbp]
    \centering
    \begin{subfigure}[b]{0.48\textwidth}{\centering\includegraphics[width=1\textwidth]{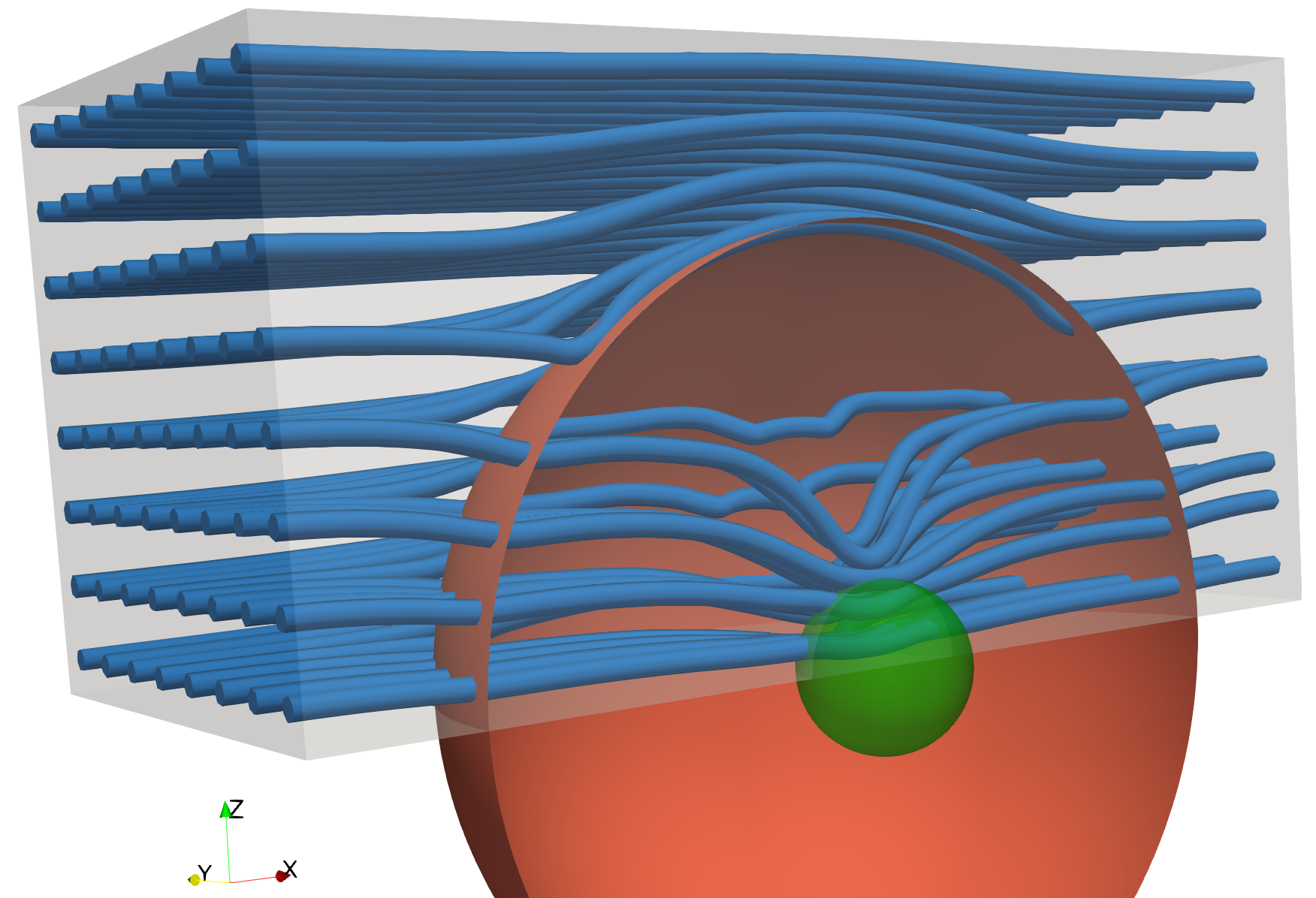}}
        \caption{Flux flow}
    \end{subfigure}\quad
    \begin{subfigure}[b]{0.34\textwidth}{\centering\includegraphics[width=0.4\textwidth]{FGM_3D_objT_1_ref2_MatModel_1_sample1_VF_views.jpg}\includegraphics[width=0.8\textwidth]{FGM_3D_objT_1_ref2_MatModel_1_sample1_VF.jpg}}
             \caption{Optimized material distribution}
    \end{subfigure}
 \caption{Optimized material distribution and flux flow for the 3D thermal concentrators. EMT model and $N_{\rm var}=129$ are considered. Optimized objective function value $J_{\rm cntr}=8.81\times 10^{-2}$ and $\varPsi_{\rm cntr}=11.36$. The thermal concentrator guides the flux streamlines towards $\mathrm{\Omega}_{\mathrm{in}}$ to concentrate them.}
 \label{fig:Chen2015case 3D cntr flux}
\end{figure}

 In this paragraph, we present the result of 3D thermal concentrators designed with two material models (EMT and Gyroid) with $N_{\rm var}=129$. 
\fref{fig:Chen2015case 3D cntr} present the optimized material distributions for both cases, which can give concentration function value of $\varPsi_{\rm cntr}=11.36$ and $\varPsi_{\rm cntr}=3.47$. Analogous to the results of the 2D thermal concentrators, which yield sector-type material distributions, two spherical cones made of $\kappa_{\rm max}$-material are found along the $x$-axis. The remaining design domain is filled with $\kappa_{\rm min}$, leaving a very small transition region for intermediate densities. The flux concentration achieved using the thermal concentrator of the EMT model is presented in \fref{fig:Chen2015case 3D cntr flux}.   

\renewcommand{\arraystretch}{1.5}   

\subsection{Thermal rotator} 
\label{sec:Thermal rotator}
 \begin{figure}[!htbp]
    \centering
    \setlength\figureheight{1\textwidth}
    \setlength\figurewidth{1\textwidth}
    \begin{subfigure}[b]{0.28\textwidth}{\centering\includegraphics[width=1\textwidth]{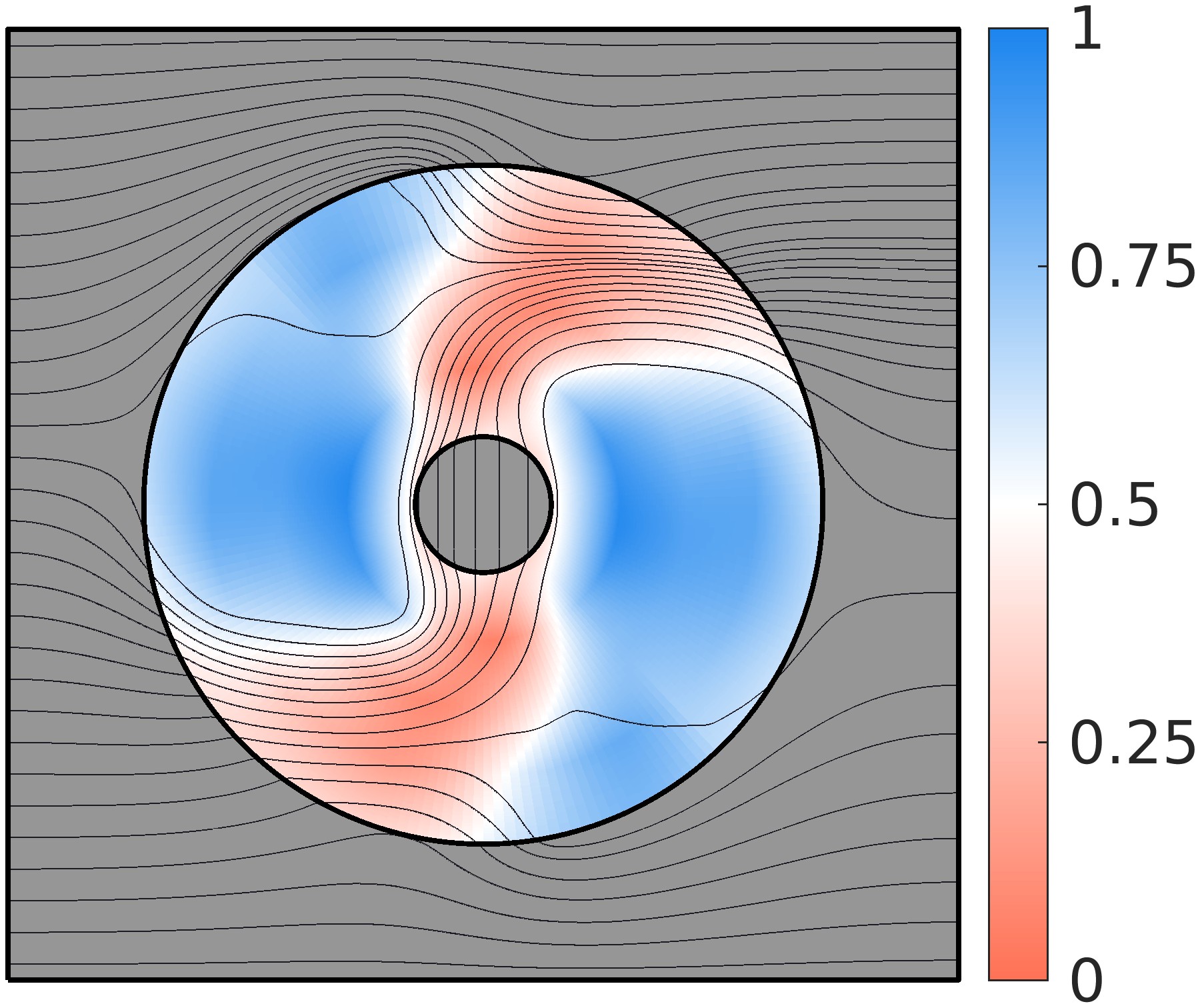}}
        \caption{$\theta=\sfrac{\pi}{2}$, $J_{\rm rtr}=9.01 \times 10^{-8}$}
    \end{subfigure}
    \begin{subfigure}[b]{0.28\textwidth}{\centering\includegraphics[width=1\textwidth]{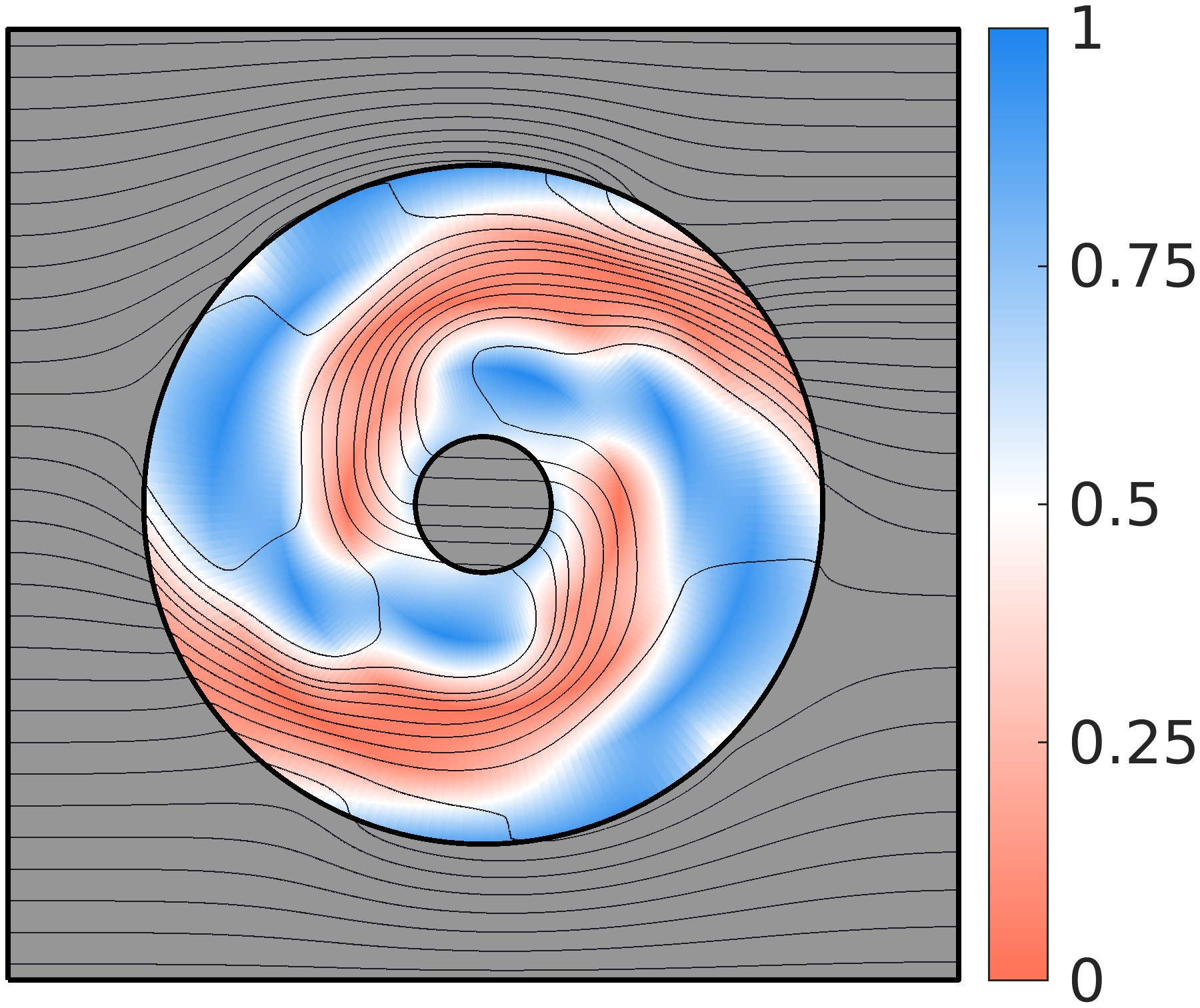}}
        \caption{$\theta={\pi}$, $J_{\rm rtr}=2.27 \times 10^{-7}$}
    \end{subfigure}
 \caption{Optimized material distributions for the thermal rotator problem. EMT model, $N_{\rm var}=100$ and two values of $\theta$, $\theta= \sfrac{\pi}{2}, {\pi}$ are considered. Optimized objective function values are $J_{\rm rtr}=9.01 \times 10^{-8}$ and $J_{\rm rtr}=2.27 \times 10^{-7}$, respectively.}
 \label{fig:Thermal rotator}
\end{figure}

\par In this subsection, we design a 2D thermal rotator. The thermal rotator was first fabricated in~\cite{Narayana2012} and later also designed for transient cases in~\cite{Guenneau2013Anisotropic} using transformation thermotics. The objective of a thermal rotator is to rotate the local direction of heat flux. In our case, we aim to rotate the flux passing through $\Omega_{\rm in}$. Accordingly, we design the objective function as:
\begin{equation}
    J_{\mathrm{rtr}}=\dfrac{1}{\widetilde{J}_{\mathrm{rtr}}} \int_{\mathrm{\Omega}_{\mathrm{in}}} \vert\vert~\mathbf{q} -\mathbf{R}~\overline{\mathbf{q}}~\vert\vert^2~d\mathrm{\Omega}, \quad \text{with} \quad  \widetilde{J}_{\mathrm{rtr}}= \int_{\mathrm{\Omega}_{\mathrm{in}}} \vert \vert ~\overline{\mathbf{q}}~\vert\vert^2~d\mathrm{\Omega}.
\end{equation}
where $\mathbf{q}$ is the flux distribution, $\overline{\mathbf{q}}$ is the flux distribution when entire $\mathrm{\Omega}$ is filled with the base material, and $\mathbf{R}$ is a 2D rotation matrix defined as: $\mathbf{R}~=~\begin{bsmallmatrix}
        \cos{\theta} & -\sin{\theta} \\
        \sin{\theta} & \cos{\theta}  
    \end{bsmallmatrix}$, with $\theta$ being the angle of rotation. 
\par Here, we take $R_{\rm in}=10$~mm, $R_{\rm out}=50$~mm, $L=140$~mm with base material filled in both $\Omega_{\rm in}$ and $\Omega_{\rm out}$. Considering the non-symmetry of the problem, we also removed the symmetry conditions for design variables. We have performed the optimization using $N_{\rm var}=100$ for two values of $\theta$, $\theta=\sfrac{\pi}{2}$ and $\theta=\pi$. The optimized material distributions with the rotated flux in $\Omega_{\rm in}$ are shown in \fref{fig:Thermal rotator}.  
\subsection{Thermal cloaked sensor}
\label{sec:Thermal cloaked sensor}
\begin{figure}[!htbp]
    \centering
    \setlength\figureheight{1\textwidth}
    \setlength\figurewidth{1\textwidth}
    \begin{subfigure}[b]{0.28\textwidth}{\centering\includegraphics[width=1\textwidth]{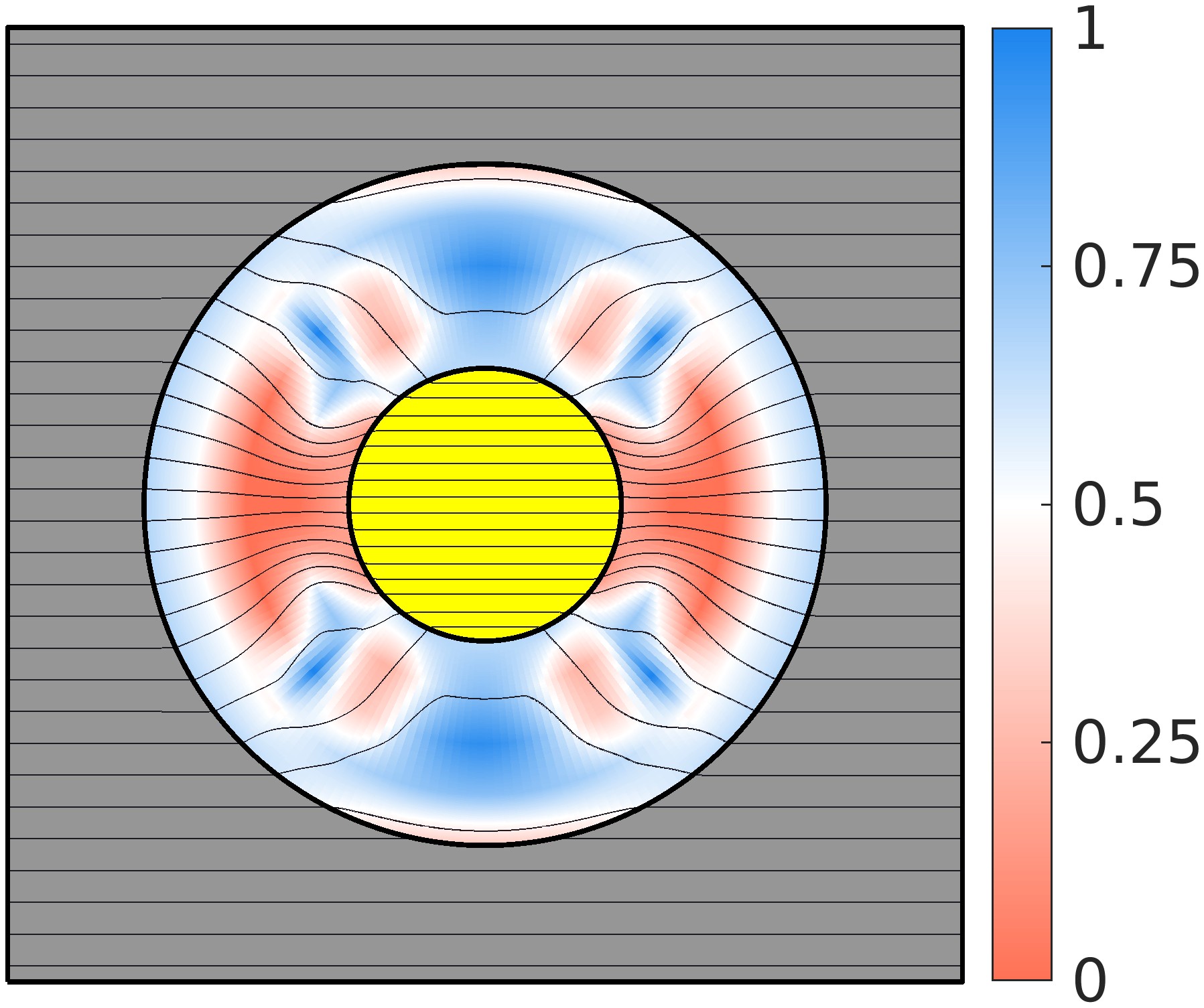}}
        \caption{\centering Optimized material distribution}
    \end{subfigure}\quad
    \begin{subfigure}[b]{0.275\textwidth}{\centering\includegraphics[width=1\textwidth]{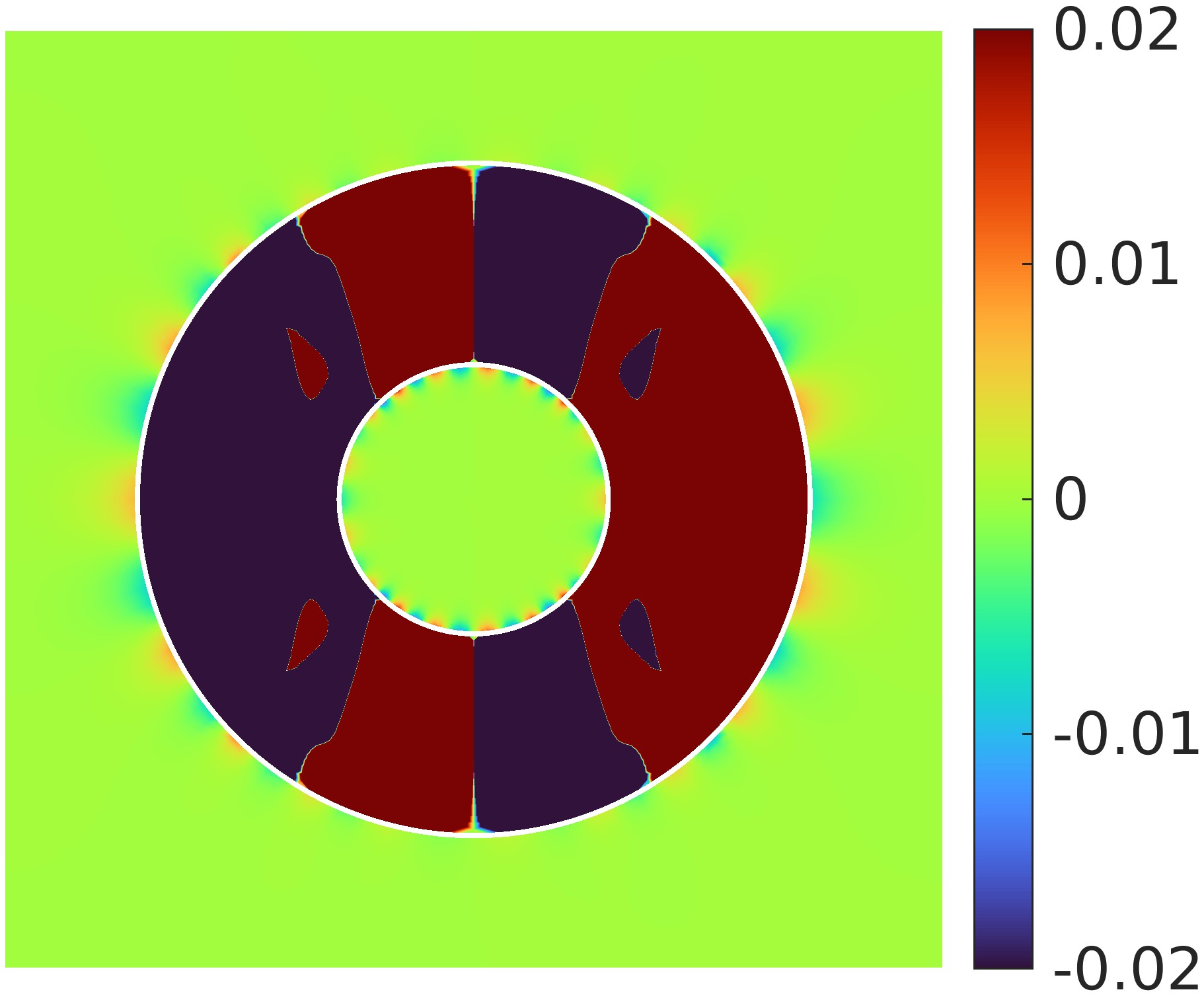}}
             \caption{\centering Temperature difference \linebreak $T-\overline{T}$}
    \end{subfigure}
 \caption{Optimized material distribution and temperature difference $T-\overline{T}$ for the thermal cloaked sensor problem. EMT model and $N_{\rm var}=25$ are considered. Optimized objective function value $J_{\rm cloaksen}=2.93 \times 10^{-7}$.  The thermal cloaked sensor keeps the flux streamlines horizontally undisturbed and diminishes the temperature disturbance in both $\mathrm{\Omega}_{\mathrm{in}}$ and $\mathrm{\Omega}_{\mathrm{out}}$.}
 \label{fig:Thermal cloaked sensor}
\end{figure}
\par Often, when a sensor is put in a physical field, the difference in properties between the background and sensor causes a disturbance around the sensor. This introduces undesirable noise in the measurement process. A thermal cloaked sensor tries to solve this issue by helping to generate the temperature profile as if the conductivities of the background and sensor are matching. As a result, the sensor has a thermal feeling of the background to measure and it can not be detected by the inline observation. Thermal cloaking sensors are already designed using other approaches in~\cite{Yang2015Invisible,JIN2020Making,SHA2022Topology}. The objective function is defined as follows:
\begin{equation}
    J_{\mathrm{cloaksen}}=\dfrac{1}{\widetilde{J}_{\mathrm{cloaksen}}} \int_{\mathrm{\Omega}_{\mathrm{in}}\cup \mathrm{\Omega}_{\mathrm{out}}} ( T - \overline{T} )^2~d\mathrm{\Omega}, \quad \text{with} \quad  \widetilde{J}_{\mathrm{cloaksen}}= \int_{\mathrm{\Omega}_{\mathrm{in}}\cup \mathrm{\Omega}_{\mathrm{out}}} ( \widetilde{T} - \overline{T} )^2~d\mathrm{\Omega},
\end{equation}
where $\overline{T}$ is the temperature when entire $\mathrm{\Omega}$ is filled with the base material, $\widetilde{T}$ is the temperature when entire $\mathrm{\Omega}_{\rm design} \cup \mathrm{\Omega}_{\mathrm{out}}$ is filled with the base material. 
\par Here, we take $R_{\rm in}=20$~mm, $R_{\rm out}=50$~mm, and $L=140$~mm. The base material is filled in $\Omega_{\rm out}$. $\Omega_{\rm in}$ represents an isotropic thermal sensor. We considered the thermal conductivity for sensor, $\kappa_{\rm sen}=130$~W/mK. The material distributions obtained by the optimization with $N_{\rm var}=25$ are shown in \fref{fig:Thermal cloaked sensor}. As the sensor has a higher conductivity than the base material, it needs more flux to create the same temperature profile as the base material. Therefore, the optimized material distribution has two sickle-shaped $\kappa_{\rm max}$-material structures along $x$-axis, which helps to concentrate more flux in $\Omega_{\rm in}$. Nevertheless, these structures remain unattached to the outer perimeter to prevent excessive streamline convergence, which could lead to temperature disturbances in $\Omega_{\rm out}$. Additionally, two dovetail-shaped $\kappa_{\rm min}$-material structures along the $y$-axis help to maintain the streamlines in a uniformly horizontal pattern inside $\Omega_{\rm in}$. We can see that the temperature difference in both $\Omega_{\rm in}$ and $\Omega_{\rm out}$ is negligible, with the objective function $J_{\rm cloaksen}=2.93 \times 10^{-7}$.  
\subsection{Thermal cloak-concentrator}
\label{sec:Thermal cloak-concentrator}
 \begin{figure}[!htbp]
    \centering
    \setlength\figureheight{1\textwidth}
    \setlength\figurewidth{1\textwidth}
    \begin{subfigure}[b]{0.27\textwidth}{\centering\includegraphics[width=1\textwidth]{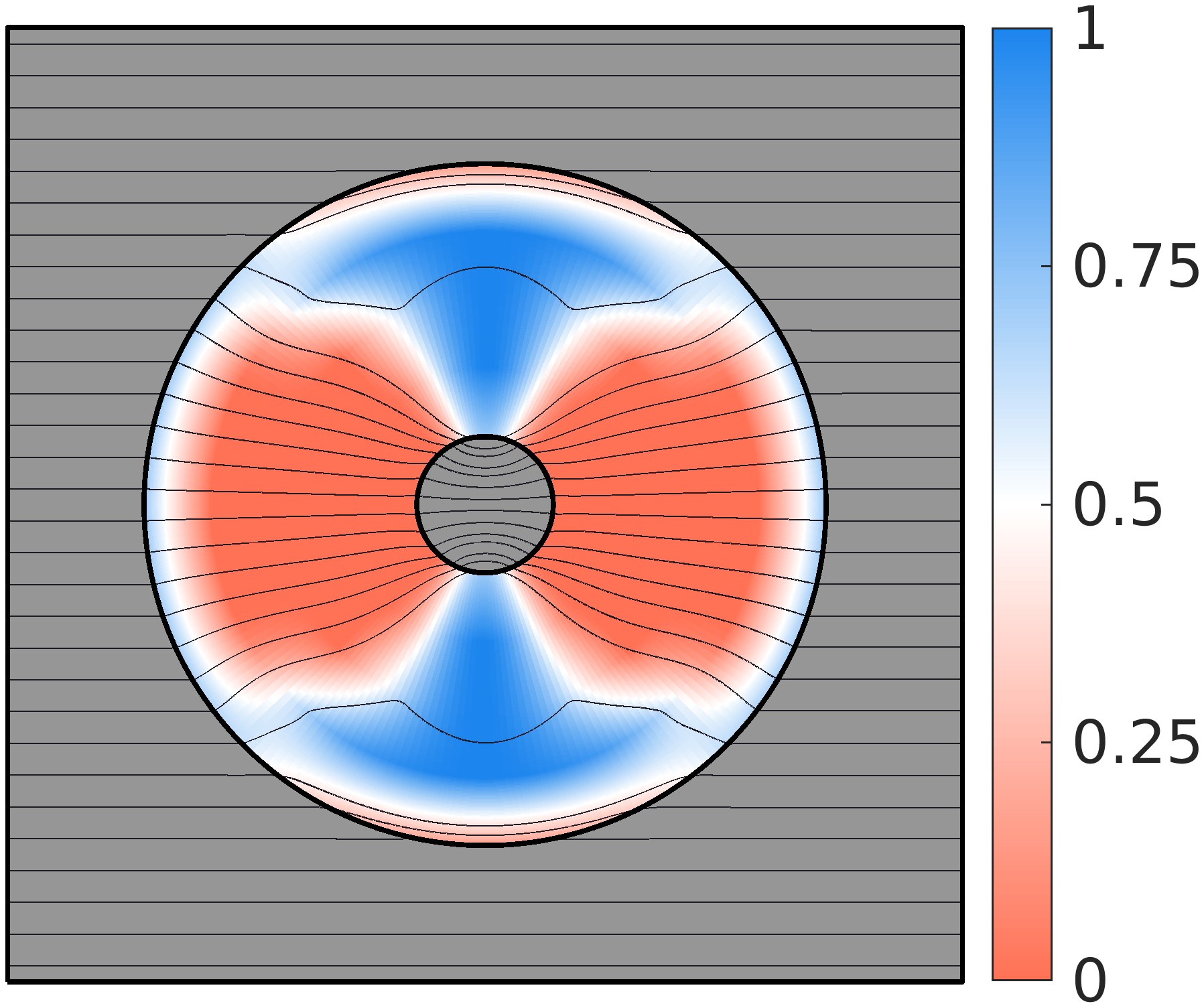}}
        \caption{\centering  Optimized material distribution}
    \end{subfigure}\quad
    \begin{subfigure}[b]{0.27\textwidth}{\centering\includegraphics[width=1\textwidth]{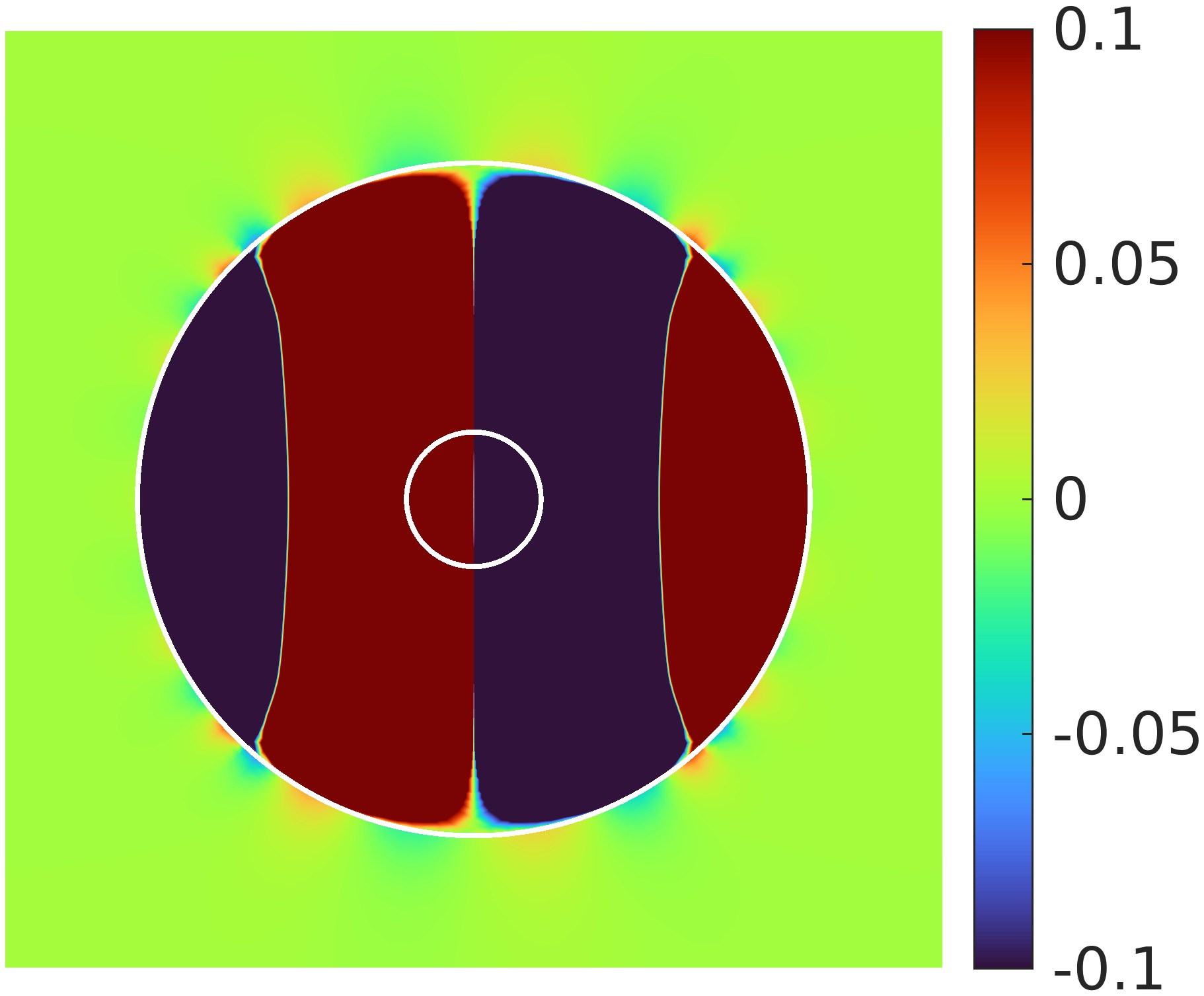}}
             \caption{\centering Temperature difference $T-\overline{T}$}
    \end{subfigure}
 \caption{Optimized material distribution and temperature difference $T-\overline{T}$ for the thermal cloak-concentrator problem. EMT model and $N_{\rm var}=25$ are considered.  Optimized objective function value $J_{\rm cloakcntr}=8.21 \times 10^{-3}$ with $J_{\rm cloak}=2.97 \times 10^{-4}$ \& $\varPsi_{\rm cntr}=3.35$. The thermal cloak-concentrator effectively achieves dual functionality—cloaking and concentrating— by balancing on each function.}
 \label{fig:Thermal cloaked concentrator}
\end{figure}

\par In this manipulator we designed a multi-functional thermal meta-structure. The cloak-concentrator performs the combined task of concentrating the flux in $\Omega_{\rm in}$ and cloaking $\Omega_{\rm in}$ as well. Thermal cloak-concentrators are constructed in articles~\cite{shen2016thermal,fujiiCloakingConcentratorThermal2020,jansari2022design} using other methods. The total objective function is defined as follows:
\begin{equation} \label{eq:thermal cloak-concentrator fn}
J_{\mathrm{cloakcntr}}=J_{\mathrm{cloak}}+\dfrac{1}{\varPsi_{\rm cntr}^4}.
\end{equation}
where $J_{\mathrm{cloak}}$ \& $\varPsi_{\rm cntr}$ are functions for cloaking and concentrating as defined in \erefs{eq:cloaking fn} -(\ref{eq:concentrating fn}). 
\par Here, we take $R_{\rm in}=10$~mm, $R_{\rm out}=50$~mm, and $L=140$~mm. The base material is filled in both $\Omega_{\rm in}$ and $\Omega_{\rm out}$. The optimization tries to find the balance between both objectives according to their relative weightage as given in \eref{eq:thermal cloak-concentrator fn}. The material distribution obtained by the optimization with $N_{\rm var}=25$ is shown in \fref{fig:Thermal cloaked concentrator}. Corresponding total objective function $J_{\rm cloakcntr}=6.35 \times 10^{-3}$ with $J_{\rm cloak}=6.12 \times 10^{-4}$ and $\varPsi_{\rm cntr}=3.63$. The optimized material distribution resembles one with a concentrator. However, the conductivities near $R_{\rm out}$ in $\Omega_{\rm design}$ vary to accommodate the cloaking function. These conductivities must be adjusted to ensure that the streamlines do not converge prematurely in $\Omega_{\rm out}$ just before entering $\Omega_{\rm design}$.

\subsection{Thermal horizontal concentrator-vertical cloak}
\label{sec:Thermal horizontal concentrator-vertical cloak}
 \begin{figure}[!htbp]
    \centering
    \setlength\figureheight{1\textwidth}
    \setlength\figurewidth{1\textwidth}
    \begin{subfigure}[t]{0.28\textwidth}{\centering\includegraphics[width=1\textwidth]{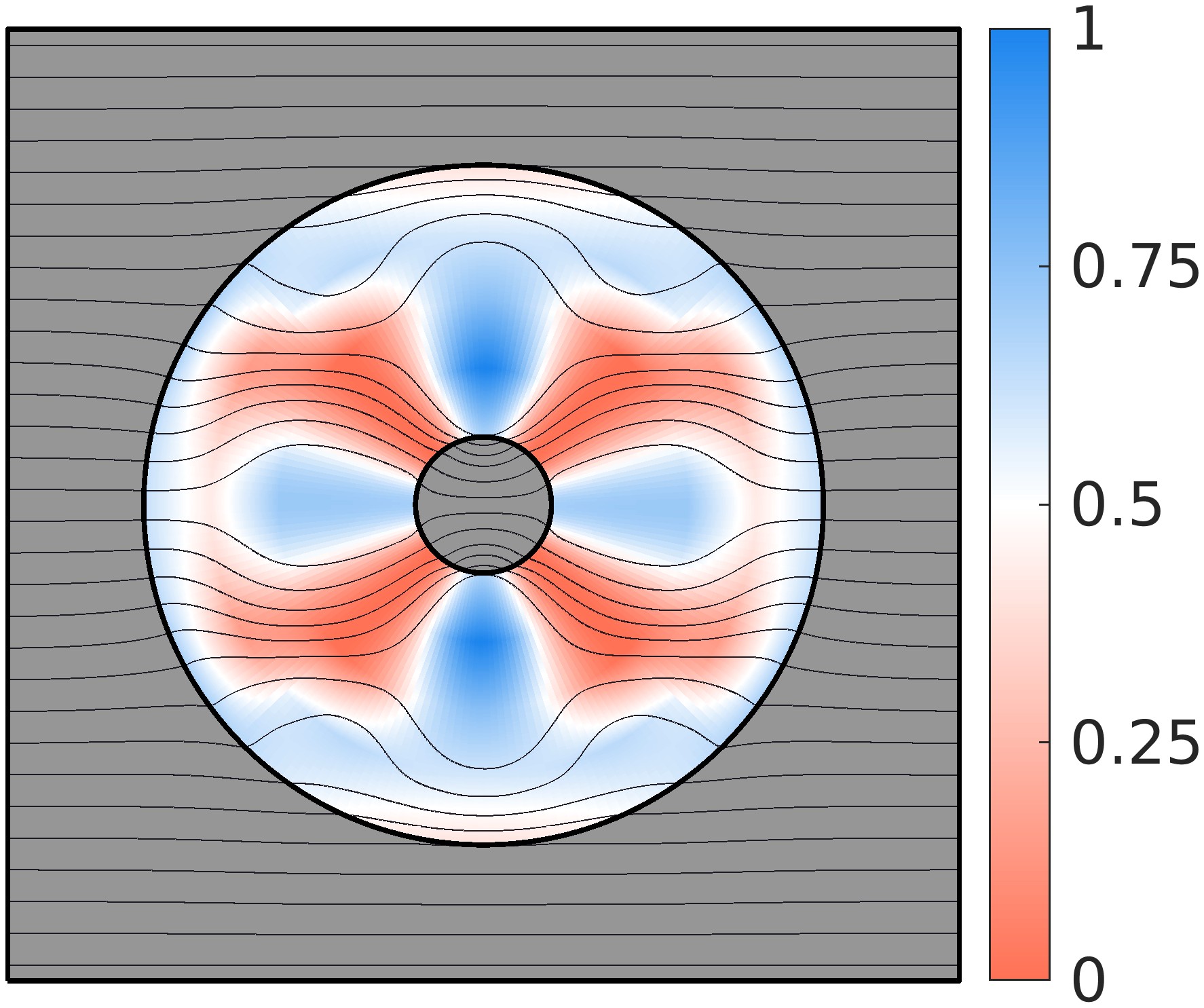}}
        \caption{Optimized material distribution and flux flow for the horizontal applied flux}
    \end{subfigure}
    \begin{subfigure}[t]{0.28\textwidth}{\centering\includegraphics[width=1\textwidth]{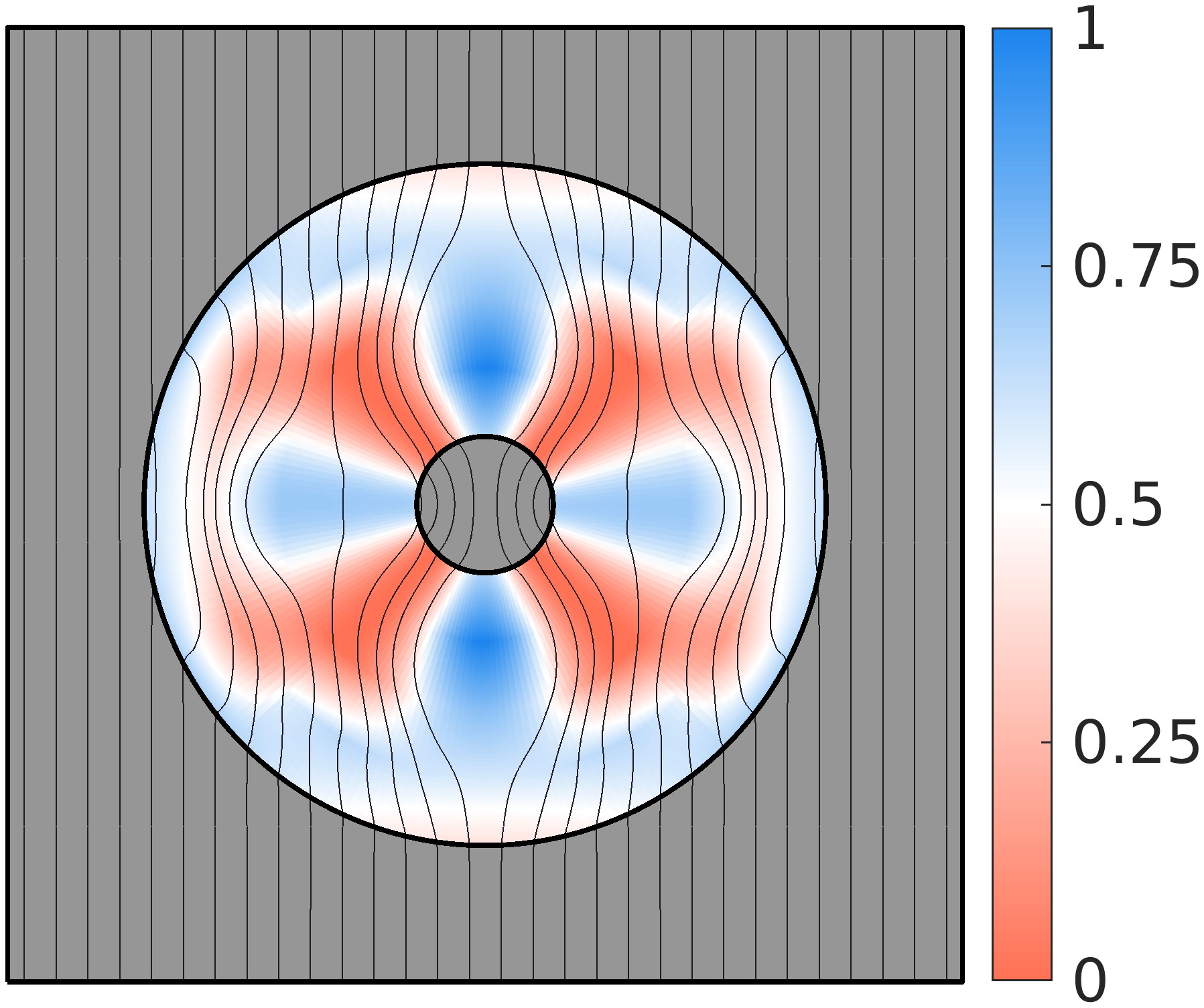}}
        \caption{Optimized material distribution and flux flow for the vertical applied flux}
    \end{subfigure}
    \begin{subfigure}[t]{0.27\textwidth}{\centering\includegraphics[width=1\textwidth]{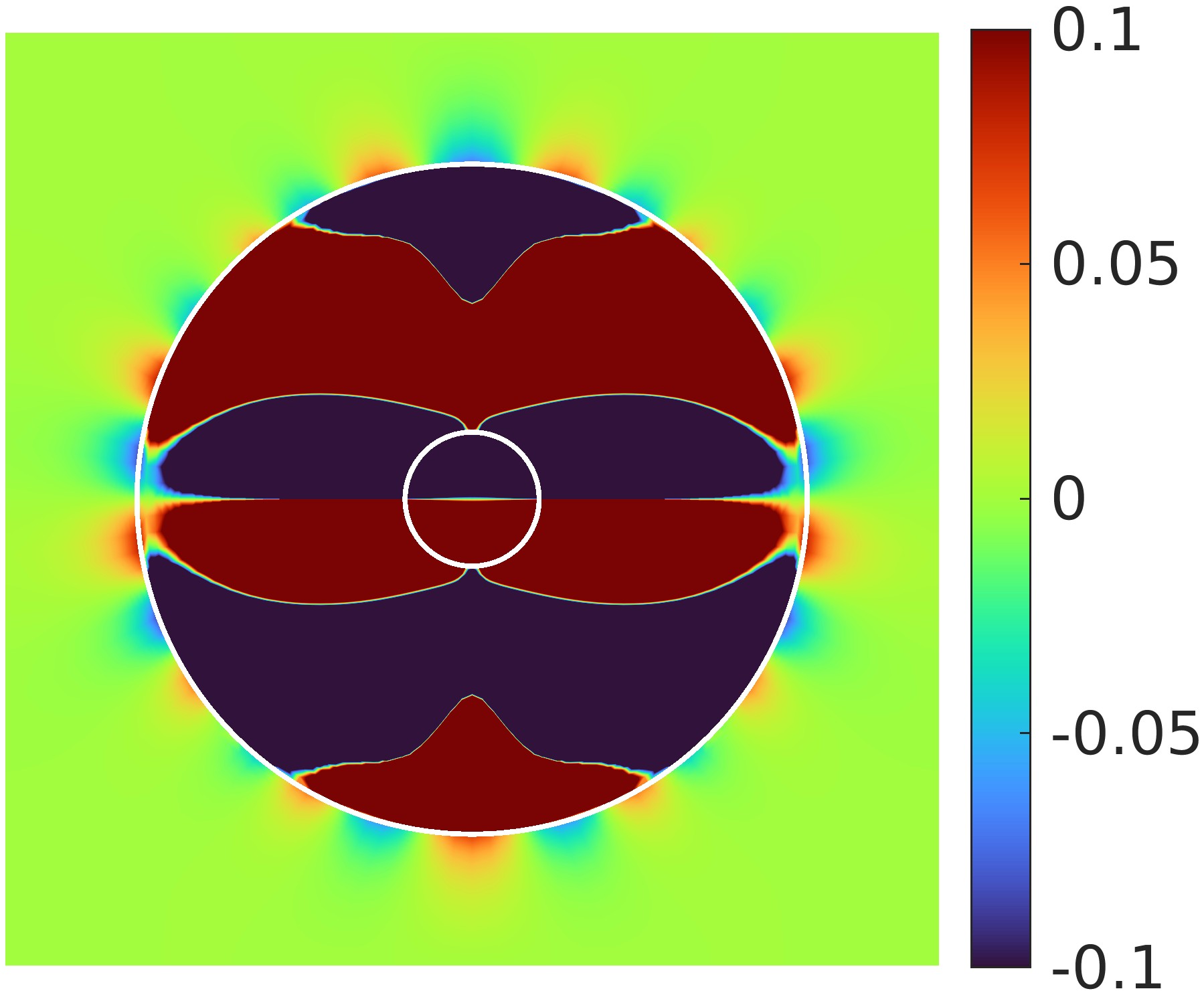}}
             \caption{\centering Temperature difference $T-\overline{T}$}
    \end{subfigure}
 \caption{Optimized material distribution (for the horizontal applied flux); optimized material distribution and temperature difference $T-\overline{T}$ (for the vertical applied flux) for thermal bi-directional thermal meta-structure (horizontal concentrator and vertical cloak). EMT model and $N_{\rm var}=25$ are considered. Optimized objective function value $J_{\rm cloakcntr}=1.75 \times 10^{-2}$ with $J_{\rm cloak}=1.31 \times 10^{-3}$ \& $\varPsi_{\rm cntr}=2.80$. }
 \label{fig:Thermal bi-direction heat manipulator}
\end{figure}
\par In this subsection, we design a bi-directional thermal meta-structure, which behaves as two different thermal meta-structures under two different sets of boundary conditions. Here, we aspire to design a thermal meta-structure that works as a concentrator for applied horizontal constant temperature difference while as a cloak for applied horizontal constant temperature difference. We use the objective function defined earlier in \eref{eq:thermal cloak-concentrator fn}, however, $J_{\mathrm{cloak}}$ \& $\varPsi_{\mathrm{cntr}}$ are calculated on two different temperature distributions based on two different sets of boundary conditions. For the cloak, we consider standard $300$~K constant temperature on the left side and $200$~K temperature on the right side, while for the concentrator the same conditions are applied on the top and bottom sides, respectively. For both cases, the remaining sides are considered as adiabatic walls. All dimensions and material allocation are the same as described in \sref{sec:Thermal cloak-concentrator}. The material distribution obtained by the optimization with $N_{\rm var}=25$ is shown in \fref{fig:Thermal bi-direction heat manipulator}. Optimized material distribution can achieve the objective function value $J_{\rm cloakcntr}=1.75 \times 10^{-2}$ with cloaking function $J_{\rm cloak}=1.31 \times 10^{-3}$ and concentrating function $\varPsi_{\rm cntr}=2.80$.

%
%
%
%
%
\section{Reconstruction of Architected Cellular Materials (ACMs)}
\label{sec:Reconstruction of ACMs}
\par From the results of the optimization of ACMs, we can reconstruct the entire structures using the density distribution and predefined unit-cells (Gyroid and TCOH in our case). We have shown here a primary reconstructed (thin-walled) Gyroid-based structure, without any manufacturing or design constraints. Note that the results obtained from the optimization provide the density field of porosity in terms of NURBS surface. Based on this surface, the density at any point on the domain can be calculated by the method of projection. 
\begin{figure}[!htbp]
    \centering
    \includegraphics[width=0.7\textwidth]{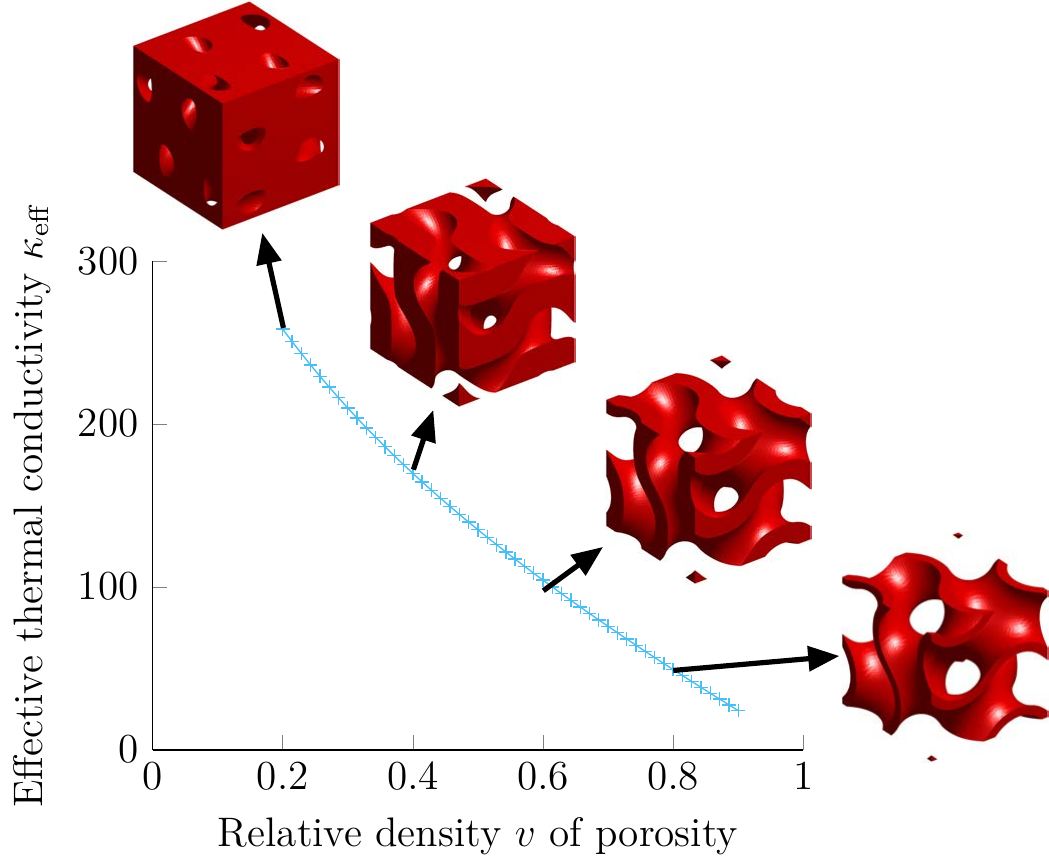}
 \caption{The relation between effective thermal conductivity and relative density $v$ of porosity. Four thin-walled Gyroid unit-cells corresponding to $v=0.2, 0.4, 0.6, 0.8$ are also shown.}
 \label{fig:Gyroid Vf}
\end{figure}
\par The unit-cell of thin-walled Gyroid is defined implicitly using Boolean operation. The volume is the intersection volume between two implicit surfaces given by:
\begin{equation} \label{eq:Gyroid surfaces}
    \cos{\left(\dfrac{2\pi}{a}x\right)} \sin{\left(\dfrac{2\pi}{a}y\right)} + \cos{\left(\dfrac{2\pi}{a}y\right)} \sin{\left(\dfrac{2\pi}{a}z\right)} + \cos{\left(\dfrac{2\pi}{a}z\right)} \sin{\left(\dfrac{2\pi}{a}x\right)} \pm t=0, 
\end{equation}
where $a$ is the side length of a cubic unit-cell and $t$ is the control parameter ($2t$ will be the thickness of the wall).  The approximate relation between the relative density $v$ of porosity and control parameter $t$ is given as~\cite{hussain2020design}:
\begin{equation} \label{eq:Gyroid vf}
    t=\dfrac{0.65}{v_m} \quad \text{with} \quad v_m=1-v,
\end{equation}
where $v_m$ and $v$ are densities of matrix material and porosity, respectively. The relation between thermal conductivity and Gyroid shape is also plotted in \fref{fig:Gyroid Vf}. 
\par ACMs are constructed by tessellating a repeating unit-cell, across the domain. There are several techniques to tessellate the unit-cell such as sweeping, meshing and trimming~\cite{aremu_voxel-based_2017}. In the swept ACMs, alignment with the boundary of the domain is enforced and this type of structure highly depends on the curvature of the boundary. Therefore, unit-cells deviate from their original shapes and properties. In the meshed ACMs, the unit-cells are mapped to the elements of the domain mesh. This mesh could be a finite element mesh used for the numerical analysis. Similar to the swept ACMs, the elements and the corresponding unit-cells follow the external geometry. Eventually, the unit-cells of the meshed ACMs deviate from their original properties too. On the contrary, the tessellation based on the trimming approach does not carry this limitation and mostly retains the properties of unit-cells. It can effectively create a complex-shaped ACM with the simple Boolean operations of the domain and tessellated unit-cells. In our case, we use the trimming approach considering our design geometries involve circular and spherical shapes. Trimmed ACMs often possess weak boundaries which lack support. A potential remedy is to create a solid skin surrounding the lattice structure~\cite{aremu_voxel-based_2017}. However, as our design domain is always surrounded by other homogeneous materials, we do not need any specific treatment for support. 
\begin{figure}[!htbp]
    \centering
    \includegraphics[width=1\textwidth]{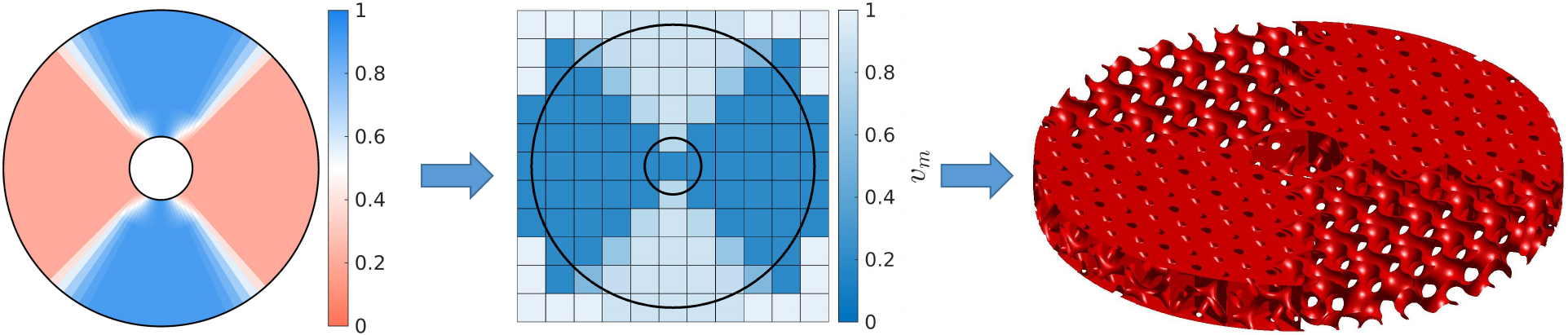}
 \caption{Reconstruction of a Gyroid-based thermal concentrator obtained by topology optimization. A very coarse $11 \times 11$ voxel mesh is considered to highlight the gradation and reconstruction features.}
 \label{fig:Reconstruction steps}
\end{figure}
\par
\begin{figure}[!htbp]
    \centering
    \setlength\figureheight{1\textwidth}
    \setlength\figurewidth{1\textwidth}
    \begin{subfigure}[b]{0.51\textwidth}{\centering\includegraphics[width=0.44\textwidth]{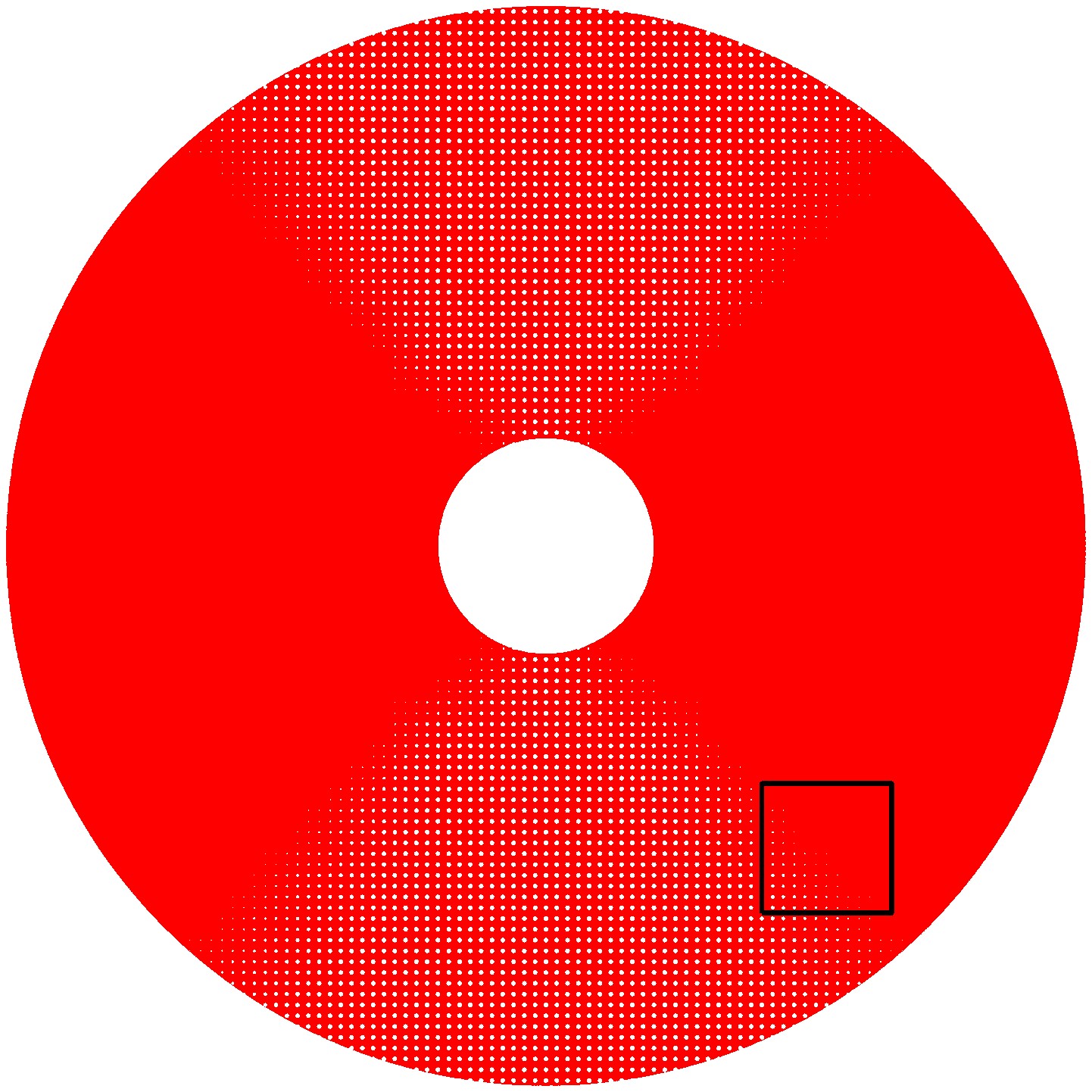}
    \includegraphics[width=0.54\textwidth]{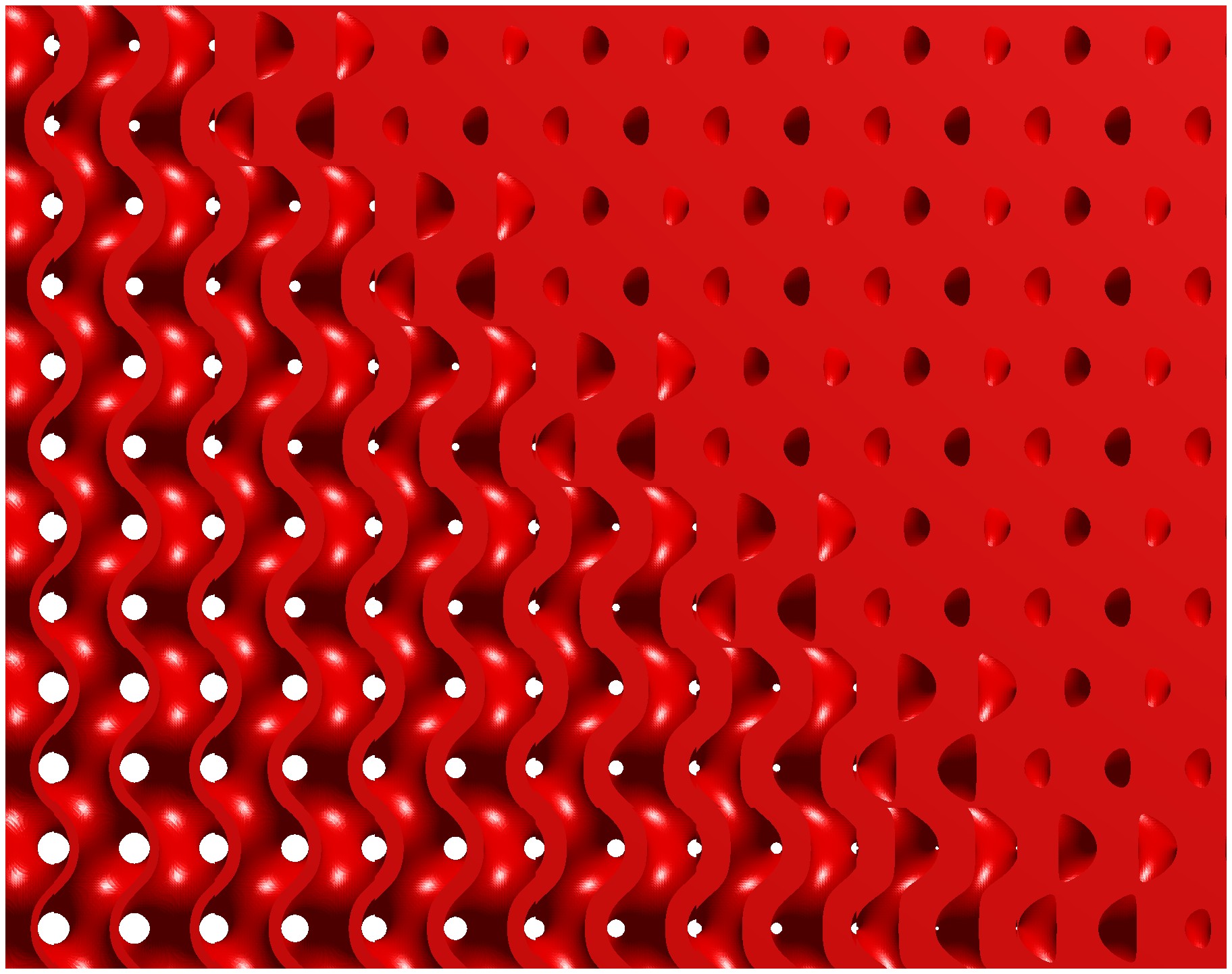}}
        \caption{2D thermal concentrator}
        \label{fig:chen2015case reconstruction a}
    \end{subfigure}\quad
     \begin{subfigure}[b]{0.46\textwidth}{\centering\includegraphics[width=0.5\textwidth]{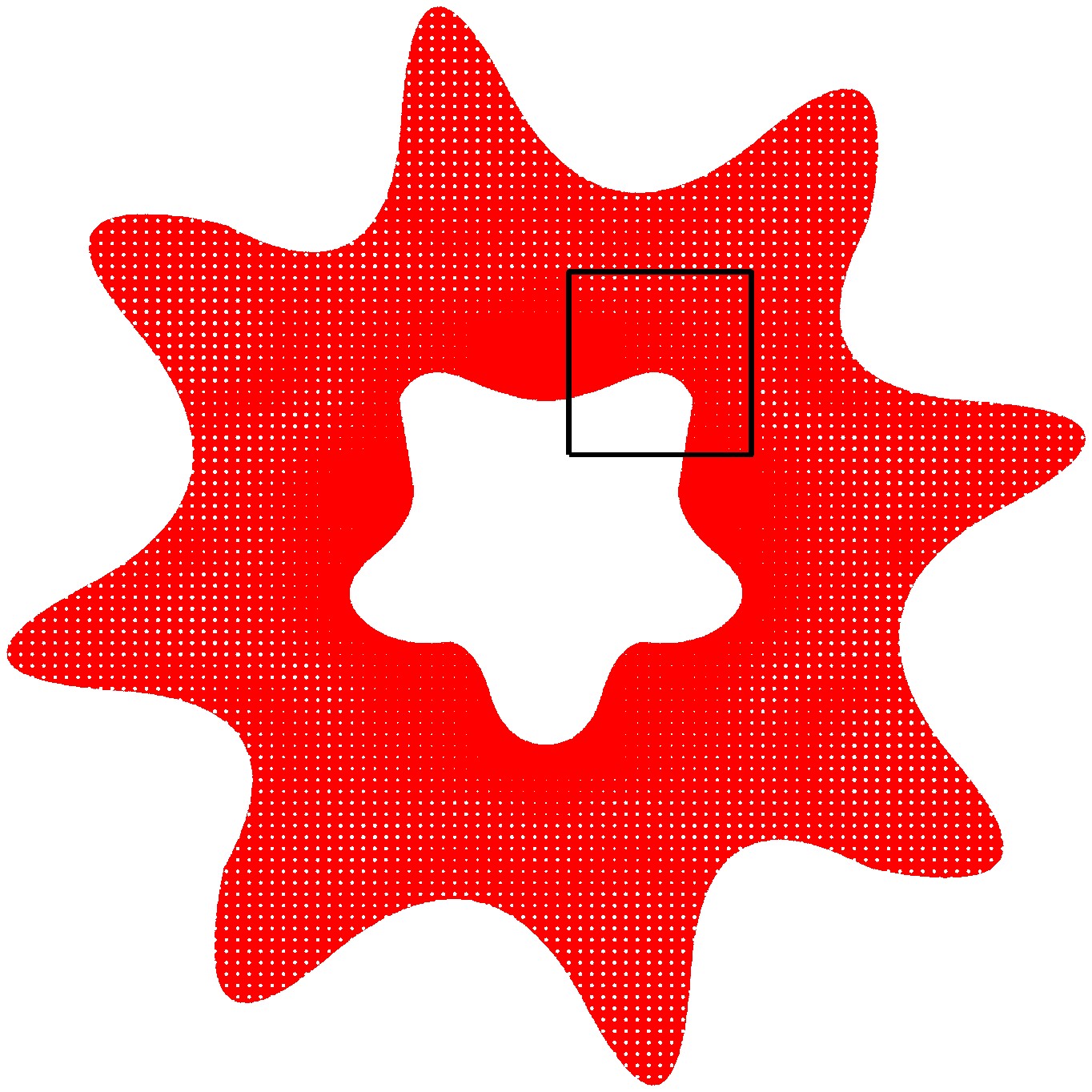}\includegraphics[width=0.5\textwidth]{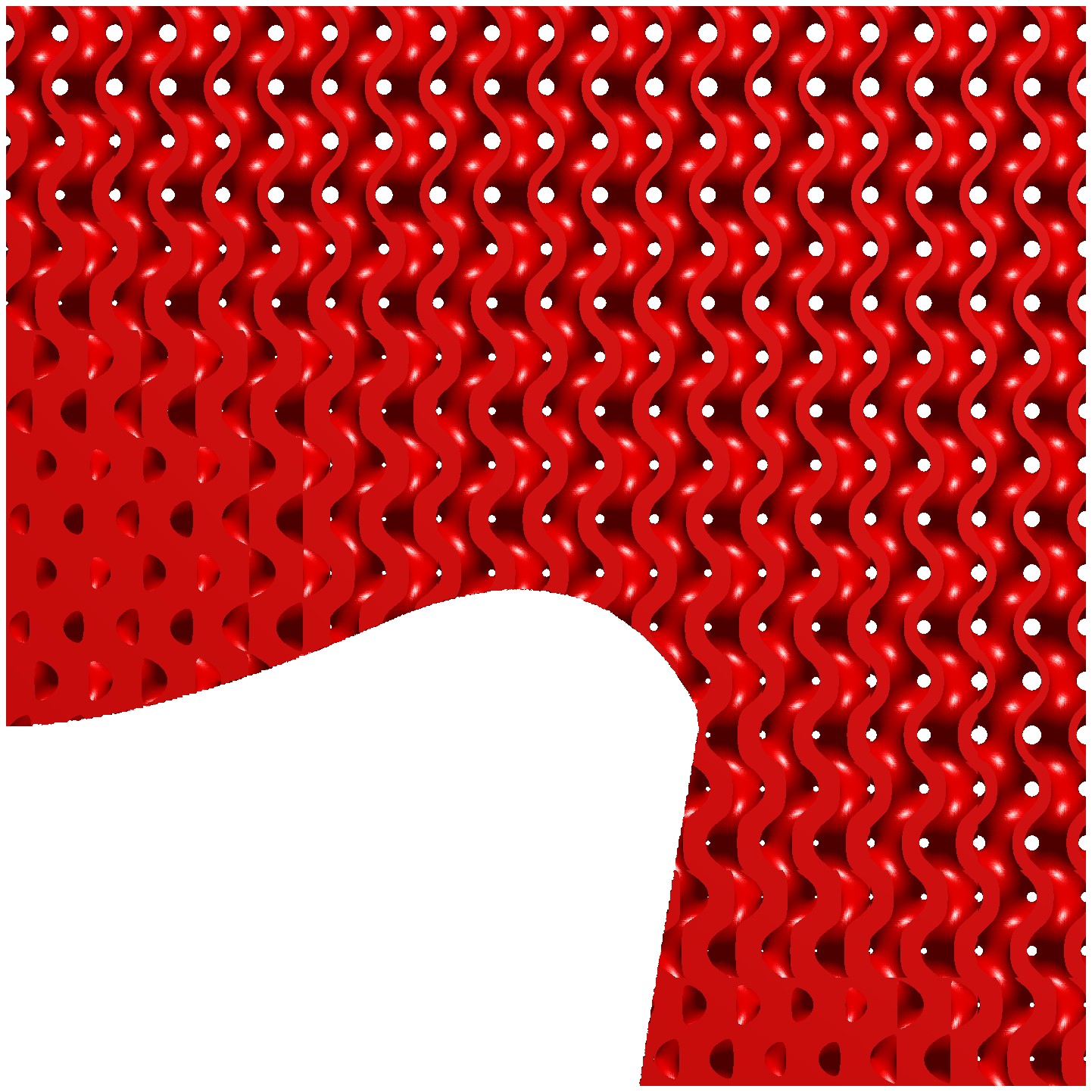}}
        \caption{2D complex star-shaped thermal cloak}
        \label{fig:chen2015case reconstruction b}
    \end{subfigure}
 \caption{Full-scale reconstruction of thermal meta-structures.}
 \label{fig:chen2015case reconstruction}
\end{figure}

 For the reconstruction, we consider a rectangular (cubical) background domain that entirely covers our actual 2D (3D) domain. Then, the background domain is voxelized, which essentially means discretizing the geometry in small rectangular (cubical) blocks~\cite{aremu_voxel-based_2017}. It is assumed that one voxel corresponds to one unit-cell and its density is calculated at the center of the voxel. The voxels that entirely outside the domain will be considered empty and given $v_m=0$, while the densities at the center of the remaining voxels will be calculated based on the NURBS entity describing the density distribution. There might be a possibility that the center of any voxel cut on the boundary might lie outside the design domain. In that case, the density value is calculated on the basis of the extrapolation of the density surface. Then, the control parameter $t$ and the corresponding Gyroid are constructed for all voxels using \erefs{eq:Gyroid vf}-(\ref{eq:Gyroid surfaces}). When all unit-cells are put together in the voxelized domain, it is referred to as the unit-cell tessellation. Finally, a Boolean intersection operation is performed between the actual domain and unit-cell tessellation to find the final trimmed ACM. The reconstruction steps are explained for a thermal concentrator in \fref{fig:Reconstruction steps}. We consider a coarse voxel mesh to highlight the characteristics of the reconstruction process. However, in actual structures, a finer mesh is required so that the scale separation hypothesis is satisfied~\cite{BERTOLINO2022twoscale}. By satisfying the scale separation hypothesis, we ensure that the homogenization rule and eventually the optimization results remain valid. A full-scale reconstructed 2D thermal concentrator and a 2D complex star-shaped thermal cloak based on a very fine voxel mesh are shown in \fref{fig:chen2015case reconstruction}.

%
%
%
%
\section{Conclusions}
\label{sec:Conclusions}
In this paper, we proposed FGM-based thermal metamaterials/meta-structures. We used the isogeometric density topology optimization method to design these meta-structures, in which the density, geometry, and solution fields are parameterized using NURBS basis functions. Following the NURBS parameterizations, IGA is utilized to solve boundary value problems. IGA gives an accurate geometric description and ease in handling higher smoothness and inter-element continuity. Additionally, NURBS-parameterized density field has a few perks in comparison to element or nodal densities, including smoother material distributions, inherent filtering against checker-boarding and straightforward calculation of the gradient of the density field. 
\par We showcase the versatility of the proposed method by designing various 2D and 3D thermal meta-structures including thermal cloaks, thermal concentrators, thermal rotators and thermal cloaked sensors. The method is robust in handling diverse material models, geometries, boundary conditions and design requirements. It can also produce alternate designs for non-convex problem by slight modification in optimization parameters or initial designs. This versatility, robustness and flexibility is one of the key benefits of our proposed tool over conventional methods (which mainly function under limited design scenarios due to their analytical nature). Additionally, the proposed method does not need any intuition-based case-dependant information. We also showed that the method can effectively design thermal meta-structures made of architected cellular materials. In order to do so, the numerical homogenization data of their unit-cells are implemented as the material law in the formulation. In the end, the full structure is generated based on the obtained density distribution and density-control parameter relation of the unit-cell.



\section*{Acknowledgements}
\label{sec:Acknowledgements}
\noindent We are grateful for the support of the University of Luxembourg. The calculations presented in this paper were carried out using the HPC facilities of the University of Luxembourg.

\appendix
\setcounter{figure}{0} 

\section{Matrix formulation of boundary value problem}
\label{sec:Appendix A}
The global stiffness matrix $\mathbf{K}$ and the global flux vector $\mathbf{F}$ (as shown in \eref{eq:Linear matrix system}) are written as:
\begin{equation} \label{eq:K}
\mathbf{K}= \mathbf{K}^b + \mathbf{K}^n+ (\mathbf{K}^n)^{\rm T}+\mathbf{K}^s + \mathbf{K}^r,
\end{equation}
\begin{equation}
\mathbf{F}= \sum_{k \in {\rm \{in,design,out \}}} \int _{\Omega_k} (\mathbf{N}^k)^{\textrm{T}}q_b~d\Omega + \int _{\Gamma_{N}} \mathbf{N}^{\textrm{T}} q_n~d\Gamma + \int _{\Gamma_{R}} \mathbf{N}^{\textrm{T}} hT_{\infty}~d\Gamma   ,
\end{equation} 
where $\mathbf{K}^b$ is the bulk stiffness matrix; $\mathbf{K}^n$ and $\mathbf{K}^s$ are the interfacial stiffness matrices; and $\mathbf{K}^r$ is the convective flux matrix (related to the robin boundary conditions). As $\Omega_{\rm in}$, $\Omega_{\rm out}$ and $\Omega_{\rm design}$ are considered separate NURBS patches, $\mathbf{K}^b$, $\mathbf{K}^n$ and $\mathbf{K}^s$ are defined as follows (following the notations used in \sref{sec:Boundary value problem}),
\begin{equation}
\mathbf{K}^b =\sum_{k \in {\rm \{in,design,out \}}} \int _{\Omega_k} (\mathbf{B}^k)^{\textrm{T}}\boldsymbol{\kappa}^k(v)\mathbf{B}^k~d\Omega,
\label{eq:Kb}
\end{equation}
\renewcommand\arraystretch{2}
\begin{equation}
\mathbf{K}^n =
\begin{bmatrix} 
-\gamma\displaystyle\int_{\Gamma_I} (\mathbf{N}^1)^{\textrm{T}}\boldsymbol{n}\boldsymbol{\kappa}^1(v)\mathbf{B}^1~d\Gamma 
& -(1-\gamma)\displaystyle\int_{\Gamma_I} (\mathbf{N}^1)^{\textrm{T}}\boldsymbol{n}\boldsymbol{\kappa}^2(v)\mathbf{B}^2~d\Gamma    \\[0.1em]
\gamma\displaystyle\int_{\Gamma_I} (\mathbf{N}^2)^{\textrm{T}}\boldsymbol{n}\boldsymbol{\kappa}^1(v)\mathbf{B}^1~d\Gamma  & (1-\gamma)\displaystyle\int_{\Gamma_I} (\mathbf{N}^2)^{\textrm{T}}\boldsymbol{n}\boldsymbol{\kappa}^2(v)\mathbf{B}^2~d\Gamma 
\end{bmatrix},
\label{eq:Kn}
\end{equation}
\begin{equation}
\mathbf{K}^s =
\begin{bmatrix} 
\beta\displaystyle\int_{\Gamma_I} (\mathbf{N}^1)^{\textrm{T}}\mathbf{N}^1~d\Gamma 
& -\beta\displaystyle\int_{\Gamma_I} (\mathbf{N}^1)^{\textrm{T}}\mathbf{N}^2~d\Gamma  \\[0.1em]
-\beta\displaystyle\int_{\Gamma_I} (\mathbf{N}^2)^{\textrm{T}}\mathbf{N}^1~d\Gamma  & \beta\displaystyle\int_{\Gamma_I} (\mathbf{N}^2)^{\textrm{T}}\mathbf{N}^2~d\Gamma 
\end{bmatrix},
\label{eq:Ks}%
\end{equation}
\begin{equation}
\mathbf{K}^r = \int_{\Gamma_R} h(\mathbf{N})^{\textrm{T}}~\mathbf{N}~d\Omega, 
\label{eq:Kr}
\end{equation}
where $\mathbf{B}$ is the matrix of basis function derivatives and $\mathbf{N}$ is the vector of basis functions. 
\par In \sref{sec:Sensitivity analysis}, the derivative of global stiffness matrix $\mathbf{K}$ with respect to relative density $v$ will be needed in the sensitivity calculation. It is defined by differentiating  \eref{eq:K} as follows: \begin{equation} \label{eq:dKdphi}
\dfrac{d\mathbf{K}}{dv}= \dfrac{d\mathbf{K}^b}{dv} + \dfrac{d\mathbf{K}^n}{dv}+ \left(\dfrac{d\mathbf{K}^n}{dv}\right)^{\rm T}+\dfrac{d\mathbf{K}^s}{dv},
\end{equation}
where
\begin{equation}\label{eq:dKbdphi}  
\dfrac{d\mathbf{K}^b}{dv}=\sum_{k \in {\rm \{in,design,out \}}}\int _{\Omega_k} (\mathbf{B}^k)^{\textrm{T}}\dfrac{d\boldsymbol{\kappa}^k(v)}{dv}\mathbf{B}^k~d\Omega,
\end{equation}
and $\dfrac{d\mathbf{K}^n}{dv}$ \& $\dfrac{d\mathbf{K}^s}{dv}$ are defined similarly by differentiating \eref{eq:Kn} \& \eref{eq:Ks}, respectively.

\section{Mesh sensitivity analysis}
\label{sec:2D cloak Mesh study}
\begin{figure}[!htbp]
    \centering
    \setlength\figureheight{1\textwidth}
    \setlength\figurewidth{1\textwidth}
    \includegraphics[width=0.7\textwidth]{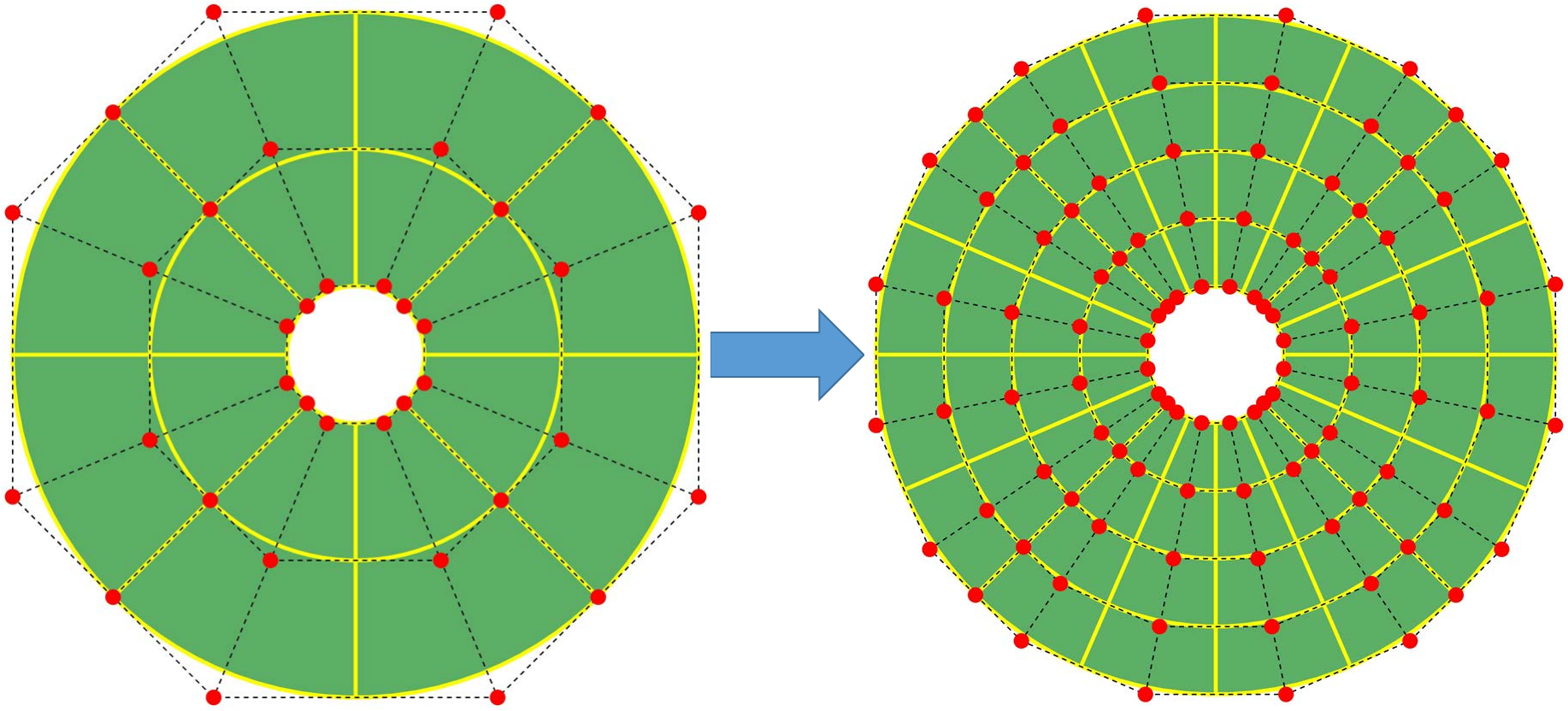}
 \caption{The solution mesh refinement strategy using the knot insertion procedure. At each stage, new knots are added at midpoints of existing knot spans in each parametric direction.}
 \label{fig:Chen2015case meshing strategy}
\end{figure}
\par At first, we conducted a mesh sensitivity analysis to find a sufficiently fine solution mesh for a given design mesh to ensure an adequate level of solution accuracy. A mesh model of a 2D NURBS patch is shown in \fref{fig:Chen2015case meshing strategy}. We consider two design meshes with $N_{\rm var}=25$ \& $N_{\rm var}=81$. For both meshes, we run the optimization problems with several stages of refinements for the solution mesh, starting with the same number of control points in the solution and design meshes. For each refinement, we insert new knots at midpoints of existing knot spans in each parametric direction as shown in \fref{fig:Chen2015case meshing strategy}. For $N_{\rm var}=25$ \& $N_{\rm var}=81$, we perform 5 and 4 stages of refinements, respectively. The last refinement solution is considered as the reference solution to calculate the relative error. 
\par
\begin{figure}[!htbp]
    \centering
    \setlength\figureheight{1\textwidth}
    \setlength\figurewidth{1\textwidth}
    \begin{subfigure}[t]{0.45\textwidth}{\centering\includegraphics[width=1\textwidth]{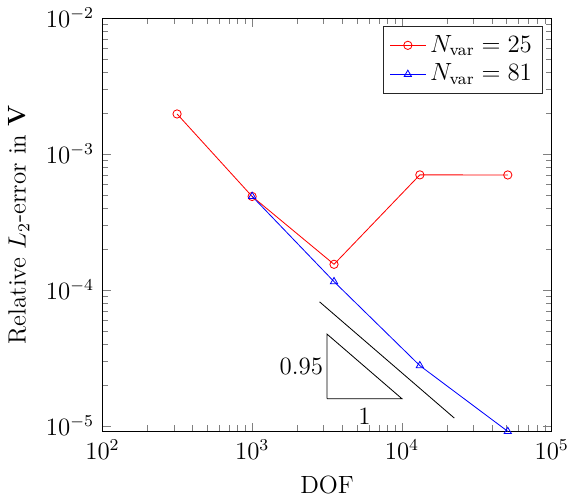}}
        \caption{2D thermal cloak}
        \label{fig:Chen2015case cloak convergence}
    \end{subfigure}\quad
    \begin{subfigure}[t]{0.45\textwidth}{\centering\includegraphics[width=1\textwidth]{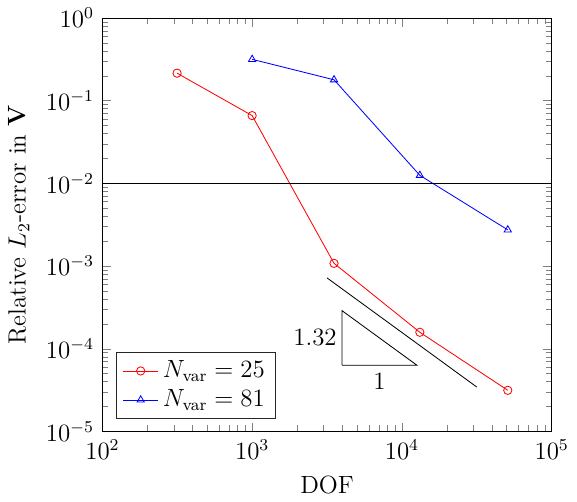}}
        \caption{2D thermal concentrator}
        \label{fig:Chen2015case cntr convergence}
    \end{subfigure}\quad
 \caption{Relative error in volume fraction $v$ value over $\Omega_{\rm design}$ with respect to the number of degrees of freedom of solution mesh for (a) 2D thermal  cloak problem and (b) 2D thermal concentrator problem. The last refinement solution is considered as the reference solution to calculate the relative error. We consider $2\%$ relative error as an acceptable error, which is represented by the black horizontal line.}
 \label{fig:Chen2015case convergence}
\end{figure}

 We perform the mesh sensitivity analysis for the 2D thermal cloak and 2D thermal concentrator. The relative error in the relative density field $v$ over $\Omega_{\rm design}$ with respect to number of degrees of freedom (DOF) is shown in \fref{fig:Chen2015case convergence}. We consider $2\%$ error as an acceptable error, which is represented by a black horizontal line in the figure. We observe that, for thermal cloak, all solution meshes and, for thermal concentrator, the solution meshes with DOF$>10^4$ satisfy the criterion. Therefore, we take a mesh with DOF=13167 as the solution mesh for both cases. 


\section*{Data availability}
\label{sec:Data availability}
\noindent Data will be made available on request.

\bibliographystyle{model1-num-names}
\bibliography{densitytop.bib}

\end{document}